\documentclass{aa} 

\usepackage[utf8]{inputenc}
\usepackage[varg]{txfonts}
\usepackage{graphicx}
\usepackage[font=small,labelfont=bf]{caption}
\usepackage{booktabs}
\usepackage{amsmath,amssymb}
\usepackage{gensymb}
\usepackage{wasysym}
\usepackage{scalerel}
\usepackage{natbib,twoopt}
\usepackage[flushleft]{threeparttable}
\usepackage{tabularx}
\usepackage{color}
\usepackage[breaklinks=true]{hyperref} 
\usepackage{adjustbox}
\usepackage{float}
\usepackage{longtable}
\usepackage[graphicx]{realboxes}
\usepackage{rotating}
\usepackage{lscape}
\usepackage{afterpage}
\usepackage{tabu}
\usepackage{multicol}
\usepackage{subfig}

\hypersetup{
      colorlinks=True,
      allcolors=blue
      }
\bibpunct{(}{)}{;}{a}{}{,} 
\makeatletter
\newcommandtwoopt{\citeads}[3][][]{\href{http://adsabs.harvard.edu/abs/#3}%
{\def\hyper@linkstart##1##2{}%
\let\hyper@linkend\@empty\citealp[#1][#2]{#3}}}
\newcommandtwoopt{\citepads}[3][][]{\href{http://adsabs.harvard.edu/abs/#3}%
{\def\hyper@linkstart##1##2{}%
\let\hyper@linkend\@empty\citep[#1][#2]{#3}}}
\newcommandtwoopt{\citetads}[3][][]{\href{http://adsabs.harvard.edu/abs/#3}%
{\def\hyper@linkstart##1##2{}%
\let\hyper@linkend\@empty\citet[#1][#2]{#3}}}
\newcommandtwoopt{\citeyearads}[3][][]%
{\href{http://adsabs.harvard.edu/abs/#3}
{\def\hyper@linkstart##1##2{}%
\let\hyper@linkend\@empty\citeyear[#1][#2]{#3}}}
\makeatother

\newcommand{\changeurlcolor}[1]{\hypersetup{urlcolor=#1}}

\def\ledd{{$\lambda_\mathrm{Edd}$}}

\def\ltsima{$\; \buildrel < \over \sim \;$}
\def\simlt{\lower.5ex\hbox{\ltsima}}

\makeatletter
\newcommand\notsotiny{\@setfontsize\notsotiny{7.5}{7}}

\makeatother

\begin{document} 

  \title{Investigating the nuclear properties of highly accreting \\ active galactic nuclei with \emph{XMM-Newton}}

  \titlerunning{Investigating the nuclear properties of highly accreting \\ active galactic nuclei with \emph{XMM-Newton}}
  
  \subtitle{}

  \author{M.~Laurenti\inst{1,2,3,4} \and F.~Tombesi\inst{1,3,4} \and F.~Vagnetti\inst{4,5} \and E.~Piconcelli\inst{3} \and M.~Guainazzi\inst{6} \and C.~Vignali\inst{7,8} \and M.~Paolillo\inst{9,10,11} \and \\ R.~Middei\inst{2,3} \and A.~Bongiorno\inst{3} \and L.~Zappacosta\inst{3}
  }
  
  \institute{INFN -- Sezione di Roma ``Tor Vergata'', Via della Ricerca Scientifica 1, I-00133 Roma, Italy \\ \email{\changeurlcolor{black}\href{mailto:marco.laurenti@roma2.infn.it}{{\fontfamily{cmtt}\selectfont
\textbf{marco.laurenti@roma2.infn.it}}
}} \and Space Science Data Center, SSDC, ASI, Via del Politecnico snc, I-00133 Roma, Italy \and INAF -- Osservatorio Astronomico di Roma, via Frascati 33, I-00078 Monte Porzio Catone, Italy \and Dipartimento di Fisica, Università degli Studi di Roma ``Tor Vergata'', via della Ricerca Scientifica 1, I-00133 Roma, Italy \and  INAF -- Istituto di Astrofisica e Planetologia Spaziali, Via del Fosso del Caveliere 100, 00133 Roma, Italy \and ESA European Space Research and Technology Centre (ESTEC), Keplerlaan 1, 2201 AZ, Noordwijk, The Netherlands \and Dipartimento di Fisica e Astronomia ‘Augusto Righi’, Università
degli Studi di Bologna, via P. Gobetti, 93/2, 40129 Bologna, Italy \and INAF -- Osservatorio di Astrofisica e Scienza dello Spazio di Bologna,
via Piero Gobetti, 93/3, I-40129 Bologna, Italy \and Dipartimento di Fisica “Ettore Pancini”, Università di Napoli Federico II, Via Cintia, 80126, Italy \and INAF – Osservatorio Astronomico di Capodimonte, Via Moiariello 16, 80131, Naples, Italy \and INFN – Unità di Napoli, via Cintia 9, 80126, Napoli, Italy}

  \date{}

  \abstract
  {Highly accreting active galactic nuclei (AGNs) have unique features that make them ideal laboratories for studying black hole accretion physics under extreme conditions. However, our understanding of the nuclear properties of these sources is hampered by the lack of a complete systematic investigation of this AGN class in terms of their main spectral and variability properties, and by the relative paucity of them in the local Universe, especially those powered by supermassive black holes with $M_\mathrm{BH} > 10^8\,M_\odot$. To overcome this limitation, we present here the X-ray spectral analysis of a new, large sample of 61 highly accreting AGNs named as the \emph{XMM-Newton} High-Eddington Serendipitous AGN Sample, or X-HESS, obtained by cross-correlating the 11th release of the \emph{XMM-Newton} serendipitous catalogue and the catalogue of spectral properties of quasars from the SDSS DR14. The X-HESS AGNs are spread across wide intervals with a redshift of $0.06<z<3.3$, a black hole mass of $6.8<\log(M_\mathrm{BH}/M_\odot)<9.8$, a bolometric luminosity of $44.7<\log(L_\mathrm{bol}/\mathrm{erg\,s}^{-1})<48.3$, and an Eddington ratio of $-0.2<\log\lambda_\mathrm{Edd}<0.5$, and more than one third of these AGNs can rely on multiple observations at different epochs, allowing us to complement their X-ray spectroscopic study with a variability analysis. 
  We find a large scatter in the $\Gamma - \lambda_\mathrm{Edd}$ distribution of the highly accreting X-HESS AGNs, in agreement with previous findings. A significant correlation is only found by considering a sample of lower-\ledd\ AGNs with $\lambda_\mathrm{Edd}\lesssim0.3$. We get hints that the $\Gamma - \lambda_\mathrm{Edd}$ relation appears to be more statistically sound for AGNs with lower $M_\mathrm{BH}$ and/or $L_\mathrm{bol}$. We investigate the possibility of transforming the $\Gamma - \lambda_\mathrm{Edd}$ plane into a fully epoch-dependent frame by calculating the Eddington ratio from the simultaneous optical/UV data from the optical monitor, $\lambda_\mathrm{Edd,O/UV}$. Interestingly, we recover a significant correlation with $\Gamma$ and a spread roughly comparable to that obtained when $L_\mathrm{bol}$ is estimated from SDSS spectra.
  Finally, we also get a mild indication of a possible anti-correlation between $\Gamma$ and the strength of the soft excess, providing hints that reflection from an ionised disc may be effective in at least a fraction of the X-HESS AGNs, though Comptonisation from hot and warm coronae cannot be ruled out as well.}


  \maketitle

\section{Introduction}\label{sec:intro}

Active galactic nuclei (AGNs) are powered by accretion processes onto a central supermassive black hole (SMBH). 
The accretion activity is usually parameterised by the Eddington ratio $\lambda_\mathrm{Edd}=L_\mathrm{bol}/L_\mathrm{Edd}$, where $L_\mathrm{bol}$ and $L_\mathrm{Edd}$ represent the bolometric and Eddington\footnote{$L_\mathrm{Edd}=1.26\times10^{38}\,(M_\mathrm{BH} / M_\odot)$ erg/s, where $M_\mathrm{BH}$ represents the mass of the SMBH.} luminosity, respectively.
Highly accreting AGNs — that is, those with $\lambda_\mathrm{Edd}\gtrsim0.6$ — are characterised by some intriguing properties that make them interesting case studies.

First, the accretion flow in the low-to-moderate \ledd\ regime envisages an optically thick, geometrically thin disc that radiates efficiently \citep[][]{shakura1973}.
The net effect of radiation pressure becomes prominent with increasing \ledd\ and, as a result, the disc thickens vertically. 
These systems are often called slim discs, as they are both optically and geometrically thick \citep[e.g][]{abramowicz1988,chen2004,sadowski2011}. Slim discs are also expected to have a low radiative efficiency because of the photon-trapping effect. 
This mechanism contributes to saturating the observed disc luminosity (${\sim}L_\mathrm{bol}$) to a limiting value of approximately 5${-}$10 $L_\mathrm{Edd}$ for steadily increasing accretion rates \citep[e.g.][]{mineshige2000}.
Although the slim disc is expected to have peculiar properties in both its spectral energy distribution and its temperature profile compared to the standard disc model, it is often difficult to detect differences between high- and low-\ledd\ AGNs \citep[e.g.][]{castello2017, liu2021, cackett2020, donnan2023}.

The high-\ledd\ accretion mode is often considered in terms of its cosmological implications. Indeed, an increasing effort has been made in recent years to detect quasars (i.e.\ AGNs with $L_\mathrm{bol}\!>\!10^{46}$ erg s$^{-1}$; QSOs hereafter) at the distant redshift of $z\sim6{-}7$, corresponding to an age when the Universe was less than ${\sim}1$ Gyr old \citep[e.g.][]{wu2015, banados2016, mazzucchelli2017,onoue2020,ighina2021}.
These QSOs do typically host massive SMBHs — that is, ones with $M_\mathrm{BH}\geq10^9\,M_\odot$ — and the physical mechanism that allowed them to grow that massive in such a relatively short period of time is still debated.
The evolution of these SMBHs via uninterrupted or rather intermittent episodes of gas accretion at a critical (or super-Eddington) rate is arguably one of the most commonly hypothesised mechanisms \citep[e.g.][]{begelman2006,inayoshi2016, valiante2017, zappacosta2020, zhang2020, lusso2023}. 

Furthermore, the launching of powerful disc winds, such as ultra-fast outflows (UFOs), is thought to occur naturally as a consequence of the high-\ledd\ accretion \citep[e.g.][]{proga2005, zubovas2013, king2015}.  
These nuclear outflows are likely to have an impact on both the SMBH growth and the evolution of the host galaxy, by depositing large amounts of energy and momentum into the interstellar medium \citep[e.g.][]{zubovas2012}, offering a possible explanation for the observed $M_\mathrm{BH}{-}\sigma$ relation \citep[e.g.][]{ferrarese2000}.
Consequently, the AGN feedback mechanism should manifest itself in full force in high-\ledd\ sources, making them the ideal laboratory for probing the impact of nuclear activity on the evolution of massive galaxies \citep[e.g.][]{reeves2009,nardini2015,tombesi2015,marziani2016,bischetti2017,bischetti2019,laurenti2021,luminari2021,middei2023}.

Despite all these remarkable features, high-\ledd\ AGNs have often been overlooked in the context of X-ray spectroscopic studies, probably due to their relative paucity in the local ($z\lesssim0.1$) Universe \citep[e.g.][]{shankar2013,shirakata2019}.
Indeed, the bulk of such works typically dealt with AGNs accreting at low-to-moderate Eddington rates \citep[e.g.][]{nandra1994, piconcelli2005, bianchi2009, liu2016, ricci2018}, with the notable exception of the Narrow Line Seyfert 1 galaxies (NLSy1s; \citealt{brandt1997, gallo2006, costantini2007, jin2013, fabian2013, waddell2020}).
However, among the high-\ledd\ systems, the NLSy1s represent a peculiar and restricted class of low-mass ($10^6{-}10^7\,M_\odot$ ) AGNs with the narrowest allowed emission lines in type-1 AGN samples (FWHM(H$\beta)<2000$ km s$^{-1}$; \citealt{osterbrock1985,marziani2018}).
In this sense, they may be regarded as a biased population, and thus enlarging the parameter space of the relations involving \ledd\ and the main X-ray spectral parameters towards AGNs hosting more massive SMBHs is of crucial importance in the light of the currently hotly debated issues that we describe in the following.

First, several authors have reported on a positive correlation between the Eddington ratio and the photon index of the power-law continuum, $\Gamma$, from which we expect that AGNs accreting close to or above the Eddington limit will present steep values of $\Gamma\geq2$ \citep[e.g.][]{shemmer2008, risaliti2009, brightman2013, kawamuro2016, huang2020}.
The general interpretation is based on the assumption that the enhancement of the optical/UV emission due to high-\ledd\ accretion increases the number of seed photons undergoing Comptonisation in the X-ray corona. As the cooling process in the corona becomes more efficient, its temperature decreases and the primary X-ray continuum becomes progressively softer.
This relation elicited immediate interest, especially for its possible applications in X-ray extragalactic surveys, since it would allow us to obtain an estimate of $M_\mathrm{BH}$ from the X-ray spectrum, including that of type-2 AGNs for which commonly used `single-epoch' estimators are not applicable.
However, recent works questioned the existence of a strong correlation between $\Gamma$ and \ledd, with some authors reporting only a weaker \citep[e.g.][]{ai2011, kamizasa2012, martocchia2017, trakhtenbrot2017, kamraj2022, trefoloni2023} or even no correlation \citep[e.g.][]{liu2016} between these two parameters, which highlights the need to study this relation in more detail.

In addition, \citet{lusso2010} claimed the existence of a positive correlation between \ledd\ and both the X-ray bolometric correction, $k_\mathrm{bol,X}=L_\mathrm{bol}/L_{2{-}10\mathrm{\,keV}}$, and the optical/UV-to-X-ray spectral index, $\alpha_\mathrm{ox}$, for a large sample of type-1 (i.e.\ unobscured) AGNs in the COSMOS survey, extending over wide intervals of redshift ($0.04<z<4.25$) and bolometric luminosity ($40.6<\log{(L_\mathrm{bol}/\mathrm{erg\,s}^{-1})}<45.3$). A more refined version of the $k_\mathrm{bol,X}-\lambda_\mathrm{Edd}$ relation has recently been provided by \citet{duras2020}, who found that $k_\mathrm{bol,X}\propto\lambda_\mathrm{Edd}^{0.61}$ for an AGN sample in which a sizeable number of highly accreting sources at $z\sim2-4$ was also included. According to these relations, one would expect that the coupling between the X-ray corona and the accretion disc is such that, for increasingly large values of \ledd, it attains a configuration leading to a dimming of the X-ray primary continuum with respect to the optical/UV emission from the disc, compared to `standard' AGNs.

However, both results and expectations based on the above relations are hampered by the limited number of high-\ledd\ AGNs, especially ones hosting more massive SMBHs, which were actually available in most of those studies. 
To overcome this limitation, in \citet{laurenti2022} we performed an X-ray spectroscopic analysis of a sample of fourteen \emph{XMM-Newton} (\citealt{jansen2001}) observations of type-1 radio-quiet QSOs accreting in a narrow interval of \ledd\ values with $\lambda_\mathrm{Edd}\gtrsim0.9$ at a redshift of $z\sim0.4-0.75$.
These AGNs were drawn from a larger sample of highly accreting sources sharing homogenous optical/UV properties described in \citet{marziani2014}, and are characterised by large black hole mass values of $M_\mathrm{BH}\sim10^{8-8.5}M_\odot$ derived from the full width at half maximum (FWHM) of the \ion{H}{$\beta$} emission line.
Despite all these QSOs possessing similar \ledd\ values, we found that $\Gamma$ was characterised by a large dispersion, with values ranging from a minimum of ${\sim}1.3$ to a maximum of ${\sim}2.5$, at odds with the expectations based on several of the previously reported $\Gamma-\lambda_\mathrm{Edd}$ relations.
Moreover, the results of our best-fit spectral analysis indicated that approximately ${\sim}30\%$ of the QSOs in our sample displayed weak X-ray emission, which could not be ascribed to cold absorption.

Since our previous findings suggest that our understanding of the high-\ledd\ accretion mode in AGN is still incomplete, it is evident that this study would benefit from a larger number of observations of highly accreting sources.
Fortunately, many of these highly accreting objects can be serendipitously found in the field of view of other X-ray sources that have been intensively and repeatedly targeted thanks to deep observational campaigns, such as those carried out with \emph{XMM-Newton}. 
The analysis of serendipitous high-\ledd\ AGNs discloses the unprecedented possibility of investigating not only their spectral but also variability properties in the X-rays in a much wider range of black hole mass, bolometric luminosity, and redshift.

For this reason, we present the spectroscopic and variability study of a large sample of high-\ledd\ AGNs named as the \emph{XMM-Newton} High-Eddington Serendipitous AGN Sample, or X-HESS, with the aim of continuing our quest to uncover the properties of such highly accreting systems.
The paper is organised as follows. We describe the procedure that leads to the definition of the sample and its general properties in Sect. \ref{sec:xhess_sample}. Section \ref{sec:data_analysis} is dedicated to the description of the X-ray spectroscopy as well as the optical/UV photometry of our sample sources.
We then present our results in Sects. \ref{sec:results} and \ref{sec:Rsp}. Finally, we draw our conclusions in \S\,\ref{sec:summary&conclusion}.
A $\Lambda$CDM cosmology with $H_0 = 70$ km s$^{-1}$ Mpc$^{-1}$, $\Omega_\mathrm{m}=0.3$ and $\Omega_\Lambda = 0.7$ is adopted throughout the paper. All errors are quoted at the $68\%$ confidence level ($\Delta{W}$-stat $=1$; \citealt{avni1976,cash1979,wachter1979}). In the following, correlations will be considered statistically significant based on the significance level of $\alpha=10^{-3}$.

\section{Sample description}\label{sec:xhess_sample}

In order to define a large sample of high-\ledd\ serendipitous AGNs, we chose to exploit the extended database included in the 11th release of the\textit{ XMM-Newton} Serendipitous Source Catalogue (4XMM-DR11; \citealt{webb2020}), which contains 12210 observations carried out between February 2000 and December 2020, in which approximately $\sim900000$ sources have been detected. 
To retrieve only those X-ray observations of spectroscopically confirmed AGNs, we considered the catalogue of spectral properties of QSOs from the Sloan Digital Sky Survey Data Release 14 Quasar Catalogue \citep[SDSS-DR14Q;][]{paris2018} described in \citet{rakshit2020}, which includes measurements of the main physical quantities of $\sim526000$ AGNs, and cross-matched it with the serendipitous \textit{XMM-Newton} catalogue within a radius of 3 arcsecs in coordinates to avoid possible spurious identifications, while maximising the completeness of the sample.
Then we imposed the additional requirements that (i) each source with multi-epoch detections must have at least one good quality observation in terms of photon counts in the broad ($E=0.2{-}12$ keV) EP8 band — that is, EP\_8\_CTS $\geq 1000$ — to ensure robust spectral results, and (ii) only those AGNs with a sufficiently high value of $\log\,$\ledd\ that we set to $\log\,$\ledd\ $\geq-0.2$ would be considered.
At this stage, we obtained a sample of 95 AGNs for a total of 217 observations. Since we are only interested in standard radio-quiet AGNs, we removed twenty sources that have been either classified as radio-loud or known to be lensed from the sample. 
Starting from this sample of 75 AGNs, we imposed the further condition that the $M_\mathrm{BH}$ estimate of each source must be derived from the FWHM of either \ion{H}{$\beta$} or \ion{Mg}{II} emission lines, which are more reliable estimators than the FWHM of \ion{C}{IV} \citep[e.g.][]{baskin2005, coatman2017, vietri2018, vietri2020}.
Two AGNs — namely, J0900+4215 and J1326--0005 — for which the sole \ion{C}{IV} black hole mass is provided in the catalogue of \citet{rakshit2020} are also included in the final sample, because they are part of the WISE/SDSS selected hyper-luminous quasar sample (WISSH; \citealt{bischetti2017}) and can benefit from refined \ion{H}{$\beta$}-based $M_\mathrm{BH}$ estimates as well as more refined measurements of $L_\mathrm{bol}$  \citep[][]{vietri2018}.
For the rest of the X-HESS AGNs, we adopted the $L_\mathrm{bol}$ values from \citet{rakshit2020} that were derived from the bolometric corrections in \citet{richards2006} involving the monochromatic luminosities at $5100\,\AA$ ($z<0.8$), $3000\,\AA$ ($0.8\leq z <1.9$), or $1350\,\AA$ ($z\geq1.9$).
Finally, we obtained a catalogue of 61 AGNs with 142 observations that we named the \textit{XMM-Newton} High-Eddington Serendipitous AGN Sample, or X-HESS.

\begin{figure}[t]
    \centering
    \includegraphics[width=\columnwidth]{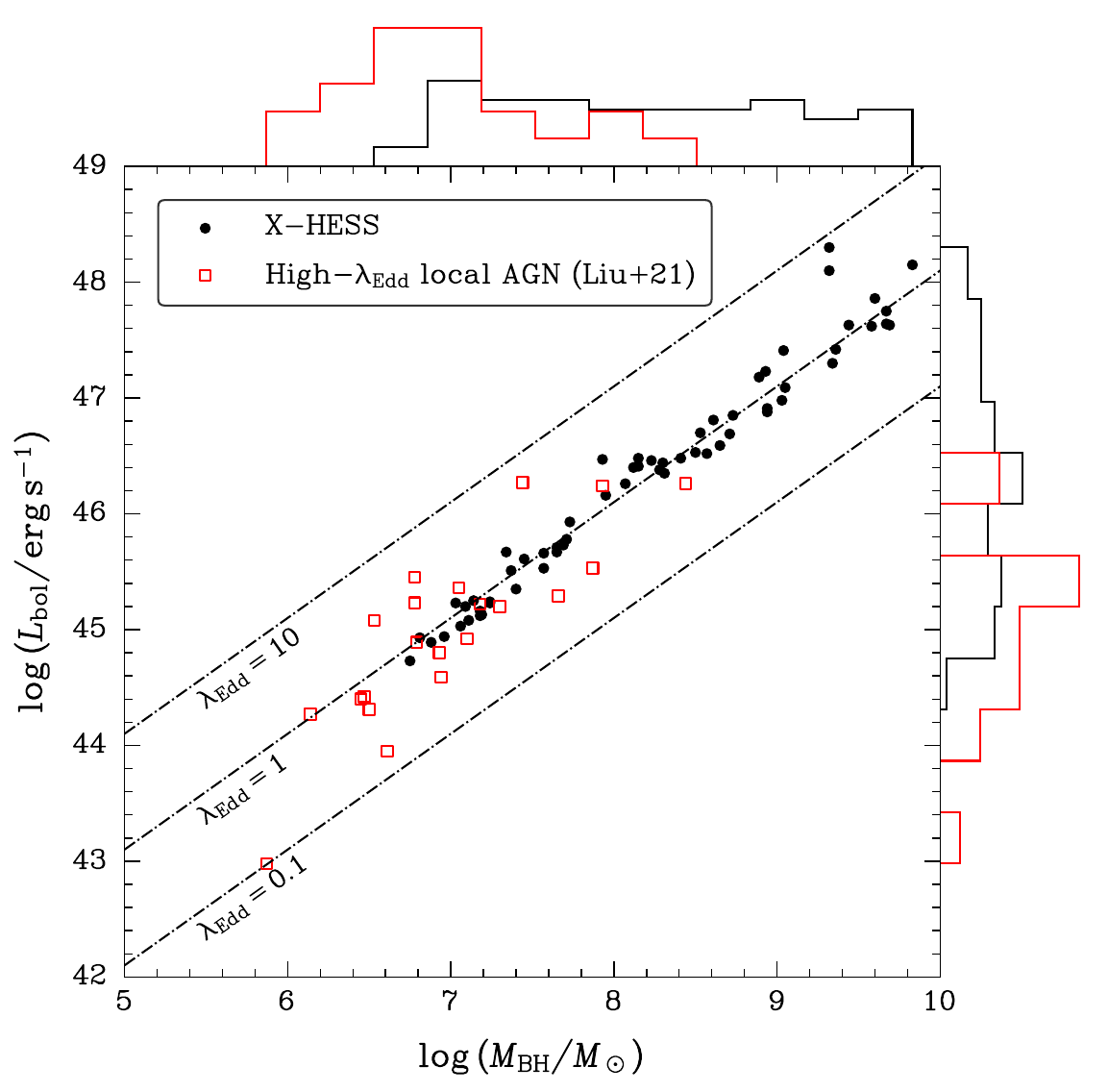}
    \caption{Distribution of the X-HESS AGNs in the $\log{L_\mathrm{bol}}{-}\log{M_\mathrm{BH}}$ plane (in black). Red squares indicate the high-\ledd\ local AGN subsample from \citet{liu2021}, as a comparison. With respect to this sample, the X-HESS AGNs extend towards larger values of both bolometric luminosity and black hole mass.
    Dash-dotted black lines describe different values of $\lambda_\mathrm{Edd}$.}
    \label{fig:LM_distro}
\end{figure}

The X-HESS AGNs are distributed over wide intervals of bolometric luminosity ($44.7<\log(L_\mathrm{bol}/\mathrm{erg\,s}^{-1}) <48.3$), black hole mass ($6.8<\log({M_\mathrm{BH}/M_\odot})<9.8$), redshift ($0.06<z<3.3$), and Eddington ratio ($-0.2< \log\lambda_\mathrm{Edd}<0.5$). 
The presence of AGNs spanning a wide range of $M_\mathrm{BH}$, $L_\mathrm{bol}$, and $z$ contributes to making X-HESS a useful tool with which to improve our understanding of high-\ledd\ AGNs in regions of the parameter space that have been poorly sampled so far.
Specifically, the broad $M_\mathrm{BH}$ and $L_\mathrm{bol}$ distributions of the X-HESS AGNs (see Fig.\ \ref{fig:LM_distro}) allow us to extend the analysis of highly accreting sources towards AGNs with both bolometric luminosity and a black hole mass substantially higher than those previously considered in other studies \citep[e.g.][]{lu2019,liu2021}.

\begin{figure}[t]
    \centering
    \includegraphics[width=\columnwidth]{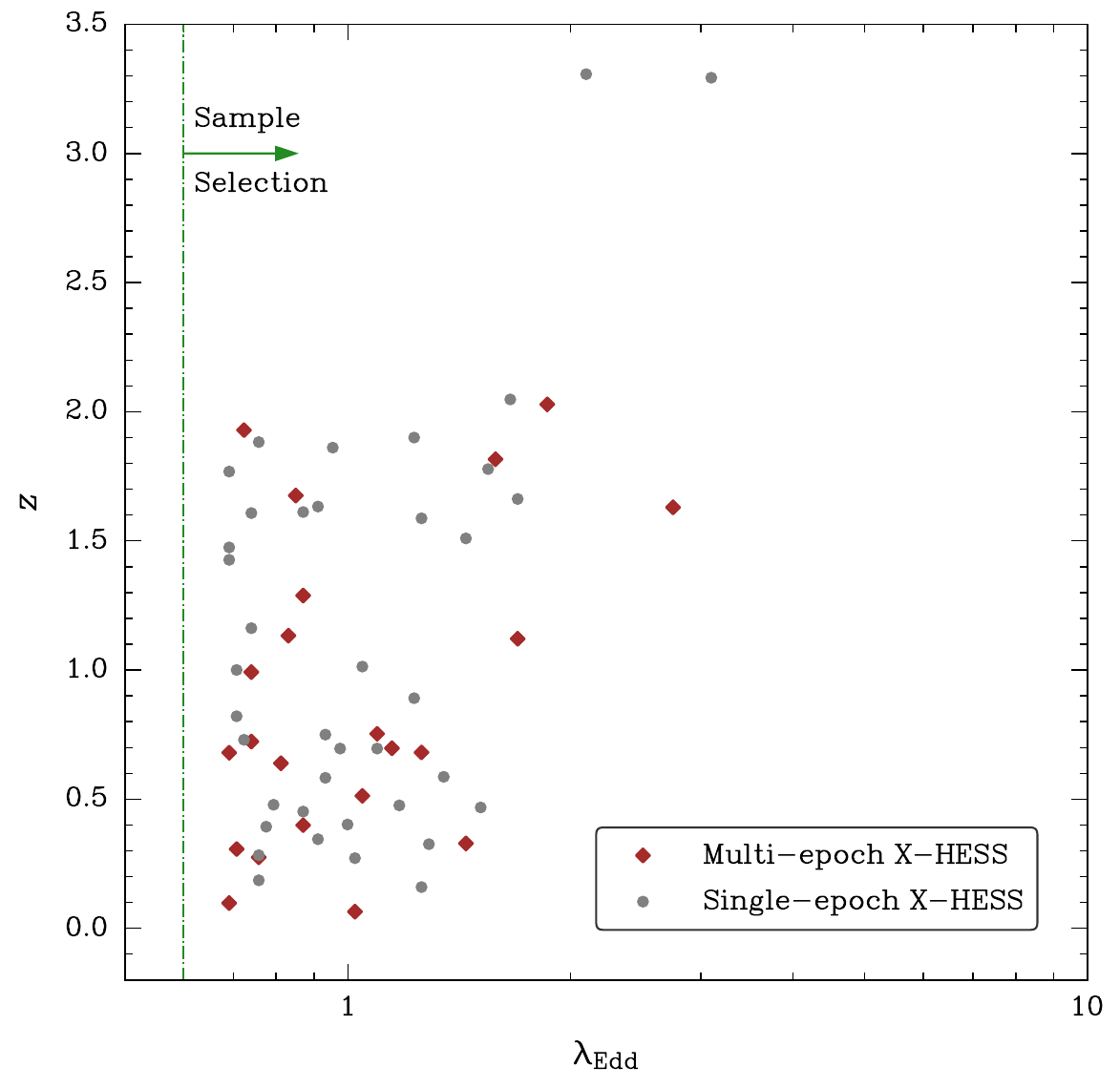}
    \caption{Distribution of the X-HESS AGNs in the $z-\lambda_\mathrm{Edd}$ plane. All the X-HESS AGNs and the subsample with multi-epoch observations are described by black dots and red diamonds, respectively. The multi-epoch subsample is well representative of the whole sample in terms of these quantities.}
    \label{fig:zLedd}
\end{figure}

Figure \ref{fig:zLedd} shows the $z-\lambda_\mathrm{Edd}$ distribution of the X-HESS sources, where AGNs with individual or multi-epoch observations are shown in the same plot. The multi-epoch subsample of X-HESS is well representative of the whole sample in terms of these quantities and the same holds for other physical parameters such as $M_\mathrm{BH}$ and $L_\mathrm{bol}$, offering the unprecedented opportunity to investigate the X-ray flux and spectral variability of 22 highly accreting sources that have been repeatedly observed by \textit{XMM-Newton} for a total of 103 times.
Furthermore, we can complement X-HESS with simultaneous optical/UV data, allowing us to investigate the interplay with the X-rays (e.g. with $\alpha_\mathrm{ox}$), by taking advantage of the observations carried out by the Optical Monitor \citep[OM;][]{mason2001} aboard \textit{XMM-Newton}. Indeed, a sizeable fraction of the X-HESS sources (${\sim}62\%$) can rely on simultaneous OM observations obtained by crossmatching X-HESS with the latest release of the XMM-OM Serendipitous Ultraviolet Source Survey Catalogue \citep[SUSS5.0;][]{page2012}.
An overview of the X-HESS AGNs, including their general properties, is provided in Appendix \ref{sec:app_agn_list}.

\section{Data analysis} \label{sec:data_analysis}
\subsection{Data reduction}\label{sec:reduction}

The raw data for each observation of the X-HESS AGNs were retrieved from the \emph{XMM-Newton} Science Archive and then processed using the \emph{XMM-Newton} Science Analysis System (SAS v21.0.0) with the latest available calibration files. We took advantage of the full potential of
\emph{XMM-Newton} in the X-ray energy interval $E = 0.3{-}10$ keV (observer frame) by collecting data from its primary instrument, the European Photon Imaging Camera (EPIC), which is equipped with three X-ray charge-coupled device cameras; namely, the pn \citep[][]{struder2001} and the two MOS detectors \citep[][]{turner2001}. Data reduction, filtering of high background periods, and spectral extraction were performed according to the standard procedures described on the SAS web page.\footnote{\url{https://www.cosmos.esa.int/web/xmm-newton/sas-threads}.} For each object, we chose a circular region with a radius of $\sim20{-}30$ arcsecs for the source extraction, corresponding to $\geq70\%$ of the encircled energy fraction for both on-axis and off-axis sources and a nearby, larger source-free circular region for the background. The spectra were binned to ensure at least one count per bin and modelled within the XSPEC (\texttt{v12.13.0g}; \citealt{arnaud1996})  package by minimising the Cash statistic with background subtraction \citep[W-stat in XSPEC;][]{cash1979, wachter1979}.

\subsection{X-ray spectroscopy}\label{sec:xray_spec}

We adopted the following procedure to analyse the X-ray spectrum of each source in our sample.
Data from EPIC-pn, MOS1, and MOS2 were always considered simultaneously and fitted together, taking into account an intercalibration constant between the three instruments that was allowed to vary within a factor of $< 20\%$. This threshold is less strict than that typically adopted in the case of dedicated on-axis pointings \citep[][]{read2014}, as in this work we deal with observations of serendipitous AGNs that are likely to be positioned off-axis with respect to the centre of the \emph{XMM-Newton} field of view (see, e.g. \citealt{mateos2009}). 
While the off-axis position leads to an increased vignetting and elongated point spread function that may have an impact on the cross-calibration of EPIC cameras \citep[e.g.][]{mateos2009,read2011}, it also implies, as an aside, that sometimes the given source can lie within a bad column or a gap between the CCDs. For this reason, for each X-HESS AGN, we checked whether any of these occurrences were actually present in the observations carried out with the EPIC cameras and then we considered only those data unaffected by such limitations to extract the X-ray spectrum.

\begin{figure}[t]
    \centering
    \includegraphics[width=\columnwidth]{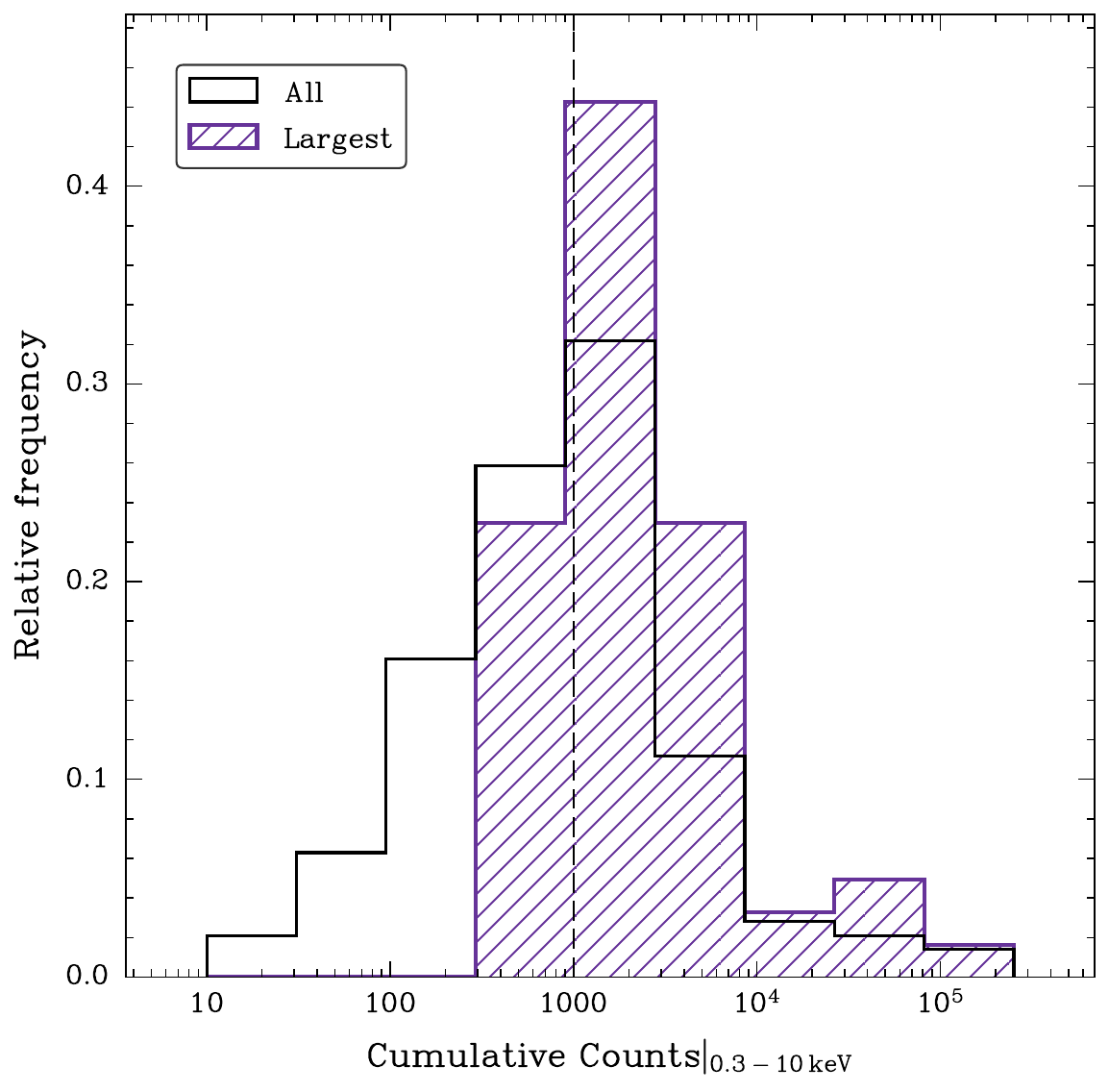}
    \caption{Relative frequency histogram showing the distribution of the total net counts from EPIC-pn, MOS1, and MOS2 in the broad $E=0.3-10$ keV observer-frame energy band for each individual observation of the X-HESS AGNs (in black). The distribution related to the observations with the largest number of counts of each X-HESS AGN is shown in purple.}
    \label{fig:Cts_vs_Exp}
\end{figure}

\begin{figure}[t]
    \centering
    \includegraphics[width=\columnwidth]{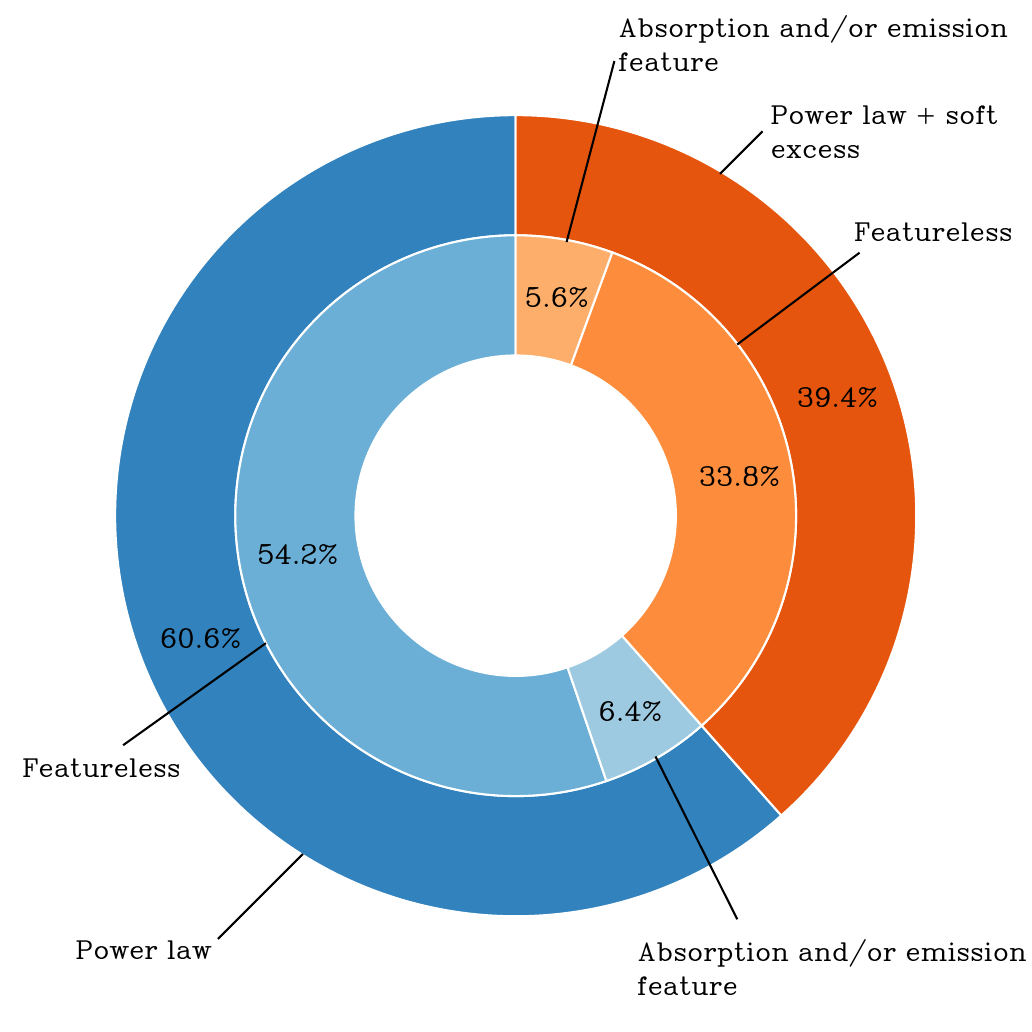}
    \caption{Pie chart describing the overall results of the spectroscopic analysis of the X-HESS sources. In most cases, the broadband X-ray continuum ($E=0.3\!-\!10$ keV) can be reproduced well by a power-law model, modified by Galactic absorption. In the remaining $\sim\!40\%$ of the observations, an excess of soft X-ray emission is found. Approximately $\sim\!12\%$ of the X-ray spectra of the X-HESS AGN reveal the presence of at least one emission and/or absorption feature due to the interaction of the primary continuum with either cold or ionised gas clouds.}
    \label{fig:pie}
\end{figure}

Figure \ref{fig:Cts_vs_Exp} provides a general description of all the 142 \emph{XMM-Newton} observations of the X-HESS AGNs in terms of the photon counts in the broad $E=0.3{-}10$ keV observer-frame energy interval, showing the relative frequency histogram of the cumulative net (i.e., background-subtracted) counts calculated by considering the joint EPIC-pn, MOS1, and MOS2 observations.
This distribution is characterised by a median value of ${\sim}900$ and is indicative of the spectral quality of the X-HESS observations, with more than $90\%$ ($40\%$) of their total having a number of counts of ${>}100$ (${>}1000$).
The comprehensive journal of observations of the X-HESS AGN can be consulted in Appendix \ref{sec:appA}, where the reader can find information about both the net exposure and counts for all the EPIC cameras.

For each spectrum, we first ignored all the data outside the $E = 0.3{-}10$ keV observer-frame energy interval and then initially considered the hard X-ray spectrum ($E>2$ keV in the rest frame, corresponding to $E>2/(1 + z)$ keV in the observer frame), to constrain the value of the photon index without any possible contamination from additional spectral components such as the soft excess -- likely to emerge at softer energies -- that could affect the $\Gamma$ measurement.
The determination of the underlying primary continuum was achieved by modelling the hard X-ray spectrum with a power law modified by Galactic absorption.
Then we extended the analysis to the broad $E=0.3{-}10$ keV interval by including the soft portion of the X-ray spectrum. Eventually, further deviations in the residuals were addressed by considering additional spectral components, which were included in the best-fit model whenever appropriate; that is, according to the statistical significance of the $\Delta{W}$-stat value with respect to the corresponding change of degrees of freedom at a 95\% confidence level for the number of interesting parameters characterising the additional component.

Though the observations with a larger number of counts are likely to return a more detailed view of the AGN spectral features, our best-fit spectral results suggest that the broadband X-ray continuum ($E=0.3{-}10$ keV) of the X-HESS sources can always be reproduced by a phenomenological model consisting of a power law modified by Galactic absorption plus, in some cases, a blackbody component to account for the excess of soft X-ray emission that characterises more than a third of all the 142 observations. Indeed, as is shown in Fig. \ref{fig:pie}, the former model accounts for the X-ray continuum emission of $\sim\!60\%$ of the observations of the X-HESS AGNs, while the latter indicates that $\sim\!40\%$ of the spectra reveals the presence of a soft excess.

Although the blackbody model provides only a phenomenological explanation of the soft excess, whose physical origin is still debated \citep[e.g.][]{sobolewska2007,fukumura2016,petrucci2018,middei2020}, it also offers the best description of the soft excess in terms of the W-statistic for all the spectra showing such a feature. Moreover, we find values of blackbody temperature, $kT_\mathrm{bb}$, that are consistent with those typically measured for type-1 AGNs \citep[e.g.][]{piconcelli2005, bianchi2009}, ranging from a minimum of $\sim100$ eV up to $\sim350$ eV.

While such basic modelling returns a satisfactory description of the vast majority of the broadband X-ray spectra ($\sim\!88\%$), as is shown in Fig. \ref{fig:pie}, we find evidence of at least one emission and/or absorption feature in approximately $\sim\!12\%$ of the observations due to the interaction of the primary continuum with either cold or ionised gas clouds. In the latter case, the absorption features are usually attributable to either warm absorbers (WAs) or UFOs. However, we note that this fraction is most likely a lower limit because of the limited S/N of several observations. We refer the reader to Appendix \ref{sec:appB} for a thorough description of the best-fit spectral results, where one can also find a few additional notes about those AGNs with relatively more complex spectra.

\subsection{Measuring the ultraviolet luminosity and $\alpha_\mathrm{ox}$}
\subsubsection{Optical Monitor data reduction}\label{sec:om}

When available, the \emph{XMM-Newton} observations were complemented with the simultaneous optical/UV data from the Optical Monitor. The raw OM data were converted to science products using the SAS task \textit{omichain}. We used the task \textit{om2pha} to convert the OM photometric points into a suitable format for XSPEC.
Galactic extinction was taken into account by considering the measurement of the colour excess from the reddening map of \citet{schlafly2011}, provided by the NASA/IPAC Infrared Science Archive (IRSA) website\footnote{\url{https://irsa.ipac.caltech.edu/applications/DUST/}.}. The spectra were then corrected for Galactic extinction according to the Milky Way reddening law of \citet{fitzpatrick1999}, assuming $R_V = 3.1$.

\subsubsection{$L_\mathrm{2500\,\AA}$ and $\alpha_\mathrm{ox}$}\label{sec:l2500_aox}

The OM measurements allowed us to estimate the monochromatic UV luminosity of $L_{2500\,\AA}$ by adopting the approach described in \citet{vagnetti2010},
which can be summarised as follows: (i) When the available rest-frame luminosity estimates from the OM filters only cover frequencies higher or lower than $2500\,\AA$, $L_{2500\,\AA}$ was calculated through curvilinear extrapolation, following the behaviour of the average optical/UV spectral energy distribution of \citet{richards2006}, computed for a large sample of type-1 AGNs with SDSS data coverage, shifted vertically to match the specific luminosity at the frequency of the nearest OM data point.
(ii) If the OM data points extend across rest-frame $\lambda{=}2500\,\AA$, $L_{2500\,\AA}$ was calculated with a linear interpolation. (iii) When only a single OM filter was available, $L_{2500\,\AA}$ is measured as in (i). We estimate the luminosity at $\lambda=4400\,\AA$, whose usage will be discussed in the \S\,\ref{sec:ep_dep}, in a similar fashion. The values of both $L_{2500\,\AA}$ and $L_{4400\,\AA}$ are listed in Table \ref{tab:bestfit}.

We estimated the possible contribution of the host galaxy starlight to the optical/UV emission by modelling the optical spectrum as a combination of AGN and galaxy components, following the procedure described in \citet{vagnetti2013}. We find that the host galaxy contribution is less than $10\%$ for the bulk of the sample and can be safely neglected for all the X-HESS sources.

\begin{figure}[t]
    \centering
    \includegraphics[width=.97\columnwidth]{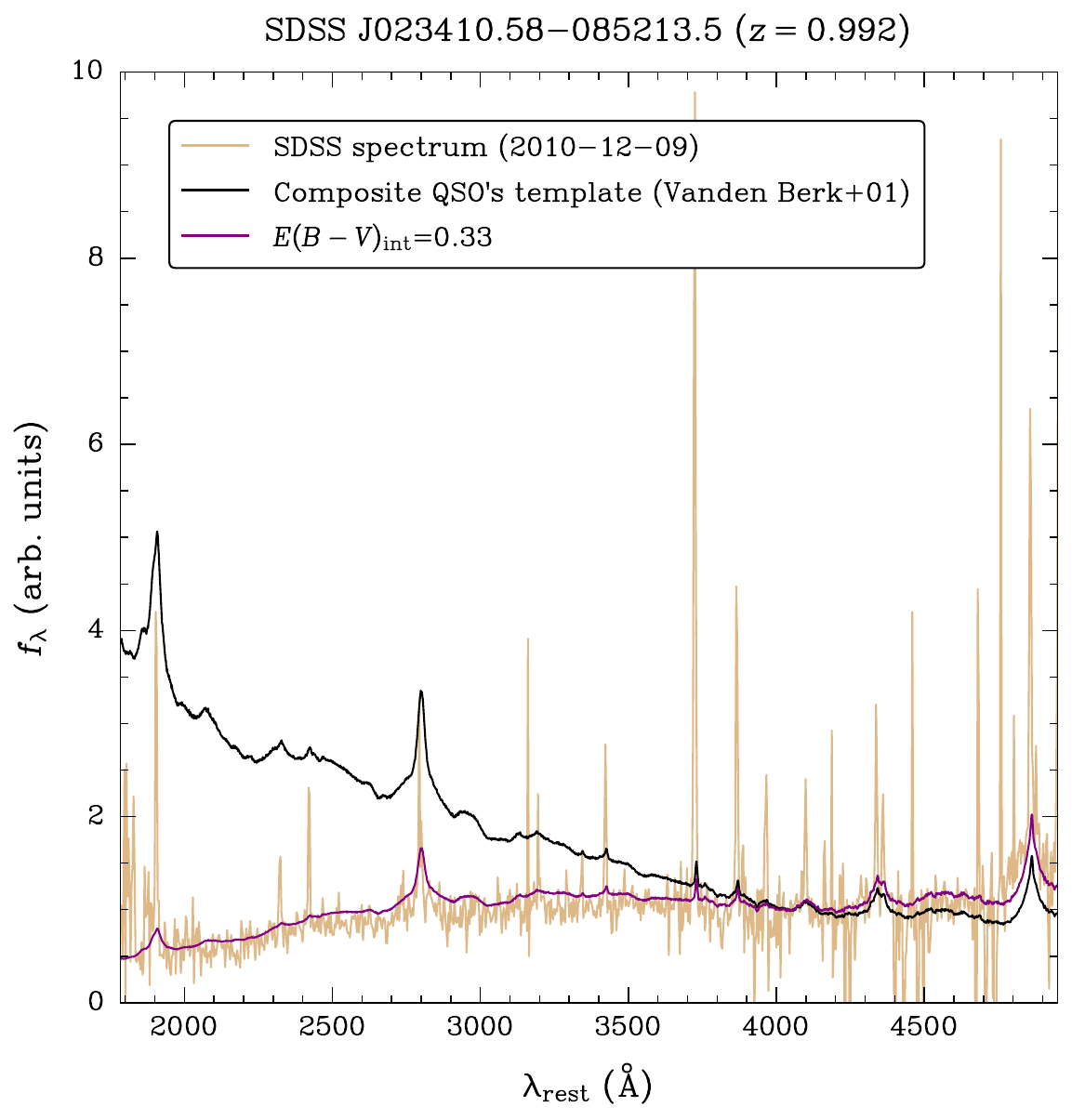}
    \caption{SDSS spectrum of J0234$-$08 at $z = 0.992$ (in gold) compared to the composite QSO's template of \citet{vandenberk2001}. The template (in black) is extinguished according to the SMC extinction law by \citet{prevot1984}, with progressively increasing values of $E(B-V)$. The value of $E(B-V)$ that minimises the distance between the reddened template (in purple) and the observed SDSS spectrum at wavelengths of $\lambda_\mathrm{obs}<8000\,\AA$ in the observer frame, corresponds to our estimate of the internal reddening of the source. J0234$-$08 has the largest $E(B-V)$ among our sources, i.e.\ $E(B-V)=0.33$.}
    \label{fig:reddening}
\end{figure}

Moreover, we measured the internal reddening of the X-HESS AGN by adopting an approach similar to that described in \citet{laurenti2022}. For each source, we compared the SDSS spectrum with the SDSS composite QSO's spectrum by \citet{vandenberk2001} and normalised them to the unit flux at rest-frame wavelengths corresponding to $\lambda_\mathrm{obs}\geq8000\,\AA$ in the observer frame.
We then applied the Small Magellanic Cloud (SMC) extinction law by \citet{prevot1984} to the template, as is shown in Fig.\ \ref{fig:reddening}. The choice of an SMC-like extinction law stems from its capability of reproducing dust reddening of QSOs at all redshifts \citep[e.g.][]{richards2003, hopkins2004, bongiorno2007, gallerani2010, krawczyk2015}.
In each case, the fiducial value of $E(B-V)$ is the one that minimises the distance between the template and the SDSS spectrum at wavelengths shorter than $\lambda_\mathrm{obs}=8000\,\AA$.
The vast majority of the sources —\ $\sim85\%$ — do have $E(B-V)\leq0.1$, while the rest of the AGNs present reddening values extending up to a maximum of $\sim0.33$. The typical uncertainty on these reddening estimates is $\sim0.01$. The values of internal reddening of the X-HESS AGNs are reported in Table \ref{tab:agn_list}.
The OM photometric data were then corrected for the effects of both Galactic and internal dust extinction to obtain a refined measurement of $L_{2500\,\AA}$.
In principle, the OM data variability for the multi-epoch AGN may be due to a combination of possible changes in the accretion rate and/or extinction, which are not easy to disentangle. Moreover, multi-epoch optical spectra would be required as well to provide deeper insights into the possible extinction changes.

Finally, the value of the monochromatic UV luminosity, $L_{2500\,\AA}$, can be used jointly with the monochromatic X-ray luminosity at 2 keV measured from the X-ray spectrum to calculate the $\alpha_\mathrm{ox}$ parameter as
\begin{equation}\label{eq:aox}
\begin{split}
  \alpha_\mathrm{ox} &= \log{\Bigg[\frac{L_\nu(2\,\mathrm{keV})}{L_\nu(2500\,\AA)}\Bigg]}\,\Bigg/\,\log{\Bigg[\frac{\nu(2\,\mathrm{keV})}{\nu(2500\,\AA)}\Bigg]} \\
  & \simeq 0.384\,\log{\Bigg[\frac{L_\nu(2\,\mathrm{keV})}{L_\nu(2500\,\AA)}\Bigg]}\,.
\end{split} 
\end{equation}

\noindent Clearly, the $\alpha_\mathrm{ox}$ parameter could only be obtained for those 38 X-HESS AGNs with OM data coverage, 14 of which have multi-epoch observations. In total, there are 82 OM observations of the X-HESS AGN and the corresponding $\alpha_\mathrm{ox}$ values are listed in Table \ref{tab:bestfit}.

\begin{figure}[t]
    \centering
    \includegraphics[width=\columnwidth]{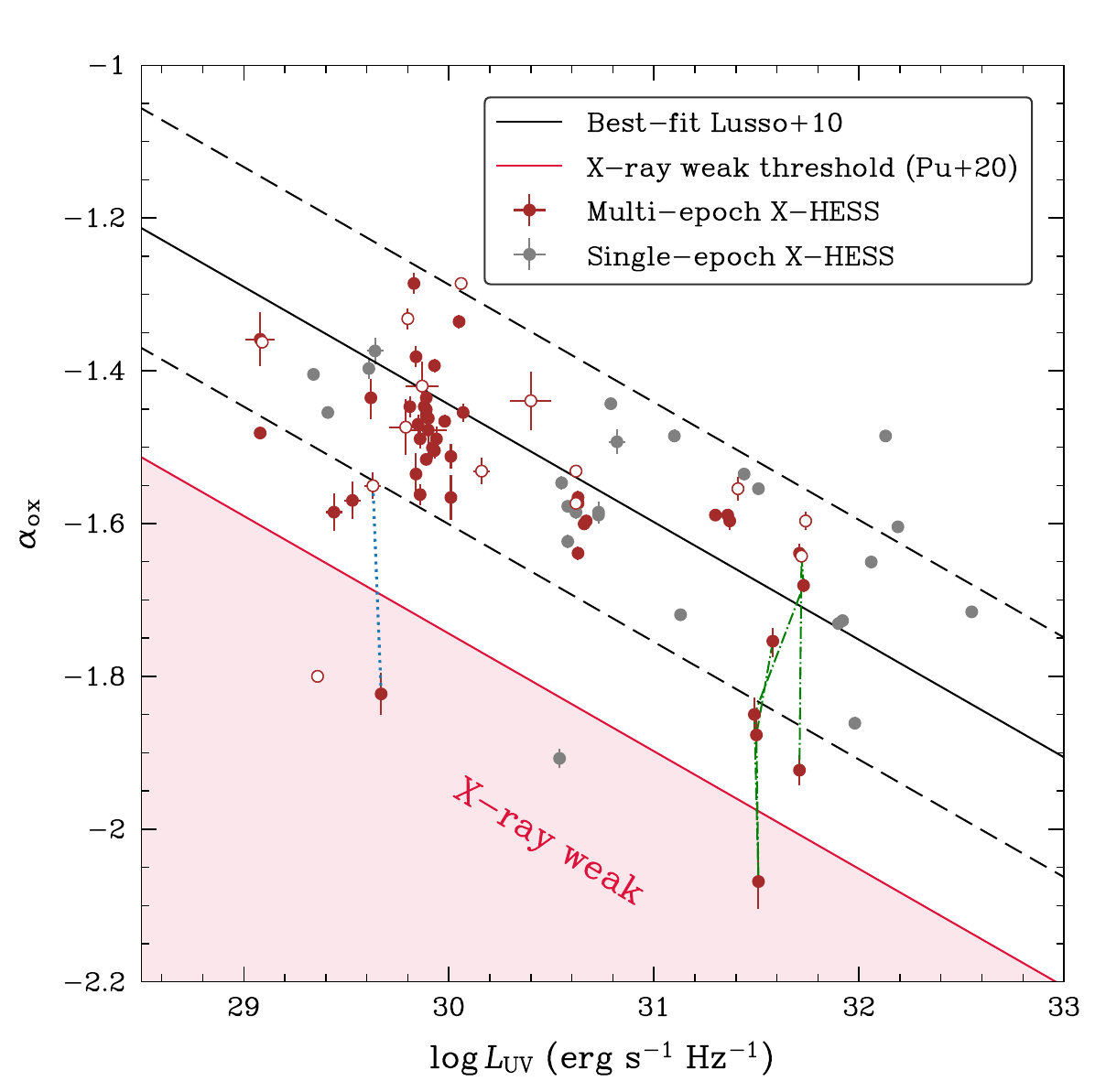}
    \caption{Distribution of the X-HESS AGNs in the $\alpha_\mathrm{ox}-\log{L_\mathrm{UV}}$ plane. Single- and multi-epoch AGNs are shown in grey and red, respectively. White circles with red edges highlight the high-flux state of the multi-epoch X-HESS AGNs. Solid and dashed black lines indicate the best-fit relation from \citet{lusso2010} and the corresponding $1\sigma$ spread. The solid red line marks the reference value for X-ray weakness, i.e. $\Delta{\alpha_\mathrm{ox}}\leq-0.3$ \citep[][]{pu2020}.  Dotted and dash-dotted blue and green lines describe the time evolution of $\alpha_\mathrm{ox}$ for the AGN X-HESS 16 and X-HESS 5, respectively. Specifically, the latter has experienced a couple of transitions between phases of standard and weak X-ray emission over the different observations.}   
    \label{fig:aox_Luv_xhess}
\end{figure}

\subsubsection{X-ray weakness}

\citet{laurenti2022} found a significant population of X-ray weak sources among optically selected QSOs with $L_\mathrm{bol}>10^{46}$ erg s$^{-1}$, comprising almost $30\%$ of their whole sample, and similar results were reported by \citet{nardini2019} and \citet{zappacosta2020}, but this is not the case for the X-HESS sources.
Figure \ref{fig:aox_Luv_xhess} shows the distribution of the X-HESS AGNs with available OM measurements in the $\alpha_\mathrm{ox}-\log{L_\mathrm{UV}}$ plane, where $\log{L_\mathrm{UV}}$ is the luminosity at $2500\,\AA$, also indicating the threshold for X-ray weakness defined by \citet{pu2020}. We recall that this threshold amounts to $\Delta{\alpha_\mathrm{ox}}\leq-0.3$, with $\Delta{\alpha_\mathrm{ox}}$ representing the difference between the observed value of $\alpha_\mathrm{ox}$ and the expected value from the reference relation which, in our case, is that from \citet{lusso2010}.
One can notice that in the vast majority of observations, the sources are distributed accordingly to the $\alpha_\mathrm{ox}-\log{L_\mathrm{UV}}$ relation from \citet{lusso2010}, while only a few observations of X-HESS AGNs included in the multi-epoch subsample are consistent with an epoch of weak X-ray emission. 
This result is not surprising, as it is probably due to the selection criteria of the X-HESS sources. Indeed, in building the sample we only considered those AGNs that have at least one observation with a sufficiently large number of photon counts in the EP8 band; that is,\ EP\_8\_CTS $\geq1000$.
In this way, we are neglecting most of the highly accreting AGNs that are also X-ray weak, as for these sources we expect photon counts to be smaller than the one we adopted as a threshold.
In order to investigate the true fraction of X-ray weak sources among the high-\ledd\ population, it would be necessary to significantly reduce such a limiting value for the photon counts \--- or better, and ideally remove it \--- though this would require a spectroscopic analysis of a huge number of AGNs, which was not feasible in this work.
Despite this limitation, an interesting result is clearly emerging in Fig. \ref{fig:aox_Luv_xhess} for SDSS J130048.10+282320.6 and SDSS J022928.41$-$051125.0; that is,\ X-HESS 5 and 16, respectively. Specifically, the time evolution of $\alpha_\mathrm{ox}$ for X-HESS 5 and 16, described by the green and blue lines in the same figure, suggests that these sources have experienced at least one transition between phases of standard and weak X-ray emission over the different observations. This result is consistent with our findings based on the ongoing \emph{Swift} monitoring of the X-ray weak AGNs described in \citet{laurenti2022}, which we shall present in a forthcoming paper (Laurenti et al., \emph{in prep.}), in which we observe similar transitions between hard- and low-flux states in some sources, and thus supports the idea that the X-ray weakness of high-\ledd\ AGNs may represent a transient phenomenon over different timescales.

In the case of X-HESS 16 ($z=0.307$), the transition towards the X-ray weak state occurs at $\sim18$ months in the source rest-frame. The soft and hard X-ray fluxes drop by a factor of approximately $\sim5$ and $\sim3$, respectively, and the photon index becomes shallower, moving from $\sim1.8$ to $\sim1.3$. For X-HESS 5 ($z=1.929$), instead, we observe two transitions from and to an X-ray weak state. In both cases, the soft and hard X-ray fluxes drop or raise by a factor of $\sim3$ and $\sim2$, respectively, and the associated timescales are much shorter than for X-HESS 16, since each transition occurs over a period of $\lesssim 1$ week in the source rest-frame. The value of the photon index, $\Gamma$, is consistent within the errors between the X-ray weak state ($\Gamma\sim1.4$) and the two contiguous higher-flux states ($\Gamma\sim1.7$), although it is slightly flatter in the former.
It is also worth mentioning that the results of our spectroscopic analysis suggest that the X-ray weak state is not directly ascribable to cold absorption. This holds not only for X-HESS 5 and 16, but also for the single-epoch X-HESS 43 caught in a low-flux state (grey dot in Fig.\ \ref{fig:aox_Luv_xhess}). However, we cannot rule out the presence of a highly ionised gas cloud with a large column density and a small covering factor along the line of sight, which would cause a decrease in the X-ray flux without heavily affecting the spectral shape (see, e.g. \citealt{laurenti2022} for a detailed discussion).
The remaining AGN lying in the X-ray weak domain is X-HESS 20, whose spectrum shows an absorption feature attributable to a warm gas cloud (see also Appendix \ref{sec:appB}). 
Unlike the variable X-ray emission, the UV luminosity at $2500\,\AA$ remains relatively stable between normal and weak X-ray emission phases for both X-HESS 5 and 16.

We note that the hypothetical presence of broad-absorption line (BAL) QSOs can potentially affect the fraction of X-ray obscured sources among high-\ledd\ AGNs. Unfortunately, we cannot quantify their impact in the present study, as we were unable to retrieve this information on the presence of BAL features (e.g. \ion{C}{IV} absorption) in the X-HESS sample.

\begin{figure}[t]
    \centering
    \includegraphics[width=\columnwidth]{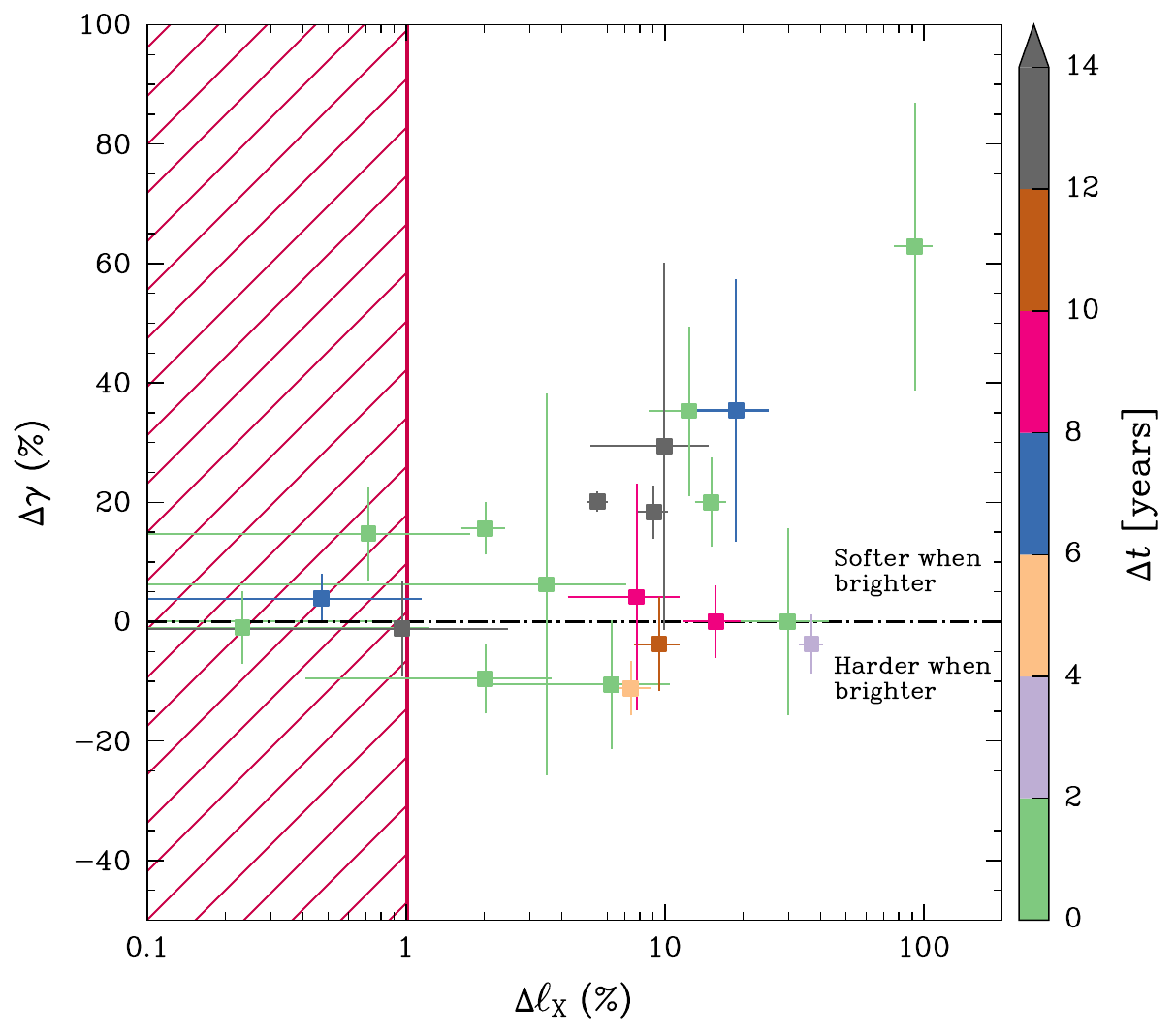}
    \caption{Spectral variation in the photon index, $\Gamma$, versus the change in $E=2-10$ keV X-ray luminosity between the lowest- and highest-flux states of the multi-epoch X-HESS AGNs. Luminosity changes of $<1\%$ occurring within the red-hatched region are considered insignificant. Most of the sources vary accordingly to a softer-when-brighter trend, while a smaller fraction do the opposite. No clear correlation is found between the amplitude of the spectral variations and the time elapsed between the two observations of interest.}
    
    \label{fig:delta_gamma}
\end{figure}

\section{Results}\label{sec:results}

\subsection{Spectral variability}\label{sec:variability}

The multi-epoch X-HESS AGNs allow us to investigate the spectral variability properties of these highly accreting sources. 
To track these variations, it is useful to consider the change in $\Gamma$ values between the observations associated with the lowest- and highest-flux states of each individual AGN, respectively. 
Specifically, we can compute the fractional variations of both the photon index and $E=2{-}10$ keV luminosity, $L_\mathrm{X}$, between the two observations of interest, $\Delta{\gamma}=(\Gamma_\mathrm{high}-\Gamma_\mathrm{low})/\Gamma_\mathrm{low}$ and $\Delta{\ell_\mathrm{X}}=(L_\mathrm{X,high}-L_\mathrm{X,low})/L_\mathrm{X,low}$, with a clear reference to the low- and high-flux states.

Figure \ref{fig:delta_gamma} shows the $\Delta{\gamma}-\Delta{\ell_\mathrm{X}}$ plane. The overall trend appears to favour positive variations in the photon index with increasingly larger changes in the hard X-ray luminosity, in agreement with a softer-when-brighter behaviour that is often reported in AGNs with a moderate-to-high Eddington ratio \citep[e.g.][]{paolillo2004, shemmer2006, sobolewska2009, gibson2012, soldi2014, vagnetti2016, serafinelli2017}, though it is not highly significant since the Spearman's correlation coefficient and the corresponding null probability are $\rho_\mathrm{S} = 0.40$ and $p(>\!|\rho_\mathrm{s}|) = 0.07$, respectively. Moreover, a minor fraction of the multi-epoch X-HESS AGNs appear to vary accordingly to the opposite trend; that is,\ harder when brighter. 

Those AGNs with $\Delta{\ell_\mathrm{X}}<1\%$ in terms of their nominal value or within their $1\sigma$ measurement error are considered as non-variable (red-shaded region in the plot). The same occurs for those sources whose $\Delta{\gamma}$ is consistent with zero.
Thus, we find that 14 of the 22 multi-epoch X-HESS AGNs ($\sim64\%$) do not show significant spectral variations between their corresponding lowest- and highest-flux observations. The eight remaining AGNs ($\sim36\%$) are characterised by significant spectral variations at the $1\sigma$ confidence level, with seven of them showing a softer-when-brighter trend, while only one source (X-HESS 11) does the opposite.
The same figure shows the time elapsed between the two observations of interest for each source, and the amplitude of the spectral variations does not show a clear dependence in terms of such a time span, although half of the significant $\Gamma$ changes apparently occur within a two-year rest-frame.
Furthermore, it is worth noting that the multi-epoch X-HESS sources have not been monitored with a uniform cadence or an equal number of times, and this could possibly affect the distribution in the $\Delta{\gamma}-\Delta{\ell_\mathrm{X}}$ plane.

\subsection{The $\Gamma-\lambda_\mathrm{Edd}$ relation}\label{sec:gam_edd}

We investigated the $\Gamma-\lambda_\mathrm{Edd}$ relation as follows. We constrained the X-ray primary continuum and derived the best-fit value of the photon index, $\Gamma$, of the X-HESS AGNs according to the procedure described in \S\,\ref{sec:xray_spec} for the X-ray spectral analysis, while the value of the Eddington ratio for each of these sources is listed in Table\ \ref{tab:agn_list}. In this section and in Sect. \ref{sec:gam_edd_mbh_lbol}, for the multi-epoch X-HESS AGNs, we only considered the value of the photon index in their highest-flux state, assuming that it most likely represents their intrinsic X-ray coronal emission if their lower-flux states were due to any kind of possible obscuration that in some cases might be potentially undetectable, as in lower S/N observations. The following results, however, still hold when jointly accounting for or averaging all the flux states of the multi-epoch X-HESS AGNs.

\begin{figure}[t]
    \centering
    \includegraphics[width=\columnwidth]{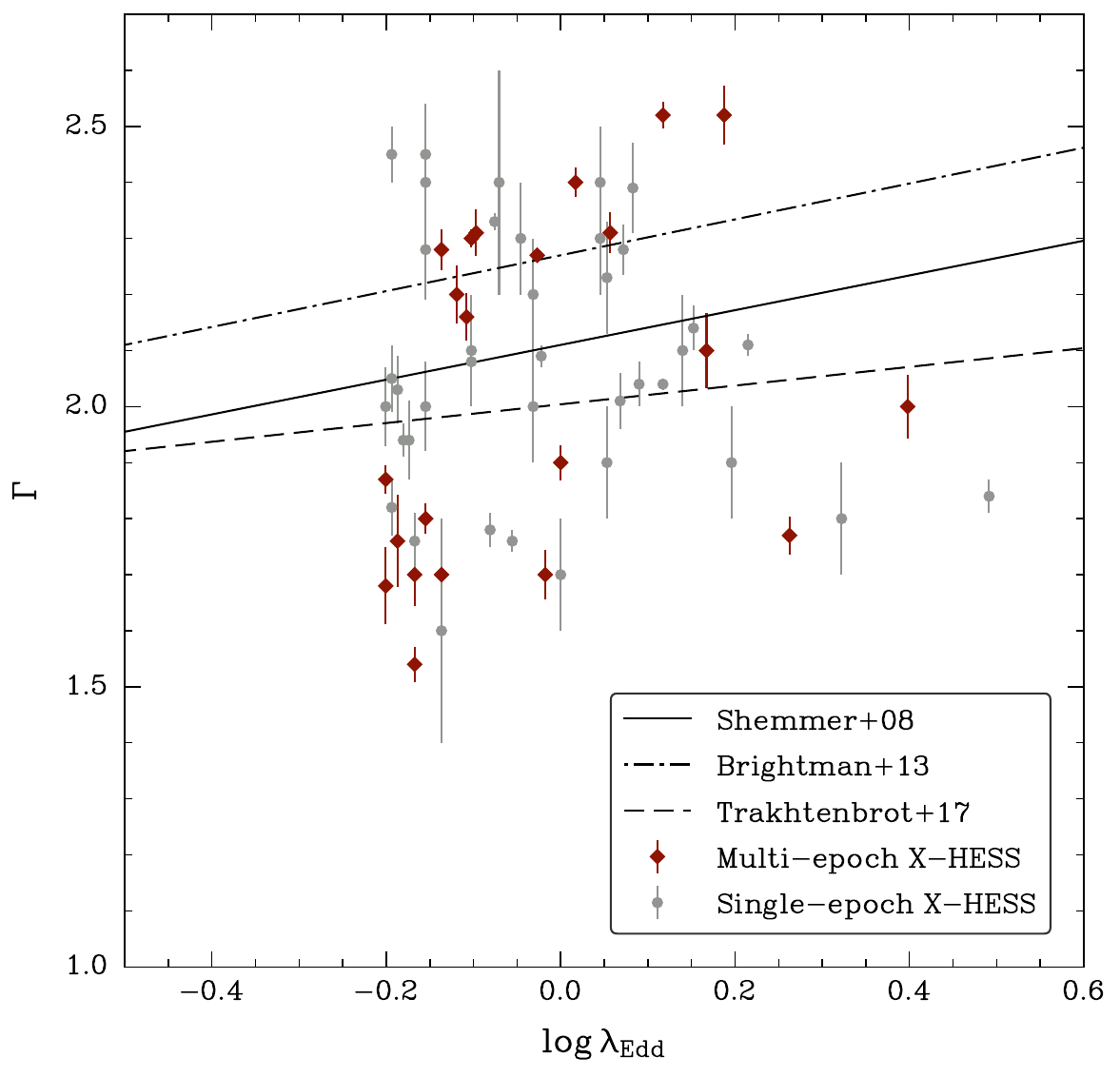}
    \caption{Distribution of the X-HESS AGNs in the $\Gamma-\log{\lambda_\mathrm{Edd}}$ plane. Single- and multi-epoch AGN are represented by grey circles and red diamonds, respectively. The best-fit relations from previous studies are reported as well. The average uncertainty on \ledd\ is 0.3 dex.}
    
    \label{fig:gamma_edd_xhess}
\end{figure}

Figure \ref{fig:gamma_edd_xhess} shows the distribution of the X-HESS AGNs in the $\Gamma-\log{\lambda_\mathrm{Edd}}$ plane. 
When accounting for both the multi- and single-epoch X-HESS subsamples, we observe a large dispersion of $\Gamma$ values spanning from a minimum of $\sim\!1.5$ to a maximum of $\sim\!2.5$, which is consistent with the range reported in \citet{laurenti2022}.
However, in that case, we studied an AGN sample comprising sources lying in a narrow interval of \ledd\ values tightly clustered around $\log\lambda_\mathrm{Edd}\sim0$, while in this case we observe the same scatter when investigating the spectral properties of AGNs with $-0.2<\log\lambda_\mathrm{Edd}<0.5$. 
This may suggest that such a dispersion is a property of high-\ledd\ AGNs as a class and is likely to emerge whenever a sufficiently large sample of highly accreting sources is considered. 
We do not find any indication of a significant correlation between $\Gamma$ and $\log\lambda_\mathrm{Edd}$, with the Spearman's correlation coefficient being $\rho_\mathrm{S}=0.21$ for a null probability of $p(>\!|\rho_\mathrm{S}|) = 0.11$. 
If we only consider the multi-epoch X-HESS AGNs, the correlation becomes weakly significant, as we find $\rho_\mathrm{S}=0.54$ and $p(>\!|\rho_\mathrm{S}|) \sim 0.01$, but it is still dominated by a large scatter.  

\begin{figure}[t]
    \centering
    \includegraphics[width=\columnwidth]{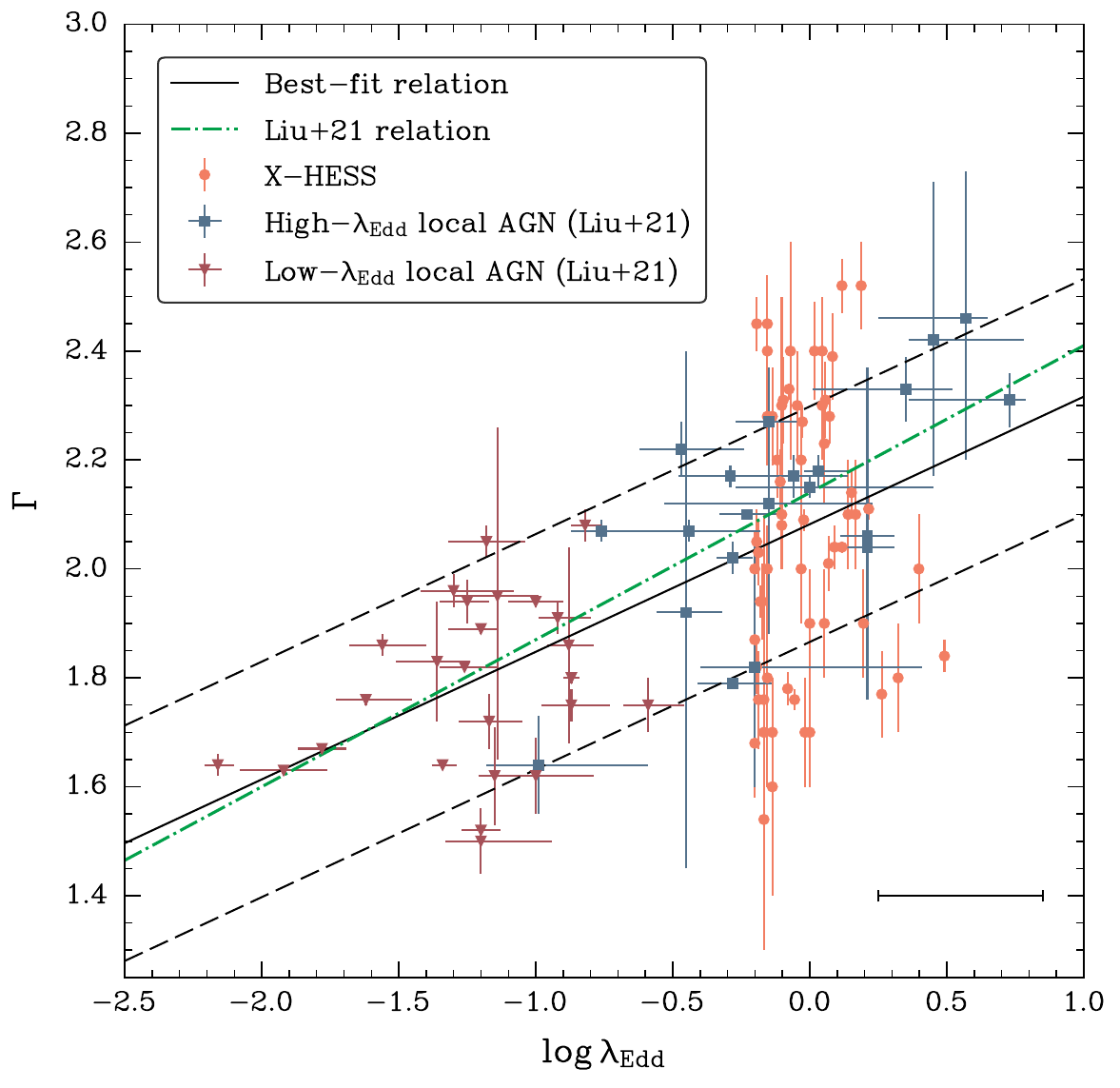}
    \caption{Distribution of the X-HESS AGNs (orange circles) and the high-\ledd\ (blue squares) and low-\ledd\ local AGNs (red triangles) from \citet{liu2021} in the $\Gamma-\log{\lambda_\mathrm{Edd}}$ plane. For the multi-epoch X-HESS AGNs, we considered the $\Gamma$ associated with their high-flux state. The black error bar in the lower right corner represents the average uncertainty on the X-HESS sources. Solid and dashed black lines describe the best-fit relation and its $1\sigma$ spread, respectively. The dash-dotted green line represents the \citealt{liu2021} best-fit relation, for comparison. }
    
    \label{fig:gamma_edd_all}
\end{figure}

To assess the relationship between $\Gamma$ and \ledd\ , one needs to extend its dynamical range by including AGNs characterised by lower Eddington ratios.
In their recent work, \citet{liu2021} did study a sample of 47 local AGNs divided into two groups consisting of 21 high-\ledd\ and 26 low-\ledd\ AGNs, respectively, in which the dividing line was set at $\lambda_\mathrm{Edd}\sim0.3$. All of these AGNs are located at a redshift of $z\lesssim0.3$ and can rely on very good quality $M_\mathrm{BH}$ measurements from reverberation mapping. Moreover, the authors only considered the highest-flux observation of their AGNs, when available, to investigate the relation between the photon index at the $E>2$ keV rest frame and $\lambda_\mathrm{Edd}$. In our case, the $\Gamma$ values reported in Table \ref{tab:bestfit} refer to the broadband $E=0.3{-}10$ keV spectral fit but, as is discussed in \S\,\ref{sec:xray_spec}, they were first constrained in the rest-frame $E>2$ keV interval. For this reason, we can safely compare our results with those in \citet{liu2021}.

Figure \ref{fig:gamma_edd_all} displays the distribution of the X-HESS AGNs and the two subsamples of local AGNs from \citet{liu2021} in the $\Gamma-\log{\lambda_\mathrm{Edd}}$ plane. 
Interestingly, if we consider both the X-HESS and high-\ledd\ local AGNs, the correlation between the two quantities is still not statistically sound, as $\rho_\mathrm{S}=0.25$ and $p(>\!|\rho_\mathrm{S}|) = 0.03$.
However, when considered altogether, the three samples show a highly significant correlation between $\Gamma$ and $\lambda_\mathrm{Edd}$, with the Spearman's  coefficient being $\rho_\mathrm{S}=0.52$ for a null probability of $p(>\!|\rho_\mathrm{S}|) < 10^{-7}$. Though this result is obviously driven by the low-\ledd\ AGN population in \citet{liu2021}, we computed the best-fit relation by adopting a bootstrap method; that is,\ by resampling $N=10000$ times the distribution in the $\Gamma-\log{\lambda_\mathrm{Edd}}$ plane, while accounting for the errors in both variables. According to this procedure, we obtain
\begin{equation}\label{eq:gam_edd}
    \Gamma = (0.23 \pm 0.04)\log{\lambda_\mathrm{Edd}} + (2.08 \pm 0.02)\,.
\end{equation}

\noindent The $1\sigma$ spread of the above relation is $\sim0.22$ and its slope is flatter -- albeit consistent within the uncertainties -- than the value of $0.27 \pm 0.04$ in \citet{liu2021} and also shallower than that reported in the bulk of previous works (e.g. $0.31\pm0.01$, \citealt{shemmer2008}; $0.31\pm0.06$, \citealt{risaliti2009}; $0.56\pm0.08$, \citealt{jin2012}; $0.32 \pm 0.05$, \citealt{brightman2013}), while it can better adhere to more conservative findings (e.g.\ $0.17\pm0.04$, \citealt{trakhtenbrot2017}; $0.16 \pm 0.03$, \citealt{trefoloni2023}) envisaging a weaker $\Gamma-\log{\lambda_\mathrm{Edd}}$ relation.

\subsection{The role of $M_\mathrm{BH}$ and $L_\mathrm{bol}$ in $\Gamma-\log{\lambda_\mathrm{Edd}}$}\label{sec:gam_edd_mbh_lbol}

The $\Gamma-\log{\lambda_\mathrm{Edd}}$ plane in Fig. \ref{fig:gamma_edd_all} is populated by a variety of sources spanning from local Seyfert-like galaxies to moderate-to-high-redshift QSOs, and thus is characterised by very different physical and observational properties. Since the Eddington ratio is proportional to both $M_\mathrm{BH}$ and $L_\mathrm{bol}$, we tried to determine whether these quantities have an impact on the observed distribution.

We considered the X-HESS and both \citet{liu2021} local AGN samples as a whole and then divided all sources into two equally populated subsamples in terms of $M_\mathrm{BH}$ and $L_\mathrm{bol}$. Specifically, we chose $\log(M_\mathrm{BH}/M_\odot)=8$ and $\log{(L_\mathrm{bol}/\mathrm{erg\,s}^{-1})}=45.5$ as the thresholds on the black hole mass and the bolometric luminosity, respectively. These thresholds are almost consistent with the transition between the average properties of NLSy1s and QSOs. 

{
\begin{table}[t]
\centering
\centering
\caption{Partial correlation analysis on the X-HESS and \citet{liu2021} subsamples with either a lower or higher black hole mass and bolometric luminosity.} 
\renewcommand{\arraystretch}{1.5}
\begin{adjustbox}{max width=\columnwidth}
\begin{threeparttable}
\begin{tabular}{c c c c c}
    \hline\hline
    Sample & Sample & $\Gamma - \lambda_\mathrm{Edd}$ & $\Gamma - ( M_\mathrm{BH} | L_\mathrm{bol}) $ & $\Gamma - ( L_\mathrm{bol} | M_\mathrm{BH}) $ \\
    & Size & $\rho_\mathrm{S}$ [  $p(>\!|\rho_\mathrm{S}|)$ ] & $\rho_\mathrm{S}$ [  $p(>\!|\rho_\mathrm{S}|)$ ] & $\rho_\mathrm{S}$ [  $p(>\!|\rho_\mathrm{S}|)$ ]  \\
    \hline 
    Low-$M_\mathrm{BH}$  & 62 & 0.62 [$<10^{-7}$] & $-0.32$ [ 0.01 ] & $0.52$ [ $2\times10^{-5}$ ] \\
    High-$M_\mathrm{BH}$ & 45 & 0.28 [0.06]       & $-0.18$  [0.24]  & $0.18$ [ 0.24 ] \\
    
    \hline
    Low-$L_\mathrm{bol}$  & 50 & 0.52 [$<10^{-4}$] &  $-0.41$  [$4\times10^{-3}$] & $0.44$  [$2\times10^{-3}$] \\
    High-$L_\mathrm{bol}$ & 57 & 0.24 [0.07]       &  $-0.21$ [ 0.12] & $0.13$ [  0.34 ]\\
    \hline
\end{tabular}

\end{threeparttable}
\label{tab:parcorr}
\end{adjustbox}
\end{table}
}

\begin{figure}[t]
    \centering
    \includegraphics[width=\columnwidth]{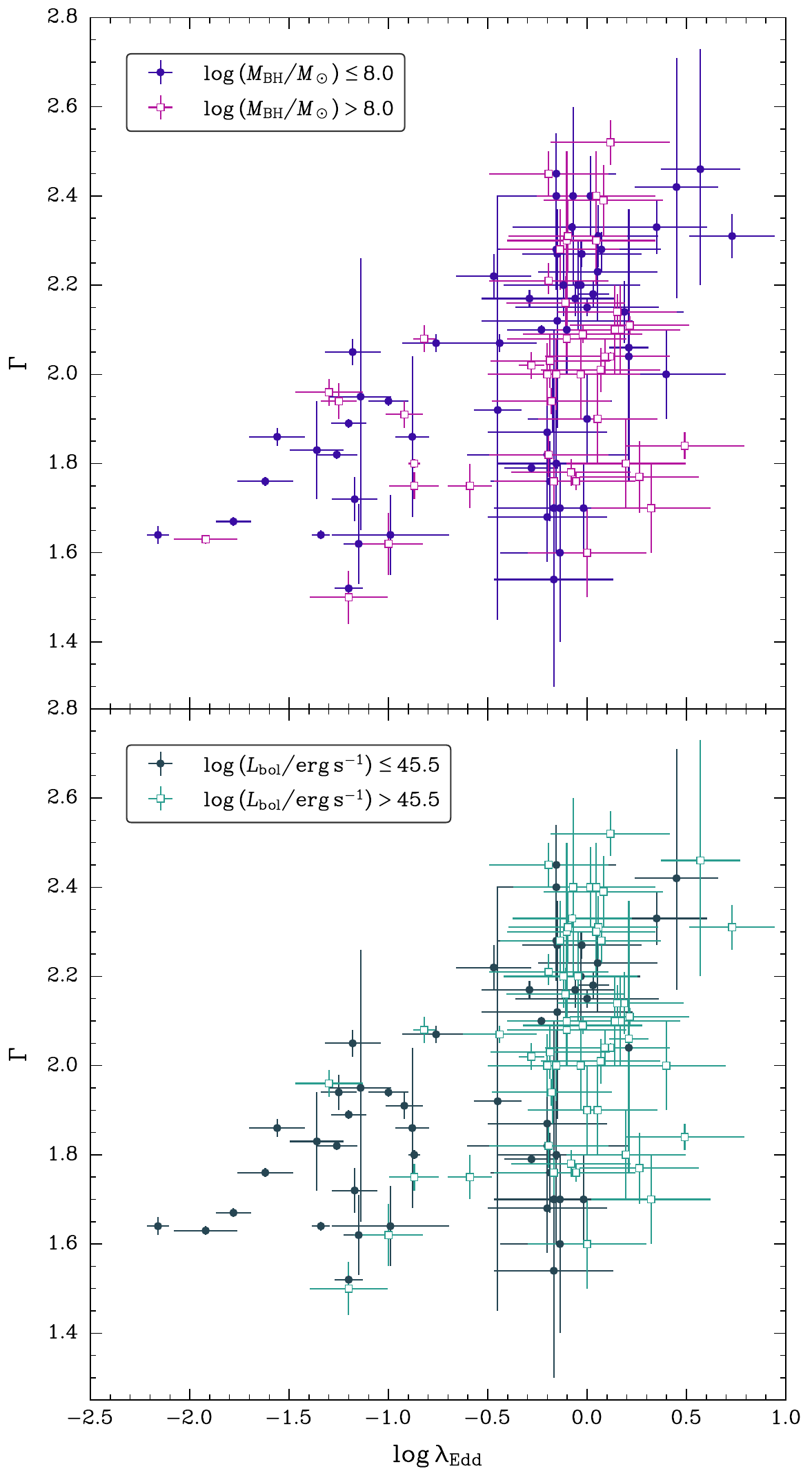}
    \caption{Distribution of the X-HESS AGNs and the high-\ledd\ and low-\ledd\ local AGNs from \citet{liu2021} in the $\Gamma-\log{\lambda_\mathrm{Edd}}$ plane. All AGNs are divided in two almost equally populated sub-intervals of black hole mass (\emph{Top panel}) and bolometric luminosity (\emph{Bottom panel}). The correlation between $\Gamma$ and \ledd\ appears to vanish for those sources with either higher $M_\mathrm{BH}$ or $L_\mathrm{bol}$. }
    
    \label{fig:gamma_edd_intervals}
\end{figure}

The distribution of all sources in the $\Gamma-\log{\lambda_\mathrm{Edd}}$ plane, highlighting the two different $M_\mathrm{BH}$ (top panel) and $L_\mathrm{bol}$ (bottom panel) subsamples, is shown in Fig. \ref{fig:gamma_edd_intervals}. We find that the lower-$M_\mathrm{BH}$ AGNs are highly correlated with the Eddington ratio ($\rho_\mathrm{S}=0.62$, $p(>\!|\rho_\mathrm{S}|) < 10^{-7}$) as opposed to the higher-$M_\mathrm{BH}$ sample that has $\rho_\mathrm{S}=0.28$ and $p(>\!|\rho_\mathrm{S}|) =0.06$. We obtain a similar result for the $L_\mathrm{bol}$ subsamples, with the lower-$L_\mathrm{bol}$ AGNs being correlated with \ledd\ ($\rho_\mathrm{S}=0.52$, $p(>\!|\rho_\mathrm{S}|) = 10^{-4}$), while the higher-$L_\mathrm{bol}$ AGNs have $\rho_\mathrm{S}=0.24$ and $p(>\!|\rho_\mathrm{S}|) =0.07$.

We also performed a partial correlation analysis on the two subsamples to disentangle the dependence on one parameter of interest while controlling for the other. The results are listed in Table \ref{tab:parcorr}. In this table, we can notice that the $\Gamma$ partial correlations with both the black hole mass and bolometric luminosity — that is, $\Gamma - ( M_\mathrm{BH} | L_\mathrm{bol}) $ and $\Gamma - ( L_\mathrm{bol} | M_\mathrm{BH}) $, respectively — become much less significant when we consider those AGNs with either higher $M_\mathrm{BH}$ or $L_\mathrm{bol}$.

\subsection{Towards an epoch-dependent \ledd\ }\label{sec:ep_dep}

Until now, we have considered a unique fixed value of \ledd\ for each X-HESS AGN independently of the observation date. However, the \ledd\ measurement reported in Table \ref{tab:agn_list} was derived by \citet{rakshit2020} as a result of the analysis of a single-epoch optical spectrum, which could have been collected at a very different time with respect to the X-ray observations of the X-HESS AGNs.
By adopting this approach, we are losing information about the specific flux state and/or variability of the sample sources.

To overcome this issue, one could leverage the simultaneous data from the OM, since the bulk of the accretion power of type-1 AGNs is actually released in the optical/UV. \citet{duras2020} have recently studied the dependence of the optical bolometric correction in an AGN sample spread over a wide interval of $L_\mathrm{bol}$. According to the authors, the optical bolometric correction, $k_\mathrm{bol,O}$, with respect to the optical $B-$band (4400$\,\AA$) luminosity is
\begin{equation}
\label{eq:kbol,opt}
k_\mathrm{bol,O}(L_\mathrm{4400\,\AA}) = 5.18 \pm 0.13.
\end{equation}

\noindent We adopted this relation and the values of $L_\mathrm{4400\,\AA}$ that we estimated for the X-HESS AGNs and the local AGN samples of \citet{liu2021} following the procedure described in \S\,\ref{sec:l2500_aox}, to obtain a simultaneous estimate of the bolometric luminosity, $L_\mathrm{bol,O/UV}$, based on the OM data. This, in turn, allowed us to translate such an $L_\mathrm{bol}$ measurement into a corresponding value of the Eddington ratio, $\lambda_\mathrm{Edd,O/UV}$, which we report in Table \ref{tab:bestfit}. This implies that, contrary to the approach adopted in \S\,\ref{sec:gam_edd}, we can now jointly consider all the flux states of the sizeable fraction ($>60\%$) of the X-HESS AGNs with simultaneous OM data coverage, without limiting the study to their highest flux state, as all of them would be characterised by different values of $\lambda_\mathrm{Edd}$. 

At the same time, we must be aware that the $L_\mathrm{bol,O/UV}$ measurements would be potentially biased if variable extinction had an impact on the X-HESS AGNs. Using optical-to-infrared light curves generated from public data such as those collected by ZTF \citep[][]{bellm2019}), Pan-STARRS \citep[][]{kaiser2002}, and NEOWISE \citep[][]{mainzer2011}, one could possibly verify whether the given AGN indeed shows strong accretion-rate variability, as is suggested by the OM data. However, this approach is beyond the scope of the present paper and is deferred to a future work.

\subsection{A new perspective on the $\Gamma$$-$\ledd\ plane}

According to the results in \S\,\ref{sec:ep_dep}, we can now rely on a useful tool that allows us to visualise the $\Gamma-\lambda_\mathrm{Edd}$ plane in a fully epoch-dependent frame, a property that is now encapsulated in both the values of the photon index, $\Gamma$, and the Eddington ratio, $\lambda_\mathrm{Edd,O/UV}$.
Of course, the same procedure can be applied to the whole set of AGNs from \citet{liu2021}. However, despite all of their sources being equipped with $\alpha_\mathrm{ox}$ measurements, not all of their optical/UV data were collected at the same time as the corresponding X-ray observation. For this reason, in this case we shall only consider those \citet{liu2021} AGNs ($\sim70\%$) disposing of simultaneous OM optical/UV observations. 

\begin{figure}[t]
    \centering
    \includegraphics[width=\columnwidth]{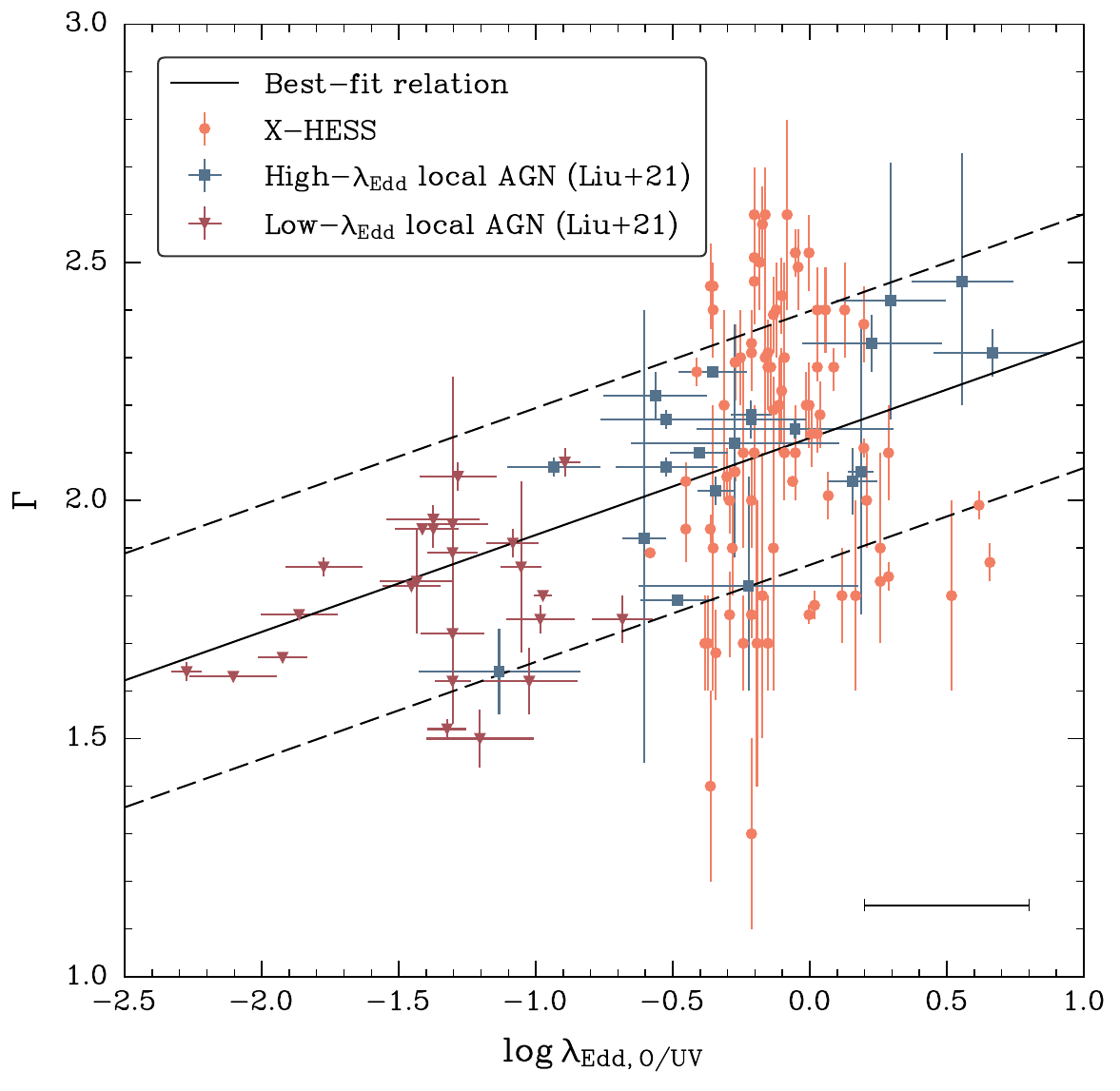}
    \caption{ Distribution of the X-HESS AGNs in the $\Gamma-\log{\lambda_\mathrm{Edd,O/UV}}$ plane. Orange circles represent the X-HESS AGNs with available OM data and, for those disposing of multi-epoch observations, we considered their high-flux state. Low-\ledd\ and high-\ledd\ local AGNs from \citet{liu2021} are shown as well with blue squares and red triangles, respectively. 
    The average error on $\log{\lambda_\mathrm{Edd,O/UV}}$ for the X-HESS AGNs is reported in the lower right corner.} 
    \label{fig:gam_vs_eddUV_all}
\end{figure}

Figure \ref{fig:gam_vs_eddUV_all} shows the distribution of the X-HESS AGNs as well as both the low-\ledd\ and high-\ledd\ local AGNs from \citet{liu2021}, which are mainly characterised by lower black hole masses compared to X-HESS, in the $\Gamma-\log\lambda_\mathrm{Edd,O/UV}$ plane. The correlation between $\Gamma$ and $\lambda_\mathrm{Edd,O/UV}$ is significant, with the Spearman's correlation coefficient being $\rho_\mathrm{S}=0.36$ for a corresponding null probability of $p(>\!|\rho_\mathrm{S}|) \simeq 8.5 \times 10^{-5}$. Thus, taking into account the specific flux state and accretion activity of each AGN at different epochs, we observe that the statistical significance of the relation between the photon index and the Eddington ratio is still present, although it is lower than previously reported in \S\,\ref{sec:gam_edd}, where only single-epoch \ledd\ determinations were considered. This effect could be ascribed to the intrinsic spread of the \citet{duras2020} relation of Eq.\ \ref{eq:kbol,opt}, which according to the authors is around $\sim0.27$ for their sample.

Following the same approach described in \S\,\ref{sec:gam_edd}, we calculated the best-fit relation by adopting a bootstrapping method. In this way, we now obtain
\begin{equation}\label{eq:gam_uvx}
    \Gamma = (0.20 \pm 0.05)\log{\lambda_\mathrm{Edd,O/UV}} + (2.13 \pm 0.03)\,,
\end{equation}

\noindent which is approximately consistent with that in Eq.\ \ref{eq:gam_edd} and has a $1\sigma$ spread of about $\sim0.26$. 
If we considered only the highest flux state for the multi-epoch X-HESS AGN in Fig. \ref{fig:gam_vs_eddUV_all}, we would obtain a similar best-fit relation, whose slope and intercept are $0.21\pm0.04$ and $2.13 \pm 0.03$, respectively. In this case, the $1\sigma$ spread is clearly smaller and amounts to $\sim0.22$.

For completeness, when dealing with the $\Gamma-\lambda_\mathrm{Edd}$ relation, we should also mention that according to recent studies, the size of the AGN broad line region calculated from the radius-luminosity relation could be significantly overestimated, affecting the $M_\mathrm{BH}$ measurements in the same way \citep[e.g.][]{du2019, gravity2024}. Moreover, in the high-\ledd\ regime, provided that the slim disc model represents a reliable description of the inner accretion flow, the optical-based $L_\mathrm{bol}$ measurements may be underestimated \citep[e.g.][]{castello2016} or, at worst, not even represent the actual accretion power due to the photon-trapping effect.

Both of these effects, when combined, might cause an overall shift in the distribution in the $\Gamma-\lambda_\mathrm{Edd}$ plane towards higher \ledd\ values. The best-fit relation could also be affected, though it is hard to quantify explicitly. In any case, the large spread of $\Gamma$ values would remain unchanged.

\section{Soft excess: The $\Gamma-R_\mathrm{S/P}$ relation}\label{sec:Rsp}

As is discussed in \S\,\ref{sec:xray_spec}, the best-fit results of our X-ray spectral analysis of the X-HESS AGNs suggest that their spectrum is characterised by an excess of soft X-ray emission that we observe in approximately $\sim40\%$ of the total number of observations.
This can be attributed to the redshift distribution of the X-HESS AGNs, since $\sim44\%$ of the sample sources is located at $z\geq1$. This implies that the observed $E=0.5{-}2$ keV range corresponds to rest-frame intervals from $E=1$ keV upward, making the detection of the soft excess very difficult in this case. On the other hand, in the rest of the cases we find that $\gtrsim70\%$ of the X-HESS AGN at $z<1$ is characterised by a soft excess.
For all the observations of these sources, we can calculate the parameter $R_\mathrm{S/P} = L^\mathrm{0.5-2\,\mathrm{keV}}_\mathrm{BB}/L^\mathrm{0.5-2\,\mathrm{keV}}_\mathrm{PL}$, which represents a proxy of the relative strength between the luminosity of the blackbody and power-law components (i.e. the soft excess and the primary continuum, respectively) in the $E=0.5\!-\!2$ keV rest-frame energy band. The $R_\mathrm{S/P}$ values of the X-HESS AGN can be found in Table \ref{tab:bestfit}.

Interestingly, we find evidence for a negative correlation between $R_\mathrm{S/P}$ and the photon index, $\Gamma$, when we consider the first source in our X-HESS 1 sample; that is,\ SDSS J094610.71+095226.3, the X-HESS AGN with the largest number of available X-ray observations (15), each showing the presence of an excess of soft X-ray emission.
This anti-correlation is significant according to Pearson, with a correlation coefficient of $r=-0.81$ for a null probability of $p(>|r|)\sim2\times10^{-4}$, while it is only marginally significant according to Spearman, with a correlation coefficient of $\rho_\mathrm{S} = -0.73$ for a null probability of $p(>|\rho_\mathrm{S}|)\sim2\times10^{-3}$.

When considered separately, each of the remaining X-HESS sources with a soft excess in their X-ray spectra dispose of less than half of the total observations of X-HESS 1, preventing us from recovering a similarly significant correlation between the two quantities.
However, when we consider the entire fraction of X-HESS AGNs with an excess of soft X-ray emission —\ $\sim40\%$ of the complete sample discussed above — we find that $\Gamma$ and $R_\mathrm{S/P}$ appear to also be anti-correlated in this case, and the anti-correlation is mildly significant according to both Pearson and Spearman, with a correlation coefficient of $r=-0.45$ and $\rho_\mathrm{S}=-0.39$ for a null probability of $p(>|r|)\sim5\times10^{-4}$ and $p(>|\rho_\mathrm{S}|)\sim 3\times 10^{-3}$, respectively. 
However, the correlation coefficient does not account for the measurement errors, which, in turn, may provide a biased result. To overcome this issue, we adopted a bootstrapping method consisting of resampling the $\Gamma$ and $R_\mathrm{S/P}$ values of the X-HESS AGNs for $N=10000$ times, while accounting for the uncertainty on both quantities (see, e.g.\ \citealt{curran2014}). In this way, we obtain a distribution of $\rho_\mathrm{S}$ values and their corresponding null probabilities, whose median values provide a less biased estimator of the Spearman's rank coefficient and $p(>|\rho_\mathrm{S}|)$. Specifically, we find $\rho_\mathrm{S}=-0.44$ and $p(>|\rho_\mathrm{S}|)\sim 7\times10^{-4}$.

\begin{figure}[t]
    \centering
    \includegraphics[width=\columnwidth]{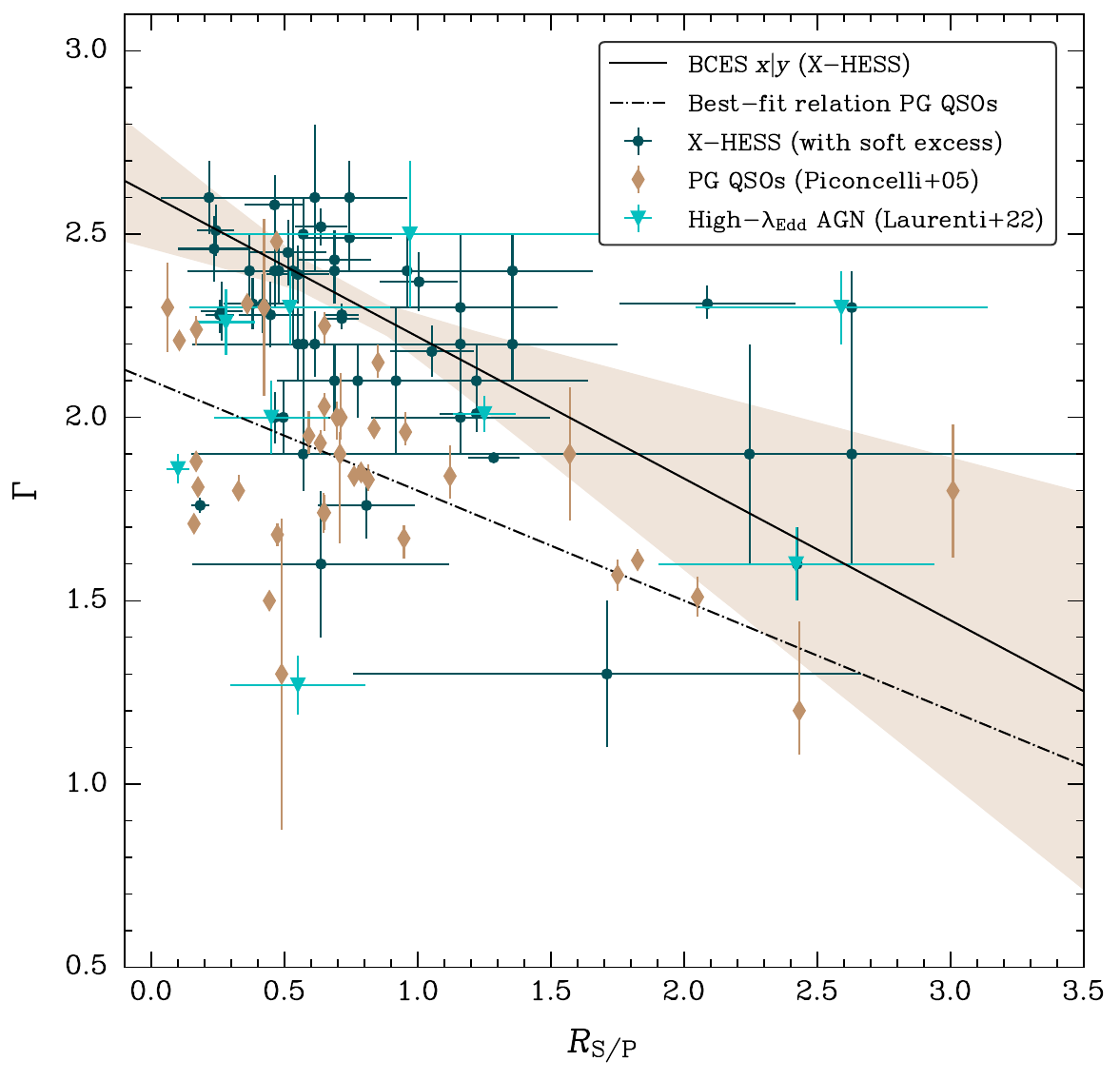}
    \caption{Distribution of the X-HESS AGNs (in green) in the $\Gamma-R_\mathrm{S/P}$ plane zoomed in on the region where $R_\mathrm{S/P}\leq3.5$. The sample of PG QSOs by \citet{piconcelli2005} and the high-\ledd\ AGNs studied by \citet{laurenti2022} are shown in brown and cyan, respectively. The dash-dotted black line indicates the best-fit relation for the PG QSOs. The solid black line describes the BCES $x|y$ best-fit relation derived from the X-HESS AGN, while the shaded area indicates the $90\%$ confidence band.}
    \label{fig:gam_rsp_all}
\end{figure}

As the total number of observations of the X-HESS AGNs with a soft excess is approximately $\sim60$, we tried to determine a best-fit relation as follows. We adopted the BCES linear regression algorithm, since it takes into account the errors in both variables. Within the BCES family of models, we chose BCES $x|y$, which assumes $y$ as the independent variable that, in our case, is represented by the slope of the power-law continuum, $\Gamma$.
The resulting best-fit relation is
\begin{equation}\label{eq:gam_rsp}
    \Gamma = (2.6 \pm 0.2) + (-0.4 \pm 0.2)\, R_\mathrm{S/P}\,.
\end{equation}
Figure \ref{fig:gam_rsp_all} shows the best-fit relation based on the X-HESS AGNs and, for comparison, the samples of Palomar-Green (PG) QSOs by \citet{piconcelli2005} and the high-\ledd\ AGNs analysed in \citet{laurenti2022} are also included. 
The X-ray spectral analysis of X-HESS can be consistently compared with those of the other two samples, as they all share the same approach. By combining the measurements of high-\ledd\ AGNs from both \citet{laurenti2022} and X-HESS, we obtain a similar relation; that is,
\begin{equation}
    \Gamma_{\mathrm{High-}\lambda_\mathrm{Edd}} = (2.53 \pm 0.09) + (-0.3 \pm 0.1)\, R_{\mathrm{S/P, High-}\lambda_\mathrm{Edd}}\,.
\end{equation}

\noindent Moreover, the previous results are in partial agreement, within the uncertainties, with what we find for the sample of $\sim\!40$ PG QSOs of \citet{piconcelli2005}. Indeed, by adopting a simple weighted least squares model to fit the data points, we obtain
\begin{equation}\label{eq:gam_rsp}
    \Gamma_\mathrm{PG} = (2.1 \pm 0.1) + (-0.3 \pm 0.1)\, R_\mathrm{S/P,\,PG}\,.
\end{equation}
However, the above relation is only marginally significant according to Pearson, with a correlation coefficient of $r=-0.47$ for a null probability of $p(>|r|)\sim4\times10^{-3}$. 

When we consider the three samples of Fig.\ \ref{fig:gam_rsp_all} altogether, we get additional evidence for the existence of a significant negative correlation between $\Gamma$ and $R_\mathrm{S/P}$, the correlation coefficients being $r=-0.49$ and $\rho_\mathrm{S}=-0.39$ for a null probability of $p(>|r|)\sim4\times10^{-5}$ and $p(>|\rho_\mathrm{S}|)\sim 10^{-3}$ according to Pearson and Spearman, respectively. Although the correlation is only mildly significant, it provides at least an indication of the possible existence of a relation between the two quantities.

Interestingly, this result appears to move in the opposite direction compared to previous findings, which suggested a putative positive correlation between the photon index and the strength of the soft excess \citep[e.g.][]{bianchi2009,boissay2016,gliozzi2020}. However, except for \citet{bianchi2009} and one of the three indicators in \citet{gliozzi2020}, the reported correlations are only marginally significant.
Although the different results may be partially affected by different models, it would be tempting to associate this trend with the high-\ledd\ nature of our AGNs, since all of the studies mentioned above contained a limited number of highly accreting AGNs.
Despite the physical origin of the soft excess being still debated, three main competing scenarios are generally hypothesised nowadays.
In the first, the soft excess arises from the enhancement of reflection from the inner regions of the accretion disc due to light-bending effects, together with a strong suppression of the primary emission \citep[e.g.][]{miniutti2004}. 
The second scenario is based on the assumption that
the soft excess can also be mimicked by absorption from a relativistically outflowing warm gas \citep[e.g.][]{gierlinski2004}.
Finally, according to the third scenario, the observed excess of soft X-ray emission may be a signature of the high-energy tail of thermal Comptonisation in a warm ($kT\sim1$ keV), optically thick ($\tau\sim10{-}20$) corona \citep[e.g.][and references therein]{petrucci2018}.

If the soft excess is due to reflection from an ionised disc, a negative correlation is expected between the photon index and the strength of the soft excess \citep[][]{boissay2016}, in agreement with the indication that we get for the X-HESS AGNs.
However, the modest S/N of some of the X-HESS observations does not allow us to constrain the properties of a putative reflection component to be included in our best-fit model. 

On the other hand, we do not find any clear evidence of reflection features even in the higher-quality X-HESS spectra. Moreover, for the X-HESS AGNs, we obtain that the optical/UV luminosity at $2500\,\AA$ and the one in the soft X-rays are highly correlated ($\rho_\mathrm{S} = 0.84$, $p(>|\rho_\mathrm{S}|) < 10^{-20}$). This would move in favour of a warm Comptonisation scenario, in which a strong link is expected between optical/UV and soft X-rays \citep[e.g.][]{mehdipour2011}, as the warm corona would lie above the surface of the inner accretion disc \citep[][]{petrucci2020}. In this case, however, we would have likely obtained a positive correlation between $\Gamma$ and the soft excess strength, $R_\mathrm{S/P}$, at least if the warm corona efficiently provides the hot corona with the bulk of soft X-ray seed photons \citep[][]{boissay2016}. All this makes a really challenging task to shed definitive light on the origin of the $\Gamma-R_\mathrm{S/P}$ relation. For the time being, our limited understanding of the interplay between the warm and hot coronae, as well as possibly different configurations and physical processes occurring in the warm corona, do not allow us to categorically rule out that the distribution in Fig.\ \ref{fig:gam_rsp_all} could be also reproduced in this scenario.

In any case, higher-quality dedicated observations, especially extending towards the hard ($E>10$ keV) X-rays, where one can find the typical spectral imprints due to reflection, are mandatory to quantify the hypothetical impact of ionised reflection in the X-HESS sample.

\section{Summary and conclusions}\label{sec:summary&conclusion}

In this paper, we have analysed a large new sample of high-\ledd\ AGNs named the \emph{XMM-Newton} High-Eddington Serendipitous AGN Sample, or X-HESS, obtained by cross-correlating the 11th release of the \emph{XMM-Newton} serendipitous catalogue and the catalogue of spectral properties of QSOs from \citet{rakshit2020}, according to the procedure described in \S\,\ref{sec:xhess_sample}. 
X-HESS contains 142 observations of 61 AGNs, and 22 of these highly accreting AGNs ($\sim36\%$ of the sample sources) have been observed at different epochs by \emph{XMM-Newton} for a total of 103 times. The X-HESS AGNs are distributed over wide intervals of bolometric luminosity ($44.7<\log(L_\mathrm{bol}/\mathrm{erg\,s}^{-1}) <48.3$), black hole mass ($6.8<\log({M_\mathrm{BH}/M_\odot})<9.8$), redshift ($0.06<z<3.3$), and the Eddington ratio ($-0.2< \log\lambda_\mathrm{Edd}<0.5$). 
Furthermore, a sizeable fraction of the X-HESS sources ($\sim62\%$) can rely on simultaneous OM observations obtained by cross-matching X-HESS with the latest release of the XMM-OM serendipitous catalogue. The main results of the analysis of the X-HESS sample can be
summarised as follows.

\begin{itemize}

    \item The multi-epoch data coverage of a large fraction of the X-HESS AGNs allows us to investigate their spectral variability across the different observations in \S\,\ref{sec:variability}. Specifically, we considered the observations associated with their lowest- and highest-flux states, respectively. The latter, especially, are also particularly interesting for their implications in the often debated $\Gamma-\log{\lambda_\mathrm{Edd}}$ relation.
    We do find that eight out of the 22 multi-epoch AGNs in X-HESS present significant spectral variations at the $1\sigma$ confidence level. The bulk of these AGNs appear to vary accordingly to a softer-when-brighter trend, whereas a single AGN follows the opposite trend. Moreover, such spectral variations do not exhibit a clear correlation with the time elapsed between the two observations of interest, which ranges from a few hours to $\sim15$ years (see Fig.\ \ref{fig:delta_gamma}).

    \item A couple of X-HESS AGNs — namely, X-HESS 5 ($z=1.929$) and 16 ($z=0.307$) — have experienced at least one transition between phases of standard-to-weak X-ray emission (see Fig.\ \ref{fig:aox_Luv_xhess}). The observed transitions occur over very different timescales in the source's rest-frame; approximately a week for the former and one and a half years for the latter. In both cases, we observe a drop in the soft and hard X-ray emission when moving from high- to low-flux states, leading to an overall hardening of the X-ray continuum. However, when accounting for the measurement errors, the corresponding spectral changes are only moderately or not at all significant for X-HESS 16 and 5, respectively. Moreover, their X-ray weak states do not appear to be due to absorption from intervening cold gas clouds.

    \item We do not find a significant correlation between the X-ray photon index, $\Gamma$, and \ledd\ for the X-HESS AGN, as is discussed in \S\,\ref{sec:gam_edd}. Not only does this result hold when accounting for (or averaging) all the flux states of the multi-epoch X-HESS AGNs, but even when we considered, where available, the sole highest-flux state of each of these AGNs as the most representative of their X-ray coronal emission, in case some hypothetical degree of obscuration is present, though mostly undetected in our case, possibly due to limited S/N observations. This approach is shared by other authors, such as \citet{liu2021}, who studied a sample of 47 local AGNs located at $z\lesssim0.3$ and divided into two groups consisting of 21 high-\ledd\ and 26 low-\ledd\ sources with $M_\mathrm{BH}$ measurements from reverberation mapping. These AGNs are mainly characterised by lower black hole masses compared to X-HESS. If we consider the X-HESS and high-\ledd\ local AGN from \citet{liu2021}, we still cannot find any significant correlation between $\Gamma$ and $\lambda_\mathrm{Edd}$ in a relatively broad interval of Eddington ratios spanning from $\sim-1$ up to $\gtrsim0.7$. On the other hand, when we consider X-HESS and both the local \citet{liu2021} AGN samples, we obtain a highly significant correlation between the two quantities. In any case, the distribution in the $\Gamma-\log{\lambda_\mathrm{Edd}}$ plane is affected by a very large spread.

    \item We have investigated the possible effects of the black hole mass and bolometric luminosity in the observed $\Gamma-\log{\lambda_\mathrm{Edd}}$ plane and found that the correlation between the photon index and the Eddington ratio is much more significant for those sources with either lower $L_\mathrm{bol}$ or $M_\mathrm{BH}$, while it basically becomes insignificant for AGN with a higher black hole mass and/or bolometric luminosity. Results from partial correlation analysis in \S\,\ref{sec:gam_edd_mbh_lbol} appear to confirm this trend and possibly suggest that such a different behaviour in the $\Gamma-\log{\lambda_\mathrm{Edd}}$ plane could likely reflect intrinsic differences between these two classes of objects, which approximately resemble the properties of local Seyfert-like galaxies and moderate-to-high-redshift QSOs, respectively.

    \item We find that the use of simultaneous optical/UV observations, such as those provided by \emph{XMM-Newton}, can offer the unprecedented possibility of obtaining measurements of the Eddington ratio, $\lambda_\mathrm{Edd,O/UV}$, at different epochs by adopting the bolometric correction in \citet{duras2020}, as is described in \S\,\ref{sec:ep_dep}. This is particularly important because it allows us to take a snapshot of the actual flux state of the given AGNs, which may differ from that associated with the single-epoch spectrum used to determine $\lambda_\mathrm{Edd}$. 
    Thanks to this approach, we re-projected the $\Gamma-\log{\lambda_\mathrm{Edd}}$ plane into its fully epoch-dependent variant, where both $\Gamma$ and $\lambda_\mathrm{Edd,O/UV}$ represent a proxy of the ongoing AGN activity (see Fig.\ \ref{fig:gam_vs_eddUV_all}). By considering the X-HESS AGNs and both the local AGN samples from \citet{liu2021}, we find that $\Gamma$ and $\lambda_\mathrm{Edd,O/UV}$ are statistically correlated and share an approximately similar spread around the best-fit relation compared to what previously found for the `standard' single-epoch \ledd\ estimates.
    In both cases, due to the large scatter in the corresponding planes, 
    we warmly recommend caution when using this relation to estimate the black hole mass. However, ideally, the adoption of a fully epoch-dependent approach would likely represent the fairest description of the actual distribution in the $\Gamma-\lambda_\mathrm{Edd}$ plane.
    Furthermore, this result shows the crucial importance for the X-ray observatories of collecting simultaneous optical/UV data, as in the case of \emph{XMM-Newton} or \emph{Swift}, to assess the time-dependent accretion properties of AGNs.

\item Approximately $\sim40\%$ of the total observations of the X-HESS AGNs, corresponding to $\gtrsim70\%$ of all the AGNs at $z<1$ for which the rest-frame $E=0.5{-}2$ keV interval is broadly visible, indicates the presence of an excess of soft X-ray emission that we modelled with a blackbody component to provide a phenomenological description. For all these observations, we calculated the parameter $R_\mathrm{S/P} = L^\mathrm{0.5-2\,\mathrm{keV}}_\mathrm{BB}/L^\mathrm{0.5-2\,\mathrm{keV}}_\mathrm{PL}$, which quantifies the relative strength of the soft excess with respect to the underlying power-law continuum in the $E=0.5{-}2$ keV observer-frame energy band. We get a mild indication of a negative correlation between $R_\mathrm{S/P}$ and the photon index, $\Gamma$, when we consider the AGN with the largest number of observations —\ X-HESS 1 — as well as when we consider the whole subsample of X-HESS AGNs characterised by a soft excess (see \S\,\ref{sec:Rsp}). This result still holds if we consider the X-HESS sources and the AGNs in the \citet{piconcelli2005} and \citet{laurenti2022} samples altogether. Interestingly, according to \citet{boissay2016}, a similar anti-correlation is expected if the soft excess is due to reflection from an ionised disc.

\end{itemize}

\noindent In the near future, we plan to update X-HESS with more recent releases of the \emph{XMM-Newton} serendipitous catalogue, which has recently reached its thirteenth version (4XMM-DR13\footnote{\url{http://xmmssc.irap.omp.eu/Catalogue/4XMM-DR13/4XMM_DR13.html}.}), to increase the number of high-\ledd\ AGNs in the X-HESS sample and their corresponding observations.
To this extent, we also note that, given the relatively bright $E=0.5{-}2$ keV fluxes of the X-HESS AGNs, we can expect that the vast majority of our highly accreting AGNs will be detected by eROSITA.   

Finally, in order to test any physical scenario to explain the observed properties of high-\ledd\ AGNs, it would be necessary to expand their study by adopting a multi-wavelength treatment. This is particularly feasible in the optical/UV band, where a large number of facilities, such as ZTF and Pan-STARRS, are continuously collecting photometric data in multiple wavelengths.
Specifically, a major leap in this quest will be provided by the forthcoming \textit{Rubin} Observatory Legacy Survey of Space and Time (LSST; \citealt{ivezic2019}), which is expected to be fully operational in 2024.
The unprecedented quality of LSST photometry will hopefully allow us, for example, to probe whether the slim disc, likely originating in highly accreting AGNs, presents peculiar optical/UV emission properties with respect to AGNs accreting in the low-to-moderate regime, and compare the X-ray and optical/UV characteristic flux variability to test different competing scenarios envisaging the presence (or lack) of correlated variations between the two bands.

\begin{acknowledgements} 

We thank the anonymous referee for the useful comments that helped us improve our manuscript. This work is based on observations obtained with \emph{XMM-Newton}, an ESA science mission with instruments and contributions directly funded by ESA member states and the USA (NASA). ML is thankful to ESA, as part of this work has been carried out at ESA-ESTEC in the context of the Archival Research Visitor Programme 2021. ML, FT and EP acknowledge funding from the European Union - Next Generation EU, PRIN/MUR 2022 (2022K9N5B4). EP acknowledges support from PRIN MIUR project "Black Hole winds and the Baryon Life Cycle of Galaxies:
the stone-guest at the galaxy evolution supper", contract no.\ 2017PH3WAT.
EP and LZ acknowledge financial support from the Bando Ricerca Fondamentale INAF 2022 Large Grant ``Toward an holistic view of the Titans: multi-band observations of $z>6$ QSOs powered by greedy supermassive black-holes''. 

\end{acknowledgements}

\bibliographystyle{aa} 
\bibliography{xhess.bib}

\begin{thebibliography}{131}
\expandafter\ifx\csname natexlab\endcsname\relax\def\natexlab#1{#1}\fi

\bibitem[{{Abramowicz} {et~al.}(1988){Abramowicz}, {Czerny}, {Lasota}, \&
  {Szuszkiewicz}}]{abramowicz1988}
{Abramowicz}, M.~A., {Czerny}, B., {Lasota}, J.~P., \& {Szuszkiewicz}, E. 1988,
  \apj, 332, 646

\bibitem[{{Ai} {et~al.}(2011){Ai}, {Yuan}, {Zhou}, {Wang}, \& {Zhang}}]{ai2011}
{Ai}, Y.~L., {Yuan}, W., {Zhou}, H.~Y., {Wang}, T.~G., \& {Zhang}, S.~H. 2011,
  \apj, 727, 31

\bibitem[{{Arnaud}(1996)}]{arnaud1996}
{Arnaud}, K.~A. 1996, in Astronomical Society of the Pacific Conference Series,
  Vol. 101, Astronomical Data Analysis Software and Systems V, ed. G.~H.
  {Jacoby} \& J.~{Barnes}, 17

\bibitem[{{Avni}(1976)}]{avni1976}
{Avni}, Y. 1976, \apj, 210, 642

\bibitem[{{Ba{\~n}ados} {et~al.}(2016){Ba{\~n}ados}, {Venemans}, {Decarli},
  {Farina}, {Mazzucchelli}, {Walter}, {Fan}, {Stern}, {Schlafly}, {Chambers},
  {Rix}, {Jiang}, {McGreer}, {Simcoe}, {Wang}, {Yang}, {Morganson}, {De Rosa},
  {Greiner}, {Balokovi{\'c}}, {Burgett}, {Cooper}, {Draper}, {Flewelling},
  {Hodapp}, {Jun}, {Kaiser}, {Kudritzki}, {Magnier}, {Metcalfe}, {Miller},
  {Schindler}, {Tonry}, {Wainscoat}, {Waters}, \& {Yang}}]{banados2016}
{Ba{\~n}ados}, E., {Venemans}, B.~P., {Decarli}, R., {et~al.} 2016, \apjs, 227,
  11

\bibitem[{{Baskin} \& {Laor}(2005)}]{baskin2005}
{Baskin}, A. \& {Laor}, A. 2005, \mnras, 356, 1029

\bibitem[{{Begelman} {et~al.}(2006){Begelman}, {Volonteri}, \&
  {Rees}}]{begelman2006}
{Begelman}, M.~C., {Volonteri}, M., \& {Rees}, M.~J. 2006, \mnras, 370, 289

\bibitem[{{Bellm} {et~al.}(2019){Bellm}, {Kulkarni}, {Graham}, {Dekany},
  {Smith}, {Riddle}, {Masci}, {Helou}, {Prince}, {Adams}, {Barbarino},
  {Barlow}, {Bauer}, {Beck}, {Belicki}, {Biswas}, {Blagorodnova}, {Bodewits},
  {Bolin}, {Brinnel}, {Brooke}, {Bue}, {Bulla}, {Burruss}, {Cenko}, {Chang},
  {Connolly}, {Coughlin}, {Cromer}, {Cunningham}, {De}, {Delacroix}, {Desai},
  {Duev}, {Eadie}, {Farnham}, {Feeney}, {Feindt}, {Flynn}, {Franckowiak},
  {Frederick}, {Fremling}, {Gal-Yam}, {Gezari}, {Giomi}, {Goldstein},
  {Golkhou}, {Goobar}, {Groom}, {Hacopians}, {Hale}, {Henning}, {Ho}, {Hover},
  {Howell}, {Hung}, {Huppenkothen}, {Imel}, {Ip}, {Ivezi{\'c}}, {Jackson},
  {Jones}, {Juric}, {Kasliwal}, {Kaspi}, {Kaye}, {Kelley}, {Kowalski},
  {Kramer}, {Kupfer}, {Landry}, {Laher}, {Lee}, {Lin}, {Lin}, {Lunnan},
  {Giomi}, {Mahabal}, {Mao}, {Miller}, {Monkewitz}, {Murphy}, {Ngeow},
  {Nordin}, {Nugent}, {Ofek}, {Patterson}, {Penprase}, {Porter}, {Rauch},
  {Rebbapragada}, {Reiley}, {Rigault}, {Rodriguez}, {van Roestel}, {Rusholme},
  {van Santen}, {Schulze}, {Shupe}, {Singer}, {Soumagnac}, {Stein}, {Surace},
  {Sollerman}, {Szkody}, {Taddia}, {Terek}, {Van Sistine}, {van Velzen},
  {Vestrand}, {Walters}, {Ward}, {Ye}, {Yu}, {Yan}, \& {Zolkower}}]{bellm2019}
{Bellm}, E.~C., {Kulkarni}, S.~R., {Graham}, M.~J., {et~al.} 2019, \pasp, 131,
  018002

\bibitem[{{Bianchi} {et~al.}(2009){Bianchi}, {Guainazzi}, {Matt}, {Fonseca
  Bonilla}, \& {Ponti}}]{bianchi2009}
{Bianchi}, S., {Guainazzi}, M., {Matt}, G., {Fonseca Bonilla}, N., \& {Ponti},
  G. 2009, \aap, 495, 421

\bibitem[{{Bischetti} {et~al.}(2019){Bischetti}, {Piconcelli}, {Feruglio},
  {Fiore}, {Carniani}, {Brusa}, {Cicone}, {Vignali}, {Bongiorno}, {Cresci},
  {Mainieri}, {Maiolino}, {Marconi}, {Nardini}, \&
  {Zappacosta}}]{bischetti2019}
{Bischetti}, M., {Piconcelli}, E., {Feruglio}, C., {et~al.} 2019, \aap, 628,
  A118

\bibitem[{{Bischetti} {et~al.}(2017){Bischetti}, {Piconcelli}, {Vietri},
  {Bongiorno}, {Fiore}, {Sani}, {Marconi}, {Duras}, {Zappacosta}, {Brusa},
  {Comastri}, {Cresci}, {Feruglio}, {Giallongo}, {La Franca}, {Mainieri},
  {Mannucci}, {Martocchia}, {Ricci}, {Schneider}, {Testa}, \&
  {Vignali}}]{bischetti2017}
{Bischetti}, M., {Piconcelli}, E., {Vietri}, G., {et~al.} 2017, \aap, 598, A122

\bibitem[{{Boissay} {et~al.}(2016){Boissay}, {Ricci}, \&
  {Paltani}}]{boissay2016}
{Boissay}, R., {Ricci}, C., \& {Paltani}, S. 2016, \aap, 588, A70

\bibitem[{{Bongiorno} {et~al.}(2007){Bongiorno}, {Zamorani}, {Gavignaud},
  {Marano}, {Paltani}, {Mathez}, {M{\o}ller}, {Picat}, {Cirasuolo},
  {Lamareille}, {Bottini}, {Garilli}, {Le Brun}, {Le F{\`e}vre}, {Maccagni},
  {Scaramella}, {Scodeggio}, {Tresse}, {Vettolani}, {Zanichelli}, {Adami},
  {Arnouts}, {Bardelli}, {Bolzonella}, {Cappi}, {Charlot}, {Ciliegi},
  {Contini}, {Foucaud}, {Franzetti}, {Guzzo}, {Ilbert}, {Iovino}, {McCracken},
  {Marinoni}, {Mazure}, {Meneux}, {Merighi}, {Pell{\`o}}, {Pollo}, {Pozzetti},
  {Radovich}, {Zucca}, {Hatziminaoglou}, {Polletta}, {Bondi}, {Brinchmann},
  {Cucciati}, {de la Torre}, {Gregorini}, {Mellier}, {Merluzzi}, {Temporin},
  {Vergani}, \& {Walcher}}]{bongiorno2007}
{Bongiorno}, A., {Zamorani}, G., {Gavignaud}, I., {et~al.} 2007, \aap, 472, 443

\bibitem[{{Brandt} {et~al.}(1997){Brandt}, {Mathur}, \& {Elvis}}]{brandt1997}
{Brandt}, W.~N., {Mathur}, S., \& {Elvis}, M. 1997, \mnras, 285, L25

\bibitem[{{Brightman} {et~al.}(2013){Brightman}, {Silverman}, {Mainieri},
  {Ueda}, {Schramm}, {Matsuoka}, {Nagao}, {Steinhardt}, {Kartaltepe},
  {Sanders}, {Treister}, {Shemmer}, {Brandt}, {Brusa}, {Comastri}, {Ho},
  {Lanzuisi}, {Lusso}, {Nandra}, {Salvato}, {Zamorani}, {Akiyama}, {Alexander},
  {Bongiorno}, {Capak}, {Civano}, {Del Moro}, {Doi}, {Elvis}, {Hasinger},
  {Laird}, {Masters}, {Mignoli}, {Ohta}, {Schawinski}, \&
  {Taniguchi}}]{brightman2013}
{Brightman}, M., {Silverman}, J.~D., {Mainieri}, V., {et~al.} 2013, \mnras,
  433, 2485

\bibitem[{{Cackett} {et~al.}(2020){Cackett}, {Gelbord}, {Li}, {Horne}, {Wang},
  {Barth}, {Bai}, {Bian}, {Carroll}, {Du}, {Edelson}, {Goad}, {Ho}, {Hu},
  {Khatu}, {Luo}, {Miller}, \& {Yuan}}]{cackett2020}
{Cackett}, E.~M., {Gelbord}, J., {Li}, Y.-R., {et~al.} 2020, \apj, 896, 1

\bibitem[{{Cash}(1979)}]{cash1979}
{Cash}, W. 1979, \apj, 228, 939

\bibitem[{{Castell{\'o}-Mor} {et~al.}(2017){Castell{\'o}-Mor}, {Kaspi},
  {Netzer}, {Du}, {Hu}, {Ho}, {Bai}, {Bian}, {Yuan}, \& {Wang}}]{castello2017}
{Castell{\'o}-Mor}, N., {Kaspi}, S., {Netzer}, H., {et~al.} 2017, \mnras, 467,
  1209

\bibitem[{{Castell{\'o}-Mor} {et~al.}(2016){Castell{\'o}-Mor}, {Netzer}, \&
  {Kaspi}}]{castello2016}
{Castell{\'o}-Mor}, N., {Netzer}, H., \& {Kaspi}, S. 2016, \mnras, 458, 1839

\bibitem[{{Chen} \& {Wang}(2004)}]{chen2004}
{Chen}, L.-H. \& {Wang}, J.-M. 2004, \apj, 614, 101

\bibitem[{{Coatman} {et~al.}(2017){Coatman}, {Hewett}, {Banerji}, {Richards},
  {Hennawi}, \& {Prochaska}}]{coatman2017}
{Coatman}, L., {Hewett}, P.~C., {Banerji}, M., {et~al.} 2017, \mnras, 465, 2120

\bibitem[{{Costantini} {et~al.}(2007){Costantini}, {Gallo}, {Brandt}, {Fabian},
  \& {Boller}}]{costantini2007}
{Costantini}, E., {Gallo}, L.~C., {Brandt}, W.~N., {Fabian}, A.~C., \&
  {Boller}, T. 2007, \mnras, 378, 873

\bibitem[{{Crummy} {et~al.}(2005){Crummy}, {Fabian}, {Brandt}, \&
  {Boller}}]{crummy2005}
{Crummy}, J., {Fabian}, A.~C., {Brandt}, W.~N., \& {Boller}, T. 2005, \mnras,
  361, 1197

\bibitem[{{Curran}(2014)}]{curran2014}
{Curran}, P.~A. 2014, arXiv e-prints, arXiv:1411.3816

\bibitem[{{Dasgupta} {et~al.}(2005){Dasgupta}, {Rao}, {Dewangan}, \&
  {Agrawal}}]{dasgupta2005}
{Dasgupta}, S., {Rao}, A.~R., {Dewangan}, G.~C., \& {Agrawal}, V.~K. 2005,
  \apjl, 618, L87

\bibitem[{{Donnan} {et~al.}(2023){Donnan}, {Hern{\'a}ndez Santisteban},
  {Horne}, {Hu}, {Du}, {Li}, {Xiao}, {Ho}, {Aceituno}, {Wang}, {Guo}, {Yang},
  {Jiang}, \& {Yao}}]{donnan2023}
{Donnan}, F.~R., {Hern{\'a}ndez Santisteban}, J.~V., {Horne}, K., {et~al.}
  2023, \mnras, 523, 545

\bibitem[{{Du} \& {Wang}(2019)}]{du2019}
{Du}, P. \& {Wang}, J.-M. 2019, \apj, 886, 42

\bibitem[{{Duras} {et~al.}(2020){Duras}, {Bongiorno}, {Ricci}, {Piconcelli},
  {Shankar}, {Lusso}, {Bianchi}, {Fiore}, {Maiolino}, {Marconi}, {Onori},
  {Sani}, {Schneider}, {Vignali}, \& {La Franca}}]{duras2020}
{Duras}, F., {Bongiorno}, A., {Ricci}, F., {et~al.} 2020, \aap, 636, A73

\bibitem[{{Fabian} {et~al.}(2013){Fabian}, {Kara}, {Walton}, {Wilkins}, {Ross},
  {Lozanov}, {Uttley}, {Gallo}, {Zoghbi}, {Miniutti}, {Boller}, {Brandt},
  {Cackett}, {Chiang}, {Dwelly}, {Malzac}, {Miller}, {Nardini}, {Ponti},
  {Reis}, {Reynolds}, {Steiner}, {Tanaka}, \& {Young}}]{fabian2013}
{Fabian}, A.~C., {Kara}, E., {Walton}, D.~J., {et~al.} 2013, \mnras, 429, 2917

\bibitem[{{Ferrarese} \& {Merritt}(2000)}]{ferrarese2000}
{Ferrarese}, L. \& {Merritt}, D. 2000, \apjl, 539, L9

\bibitem[{{Fitzpatrick}(1999)}]{fitzpatrick1999}
{Fitzpatrick}, E.~L. 1999, \pasp, 111, 63

\bibitem[{{Fukumura} {et~al.}(2016){Fukumura}, {Hendry}, {Clark}, {Tombesi}, \&
  {Takahashi}}]{fukumura2016}
{Fukumura}, K., {Hendry}, D., {Clark}, P., {Tombesi}, F., \& {Takahashi}, M.
  2016, \apj, 827, 31

\bibitem[{{Gallerani} {et~al.}(2010){Gallerani}, {Maiolino}, {Juarez}, {Nagao},
  {Marconi}, {Bianchi}, {Schneider}, {Mannucci}, {Oliva}, {Willott}, {Jiang},
  \& {Fan}}]{gallerani2010}
{Gallerani}, S., {Maiolino}, R., {Juarez}, Y., {et~al.} 2010, \aap, 523, A85

\bibitem[{{Gallo}(2006)}]{gallo2006}
{Gallo}, L.~C. 2006, \mnras, 368, 479

\bibitem[{{Gibson} \& {Brandt}(2012)}]{gibson2012}
{Gibson}, R.~R. \& {Brandt}, W.~N. 2012, \apj, 746, 54

\bibitem[{{Gierli{\'n}ski} \& {Done}(2004)}]{gierlinski2004}
{Gierli{\'n}ski}, M. \& {Done}, C. 2004, \mnras, 349, L7

\bibitem[{{Gliozzi} \& {Williams}(2020)}]{gliozzi2020}
{Gliozzi}, M. \& {Williams}, J.~K. 2020, \mnras, 491, 532

\bibitem[{{GRAVITY Collaboration} {et~al.}(2024){GRAVITY Collaboration},
  {Amorim}, {Bourdarot}, {Brandner}, {Cao}, {Cl{\'e}net}, {Davies}, {de Zeeuw},
  {Dexter}, {Drescher}, {Eckart}, {Eisenhauer}, {Fabricius}, {Feuchtgruber},
  {F{\"o}rster Schreiber}, {Garcia}, {Genzel}, {Gillessen}, {Gratadour},
  {H{\"o}nig}, {Kishimoto}, {Lacour}, {Lutz}, {Millour}, {Netzer}, {Ott},
  {Paumard}, {Perraut}, {Perrin}, {Peterson}, {Petrucci}, {Pfuhl}, {Prieto},
  {Rabien}, {Rouan}, {Santos}, {Shangguan}, {Shimizu}, {Sternberg},
  {Straubmeier}, {Sturm}, {Tacconi}, {Tristram}, {Widmann}, \&
  {Woillez}}]{gravity2024}
{GRAVITY Collaboration}, {Amorim}, A., {Bourdarot}, G., {et~al.} 2024, \aap,
  684, A167

\bibitem[{{HI4PI Collaboration} {et~al.}(2016){HI4PI Collaboration}, {Ben
  Bekhti}, {Fl{\"o}er}, {Keller}, {Kerp}, {Lenz}, {Winkel}, {Bailin},
  {Calabretta}, {Dedes}, {Ford}, {Gibson}, {Haud}, {Janowiecki}, {Kalberla},
  {Lockman}, {McClure-Griffiths}, {Murphy}, {Nakanishi}, {Pisano}, \&
  {Staveley-Smith}}]{bekhti2016}
{HI4PI Collaboration}, {Ben Bekhti}, N., {Fl{\"o}er}, L., {et~al.} 2016, \aap,
  594, A116

\bibitem[{{Hopkins} {et~al.}(2004){Hopkins}, {Strauss}, {Hall}, {Richards},
  {Cooper}, {Schneider}, {Vanden Berk}, {Jester}, {Brinkmann}, \&
  {Szokoly}}]{hopkins2004}
{Hopkins}, P.~F., {Strauss}, M.~A., {Hall}, P.~B., {et~al.} 2004, \aj, 128,
  1112

\bibitem[{{Huang} {et~al.}(2020){Huang}, {Luo}, {Du}, {Hu}, {Wang}, \&
  {Li}}]{huang2020}
{Huang}, J., {Luo}, B., {Du}, P., {et~al.} 2020, \apj, 895, 114

\bibitem[{{Ighina} {et~al.}(2021){Ighina}, {Belladitta}, {Caccianiga},
  {Broderick}, {Drouart}, {Moretti}, \& {Seymour}}]{ighina2021}
{Ighina}, L., {Belladitta}, S., {Caccianiga}, A., {et~al.} 2021, \aap, 647, L11

\bibitem[{{Inayoshi} {et~al.}(2016){Inayoshi}, {Haiman}, \&
  {Ostriker}}]{inayoshi2016}
{Inayoshi}, K., {Haiman}, Z., \& {Ostriker}, J.~P. 2016, \mnras, 459, 3738

\bibitem[{{Inoue} {et~al.}(2007){Inoue}, {Terashima}, \& {Ho}}]{inoue2007}
{Inoue}, H., {Terashima}, Y., \& {Ho}, L.~C. 2007, \apj, 662, 860

\bibitem[{{Ivezi{\'c}} {et~al.}(2019){Ivezi{\'c}}, {Kahn}, {Tyson}, {Abel},
  {Acosta}, {Allsman}, {Alonso}, {AlSayyad}, {Anderson}, {Andrew}, {Angel},
  {Angeli}, {Ansari}, {Antilogus}, {Araujo}, {Armstrong}, {Arndt}, {Astier},
  {Aubourg}, {Auza}, {Axelrod}, {Bard}, {Barr}, {Barrau}, {Bartlett}, {Bauer},
  {Bauman}, {Baumont}, {Bechtol}, {Bechtol}, {Becker}, {Becla}, {Beldica},
  {Bellavia}, {Bianco}, {Biswas}, {Blanc}, {Blazek}, {Blandford}, {Bloom},
  {Bogart}, {Bond}, {Booth}, {Borgland}, {Borne}, {Bosch}, {Boutigny},
  {Brackett}, {Bradshaw}, {Brandt}, {Brown}, {Bullock}, {Burchat}, {Burke},
  {Cagnoli}, {Calabrese}, {Callahan}, {Callen}, {Carlin}, {Carlson},
  {Chandrasekharan}, {Charles-Emerson}, {Chesley}, {Cheu}, {Chiang}, {Chiang},
  {Chirino}, {Chow}, {Ciardi}, {Claver}, {Cohen-Tanugi}, {Cockrum}, {Coles},
  {Connolly}, {Cook}, {Cooray}, {Covey}, {Cribbs}, {Cui}, {Cutri}, {Daly},
  {Daniel}, {Daruich}, {Daubard}, {Daues}, {Dawson}, {Delgado}, {Dellapenna},
  {de Peyster}, {de Val-Borro}, {Digel}, {Doherty}, {Dubois},
  {Dubois-Felsmann}, {Durech}, {Economou}, {Eifler}, {Eracleous}, {Emmons},
  {Fausti Neto}, {Ferguson}, {Figueroa}, {Fisher-Levine}, {Focke}, {Foss},
  {Frank}, {Freemon}, {Gangler}, {Gawiser}, {Geary}, {Gee}, {Geha}, {Gessner},
  {Gibson}, {Gilmore}, {Glanzman}, {Glick}, {Goldina}, {Goldstein}, {Goodenow},
  {Graham}, {Gressler}, {Gris}, {Guy}, {Guyonnet}, {Haller}, {Harris},
  {Hascall}, {Haupt}, {Hernandez}, {Herrmann}, {Hileman}, {Hoblitt}, {Hodgson},
  {Hogan}, {Howard}, {Huang}, {Huffer}, {Ingraham}, {Innes}, {Jacoby}, {Jain},
  {Jammes}, {Jee}, {Jenness}, {Jernigan}, {Jevremovi{\'c}}, {Johns}, {Johnson},
  {Johnson}, {Jones}, {Juramy-Gilles}, {Juri{\'c}}, {Kalirai}, {Kallivayalil},
  {Kalmbach}, {Kantor}, {Karst}, {Kasliwal}, {Kelly}, {Kessler}, {Kinnison},
  {Kirkby}, {Knox}, {Kotov}, {Krabbendam}, {Krughoff}, {Kub{\'a}nek},
  {Kuczewski}, {Kulkarni}, {Ku}, {Kurita}, {Lage}, {Lambert}, {Lange},
  {Langton}, {Le Guillou}, {Levine}, {Liang}, {Lim}, {Lintott}, {Long},
  {Lopez}, {Lotz}, {Lupton}, {Lust}, {MacArthur}, {Mahabal}, {Mandelbaum},
  {Markiewicz}, {Marsh}, {Marshall}, {Marshall}, {May}, {McKercher}, {McQueen},
  {Meyers}, {Migliore}, {Miller}, {Mills}, {Miraval}, {Moeyens}, {Moolekamp},
  {Monet}, {Moniez}, {Monkewitz}, {Montgomery}, {Morrison}, {Mueller},
  {Muller}, {Mu{\~n}oz Arancibia}, {Neill}, {Newbry}, {Nief}, {Nomerotski},
  {Nordby}, {O'Connor}, {Oliver}, {Olivier}, {Olsen}, {O'Mullane}, {Ortiz},
  {Osier}, {Owen}, {Pain}, {Palecek}, {Parejko}, {Parsons}, {Pease},
  {Peterson}, {Peterson}, {Petravick}, {Libby Petrick}, {Petry},
  {Pierfederici}, {Pietrowicz}, {Pike}, {Pinto}, {Plante}, {Plate}, {Plutchak},
  {Price}, {Prouza}, {Radeka}, {Rajagopal}, {Rasmussen}, {Regnault}, {Reil},
  {Reiss}, {Reuter}, {Ridgway}, {Riot}, {Ritz}, {Robinson}, {Roby}, {Roodman},
  {Rosing}, {Roucelle}, {Rumore}, {Russo}, {Saha}, {Sassolas}, {Schalk},
  {Schellart}, {Schindler}, {Schmidt}, {Schneider}, {Schneider}, {Schoening},
  {Schumacher}, {Schwamb}, {Sebag}, {Selvy}, {Sembroski}, {Seppala}, {Serio},
  {Serrano}, {Shaw}, {Shipsey}, {Sick}, {Silvestri}, {Slater}, {Smith},
  {Smith}, {Sobhani}, {Soldahl}, {Storrie-Lombardi}, {Stover}, {Strauss},
  {Street}, {Stubbs}, {Sullivan}, {Sweeney}, {Swinbank}, {Szalay}, {Takacs},
  {Tether}, {Thaler}, {Thayer}, {Thomas}, {Thornton}, {Thukral}, {Tice},
  {Trilling}, {Turri}, {Van Berg}, {Vanden Berk}, {Vetter}, {Virieux},
  {Vucina}, {Wahl}, {Walkowicz}, {Walsh}, {Walter}, {Wang}, {Wang}, {Warner},
  {Wiecha}, {Willman}, {Winters}, {Wittman}, {Wolff}, {Wood-Vasey}, {Wu},
  {Xin}, {Yoachim}, \& {Zhan}}]{ivezic2019}
{Ivezi{\'c}}, {\v{Z}}., {Kahn}, S.~M., {Tyson}, J.~A., {et~al.} 2019, \apj,
  873, 111

\bibitem[{{Jansen} {et~al.}(2001){Jansen}, {Lumb}, {Altieri}, {Clavel}, {Ehle},
  {Erd}, {Gabriel}, {Guainazzi}, {Gondoin}, {Much}, {Munoz}, {Santos},
  {Schartel}, {Texier}, \& {Vacanti}}]{jansen2001}
{Jansen}, F., {Lumb}, D., {Altieri}, B., {et~al.} 2001, \aap, 365, L1

\bibitem[{{Jin} {et~al.}(2013){Jin}, {Done}, {Middleton}, \& {Ward}}]{jin2013}
{Jin}, C., {Done}, C., {Middleton}, M., \& {Ward}, M. 2013, \mnras, 436, 3173

\bibitem[{{Jin} {et~al.}(2012){Jin}, {Ward}, \& {Done}}]{jin2012}
{Jin}, C., {Ward}, M., \& {Done}, C. 2012, \mnras, 425, 907

\bibitem[{{Kaiser} {et~al.}(2002){Kaiser}, {Aussel}, {Burke}, {Boesgaard},
  {Chambers}, {Chun}, {Heasley}, {Hodapp}, {Hunt}, {Jedicke}, {Jewitt},
  {Kudritzki}, {Luppino}, {Maberry}, {Magnier}, {Monet}, {Onaka}, {Pickles},
  {Rhoads}, {Simon}, {Szalay}, {Szapudi}, {Tholen}, {Tonry}, {Waterson}, \&
  {Wick}}]{kaiser2002}
{Kaiser}, N., {Aussel}, H., {Burke}, B.~E., {et~al.} 2002, in Society of
  Photo-Optical Instrumentation Engineers (SPIE) Conference Series, Vol. 4836,
  Survey and Other Telescope Technologies and Discoveries, ed. J.~A. {Tyson} \&
  S.~{Wolff}, 154--164

\bibitem[{{Kamizasa} {et~al.}(2012){Kamizasa}, {Terashima}, \&
  {Awaki}}]{kamizasa2012}
{Kamizasa}, N., {Terashima}, Y., \& {Awaki}, H. 2012, \apj, 751, 39

\bibitem[{{Kamraj} {et~al.}(2022){Kamraj}, {Brightman}, {Harrison}, {Stern},
  {Garc{\'\i}a}, {Balokovi{\'c}}, {Ricci}, {Koss}, {Mej{\'\i}a-Restrepo}, {Oh},
  {Powell}, \& {Urry}}]{kamraj2022}
{Kamraj}, N., {Brightman}, M., {Harrison}, F.~A., {et~al.} 2022, \apj, 927, 42

\bibitem[{{Kawamuro} {et~al.}(2016){Kawamuro}, {Ueda}, {Tazaki}, {Ricci}, \&
  {Terashima}}]{kawamuro2016}
{Kawamuro}, T., {Ueda}, Y., {Tazaki}, F., {Ricci}, C., \& {Terashima}, Y. 2016,
  \apjs, 225, 14

\bibitem[{{King} \& {Pounds}(2015)}]{king2015}
{King}, A. \& {Pounds}, K. 2015, \araa, 53, 115

\bibitem[{{Kosec} {et~al.}(2020){Kosec}, {Zoghbi}, {Walton}, {Pinto}, {Fabian},
  {Parker}, \& {Reynolds}}]{kosec2020}
{Kosec}, P., {Zoghbi}, A., {Walton}, D.~J., {et~al.} 2020, \mnras, 495, 4769

\bibitem[{{Krawczyk} {et~al.}(2015){Krawczyk}, {Richards}, {Gallagher},
  {Leighly}, {Hewett}, {Ross}, \& {Hall}}]{krawczyk2015}
{Krawczyk}, C.~M., {Richards}, G.~T., {Gallagher}, S.~C., {et~al.} 2015, \aj,
  149, 203

\bibitem[{{Laurenti} {et~al.}(2021){Laurenti}, {Luminari}, {Tombesi},
  {Vagnetti}, {Middei}, \& {Piconcelli}}]{laurenti2021}
{Laurenti}, M., {Luminari}, A., {Tombesi}, F., {et~al.} 2021, \aap, 645, A118

\bibitem[{{Laurenti} {et~al.}(2022){Laurenti}, {Piconcelli}, {Zappacosta},
  {Tombesi}, {Vignali}, {Bianchi}, {Marziani}, {Vagnetti}, {Bongiorno},
  {Bischetti}, {del Olmo}, {Lanzuisi}, {Luminari}, {Middei}, {Perri}, {Ricci},
  \& {Vietri}}]{laurenti2022}
{Laurenti}, M., {Piconcelli}, E., {Zappacosta}, L., {et~al.} 2022, \aap, 657,
  A57

\bibitem[{{Liu} {et~al.}(2021){Liu}, {Luo}, {Brandt}, {Brotherton},
  {Gallagher}, {Ni}, {Shemmer}, \& {Timlin}}]{liu2021}
{Liu}, H., {Luo}, B., {Brandt}, W.~N., {et~al.} 2021, \apj, 910, 103

\bibitem[{{Liu} {et~al.}(2016){Liu}, {Merloni}, {Georgakakis}, {Menzel},
  {Buchner}, {Nandra}, {Salvato}, {Shen}, {Brusa}, \& {Streblyanska}}]{liu2016}
{Liu}, Z., {Merloni}, A., {Georgakakis}, A., {et~al.} 2016, \mnras, 459, 1602

\bibitem[{{Lu} {et~al.}(2019){Lu}, {Huang}, {Zhang}, {Wang}, {Du}, {Hu},
  {Xiao}, {Li}, {Bai}, {Bian}, {Yuan}, {Ho}, {Wang}, \& {SEAMBH
  Collaboration}}]{lu2019}
{Lu}, K.-X., {Huang}, Y.-K., {Zhang}, Z.-X., {et~al.} 2019, \apj, 877, 23

\bibitem[{{Luminari} {et~al.}(2021){Luminari}, {Nicastro}, {Elvis},
  {Piconcelli}, {Tombesi}, {Zappacosta}, \& {Fiore}}]{luminari2021}
{Luminari}, A., {Nicastro}, F., {Elvis}, M., {et~al.} 2021, \aap, 646, A111

\bibitem[{{Lusso} {et~al.}(2010){Lusso}, {Comastri}, {Vignali}, {Zamorani},
  {Brusa}, {Gilli}, {Iwasawa}, {Salvato}, {Civano}, {Elvis}, {Merloni},
  {Bongiorno}, {Trump}, {Koekemoer}, {Schinnerer}, {Le Floc'h}, {Cappelluti},
  {Jahnke}, {Sargent}, {Silverman}, {Mainieri}, {Fiore}, {Bolzonella}, {Le
  F{\`e}vre}, {Garilli}, {Iovino}, {Kneib}, {Lamareille}, {Lilly}, {Mignoli},
  {Scodeggio}, \& {Vergani}}]{lusso2010}
{Lusso}, E., {Comastri}, A., {Vignali}, C., {et~al.} 2010, \aap, 512, A34

\bibitem[{{Lusso} {et~al.}(2023){Lusso}, {Valiante}, \& {Vito}}]{lusso2023}
{Lusso}, E., {Valiante}, R., \& {Vito}, F. 2023, in Handbook of X-ray and
  Gamma-ray Astrophysics. Edited by Cosimo Bambi and Andrea Santangelo, 122

\bibitem[{{Mainzer} {et~al.}(2011){Mainzer}, {Bauer}, {Grav}, {Masiero},
  {Cutri}, {Dailey}, {Eisenhardt}, {McMillan}, {Wright}, {Walker}, {Jedicke},
  {Spahr}, {Tholen}, {Alles}, {Beck}, {Brandenburg}, {Conrow}, {Evans},
  {Fowler}, {Jarrett}, {Marsh}, {Masci}, {McCallon}, {Wheelock}, {Wittman},
  {Wyatt}, {DeBaun}, {Elliott}, {Elsbury}, {Gautier}, {Gomillion}, {Leisawitz},
  {Maleszewski}, {Micheli}, \& {Wilkins}}]{mainzer2011}
{Mainzer}, A., {Bauer}, J., {Grav}, T., {et~al.} 2011, \apj, 731, 53

\bibitem[{{Mallick} \& {Dewangan}(2018)}]{mallick2018}
{Mallick}, L. \& {Dewangan}, G.~C. 2018, \apj, 863, 178

\bibitem[{{Martocchia} {et~al.}(2017){Martocchia}, {Piconcelli}, {Zappacosta},
  {Duras}, {Vietri}, {Vignali}, {Bianchi}, {Bischetti}, {Bongiorno}, {Brusa},
  {Lanzuisi}, {Marconi}, {Mathur}, {Miniutti}, {Nicastro}, {Bruni}, \&
  {Fiore}}]{martocchia2017}
{Martocchia}, S., {Piconcelli}, E., {Zappacosta}, L., {et~al.} 2017, \aap, 608,
  A51

\bibitem[{{Marziani} {et~al.}(2018){Marziani}, {del Olmo}, {D'Onofrio},
  {Dultzin}, {Negrete}, {Mart{\'\i}nez-Aldama}, {Bon}, {Bon}, \&
  {Stirpe}}]{marziani2018}
{Marziani}, P., {del Olmo}, A., {D'Onofrio}, M., {et~al.} 2018, in Revisiting
  Narrow-Line Seyfert 1 Galaxies and their Place in the Universe, 2

\bibitem[{{Marziani} {et~al.}(2016){Marziani}, {Mart{\'\i}nez Carballo},
  {Sulentic}, {Del Olmo}, {Stirpe}, \& {Dultzin}}]{marziani2016}
{Marziani}, P., {Mart{\'\i}nez Carballo}, M.~A., {Sulentic}, J.~W., {et~al.}
  2016, \apss, 361, 29

\bibitem[{{Marziani} \& {Sulentic}(2014)}]{marziani2014}
{Marziani}, P. \& {Sulentic}, J.~W. 2014, \mnras, 442, 1211

\bibitem[{{Mason} {et~al.}(2001){Mason}, {Breeveld}, {Much}, {Carter},
  {Cordova}, {Cropper}, {Fordham}, {Huckle}, {Ho}, {Kawakami}, {Kennea},
  {Kennedy}, {Mittaz}, {Pandel}, {Priedhorsky}, {Sasseen}, {Shirey}, {Smith},
  \& {Vreux}}]{mason2001}
{Mason}, K.~O., {Breeveld}, A., {Much}, R., {et~al.} 2001, \aap, 365, L36

\bibitem[{{Mateos} {et~al.}(2009){Mateos}, {Saxton}, {Read}, \&
  {Sembay}}]{mateos2009}
{Mateos}, S., {Saxton}, R.~D., {Read}, A.~M., \& {Sembay}, S. 2009, \aap, 496,
  879

\bibitem[{{Mazzucchelli} {et~al.}(2017){Mazzucchelli}, {Ba{\~n}ados},
  {Venemans}, {Decarli}, {Farina}, {Walter}, {Eilers}, {Rix}, {Simcoe},
  {Stern}, {Fan}, {Schlafly}, {De Rosa}, {Hennawi}, {Chambers}, {Greiner},
  {Burgett}, {Draper}, {Kaiser}, {Kudritzki}, {Magnier}, {Metcalfe}, {Waters},
  \& {Wainscoat}}]{mazzucchelli2017}
{Mazzucchelli}, C., {Ba{\~n}ados}, E., {Venemans}, B.~P., {et~al.} 2017, \apj,
  849, 91

\bibitem[{{Mehdipour} {et~al.}(2011){Mehdipour}, {Branduardi-Raymont},
  {Kaastra}, {Petrucci}, {Kriss}, {Ponti}, {Blustin}, {Paltani}, {Cappi},
  {Detmers}, \& {Steenbrugge}}]{mehdipour2011}
{Mehdipour}, M., {Branduardi-Raymont}, G., {Kaastra}, J.~S., {et~al.} 2011,
  \aap, 534, A39

\bibitem[{{Middei} {et~al.}(2023){Middei}, {Nardini}, {Matzeu}, {Bianchi},
  {Braito}, {Perri}, \& {Puccetti}}]{middei2023}
{Middei}, R., {Nardini}, E., {Matzeu}, G.~A., {et~al.} 2023, \aap, 680, A50

\bibitem[{{Middei} {et~al.}(2020){Middei}, {Petrucci}, {Bianchi}, {Ursini},
  {Cappi}, {Clavel}, {De Rosa}, {Marinucci}, {Matt}, \& {Tortosa}}]{middei2020}
{Middei}, R., {Petrucci}, P.~O., {Bianchi}, S., {et~al.} 2020, \aap, 640, A99

\bibitem[{{Mineshige} {et~al.}(2000){Mineshige}, {Kawaguchi}, {Takeuchi}, \&
  {Hayashida}}]{mineshige2000}
{Mineshige}, S., {Kawaguchi}, T., {Takeuchi}, M., \& {Hayashida}, K. 2000,
  \pasj, 52, 499

\bibitem[{{Miniutti} \& {Fabian}(2004)}]{miniutti2004}
{Miniutti}, G. \& {Fabian}, A.~C. 2004, \mnras, 349, 1435

\bibitem[{{Nandra} \& {Pounds}(1994)}]{nandra1994}
{Nandra}, K. \& {Pounds}, K.~A. 1994, \mnras, 268, 405

\bibitem[{{Nardini} {et~al.}(2019){Nardini}, {Lusso}, \&
  {Bisogni}}]{nardini2019}
{Nardini}, E., {Lusso}, E., \& {Bisogni}, S. 2019, \mnras, 482, L134

\bibitem[{{Nardini} {et~al.}(2015){Nardini}, {Reeves}, {Gofford}, {Harrison},
  {Risaliti}, {Braito}, {Costa}, {Matzeu}, {Walton}, {Behar}, {Boggs},
  {Christensen}, {Craig}, {Hailey}, {Matt}, {Miller}, {O'Brien}, {Stern},
  {Turner}, \& {Ward}}]{nardini2015}
{Nardini}, E., {Reeves}, J.~N., {Gofford}, J., {et~al.} 2015, Science, 347, 860

\bibitem[{{Onoue} {et~al.}(2020){Onoue}, {Ba{\~n}ados}, {Mazzucchelli},
  {Venemans}, {Schindler}, {Walter}, {Hennawi}, {Andika}, {Davies}, {Decarli},
  {Farina}, {Jahnke}, {Nagao}, {Tominaga}, \& {Wang}}]{onoue2020}
{Onoue}, M., {Ba{\~n}ados}, E., {Mazzucchelli}, C., {et~al.} 2020, \apj, 898,
  105

\bibitem[{{Osterbrock} \& {Pogge}(1985)}]{osterbrock1985}
{Osterbrock}, D.~E. \& {Pogge}, R.~W. 1985, \apj, 297, 166

\bibitem[{{Page} {et~al.}(2012){Page}, {Brindle}, {Talavera}, {Still}, {Rosen},
  {Yershov}, {Ziaeepour}, {Mason}, {Cropper}, {Breeveld}, {Loiseau}, {Mignani},
  {Smith}, \& {Murdin}}]{page2012}
{Page}, M.~J., {Brindle}, C., {Talavera}, A., {et~al.} 2012, \mnras, 426, 903

\bibitem[{{Paolillo} {et~al.}(2004){Paolillo}, {Schreier}, {Giacconi},
  {Koekemoer}, \& {Grogin}}]{paolillo2004}
{Paolillo}, M., {Schreier}, E.~J., {Giacconi}, R., {Koekemoer}, A.~M., \&
  {Grogin}, N.~A. 2004, \apj, 611, 93

\bibitem[{{P{\^a}ris} {et~al.}(2018){P{\^a}ris}, {Petitjean}, {Aubourg},
  {Myers}, {Streblyanska}, {Lyke}, {Anderson}, {Armengaud}, {Bautista},
  {Blanton}, {Blomqvist}, {Brinkmann}, {Brownstein}, {Brandt}, {Burtin},
  {Dawson}, {de la Torre}, {Georgakakis}, {Gil-Mar{\'\i}n}, {Green}, {Hall},
  {Kneib}, {LaMassa}, {Le Goff}, {MacLeod}, {Mariappan}, {McGreer}, {Merloni},
  {Noterdaeme}, {Palanque-Delabrouille}, {Percival}, {Ross}, {Rossi},
  {Schneider}, {Seo}, {Tojeiro}, {Weaver}, {Weijmans}, {Y{\`e}che}, {Zarrouk},
  \& {Zhao}}]{paris2018}
{P{\^a}ris}, I., {Petitjean}, P., {Aubourg}, {\'E}., {et~al.} 2018, \aap, 613,
  A51

\bibitem[{{Petrucci} {et~al.}(2020){Petrucci}, {Gronkiewicz}, {Rozanska},
  {Belmont}, {Bianchi}, {Czerny}, {Matt}, {Malzac}, {Middei}, {De Rosa},
  {Ursini}, \& {Cappi}}]{petrucci2020}
{Petrucci}, P.~O., {Gronkiewicz}, D., {Rozanska}, A., {et~al.} 2020, \aap, 634,
  A85

\bibitem[{{Petrucci} {et~al.}(2018){Petrucci}, {Ursini}, {De Rosa}, {Bianchi},
  {Cappi}, {Matt}, {Dadina}, \& {Malzac}}]{petrucci2018}
{Petrucci}, P.~O., {Ursini}, F., {De Rosa}, A., {et~al.} 2018, \aap, 611, A59

\bibitem[{{Piconcelli} {et~al.}(2005){Piconcelli}, {Jimenez-Bail{\'o}n},
  {Guainazzi}, {Schartel}, {Rodr{\'\i}guez-Pascual}, \&
  {Santos-Lle{\'o}}}]{piconcelli2005}
{Piconcelli}, E., {Jimenez-Bail{\'o}n}, E., {Guainazzi}, M., {et~al.} 2005,
  \aap, 432, 15

\bibitem[{{Prevot} {et~al.}(1984){Prevot}, {Lequeux}, {Maurice}, {Prevot}, \&
  {Rocca-Volmerange}}]{prevot1984}
{Prevot}, M.~L., {Lequeux}, J., {Maurice}, E., {Prevot}, L., \&
  {Rocca-Volmerange}, B. 1984, \aap, 132, 389

\bibitem[{{Proga}(2005)}]{proga2005}
{Proga}, D. 2005, \apjl, 630, L9

\bibitem[{{Pu} {et~al.}(2020){Pu}, {Luo}, {Brandt}, {Timlin}, {Liu}, {Ni}, \&
  {Wu}}]{pu2020}
{Pu}, X., {Luo}, B., {Brandt}, W.~N., {et~al.} 2020, \apj, 900, 141

\bibitem[{{Rakshit} {et~al.}(2020){Rakshit}, {Stalin}, \&
  {Kotilainen}}]{rakshit2020}
{Rakshit}, S., {Stalin}, C.~S., \& {Kotilainen}, J. 2020, \apjs, 249, 17

\bibitem[{{Read} {et~al.}(2014){Read}, {Guainazzi}, \& {Sembay}}]{read2014}
{Read}, A.~M., {Guainazzi}, M., \& {Sembay}, S. 2014, \aap, 564, A75

\bibitem[{{Read} {et~al.}(2011){Read}, {Rosen}, {Saxton}, \&
  {Ramirez}}]{read2011}
{Read}, A.~M., {Rosen}, S.~R., {Saxton}, R.~D., \& {Ramirez}, J. 2011, \aap,
  534, A34

\bibitem[{{Reeves} {et~al.}(2023){Reeves}, {Braito}, {Porquet}, {Laurenti},
  {Lobban}, \& {Matzeu}}]{reeves2023}
{Reeves}, J.~N., {Braito}, V., {Porquet}, D., {et~al.} 2023, \apj, 952, 52

\bibitem[{{Reeves} {et~al.}(2009){Reeves}, {O'Brien}, {Braito}, {Behar},
  {Miller}, {Turner}, {Fabian}, {Kaspi}, {Mushotzky}, \& {Ward}}]{reeves2009}
{Reeves}, J.~N., {O'Brien}, P.~T., {Braito}, V., {et~al.} 2009, \apj, 701, 493

\bibitem[{{Ricci} {et~al.}(2018){Ricci}, {Ho}, {Fabian}, {Trakhtenbrot},
  {Koss}, {Ueda}, {Lohfink}, {Shimizu}, {Bauer}, {Mushotzky}, {Schawinski},
  {Paltani}, {Lamperti}, {Treister}, \& {Oh}}]{ricci2018}
{Ricci}, C., {Ho}, L.~C., {Fabian}, A.~C., {et~al.} 2018, \mnras, 480, 1819

\bibitem[{{Richards} {et~al.}(2003){Richards}, {Hall}, {Vanden Berk},
  {Strauss}, {Schneider}, {Weinstein}, {Reichard}, {York}, {Knapp}, {Fan},
  {Ivezi{\'c}}, {Brinkmann}, {Budav{\'a}ri}, {Csabai}, \&
  {Nichol}}]{richards2003}
{Richards}, G.~T., {Hall}, P.~B., {Vanden Berk}, D.~E., {et~al.} 2003, \aj,
  126, 1131

\bibitem[{{Richards} {et~al.}(2006){Richards}, {Lacy}, {Storrie-Lombardi},
  {Hall}, {Gallagher}, {Hines}, {Fan}, {Papovich}, {Vanden Berk}, {Trammell},
  {Schneider}, {Vestergaard}, {York}, {Jester}, {Anderson}, {Budav{\'a}ri}, \&
  {Szalay}}]{richards2006}
{Richards}, G.~T., {Lacy}, M., {Storrie-Lombardi}, L.~J., {et~al.} 2006, \apjs,
  166, 470

\bibitem[{{Risaliti} {et~al.}(2009){Risaliti}, {Young}, \&
  {Elvis}}]{risaliti2009}
{Risaliti}, G., {Young}, M., \& {Elvis}, M. 2009, \apjl, 700, L6

\bibitem[{{Schlafly} \& {Finkbeiner}(2011)}]{schlafly2011}
{Schlafly}, E.~F. \& {Finkbeiner}, D.~P. 2011, \apj, 737, 103

\bibitem[{{Serafinelli} {et~al.}(2017){Serafinelli}, {Vagnetti}, \&
  {Middei}}]{serafinelli2017}
{Serafinelli}, R., {Vagnetti}, F., \& {Middei}, R. 2017, \aap, 600, A101

\bibitem[{{Shakura} \& {Sunyaev}(1973)}]{shakura1973}
{Shakura}, N.~I. \& {Sunyaev}, R.~A. 1973, \aap, 24, 337

\bibitem[{{Shankar} {et~al.}(2013){Shankar}, {Weinberg}, \&
  {Miralda-Escud{\'e}}}]{shankar2013}
{Shankar}, F., {Weinberg}, D.~H., \& {Miralda-Escud{\'e}}, J. 2013, \mnras,
  428, 421

\bibitem[{{Shemmer} {et~al.}(2006){Shemmer}, {Brandt}, {Netzer}, {Maiolino}, \&
  {Kaspi}}]{shemmer2006}
{Shemmer}, O., {Brandt}, W.~N., {Netzer}, H., {Maiolino}, R., \& {Kaspi}, S.
  2006, \apjl, 646, L29

\bibitem[{{Shemmer} {et~al.}(2008){Shemmer}, {Brandt}, {Netzer}, {Maiolino}, \&
  {Kaspi}}]{shemmer2008}
{Shemmer}, O., {Brandt}, W.~N., {Netzer}, H., {Maiolino}, R., \& {Kaspi}, S.
  2008, \apj, 682, 81

\bibitem[{{Shirakata} {et~al.}(2019){Shirakata}, {Kawaguchi}, {Oogi},
  {Okamoto}, \& {Nagashima}}]{shirakata2019}
{Shirakata}, H., {Kawaguchi}, T., {Oogi}, T., {Okamoto}, T., \& {Nagashima}, M.
  2019, \mnras, 487, 409

\bibitem[{{S{\k{a}}dowski} {et~al.}(2011){S{\k{a}}dowski}, {Abramowicz},
  {Bursa}, {Klu{\'z}niak}, {Lasota}, \& {R{\'o}{\.z}a{\'n}ska}}]{sadowski2011}
{S{\k{a}}dowski}, A., {Abramowicz}, M., {Bursa}, M., {et~al.} 2011, \aap, 527,
  A17

\bibitem[{{Sobolewska} \& {Done}(2007)}]{sobolewska2007}
{Sobolewska}, M.~A. \& {Done}, C. 2007, \mnras, 374, 150

\bibitem[{{Sobolewska} \& {Papadakis}(2009)}]{sobolewska2009}
{Sobolewska}, M.~A. \& {Papadakis}, I.~E. 2009, \mnras, 399, 1597

\bibitem[{{Soldi} {et~al.}(2014){Soldi}, {Beckmann}, {Baumgartner}, {Ponti},
  {Shrader}, {Lubi{\'n}ski}, {Krimm}, {Mattana}, \& {Tueller}}]{soldi2014}
{Soldi}, S., {Beckmann}, V., {Baumgartner}, W.~H., {et~al.} 2014, \aap, 563,
  A57

\bibitem[{{Str{\"u}der} {et~al.}(2001){Str{\"u}der}, {Briel}, {Dennerl},
  {Hartmann}, {Kendziorra}, {Meidinger}, {Pfeffermann}, {Reppin}, {Aschenbach},
  {Bornemann}, {Br{\"a}uninger}, {Burkert}, {Elender}, {Freyberg}, {Haberl},
  {Hartner}, {Heuschmann}, {Hippmann}, {Kastelic}, {Kemmer}, {Kettenring},
  {Kink}, {Krause}, {M{\"u}ller}, {Oppitz}, {Pietsch}, {Popp}, {Predehl},
  {Read}, {Stephan}, {St{\"o}tter}, {Tr{\"u}mper}, {Holl}, {Kemmer}, {Soltau},
  {St{\"o}tter}, {Weber}, {Weichert}, {von Zanthier}, {Carathanassis}, {Lutz},
  {Richter}, {Solc}, {B{\"o}ttcher}, {Kuster}, {Staubert}, {Abbey}, {Holland},
  {Turner}, {Balasini}, {Bignami}, {La Palombara}, {Villa}, {Buttler},
  {Gianini}, {Lain{\'e}}, {Lumb}, \& {Dhez}}]{struder2001}
{Str{\"u}der}, L., {Briel}, U., {Dennerl}, K., {et~al.} 2001, \aap, 365, L18

\bibitem[{{Tombesi} {et~al.}(2015){Tombesi}, {Mel{\'e}ndez}, {Veilleux},
  {Reeves}, {Gonz{\'a}lez-Alfonso}, \& {Reynolds}}]{tombesi2015}
{Tombesi}, F., {Mel{\'e}ndez}, M., {Veilleux}, S., {et~al.} 2015, \nat, 519,
  436

\bibitem[{{Trakhtenbrot} {et~al.}(2017){Trakhtenbrot}, {Ricci}, {Koss},
  {Schawinski}, {Mushotzky}, {Ueda}, {Veilleux}, {Lamperti}, {Oh}, {Treister},
  {Stern}, {Harrison}, {Balokovi{\'c}}, \& {Gehrels}}]{trakhtenbrot2017}
{Trakhtenbrot}, B., {Ricci}, C., {Koss}, M.~J., {et~al.} 2017, \mnras, 470, 800

\bibitem[{{Trefoloni} {et~al.}(2023){Trefoloni}, {Lusso}, {Nardini},
  {Risaliti}, {Bargiacchi}, {Bisogni}, {Civano}, {Elvis}, {Fabbiano}, {Gilli},
  {Marconi}, {Richards}, {Sacchi}, {Salvestrini}, {Signorini}, \&
  {Vignali}}]{trefoloni2023}
{Trefoloni}, B., {Lusso}, E., {Nardini}, E., {et~al.} 2023, \aap, 677, A111

\bibitem[{{Turner} {et~al.}(2001){Turner}, {Abbey}, {Arnaud}, {Balasini},
  {Barbera}, {Belsole}, {Bennie}, {Bernard}, {Bignami}, {Boer}, {Briel},
  {Butler}, {Cara}, {Chabaud}, {Cole}, {Collura}, {Conte}, {Cros}, {Denby},
  {Dhez}, {Di Coco}, {Dowson}, {Ferrando}, {Ghizzardi}, {Gianotti}, {Goodall},
  {Gretton}, {Griffiths}, {Hainaut}, {Hochedez}, {Holland}, {Jourdain},
  {Kendziorra}, {Lagostina}, {Laine}, {La Palombara}, {Lortholary}, {Lumb},
  {Marty}, {Molendi}, {Pigot}, {Poindron}, {Pounds}, {Reeves}, {Reppin},
  {Rothenflug}, {Salvetat}, {Sauvageot}, {Schmitt}, {Sembay}, {Short},
  {Spragg}, {Stephen}, {Str{\"u}der}, {Tiengo}, {Trifoglio}, {Tr{\"u}mper},
  {Vercellone}, {Vigroux}, {Villa}, {Ward}, {Whitehead}, \&
  {Zonca}}]{turner2001}
{Turner}, M.~J.~L., {Abbey}, A., {Arnaud}, M., {et~al.} 2001, \aap, 365, L27

\bibitem[{{Vagnetti} {et~al.}(2013){Vagnetti}, {Antonucci}, \&
  {Trevese}}]{vagnetti2013}
{Vagnetti}, F., {Antonucci}, M., \& {Trevese}, D. 2013, \aap, 550, A71

\bibitem[{{Vagnetti} {et~al.}(2016){Vagnetti}, {Middei}, {Antonucci},
  {Paolillo}, \& {Serafinelli}}]{vagnetti2016}
{Vagnetti}, F., {Middei}, R., {Antonucci}, M., {Paolillo}, M., \&
  {Serafinelli}, R. 2016, \aap, 593, A55

\bibitem[{{Vagnetti} {et~al.}(2010){Vagnetti}, {Turriziani}, {Trevese}, \&
  {Antonucci}}]{vagnetti2010}
{Vagnetti}, F., {Turriziani}, S., {Trevese}, D., \& {Antonucci}, M. 2010, \aap,
  519, A17

\bibitem[{{Valiante} {et~al.}(2017){Valiante}, {Agarwal}, {Habouzit}, \&
  {Pezzulli}}]{valiante2017}
{Valiante}, R., {Agarwal}, B., {Habouzit}, M., \& {Pezzulli}, E. 2017, \pasa,
  34, e031

\bibitem[{{Vanden Berk} {et~al.}(2001){Vanden Berk}, {Richards}, {Bauer},
  {Strauss}, {Schneider}, {Heckman}, {York}, {Hall}, {Fan}, {Knapp},
  {Anderson}, {Annis}, {Bahcall}, {Bernardi}, {Briggs}, {Brinkmann}, {Brunner},
  {Burles}, {Carey}, {Castander}, {Connolly}, {Crocker}, {Csabai}, {Doi},
  {Finkbeiner}, {Friedman}, {Frieman}, {Fukugita}, {Gunn}, {Hennessy},
  {Ivezi{\'c}}, {Kent}, {Kunszt}, {Lamb}, {Leger}, {Long}, {Loveday}, {Lupton},
  {Meiksin}, {Merelli}, {Munn}, {Newberg}, {Newcomb}, {Nichol}, {Owen}, {Pier},
  {Pope}, {Rockosi}, {Schlegel}, {Siegmund}, {Smee}, {Snir}, {Stoughton},
  {Stubbs}, {SubbaRao}, {Szalay}, {Szokoly}, {Tremonti}, {Uomoto}, {Waddell},
  {Yanny}, \& {Zheng}}]{vandenberk2001}
{Vanden Berk}, D.~E., {Richards}, G.~T., {Bauer}, A., {et~al.} 2001, \aj, 122,
  549

\bibitem[{{Vietri} {et~al.}(2020){Vietri}, {Mainieri}, {Kakkad}, {Netzer},
  {Perna}, {Circosta}, {Harrison}, {Zappacosta}, {Husemann}, {Padovani},
  {Bischetti}, {Bongiorno}, {Brusa}, {Carniani}, {Cicone}, {Comastri},
  {Cresci}, {Feruglio}, {Fiore}, {Lanzuisi}, {Mannucci}, {Marconi},
  {Piconcelli}, {Puglisi}, {Salvato}, {Schramm}, {Schulze}, {Scholtz},
  {Vignali}, \& {Zamorani}}]{vietri2020}
{Vietri}, G., {Mainieri}, V., {Kakkad}, D., {et~al.} 2020, \aap, 644, A175

\bibitem[{{Vietri} {et~al.}(2018){Vietri}, {Piconcelli}, {Bischetti}, {Duras},
  {Martocchia}, {Bongiorno}, {Marconi}, {Zappacosta}, {Bisogni}, {Bruni},
  {Brusa}, {Comastri}, {Cresci}, {Feruglio}, {Giallongo}, {La Franca},
  {Mainieri}, {Mannucci}, {Ricci}, {Sani}, {Testa}, {Tombesi}, {Vignali}, \&
  {Fiore}}]{vietri2018}
{Vietri}, G., {Piconcelli}, E., {Bischetti}, M., {et~al.} 2018, \aap, 617, A81

\bibitem[{{Wachter} {et~al.}(1979){Wachter}, {Leach}, \&
  {Kellogg}}]{wachter1979}
{Wachter}, K., {Leach}, R., \& {Kellogg}, E. 1979, \apj, 230, 274

\bibitem[{{Waddell} \& {Gallo}(2020)}]{waddell2020}
{Waddell}, S.~G.~H. \& {Gallo}, L.~C. 2020, \mnras, 498, 5207

\bibitem[{{Webb} {et~al.}(2020){Webb}, {Coriat}, {Traulsen}, {Ballet}, {Motch},
  {Carrera}, {Koliopanos}, {Authier}, {de la Calle}, {Ceballos}, {Colomo},
  {Chuard}, {Freyberg}, {Garcia}, {Kolehmainen}, {Lamer}, {Lin}, {Maggi},
  {Michel}, {Page}, {Page}, {Perea-Calderon}, {Pineau}, {Rodriguez}, {Rosen},
  {Santos Lleo}, {Saxton}, {Schwope}, {Tom{\'a}s}, {Watson}, \&
  {Zakardjian}}]{webb2020}
{Webb}, N.~A., {Coriat}, M., {Traulsen}, I., {et~al.} 2020, \aap, 641, A136

\bibitem[{{Wu} {et~al.}(2015){Wu}, {Wang}, {Fan}, {Yi}, {Zuo}, {Bian}, {Jiang},
  {McGreer}, {Wang}, {Yang}, {Yang}, {Thompson}, \& {Beletsky}}]{wu2015}
{Wu}, X.-B., {Wang}, F., {Fan}, X., {et~al.} 2015, \nat, 518, 512

\bibitem[{{Zappacosta} {et~al.}(2020){Zappacosta}, {Piconcelli}, {Giustini},
  {Vietri}, {Duras}, {Miniutti}, {Bischetti}, {Bongiorno}, {Brusa},
  {Chiaberge}, {Comastri}, {Feruglio}, {Luminari}, {Marconi}, {Ricci},
  {Vignali}, \& {Fiore}}]{zappacosta2020}
{Zappacosta}, L., {Piconcelli}, E., {Giustini}, M., {et~al.} 2020, \aap, 635,
  L5

\bibitem[{{Zhang} {et~al.}(2020){Zhang}, {Lu}, \& {Fang}}]{zhang2020}
{Zhang}, X., {Lu}, Y., \& {Fang}, T. 2020, \apjl, 903, L18

\bibitem[{{Zubovas} \& {King}(2012)}]{zubovas2012}
{Zubovas}, K. \& {King}, A. 2012, \apjl, 745, L34

\bibitem[{{Zubovas} \& {King}(2013)}]{zubovas2013}
{Zubovas}, K. \& {King}, A. 2013, \apj, 769, 51

\end{thebibliography}


\clearpage

\appendix

\section{The X-HESS sample}\label{sec:app_agn_list}

The AGN included in the X-HESS sample are listed in Table \ref{tab:agn_list}, where the values of redshift, black hole mass and bolometric luminosity are drawn from the \citet{rakshit2020} catalogue, unless otherwise stated. 
The values of $\lambda_\mathrm{Edd}$ reported in Table \ref{tab:agn_list} are calculated as the ratio of the bolometric to Eddington luminosity, $L_\mathrm{Edd} = 1.26 \times 10^{38}\,(M_\mathrm{BH}/M_\odot)$ erg/s, where $M_\mathrm{BH}$ represents the  fiducial estimate of the black hole mass based on the FWHM of either \ion{H}{$\beta$} or \ion{Mg}{II}.

\section{List of \emph{XMM-Newton} observations}\label{sec:appA}

In Table \ref{tab:longjournal} is the comprehensive journal of all \emph{XMM-Newton} observations of the X-HESS AGN. For each source, we report their observation(s) with the corresponding X-HESS and \emph{XMM-Newton} identifier, the starting date, as well as both the net exposure and counts in the broadband $E=0.3-10$ keV observer-frame energy interval. 

The hyphens reported in the same table indicate that the corresponding EPIC pn/MOS1/MOS2 observations are either unavailable or excluded from the analysis according to the procedure described in \S\,\ref{sec:xray_spec}.

\section{X-ray spectra of the X-HESS AGN}\label{sec:appB}
In this section we report a complete description of the best-fit spectral results for each AGN in the X-HESS sample. Specifically, the best-fit model and parameters of each observation are listed in Table \ref{tab:bestfit}, while the corresponding X-ray spectrum can be visualised in Fig. \ref{fig:xhess_spectra}, where data from EPIC-pn, MOS1 and MOS2 are shown in black, red and blue, respectively. Solid green line indicates the best-fit model. The bottom panel in each plot describes the ratio between the data points and the best-fit model. 

As already stated in \S\,\ref{sec:xray_spec}, we find that the X-ray continuum of the X-HESS sources can be well reproduced by a phenomenological model described by a power law absorbed by the Galactic gas column density plus, in some cases, a blackbody component to account for the soft excess.
However, some of the X-HESS AGN are characterised by at least one spectrum that requires relatively more complex models, that we discuss below. In each case, the adoption of an additional model component is favoured whenever the value of $\Delta{W}$-stat with respect to the corresponding change of degrees of freedom is significant at $95\%$ confidence level for the number of interesting parameters characterising the additional component. 

\paragraph{X-HESS 7.}

This AGN is located at redshift $z\!\sim\!1$ and has been serendipitously observed by \emph{XMM-Newton} for a total of five times between July 2003 and January 2015.
Our spectroscopic analysis indicates that the broadband spectrum of SDSS J023410.58–085213.5 can be described by a simple model comprising a power-law continuum modified by Galactic absorption.
Moreover, in three spectra we find evidence for cold absorption by a gas cloud with a column density of around $\sim\!3\times10^{21}$ cm$^{-2}$ which we accounted for with the \texttt{zTBabs} component in XSPEC.
The power-law continuum is quite hard, with the photon index varying between a minimum of $\sim\!1.3$ and a maximum of $\sim\!1.6$.
According to NASA/IPAC Extragalactic Database (NED)\footnote{\url{https://ned.ipac.caltech.edu}.} this source has not been observed yet in the radio (apparently the only case among the X-HESS AGN) and since such low values of $\Gamma$ are not typically expected for a RQ, type-1 AGN, while more commonly seen in RL sources, we cannot exclude a priori that this AGN actually belongs to the class of radio-loud quasars.

\paragraph{X-HESS 9.}

The ninth X-HESS AGN is located at redshift $z\!\sim\!0.7$ and has a total of four available observations obtained in the same year, from July to August 2017.
In general, a good description of the broadband X-ray spectrum is provided by a model consisting of a power-law continuum modified by Galactic absorption.
In one spectrum, however, we find necessary to include a cold absorption component, modelled with \texttt{zTBabs} in XSPEC, to take into account a thick, neutral absorber of column width $N_\mathrm{H} \sim 10^{22}$ cm$^{-2}$.
No evidence of such an absorption structure is found in the other observations.
The power-law continuum is characterised by photon indices with values ranging from a minimum of $\sim\!1.4$ to a maximum of $\sim\!1.7$, and this AGN does not show large variations in both the soft and hard X-ray fluxes over the different observations.

\paragraph{X-HESS 13.}

This source has three available \emph{XMM-Newton} observations carried out between May 2017 and November 2018, with the first two observations being separated by only one day and thus almost coincident.
This AGN is also known as Ton 28 and, specifically, the two 2017 observations were analysed by \citet{nardini2019}, who found an absorption feature at $\sim\!6.97$ keV that they identified as blueshifted absorption by \ion{Fe}{XXVI} associated with a UFO with a velocity of $\sim\!0.25-0.30\,c$.
For these two observations we find that a good description of the hard X-ray spectrum is provided by a power law modified by Galactic absorption plus an emission feature at $\sim\!6.82-6.86$ keV, that \citet{nardini2019} ascribed to a superposition of fluorescent K$\alpha$ lines from \ion{Fe}{XXV-XXVI}, confirming the existence of highly ionised gas in the nuclear regions of Ton 28. 
In our simplified spectral modelling, even if some negative residuals around $9.2$ keV are actually visible, we cannot constrain the UFO absorption feature and including such an absorption component does not provide a better spectral fit in terms of the W-statistic. For this reason we refer the reader to the paper by \citet{nardini2019} for a detailed spectroscopic analysis of the UFO absorption feature.
We find a considerable excess of soft emission when we extend our spectral analysis to the soft X-rays. While in the first observation such soft excess can be well described by a single blackbody component with $kT_\mathrm{bb}\sim126$ eV, we need to include two blackbody components of temperatures $\sim\!112$ and $\sim\!260$ eV to account for the large soft X-ray emission. The photon index $\Gamma$ undergoes a small variation, moving from a value of $\sim\!2.3$ in the first observation to $\sim\!2.2$ in the second.
For the most recent 2018 observation, a good description of the broadband X-ray continuum is provided by a model consisting of a power law of photon index $\sim\!2.5$ modified by Galactic absorption plus a blackbody component of temperature $\sim\!126$ eV to account for the soft excess. We also find that the hard portion of the X-ray spectrum is characterised by the appearance of an emission feature at $\sim\!6.54$ keV that we model with a Gaussian of $EW\sim340$ eV and is likely ascribable to Fe K$\alpha$ emission from ionised iron. The width of the line is unconstrained, and
we fix it to a value consistent with the energy resolution of the instruments of 100 eV. 
In the 2018 observation we find no trace of the UFO absorption feature that was present in the previous observations.

\paragraph{X-HESS 20.}

This X-HESS AGN is found at redshift $z\!\sim\!0.1$ and has been observed twice by \emph{XMM-Newton} in June 2001 and January 2016, respectively.
This source is a well known NLSy1, frequently referred to as PG 1404+226, that has been extensively studied for its peculiar spectral and emission features \citep[e.g.][and references therein]{dasgupta2005, crummy2005, mallick2018}.
Concerning the 2001 \emph{XMM-Newton} observation, the authors found that the broadband X-ray spectrum of PG 1404+226 could be described by a baseline model consisting of a power-law continuum modified by Galactic absorption, plus a blackbody component to account for the unusually strong excess of soft X-ray emission.
The results from our spectroscopic analysis agree with their previous findings, and we find a value of $\Gamma\!\sim\!1.74$ and a blackbody temperature of $kT_\mathrm{bb}\!\sim\!114$ eV, both consistent within the errors to what reported in e.g. \citet{dasgupta2005}.
Moreover, the same authors did also find negative residuals at around 1 and 3 keV, respectively, that were modelled with three Gaussian absorption lines at $\sim\!1$, $\sim\!1.17$ and $\sim\!3.07$ keV.
We also find these residuals, but in our case only the absorption feature at $\sim1$ keV provides a significant improvement to the spectral fit, i.e.\ $\Delta{W} \sim 33$ for two degrees of freedom. By modelling such an absorption feature with a Gaussian component (\texttt{zgauss} in XSPEC), we find that the line centroid is at $E\sim1.01$ keV, possibily ascribable to absorption by ionised Ne or Fe, with $\sigma\sim140$ eV and an equivalent width of $EW\sim-90$ eV.
\citet{dasgupta2005} reported on a considerable intra-observation flux variability that we also observe in the EPIC light curves, and they also concluded that there might be a connection between the flux and spectral variability, as the absorption features get more pronounced in the high-flux state.
The most recent 2016 \emph{XMM-Newton} observation has been analysed in great detail by \citet{mallick2018}, who carried out both a spectral and timing study. They found a large short-term, intra-observation flux variability consisting of an exponential increase of the X-ray (0.3–8 keV) by a factor of $\sim\!7$ in about 10 ks followed by a sharp drop. A baseline model similar to the one adopted by \citet{dasgupta2005} was used to model the broadband X-ray spectrum of PG 1404+226. In this case, however, the soft X-ray emission below $\sim\!1$ keV was much more complex, probably due to the longer exposure of the 2016 observation, and revealed the presence of a WA, consisting of a highly ionised ($\xi \sim 600$ erg cm s$^{-1}$) \ion{Ne}{X} Ly$\alpha$ absorbing cloud along the line of sight with a column density of $N_\mathrm{H} \sim 5\times 10^{22}$ cm$^{-2}$, that the authors accounted for by means of an \emph{ad-hoc} model in \texttt{XSTAR}.
For our purposes, a phenomenological description of the broadband X-ray spectrum by means of a power-law continuum modified by Galactic absorption, plus a blackbody component to account for the excess of soft X-ray emission and the \texttt{zxipcf} component to deal with the ionised absorption of the WA is sufficient to provide a good spectral fit and we refer the reader to the paper by \citet{mallick2018} for a more detailed treatment.
We find that the photon index has a value of $\Gamma\sim1.68$ and the blackbody temperature is around $kT_\mathrm{bb}\sim115$ eV, consistent within the 90\% confidence level with the values reported by \citet{mallick2018}. The \texttt{zxipcf} component indicates the presence of an absorbing gas cloud with a column density of $N_\mathrm{H} \sim 36\times 10^{22}$ cm$^{-2}$ and ionisation parameter of $\log(\xi/\mathrm{erg\,cm\,s}^{-1})\sim3.1$.
In addition, we also find negative residuals around $\sim\!1$ keV that we account for by including a Gaussian absorption line, with centroid energy of $E\sim1.06$ keV, $\sigma\sim120$ eV and an equivalent width of $EW\!\sim\!-80$ eV.
Between the two observations the soft X-ray flux has dropped by a factor of $\sim\!2$, while the hard X-ray emission has maintained approximately the same level.

\paragraph{X-HESS 21.}

This AGN is a NLSy1 galaxy, also known as PG 1448+273, that has been extensively studied by means of detailed spectral analyses due to the presence of a powerful and variable UFO in its X-ray spectrum \citep[for an indepth discussion, see e.g.][]{kosec2020, laurenti2021, reeves2023}.
Here it will be sufficient to examine the differences between the phenomenological descriptions of the broadband X-ray spectrum for the two available \emph{XMM-Newton} observations in the X-HESS sample.
The 2003 observation was studied by \citet{inoue2007}, who found that the broadband X-ray spectrum of PG 1448+273 was described well by a baseline model consisting of a power-law continuum modified by Galactic absorption, plus two blackbody components two account for the complex excess of soft X-ray emission, a Gaussian emission line at $E\sim6$ keV to include the Fe K$\alpha$ emission and an edge component to deal with an absorption feature in the soft X-rays.
Our independent analysis is in partial agreement with their findings and the best-fit model is approximately the same described by \citet{inoue2007}, expect for the lack of the need to include an edge component in the soft X-ray portion of the spectrum. We find the photon index has a value of $\Gamma\sim2.27$ and the blackbody temperatures are $kT_\mathrm{bb} \sim 80$ eV and $kT_\mathrm{bb}\sim170$ eV, respectively.
Including a Gaussian emission line at around $6.4$ keV leads to a net improvement of the spectral fit, with a $\Delta W \sim 16$ for two degrees of freedom, but we need to fix the line width to the value of 200 eV returned by the fit, similarly to the approach by \citet{inoue2007}, otherwise the line energy cannot be resolved.
We find the line centroid energy at $E\sim6.49$ keV with an equivalent width of around $EW\!\sim\!190$ eV.
For the most recent 2017 observation, we find that a good description of the broadband X-ray spectrum is provided by a similar model, but in this case a single blackbody component is necessary to account for the soft X-ray emission and an additional Gaussian line is needed to model the UFO absorption feature \citep[][]{kosec2020,laurenti2021, reeves2023}.
We find that the photon index has a value of $\Gamma\sim1.9$ and the blackbody temperature is now of $kT_\mathrm{bb}\sim110$ eV. To constrain the emission line we must fix its width to 100 eV, the value returned by the fit, while the absorption line has a width of $\sigma\sim260$ eV. The line centroid energies are $E\sim6.65$ keV and $E\sim7.48$ keV, respectively, and the former has an equivalent width is $EW\sim100$ eV, while the latter has $EW\sim-340$ eV.
The soft X-ray flux has dropped by a factor of $\sim2$ between the two observations and the hard X-ray flux has diminished by around $\sim50\%$ of its original value observed in 2003.

A more recent joint \emph{XMM-Newton} and \emph{NuSTAR} observation of PG 1448+273 was carried out in 2022 and \citet{reeves2023} discovered that the UFO is persistent, significantly faster than in the previous 2017 observation and rapidly variable on timescales down to 10 ks.

\paragraph{X-HESS 25.}

This AGN is found at $z\sim0.7$ and has been observed in July 2012. We find that the broadband X-ray continuum may be described by a simple model comprising a power law with $\Gamma\sim1.9$ modified by Galactic absorption.
Positive residuals emerge below $\sim4$ keV and we include a Gaussian line to deal with them. We obtain a significant improvement of the spectral fit by including the line, being $\Delta{W}\sim15$ for two degrees of freedom. The line centroid is at $E\sim3.8$ keV, with a width of $\sigma\sim190$ eV that we must keep fixed to the best-fit value otherwise it cannot be constained, and the equivalent width is $EW\sim270$ eV. This line can be tentatively identified with a contribution from \ion{Ca}{XIX} emission at $\sim3.9$ keV.

\paragraph{X-HESS 30.}

This AGN is located at $z\sim0.3$ and has been observed in September 2016. We find that the broadband X-ray continuum may be described by a simple model comprising a power law with $\Gamma\sim2.3$ modified by Galactic absorption.
Negative residuals emerge around $\sim1$ keV and we include a Gaussian line to take them into account. The spectral fit improves by including the line, being $\Delta{W}>10$ for two degrees of freedom. The line centroid is at $E\sim1.10$ keV, with an unresolved width, and the equivalent width is $EW\sim-70$ eV. This line can be tentatively ascribed to absorption by \ion{Fe}{XVII} and/or \ion{Fe}{XVIII}.

\paragraph{X-HESS 32.}

This AGN lies at $z\sim1.50$ and has been observed in October 2014. We find that the broadband X-ray continuum may be described by a simple model comprising a power law with $\Gamma\sim1.9$ modified by Galactic absorption.
Positive residuals emerge around $\sim6.4$ keV and we deal with them by adding a Gaussian line to the baseline model. We obtain a significant improvement of the spectral fit by including the line, being $\Delta{W}\sim16$ for two degrees of freedom. The line centroid is at $E\sim6.58$ keV, the width is unresolved, and the equivalent width is $EW\sim60$ eV. This line can be tentatively identified with a contribution from Fe K$\alpha$ emission from ionised iron, likely \ion{Fe}{XXV}.

\paragraph{X-HESS 37.}

This X-HESS AGN lies at $z\sim0.16$ and has been observed by \emph{XMM-Newton} in October 2005. 
The broadband X-ray continuum may be described by a baseline model consisting of a power law modified by Galactic absorption, plus two blackbody components necessary to account for the excess of soft X-ray emission.
The photon index has a value of $\Gamma\sim2.3$ and the blackbody temperature of the two components is $kT_\mathrm{bb}\sim134$ eV and $kT_\mathrm{bb}\sim70$ eV, respectively.
We find positive residuals around $\sim1$ keV that we model with a Gaussian line. The fit improvement is significant, i.e.\ $\Delta{W}\sim25$ for three degrees of freedom, by including the emission line in the best-fit model.
The line width is $\sigma\sim100$ eV and the centroid energy is $E\sim0.94$ keV, with an equivalent width of $EW\sim50$ eV, and is likely due to emission by ionised iron.

\paragraph{X-HESS 54.}

This source is an AGN at redshift $z\sim3.3$ that has been observed by \emph{XMM-Newton} in December 2017. 
We find that the broadband X-ray spectrum can be described well by a baseline model consisting of a power-law continuum with $\Gamma\sim1.7$ modified by Galactic absorption.
The soft portion of the X-ray spectrum shows clear signatures of absorption that we try to model with different components.
The spectral fit has a significant improvement, being $\Delta{W}\sim20$ for one degree of freedom, by including absorption by a cold gas cloud (\texttt{zTBabs} in XSPEC).
We find that such absorber has a column width of around $N_\mathrm{H}\sim6\times10^{22}$ cm$^{-2}$.

\paragraph{X-HESS 56.}

This AGN is at redshift $z\sim1.6$ and has been observed by \emph{XMM-Newton} in January 2016. 
We find that a satisfactory description of the broadband X-ray continuum is provided by a baseline model consisting of a power-law with $\Gamma\sim1.8$ modified by Galactic absorption.
The hard portion of the X-ray spectrum is dominated by a broad absorption feature below $\sim10$ keV in the rest frame. 
We obtain a significant improvement of the spectral fit, being $\Delta{W}\sim40$ for two degrees of freedom, by including an absorption edge in the best-fit model with \texttt{zedge} in XSPEC, that is likely ascribable to \ion{Fe}{XXVI}.
The energy of the absorption edge is $E\sim9.3$ keV with $\tau\sim 1.1$.
The presence of this spectral feature may be indicative of resonant absorption lines due to the same ion species, thus potentially suggesting that the absorption edge can represent a broad and blueshifted absorption line likely linked to a UFO. However, if we substitute the edge with a Gaussian line in absorption, this leads to a worsening of the spectral fit and the impossibility to constrain the line energy and width.

\setcounter{table}{0} 
\setcounter{figure}{0} 
\renewcommand{\thetable}{A.\arabic{table}}
\renewcommand{\thefigure}{A.\arabic{figure}}
\renewcommand{\theHfigure}{A.\arabic{figure}}

\begin{table*}[htbp]
\centering
\renewcommand{\arraystretch}{1.5}
\caption{X-HESS: comprehensive AGN list and their general properties.}
\label{tab:agn_list}
\begin{adjustbox}{max width=\textwidth}
\begin{threeparttable}
\begin{tabular}{c c c c c c c c c c c }
\hline
  \multicolumn{1}{c}{ID} &
  \multicolumn{1}{c}{SDSS Name} &
  \multicolumn{1}{c}{RA} &
  \multicolumn{1}{c}{DEC} &
  \multicolumn{1}{c}{$N_\mathrm{H,gal}$} &
  \multicolumn{1}{c}{$E(B-V)_\mathrm{int}$} &
  \multicolumn{1}{c}{$z$} &
  \multicolumn{1}{c}{$\log(M_\mathrm{BH}/M_\odot)$} &
  \multicolumn{1}{c}{$\log(L_\mathrm{bol}/\mathrm{erg\,s}^{-1})$} &
  \multicolumn{1}{c}{$\lambda_\mathrm{Edd}$} &
  \multicolumn{1}{c}{$N_\mathrm{epo}$} \\
  
  \multicolumn{1}{c}{(1)} &
  \multicolumn{1}{c}{(2)} &
  \multicolumn{1}{c}{(3)} &
  \multicolumn{1}{c}{(4)} &
  \multicolumn{1}{c}{(5)} &
  \multicolumn{1}{c}{(6)} &
  \multicolumn{1}{c}{(7)} &
  \multicolumn{1}{c}{(8)} &
  \multicolumn{1}{c}{(9)} &
  \multicolumn{1}{c}{(10)} &
  \multicolumn{1}{c}{(11)} \\
\hline\hline
  1  & J094610.71+095226.3 & 146.54 & 9.87      & 2.36 & 0.02    & 0.697   & 7.5  & 45.6  & 1.1  & 15\\
  2  & J095847.88+690532.7 & 149.70 & 69.09     & 7.00 & <0.01   & 1.2882  & 9.1  & 47.1  & 0.9  & 13\\
  3  & J122549.87+332454.9 & 186.46 & 33.42     & 2.47 & <0.01   & 1.1329  & 7.7  & 45.7  & 0.8  & 8\\
  4  & J113233.55+273956.3 & 173.14 & 27.67     & 1.65 & <0.01   & 0.681   & 7.7  & 45.9  & 1.3  & 7\\
  5  & J130048.10+282320.6 & 195.20 & 28.39     & 1.04 & 0.01    & 1.929   & 9.3  & 47.3  & 0.7  & 7\\
  6  & J172255.24+320307.5 & 260.73 & 32.05     & 3.23 & 0.33    & 0.2752  & 6.8  & 44.7  & 0.8  & 6\\
  7  & J023410.58$-$085213.5 & 38.54 & $-$8.87  & 3.46 & 0.33    & 0.992   & 7.1  & 45.1  & 0.7  & 5\\
  8  & J005324.60+123343.5 & 13.35 & 12.56      & 4.61 & 0.12    & 0.68    & 7.2  & 45.1  & 0.7  & 4\\
  9  & J021702.01+015352.0 & 34.26 & 1.90       & 3.40 & <0.01   & 0.723   & 7.1  & 45.0  & 0.7  & 4\\
  10 & J021704.90+014935.2 & 34.27 & 1.83       & 3.39 & 0.2     & 0.639   & 6.9  & 44.9  & 0.8  & 4\\
  11 & J154530.23+484608.9 & 236.38 & 48.77     & 1.59 & <0.01   & 0.3993  & 8.3  & 46.4  & 0.9  & 4\\
  12 & J171522.17+573626.7 & 258.84 & 57.61     & 2.27 & 0.02    & 2.0284  & 9.0  & 47.4  & 1.9  & 4\\
  13 & J100402.61+285535.3 & 151.01 & 28.93     & 1.76 & <0.01   & 0.3287  & 8.2  & 46.4  & 1.4  & 3\\
  14 & J221715.18+002615.0 & 334.31 & 0.44      & 3.55 & 0.21    & 0.7532  & 7.4  & 45.5  & 1.1  & 3\\
  15 & J221738.41+001206.5 & 334.41 & 0.20      & 3.97 & <0.01   & 1.1211  & 7.3  & 45.7  & 1.7  & 2\\
  16 & J022928.41$-$051125.0 & 37.37 & $-$5.19  & 2.38 & 0.26    & 0.307   & 7.4  & 45.4  & 0.7  & 2\\
  17 & J074545.01+392700.9 & 116.44 & 39.45     & 5.69 & 0.02    & 1.6299  & 7.9  & 46.5  & 2.8  & 2\\
  18 & J114229.22+264012.4 & 175.62 & 26.67     & 1.75 & <0.01   & 1.6756  & 8.5  & 46.5  & 0.9  & 2\\
  19 & J123034.20+073305.3 & 187.64 & 7.55      & 1.63 & 0.01    & 1.816   & 8.9  & 47.2  & 1.6  & 2\\
  20 & J140621.89+222346.5 & 211.59 & 22.40     & 1.93 & <0.01   & 0.0979  & 7.2  & 45.1  & 0.7  & 2\\
  21 & J145108.76+270926.9 & 222.79 & 27.16     & 3.01 & <0.01   & 0.0645  & 7.1  & 45.3  & 1.0  & 2\\
  22 & J233317.38$-$002303.4 & 353.32 & $-$0.38 & 3.73 & 0.28    & 0.5129  & 6.8  & 44.9  & 1.0  & 2\\
  23 & J002209.69+013213.0 & 5.54  & 1.54       & 2.59 & 0.09    & 1.8826  & 8.7  & 46.7  & 0.8  & 1\\
  24 & J014634.38--093014.3 & 26.64 & --9.50    & 2.72 & <0.01   & 0.3934  & 7.2  & 45.2  & 0.8  & 1\\
  25 & J014904.48+125746.2 & 27.27 & 12.96      & 5.32 & 0.01    & 0.73    & 7.6  & 45.5  & 0.7  & 1\\
  26 & J015828.31--014810.0 & 29.62 & --1.80    & 2.44 & 0.01    & 1.7779  & 8.9  & 47.2  & 1.5  & 1\\
  27 & J020326.22--051020.5 & 30.86 & --5.17    & 1.97 & 0.06    & 0.891   & 8.1  & 46.3  & 1.2  & 1\\
  28 & J022039.48--030820.3 & 35.16 & --3.14    & 2.11 & 0.01    & 0.452   & 7.7  & 45.7  & 0.9  & 1\\
  29 & J024651.91--005930.9 & 41.72 & --0.99    & 3.64 & <0.01   & 0.4679  & 8.1  & 46.4  & 1.5  & 1\\
  30 & J034000.50--051747.1 & 55.00 & --5.30    & 4.20 & <0.01   & 0.345   & 7.7  & 45.7  & 0.9  & 1\\
  31 & J081014.48+280337.1 & 122.56 & 28.06     & 3.14 & <0.01   & 0.821   & 8.6  & 46.5  & 0.7  & 1\\
  32 & J081331.28+254503.0 & 123.38 & 25.75     & 3.82 & 0.08    & 1.5096  & 9.6  & 47.9  & 1.4  & 1\\
  33 & J083850.15+261105.4 & 129.71 & 26.18     & 3.18 & <0.01   & 1.6116  & 9.6  & 47.6  & 0.9  & 1\\
  34 & J084153.99+194303.1 & 130.47 & 19.72     & 2.56 & 0.13    & 0.4784  & 7.2  & 45.2  & 0.8  & 1\\
  35 & J085626.07+380519.6 & 134.11 & 38.09     & 3.17 & <0.01   & 0.283   & 7.0  & 44.9  & 0.8  & 1\\
  36 & J090033.50+421547.0 & 135.14 & 42.26     & 2.37 & 0.02    & 3.293   & 9.3$^{(a)}$ & 48.1$^{(b)}$ & 3.1$^{(b)}$ & 1\\
  37 & J092247.03+512038.0 & 140.70 & 51.34     & 1.36 & <0.01   & 0.1597  & 7.0 & 45.2 & 1.3 & 1\\
  38 & J092943.41+004127.3 & 142.43 & 0.69      & 3.36 & <0.01   & 0.5866  & 8.2 & 46.5 & 1.3 & 1\\
  39 & J093922.89+370944.0 & 144.85 & 37.16     & 1.11 & <0.01   & 0.1859  & 7.2 & 45.2 & 0.8 & 1\\
  
\hline\end{tabular}

\end{threeparttable}
\end{adjustbox}
\end{table*}

\begin{table*}[htbp]
\centering
\renewcommand{\arraystretch}{1.5}
\caption*{Table \ref{tab:agn_list} (Continued).}
\begin{adjustbox}{max width=\textwidth}
\begin{threeparttable}
\begin{tabular}{c c c c c c c c c c c }
\hline
  \multicolumn{1}{c}{ID} &
  \multicolumn{1}{c}{SDSS Name} &
  \multicolumn{1}{c}{RA} &
  \multicolumn{1}{c}{DEC} &
  \multicolumn{1}{c}{$N_\mathrm{H,gal}$} &
  \multicolumn{1}{c}{$E(B-V)_\mathrm{int}$} &
  \multicolumn{1}{c}{$z$} &
  \multicolumn{1}{c}{$\log(M_\mathrm{BH}/M_\odot)$} &
  \multicolumn{1}{c}{$\log(L_\mathrm{bol}/\mathrm{erg\,s}^{-1})$} &
  \multicolumn{1}{c}{$\lambda_\mathrm{Edd}$} &
  \multicolumn{1}{c}{$N_\mathrm{epo}$} \\
  
  \multicolumn{1}{c}{(1)} &
  \multicolumn{1}{c}{(2)} &
  \multicolumn{1}{c}{(3)} &
  \multicolumn{1}{c}{(4)} &
  \multicolumn{1}{c}{(5)} &
  \multicolumn{1}{c}{(6)} &
  \multicolumn{1}{c}{(7)} &
  \multicolumn{1}{c}{(8)} &
  \multicolumn{1}{c}{(9)} &
  \multicolumn{1}{c}{(10)} &
  \multicolumn{1}{c}{(11)} \\
\hline\hline
  
  40 & J094033.75+462315.0 & 145.14 & 46.39   & 1.06 & <0.01  &  0.696  & 8.3$^{(c)}$ \iffalse8.12\fi & 46.4$^{(c)}$\iffalse46.47\fi & 1.1$^{(c)}$\iffalse1.61\fi & 1\\
  41 & J103928.14+392342.1 & 159.87 & 39.40   & 1.60 & <0.01  &  1.0134 & 8.7 & 46.9 & 1.0 & 1\\
  42 & J110035.00+101027.4 & 165.15 & 10.17   & 2.44 & <0.01  &  1.5873 & 8.6 & 46.8 & 1.3 & 1\\
  43 & J110312.93+414154.9 & 165.80 & 41.70   & 0.81 & <0.01  &  0.402  & 8.3$^{(c)}$ \iffalse8.15\fi & 46.4$^{(c)}$ \iffalse46.28\fi & 1.0$^{(c)}$ \iffalse0.96\fi & 1\\
  44 & J112306.33+013749.6 & 170.78 & 1.63    & 3.61 & 0.01   &  0.696  & 7.6 & 45.7 & 1.0 & 1\\
  45 & J112317.51+051804.0 & 170.82 & 5.30    & 3.23 & <0.01  &  1.0002 & 9.0 & 47.0 & 0.7 & 1\\
  46 & J112405.15+061248.8 & 171.02 & 6.21    & 3.91 & 0.02   &  0.2717 & 7.1 & 45.2 & 1.0 & 1\\
  47 & J112818.49+240217.4 & 172.08 & 24.04   & 1.13 & 0.02   &  1.6077 & 8.9 & 46.9 & 0.7 & 1\\
  48 & J120734.62+150643.7 & 181.89 & 15.11   & 2.26 & 0.14   &  0.75   & 8.4$^{(c)}$ \iffalse8.55\fi & 46.5$^{(c)}$ \iffalse46.63\fi & 0.9$^{(c)}$ \iffalse0.87\fi & 1\\
  49 & J120858.01+454035.4 & 182.24 & 45.68   & 1.28 & <0.01  &  1.162  & 9.7 & 47.6 & 0.7 & 1\\
  50 & J124615.77+673032.7 & 191.57 & 67.51   & 1.55 & 0.05   &  1.7684 & 8.9 & 46.9 & 0.7 & 1\\
  51 & J125005.72+263107.5 & 192.52 & 26.52   & 0.94 & <0.01  &  2.0476 & 9.8 & 48.2 & 1.7 & 1\\
  52 & J125216.58+052737.7 & 193.07 & 5.46    & 1.87 & <0.01  &  1.9    & 9.4 & 47.6 & 1.2 & 1\\
  53 & J130112.91+590206.6 & 195.30 & 59.04   & 1.40 & <0.01  &  0.476  & 8.5$^{(c)}$ \iffalse8.67\fi & 46.7$^{(c)}$ \iffalse46.75\fi & 1.2$^{(c)}$ \iffalse0.87\fi & 1\\
  54 & J132654.95--000530.1 & 201.72 & --0.09 & 1.69 & 0.18   &  3.307  & 9.3$^{(a)}$ & 48.3$^{(b)}$ & 2.1$^{(b)}$ & 1\\
  55 & J133618.53+101742.6 & 204.08 & 10.30   & 2.22 & <0.01  &  0.5827 & 7.7 & 45.8 & 0.9 & 1\\
  56 & J135306.34+113804.7 & 208.28 & 11.63   & 1.69 & 0.03   &  1.633  & 9.4 & 47.4 & 0.9 & 1\\
  57 & J144741.76--020339.1 & 221.92 & --2.06 & 4.16 & <0.01  &  1.4266 & 8.7 & 46.6 & 0.7 & 1\\
  58 & J161434.67+470420.0 & 243.64 & 47.07   & 0.95 & 0.05   &  1.861  & 9.7 & 47.8 & 1.0 & 1\\
  59 & J162617.23+142130.8 & 246.57 & 14.36   & 3.93 & 0.04   &  1.6622 & 8.2 & 46.5 & 1.7 & 1\\
  60 & J163201.11+373749.9 & 248.00 & 37.63   & 0.83 & <0.01  &  1.4747 & 9.7 & 47.6 & 0.7 & 1\\
  61 & J223607.68+134355.3 & 339.03 & 13.73   & 4.76 & <0.01  &  0.3258 & 8.0 & 46.2 & 1.3 & 1\\
\hline\end{tabular}

\tablefoot{(1) Unique X-HESS identifier; (2) SDSS IAU name; (3) right ascension; (4) declination; (5) Galactic column density ($10^{20}$ cm$^{-2}$) from the full-sky H{\,\scriptsize{I}} map by \cite{bekhti2016}; (6) internal reddening measured according to the procedure in \S\,\ref{sec:l2500_aox} (the typical uncertainty is $\sim0.01$); (7) redshift; (8) black hole mass; (9) bolometric luminosity; (10) Eddington ratio; (11) total number of available epochs.
$^{(a)}$ \ion{H}{$\beta$}-based black hole mass measurement from \citet{bischetti2017}. 
$^{(b)}$ bolometric luminosity corrected for the orientation effects and Eddington ratio from \citet{vietri2018}.
$^{(c)}$ refined measurements of \ion{H}{$\beta$}-based black hole mass, bolometric luminosity and Eddington ratio from \citet{marziani2014}.}

\end{threeparttable}
\end{adjustbox}
\end{table*}

\setcounter{table}{0} 
\setcounter{figure}{0} 
\renewcommand{\thetable}{B.\arabic{table}}
\renewcommand{\thefigure}{B.\arabic{figure}}
\renewcommand{\theHfigure}{B.\arabic{figure}}

\clearpage
\onecolumn
\renewcommand{\arraystretch}{1.2}
\begin{longtable}[c]{ c  c  c  c  c  c } \\
\caption{Journal of observations.\label{tab:longjournal}} \\
 
 \multicolumn{6}{c}{} \\
 \hline\hline
 ID & Obs.\ No. & Obs.\ ID & Start Date & Exp &  Cts$\lvert_\mathrm{tot}$  \\
    (1) & (2) & (3) & (4)& (5)  & (6) \\
 \hline
 \endfirsthead

 \multicolumn{6}{c}{Continuation of Table \ref{tab:longjournal}}\\
 \hline\hline
 ID & Obs.\ No. & Obs.\ ID & Start Date & Exp &  Cts$\lvert_\mathrm{tot}$  \\
    (1) & (2) & (3) & (4)& (5)  & (6) \\
 \hline
 \endhead

 \hline
 \endfoot

 \multicolumn{6}{ c }{}\\
 
 \endlastfoot
   
    1 & 1  & 0150970101 & 2003-05-07 & --/9.7/9.7      & --/69/53      \\
    1 & 2  & 0150970301 & 2003-12-02 & 17.0/23.4/23.4 & 577/192/203  \\
    1 & 3  & 0671540201 & 2011-11-04 & 13.5/21.0/21.0 & 398/159/123  \\
    1 & 4  & 0671540301 & 2011-11-06 & 20.8/25.0/25.1 & 1150/408/362 \\
    1 & 5  & 0671540401 & 2011-11-21 & 17.5/23.4/24.0 & 610/178/182  \\
    1 & 6  & 0671540501 & 2011-11-27 & 17.4/22.0/22.0 & 518/184/188  \\
    1 & 7  & 0671540601 & 2011-12-01 & 13.2/19.7/20.2 & 582/181/178  \\
    1 & 8  & 0671540701 & 2011-12-04 & 2.7/6.6/9.4    & 69/32/46     \\
    1 & 9  & 0743950101 & 2014-10-31 & 23.9/64.3/62.2 & 1348/952/750 \\
    1 & 10 & 0743950201 & 2014-11-03 & 32.6/57.7/56.4 & 1020/462/384 \\
    1 & 11 & 0743950301 & 2014-11-05 & 20.7/33.5/31.2 & 954/422/369  \\
    1 & 12 & 0743950401 & 2014-11-21 & 38.6/53.3/53.3 & 894/288/233  \\
    1 & 13 & 0743950501 & 2014-11-23 & 39.3/57.1/--   & 682/251/--    \\
    1 & 14 & 0743950601 & 2014-11-25 & 46.9/60.7/60.8 & 1743/499/391 \\
    1 & 15 & 0743950701 & 2014-11-27 & 44.1/61.0/58.6 & 2302/655/525 \\
    2 & 1  & 0112521001 & 2002-04-10 & --/10.0/10.0     & --/143/154      \\
    2 & 2  & 0112521101 & 2002-04-16 & 7.6/9.7/9.6      & 409/102/136    \\
    2 & 3  & 0200980101 & 2004-09-26 & 70.3/102.3/103.0 & 3837/1488/1773 \\
    2 & 4  & 0657802001 & 2011-03-24 & 3.2/5.6/8.3      & 45/36/22       \\
    2 & 5  & 0657801601 & 2011-04-17 & 2.1/4.3/4.8      & 61/48/34       \\
    2 & 6  & 0657801801 & 2011-09-26 & 12.4/--/15.7     & 342/--/137      \\
    2 & 7  & 0657802201 & 2011-11-23 & 12.9/15.9/15.9   & 234/109/101    \\
    2 & 8  & 0693850801 & 2012-10-23 & 6.0/8.4/8.7      & 193/88/97      \\
    2 & 9  & 0693850901 & 2012-10-25 & 7.1/13.6/13.6    & 316/174/179    \\
    2 & 10 & 0693851001 & 2012-10-27 & 3.9/10.2/11.1    & 109/82/97      \\
    2 & 11 & 0693851701 & 2012-11-12 & --/9.5/9.4       & --/62/63        \\
    2 & 12 & 0693851801 & 2012-11-14 & 6.8/8.8/8.8      & 201/69/85      \\
    2 & 13 & 0693851101 & 2012-11-16 & 2.9/4.9/5.0      & 66/48/56       \\
    3 & 1  & 0112521901 & 2002-05-31 & 10.3/15.0/14.9   & 30/12/14       \\
    3 & 2  & 0142830101 & 2003-11-30 & 90.0/101.8/102.3 & 387/83/115     \\
    3 & 3  & 0744010101 & 2014-12-28 & --/--/51.1       & --/--/35         \\
    3 & 4  & 0744010201 & 2014-12-30 & --/--/20.1       & --/--/17         \\
    3 & 5  & 0824610101 & 2018-12-13 & 83.5/--/104.2    & 130/--/47       \\
    3 & 6  & 0824610201 & 2018-12-19 & 59.9/--/--       & 100/--/--        \\
    3 & 7  & 0824610301 & 2018-12-31 & --/--/81.9       & --/--/26         \\
    3 & 8  & 0824610401 & 2019-01-02 & --/--/110.2      & --/--/42         \\
    4 & 1  & 0693380301 & 2013-05-13 & --/28.3/28.4     & --/240/273    \\
    4 & 2  & 0693380501 & 2013-06-20 & 13.2/16.2/16.2 & 743/244/224  \\
    4 & 3  & 0727960701 & 2013-11-14 & 15.2/25.3/25.1 & 1357/624/652 \\
    4 & 4  & 0727960801 & 2013-11-16 & 11.1/18.4/18.2 & 1659/708/594 \\
    4 & 5  & 0764850201 & 2015-05-31 & 46.9/59.7/59.3 & 1597/483/542 \\
    4 & 6  & 0764850301 & 2015-12-12 & 23.5/33.1/31.8 & 796/309/254  \\
    4 & 7  & 0764850401 & 2015-12-24 & 9.5/14.2/14.2  & 248/99/88    \\
    5 & 1  & 0124712501 & 2002-06-07 & --/27.7/27.7    & --/143/127    \\
    5 & 2  & 0204040101 & 2004-06-06 & --/77.4/77.2    & --/74/97      \\
    5 & 3  & 0204040201 & 2004-06-18 & --/--/54.2      & --/--/78       \\
    5 & 4  & 0204040301 & 2004-07-12 & 46.6/--/--      & 178/--/--      \\
    5 & 5  & 0304320301 & 2005-06-27 & 26.1/--/--       & 562/--/--      \\
    5 & 6  & 0304320201 & 2005-06-28 & 58.7/--/--       & 1542/--/--     \\
    5 & 7  & 0304320801 & 2006-06-06 & 35.9/--/--       & 161/--/--      \\
    6 & 1  & 0093030301 & 2003-08-15 & --/4.9/5.0   & --/34/46  \\
    6 & 2  & 0093031001 & 2003-08-29 & 1.9/5.1/4.3 & 34/23/24 \\
    6 & 3  & 0093031101 & 2003-09-06 & --/2.7/2.8   & --/23/27  \\
    6 & 4  & 0093031401 & 2003-10-08 & --/3.1/3.2   & --/61/92  \\
    6 & 5  & 0093031501 & 2003-10-10 & 1.6/4.2/4.3 & 63/86/75 \\
    6 & 6  & 0693180901 & 2012-08-07 & 24.2/--/--    & 2557/--/-- \\
    7 & 1  & 0150470601 & 2003-07-15 & --/37.9/25.7    & --/248/125  \\
    7 & 2  & 0690870101 & 2013-07-20 & 15.9/19.3/19.3 & 197/78/78  \\
    7 & 3  & 0690870501 & 2013-08-10 & --/101.3/101.3  & --/451/381  \\
    7 & 4  & 0743830501 & 2015-01-13 & --/117.1/116.4  & --/617/595  \\
    7 & 5  & 0743830601 & 2015-01-25 & --/110.2/110.4  & --/661/659  \\
    8 & 1  & 0110890301 & 2002-06-22 & 17.9/20.9/21.0 & 1009/365/363  \\
    8 & 2  & 0300470101 & 2005-07-18 & 79.1/--/--       & 704/--/--       \\
    8 & 3  & 0743050301 & 2015-01-19 & 134.9/--/--      & 982/--/--      \\
    8 & 4  & 0743050801 & 2015-01-21 & 111.3/--/--      & 876/--/--       \\
    9 & 1  & 0763450301 & 2017-07-18 & 114.6/--/--      & 477/--/--       \\
    9 & 2  & 0784030101 & 2017-07-21 & 17.3/21.3/21.4 & 78/26/20      \\
    9 & 3  & 0784030501 & 2017-08-04 & 17.4/21.7/20.9 & 72/17/20      \\
    9 & 4  & 0784030601 & 2017-08-15 & --/22.3/22.2    & --/30/8        \\
    10& 1  & 0763450301 & 2017-07-18 & 114.5/--/--      & 541/--/--   \\
    10& 2  & 0784030101 & 2017-07-21 & 17.3/21.3/21.4 & 100/29/21 \\
    10& 3  & 0784030501 & 2017-08-04 & 17.4/21.7/20.9 & 84/18/14  \\
    10& 4  & 0784030601 & 2017-08-15 & 17.7/22.3/22.2 & 84/44/43  \\
    11& 1  & 0153220401 & 2003-02-08 & 2.4/8.6/8.9  & 661/628/678  \\
    11& 2  & 0505050201 & 2007-06-09 & 4.5/9.1/9.3  & 1685/933/864 \\
    11& 3  & 0505050701 & 2007-06-15 & 6.2/10.2/9.3 & 2255/974/884 \\
    11& 4  & 0505050301 & 2007-06-17 & --/14.7/14.4  & --/1157/1277  \\
    12& 1  & 0305750401 & 2005-06-23 & 2.7/--/6.0       & 23/--/16      \\
    12& 2  & 0741580101 & 2014-12-04 & 9.4/12.1/12.1   & 269/97/106   \\
    12& 3  & 0764910201 & 2015-10-23 & 33.1/51.4/48.8  & 430/222/191  \\
    12& 4  & 0792790101 & 2016-10-12 & 48.0/--/68.0     & 961/--/446    \\
    13& 1  & 0804560101 & 2017-05-14 & 21.5/30.0/32.5 & 18706/5356/6060  \\
    13& 2  & 0804560201 & 2017-05-15 & 17.0/23.4/23.4 & 577/192/203      \\
    13& 3  & 0823090301 & 2018-11-21 & 6.7/9.4/10.1   & 8801/2818/2774   \\
    14& 1  & 0094310101 & 2002-11-18 & 56.7/64.6/64.3  & 533/230/221  \\
    14& 2  & 0094310201 & 2002-12-15 & 59.9/71.5/72.2  & 803/393/365  \\
    14& 3  & 0673000142 & 2011-12-07 & 2.4/4.2/4.2     & 99/51/32     \\
    15& 1  & 0094310101 & 2002-11-18 & 57.0/64.5/64.3  & 457/118/128  \\
    15& 2  & 0094310201 & 2002-12-15 & 60.2/71.4/72.1  & 445/119/143  \\
    16& 1  & 0677590132 & 2011-07-23 & 11.0/12.4/12.4  & 893/285/252  \\
    16& 2  & 0820080401 & 2019-01-11 & 9.2/--/--         & 146/--/--      \\
    17& 1  & 0551850101 & 2008-10-17 & 34.2/51.9/53.4  & 380/157/209 \\
    17& 2  & 0551851201 & 2009-03-21 & --/79.1/79.0     & --/184/212   \\
    18& 1  & 0556560101 & 2008-12-10 & 26.9/--/--  &  946/--/-- \\
    18& 2  & 0764100501 & 2015-11-21 & 20.3/--/--  &  676/--/-- \\
    19& 1  & 0722670301 & 2014-01-09 & 39.7/--/--  &  379/--/-- \\
    19& 2  & 0722670601 & 2014-01-13 & 27.9/--/--  &  330/--/-- \\
    20& 1  & 0051760201 & 2001-06-18 & 13.4/17.8/17.1  &  5915/1886/2048  \\
    20& 2  & 0763480101 & 2016-01-25 & 39.8/60.3/60.1  &  12120/3387/3092 \\
    21& 1  & 0152660101 & 2003-02-08 & 17.8/21.1/21.1   &  66454/19520/19243   \\
    21& 2  & 0781430101 & 2017-01-24 & 76.3/107.2/110.0 &  149871/46126/46036  \\
    22& 1  & 0673002332 & 2012-05-25 & 3.7/4.2/4.2  &  529/216/188  \\
    22& 2  & 0673002333 & 2012-05-25 & --/4.2/4.2    &  --/97/106     \\
    23& 1  & 0553230201 & 2008-12-30 & 51.0/60.8/60.5  &  333/120/92  \\
    24& 1  & 0673750101 & 2011-12-24 & 18.1/22.4/22.4  &  1863/593/563  \\
    25& 1  & 0673770301 & 2012-07-02 & 28.6/39.0/38.9  &  407/153/144  \\
    26& 1  & 0762870301 & 2016-02-05 & 19.0/25.2/26.9  &  1056/389/365  \\
    27& 1  & 0677640138 & 2012-01-12 & 10.9/12.4/12.4  &  702/190/210  \\
    28& 1  & 0037982701 & 2002-08-15 & 11.9/17.5/17.8  &  765/262/278  \\
    29& 1  & 0402320101 & 2007-02-02 & 4.9/6.2/6.4  &  1207/366/367  \\
    30& 1  & 0780320101 & 2016-09-02 & --/15.4/16.4  &  --/261/298  \\
    31& 1  & 0152530101 & 2002-10-05 & 18.8/27.7/28.0  &  1475/665/594  \\
    32& 1  & 0728990101 & 2014-10-04 & 45.5/55.0/55.0  &  9109/2909/2914  \\
    33& 1  & 0821730301 & 2019-04-27 & 17.6/22.2/22.6  &  1162/441/445  \\
    34& 1  & 0742570101 & 2015-04-09 & 54.0/--/62.4  &  793/--/231  \\
    35& 1  & 0302581801 & 2005-10-10 & 19.2/--/28.0  &  2163/--/707  \\
    36& 1  & 0803950601 & 2017-11-17 & 15.2/19.7/21.9  &  1703/559/699  \\
    37& 1  & 0300910301 & 2005-10-08 & 22.0/29.5/30.5  &  22801/6724/8556  \\
    38& 1  & 0802220601 & 2017-12-04 & 9.6/12.5/12.5  &  1444/449/479  \\
    39& 1  & 0411980301 & 2006-11-01 & 4.2/6.5/6.5  &  2237/719/809  \\
    40& 1  & 0843830301 & 2019-05-14 & 13.0/19.4/19.5  &  1831/505/469  \\
    41& 1  & 0783520201 & 2016-05-04 & 17.6/21.4/21.4  &  3797/1251/1216  \\
    42& 1  & 0802200801 & 2017-11-28 & 23.6/--/34.6  &  578/--/264  \\
    43& 1  & 0843830501 & 2019-05-05 & 13.7/16.8/16.8  &  999/269/270  \\
    44& 1  & 0145750101 & 2003-06-23 & 22.0/29.7/29.8  &  902/280/289  \\
    45& 1  & 0083000301 & 2001-12-15 & --/30.6/30.8  &  --/387/380  \\
    46& 1  & 0103863201 & 2002-12-19 & 4.5/8.4/8.4  &  576/275/290  \\
    47& 1  & 0822530201 & 2018-05-28 & 29.3/35.1/35.1  &  1568/396/477  \\
    48& 1  & 0843830901 & 2019-07-05 & 10.8/16.4/16.4  &  948/321/331  \\
    49& 1  & 0033540601 & 2002-05-11 & 5.9/7.4/7.4  &  574/159/168  \\
    50& 1  & 0306680201 & 2005-11-29 & 4.2/5.0/5.0  &  616/233/239  \\
    51& 1  & 0143150201 & 2003-06-18 & 20.2/27.2/27.3  &  4093/1516/1496  \\
    52& 1  & 0801790601 & 2017-06-18 & 22.0/27.0/27.2  &  1307/384/424  \\
    53& 1  & 0304570101 & 2006-04-20 & 8.5/11.1/11.2  &  4037/1336/1297  \\
    54& 1  & 0804480101 & 2017-12-30 & 35.6/45.4/45.1  &  373/154/151  \\
    55& 1  & 0761590701 & 2015-06-20 & 18.7/--/22.4  &  680/--/251  \\
    56& 1  & 0762520101 & 2016-01-13 & 26.8/36.2/35.9  &  1846/719/772  \\
    57& 1  & 0782360201 & 2016-07-19 & 30.6/35.6/36.0  &  823/236/223  \\
    58& 1  & 0803910101 & 2017-08-05 & 69.3/85.7/87.1  &  17087/5133/5637  \\
    59& 1  & 0783520301 & 2016-08-26 & --/60.9/60.9  &  --/153/144  \\
    60& 1  & 0033540901 & 2002-01-06 & 11.4/15.3/15.2  &  1414/541/476  \\
    61& 1  & 0153220601 & 2003-05-28 & 7.0/7.5/8.7  &  3975/1171/1301  \\
    \hline

\end{longtable}

\begin{minipage}{\linewidth}
\renewcommand{\footnoterule}{}
\textbf{Notes:} (1) Unique X-HESS identifier; (2) Observation number; (3) \emph{XMM-Newton} observation ID; (4) Start date of the observation; (5) Net exposure (in ks) for the pn/MOS1/MOS2 cameras, respectively; (6) net counts for the pn/MOS1/MOS2 cameras in the broad ($E=0.3-10$ keV) observer-frame energy band.

\end{minipage}


\setcounter{table}{0} 
\setcounter{figure}{0} 
\renewcommand{\thetable}{C.\arabic{table}}
\renewcommand{\thefigure}{C.\arabic{figure}}
\renewcommand{\theHfigure}{C.\arabic{figure}}

\onecolumn
\begin{landscape}

\renewcommand{\arraystretch}{1.5}
\setlength{\tabcolsep}{2.5pt}
\notsotiny
\setlength\LTleft{0pt}              
\setlength\LTright{0pt}             

\begin{longtable}[c]{ c c >{\centering\arraybackslash}p{3cm} c c c c c c c c c c c c c c }\\
\caption{Best fitting spectral parameters for the broadband ($E = 0.3-10$ keV) X-ray spectra of the X-HESS AGN, along with derived quantities from EPIC and OM data.\label{tab:bestfit}} \\

 \multicolumn{17}{c}{}\\
 \hline\hline
 
 ID & \# & Model & W-stat/dof (dof) &  $\Gamma$ & $kT_\mathrm{bb}$ & $F_{0.5-2\,\mathrm{keV}}$ & $F_{2-10\,\mathrm{keV}}$ & $\log{L_{0.5-2\,\mathrm{keV}}}$ & $\log{L_{2-10\,\mathrm{keV}}}$  & $k_\mathrm{bol,X}$ &  $\log{L_{2\mathrm{\,kev}}}$ & $\log{L_{2500\mathrm{\AA}}}$ & $\log{L_{4400\mathrm{\AA}}}$ & $\alpha_\mathrm{ox}$ & $\log{\lambda_\mathrm{Edd,O/UV}}$ & $R_\mathrm{S/P}$ \\
    (1) & (2) & (3) & (4) & (5) & (6) & (7) & (8) & (9) & (10) & (11) & (12) & (13) & (14) & (15) & (16) & (17)  \\
    \hline 
 \endfirsthead

 \multicolumn{17}{c}{Continuation of Table \ref{tab:bestfit}}\\
 \hline\hline
 ID & \# & Model & W-stat/dof (dof) &  $\Gamma$ & $kT_\mathrm{bb}$ & $F_{0.5-2\,\mathrm{keV}}$ & $F_{2-10\,\mathrm{keV}}$ & $\log{L_{0.5-2\,\mathrm{keV}}}$ & $\log{L_{2-10\,\mathrm{keV}}}$  & $k_\mathrm{bol,X}$ &  $\log{L_{2\mathrm{\,kev}}}$ & $\log{L_{2500\mathrm{\AA}}}$ & $\log{L_{4400\mathrm{\AA}}}$ & $\alpha_\mathrm{ox}$ & $\log{\lambda_\mathrm{Edd,O/UV}}$ & $R_\mathrm{S/P}$ \\
    (1) & (2) & (3) & (4) & (5) & (6) & (7) & (8) & (9) & (10) & (11) & (12) & (13) & (14) & (15) & (16) & (17)  \\
    \hline 
 \endhead

 \hline
 \endfoot

 \multicolumn{17}{c}{}\\
 
 \endlastfoot

1       &       1       &         const*TBabs*(zbb + zpo)       &        1.10 (34)       &        $1.9 \pm 0.3$                  &        $100 \pm 20$             &        $2.6_{-0.6}^{+0.2}$    &        $3.1_{-0.9}^{+1.0}$         &         $44.16_{-0.07}^{+0.08}$       &        $43.80 \pm 0.08$         &        $60 \pm 30$    &         $25.88_{-0.07}^{+0.06}$       &         $29.62 \pm 0.02$       &       $29.65 \pm 0.02$        &          $-1.44_{-0.03}^{+0.02}$         &       $-0.4$  &        $2.2 \pm 2.1$  \\
1       &       2       &         const*TBabs*(zbb + zpo)       &        1.05 (95)       &        $2.2 \pm 0.1$                  &        $130 \pm 10$             &        $3.9_{-0.2}^{+0.1}$    &        $2.5 \pm 0.3$           &         $44.26 \pm 0.02$              &        $43.78 \pm 0.04$         &        $70 \pm 30$    &         $25.98 \pm 0.03$              &         $29.86 \pm 0.01$       &       $29.89 \pm 0.01$        &          $-1.49 \pm 0.01$       &       $-0.1$  &        $1.4 \pm 0.4$  \\
1       &       3       &         const*TBabs*(zbb + zpo)       &        1.00 (108)      &        $2.1 \pm 0.1$                  &        $110 \pm 10$             &        $3.6_{-0.3}^{+0.1}$    &        $3.1 \pm 0.5$           &         $44.23 \pm 0.03$              &        $43.86 \pm 0.04$         &        $60 \pm 30$    &         $26.01 \pm 0.03$              &         $29.92 \pm 0.02$       &       $29.95 \pm 0.02$        &          $-1.50 \pm 0.01$       &       $-0.1$  &        $1.2 \pm 0.4$  \\
\hline

 \end{longtable}

\begin{minipage}{0.95\linewidth}
\renewcommand{\footnoterule}{}
\textbf{Notes:} (1) Unique X-HESS identifier; (2) Obs.\ Number; (3) best-fit model; (4) ratio between the W-stat value of the spectral fit and the degrees of freedom; (5) photon index; (6) blackbody temperature (eV); (7) absorbed $E=0.5-2$ keV flux in units of $10^{-14}$ erg cm$^{-2}$ s$^{-1}$; (8) absorbed $E=2-10$ keV flux in units of $10^{-14}$ erg cm$^{-2}$ s$^{-1}$; (9) intrinsic luminosity in the $E=0.5-2$ keV band; (10) intrinsic luminosity in the $E=2-10$ keV band; (11) X-ray bolometric correction calculated from $L_\mathrm{bol}$ in \citet{rakshit2020} and the hard X-ray luminosity in column (10); (12) 2 keV rest-frame luminosity (erg s$^{-1}$ Hz$^{-1}$); (13) 2500 \AA\ rest-frame luminosity (erg s$^{-1}$ Hz$^{-1}$); (14) 4400 \AA\ rest-frame luminosity (erg s$^{-1}$ Hz$^{-1}$); (15) $\alpha_\mathrm{ox}$ parameter; (16) Eddington ratio from the simultaneous optical/UV OM data and \citet{duras2020} optical bolometric correction (typical uncertainty of $\sim0.3$ dex); (17) $R_\mathrm{S/P}$ parameter, a proxy of the relative strength between the luminosity of the blackbody and power-law components in the $E=0.5\!-\!2$ keV energy band.
$^{(a)}$ The column density of the cold gas cloud is $N_\mathrm{H} = (3 \pm 2) \times 10^{21}$ cm$^{-2}$; $^{(b)}$ The column density of the cold gas cloud is $N_\mathrm{H} = (2 \pm 1) \times 10^{21}$ cm$^{-2}$; $^{(c)}$ The column density of the cold gas cloud is $N_\mathrm{H} = (3 \pm 1) \times 10^{21}$ cm$^{-2}$; $^{(d)}$ Photoelectric absorption by a gas cloud with $N_\mathrm{H} = 1.3_{-0.7}^{+0.8} \times 10^{22}$ cm$^{-2}$; $^{(e)}$ Gaussian emission line at $E=6.82_{-0.03}^{+0.06}$ keV, with unconstrained width and equivalent width of $EW = 160 \pm 50$ eV ; $^{(f)}$ Additional blackbody component with $kT = 260 \pm 40$ eV; $^{(g)}$ Gaussian emission line at $6.88 \pm 0.02$, with unconstrained width and an equivalent width of $EW = 200 \pm 40$ eV; $^{(h)}$ Gaussian emission line at $6.54 \pm 0.06$, with unconstrained width and an equivalent width of $EW = 340 \pm 110$ eV; $^{(j)}$ Gaussian absorption line at $E=1.01_{-0.07}^{+0.04}$ keV, with a width of $\sigma=140_{-20}^{+40}$ eV and an equivalent width of $\mathrm{FWHM} = -90_{-10}^{+30}$ eV; $^{(k)}$ Ionised gas cloud with $N_\mathrm{H}\,(10^{22}\,\mathrm{cm}^{-2}) = 36 \pm 8$, $\log\xi = 3.15_{-0.05}^{+0.07}$ erg cm$^{-1}$ s $^{-1}$ and fixed covering factor $C_\mathrm{f}=1$; $^{(l)}$ Gaussian absorption line at $E=1.06_{-0.04}^{+0.03}$ keV, with a width of $\sigma=120_{-30}^{+40}$ eV and an equivalent width of $\mathrm{FWHM} = -80 \pm 10$ eV; $^{(m)}$ Gaussian emission line at $E=6.49_{-0.07}^{+0.08}$ keV, with a fixed width of $\sigma=200$ eV and equivalent width of $EW = 190 \pm 60$ eV; $^{(n)}$ Additional blackbody component with $kT_\mathrm{bb}=80\pm3$ eV; $^{(o)}$ Gaussian emission line with fixed width $\sigma=100$ eV at $E=6.65_{-0.04}^{+0.05}$ keV, with an equivalent width of $EW=100\pm20$ eV; $^{(p)}$ Gaussian absorption line at $E=7.48 \pm 0.04$ keV, with a width of $\sigma=260_{-40}^{+50}$ eV and an equivalent width of $EW=-340\pm40$ eV; $^{(q)}$ Gaussian emission line at $E=3.80_{-0.03}^{+0.04}$ keV, with unresolved width and an equivalent width of $EW=270\pm90$ eV; $^{(r)}$ Gaussian absorption line at $E=1.10 \pm 0.01$ keV, with an unresolved width and an equivalent width of $EW = -70_{-10}^{+20}$ eV; $^{(s)}$ Gaussian emission line at $E=6.58_{-0.04}^{+0.03}$, with an unresolved width and an equivalent width of $EW = 60 \pm 20$ eV; $^{(t)}$ Gaussian emission line at $E=0.94_{-0.03}^{+0.02}$ eV, with a width of $\sigma=100 \pm 20$ eV and an equivalent width of $EW=50 \pm 20$ eV; $^{(u)}$ Additional blackbody component with $kT_\mathrm{bb} = 200_{-50}^{+40}$ eV; $^{(v)}$ Cold absorbing gas cloud with column density of $N_\mathrm{H} (10^{22} \mathrm{cm}^{-2}) = 5.6_{-1.5}^{+1.6}$; $^{(w)}$ Absorption edge at $E = 9.3 \pm 0.2$ keV and depth $\tau = 1.1 \pm 0.2$.

\end{minipage}
 
\end{landscape}

\begin{figure}[h]
     \vspace{-0.5cm}
     \centering
     \renewcommand{\arraystretch}{2}
     \begin{tabular}{ccc}
     \subfloat{\includegraphics[width = 2.1in]{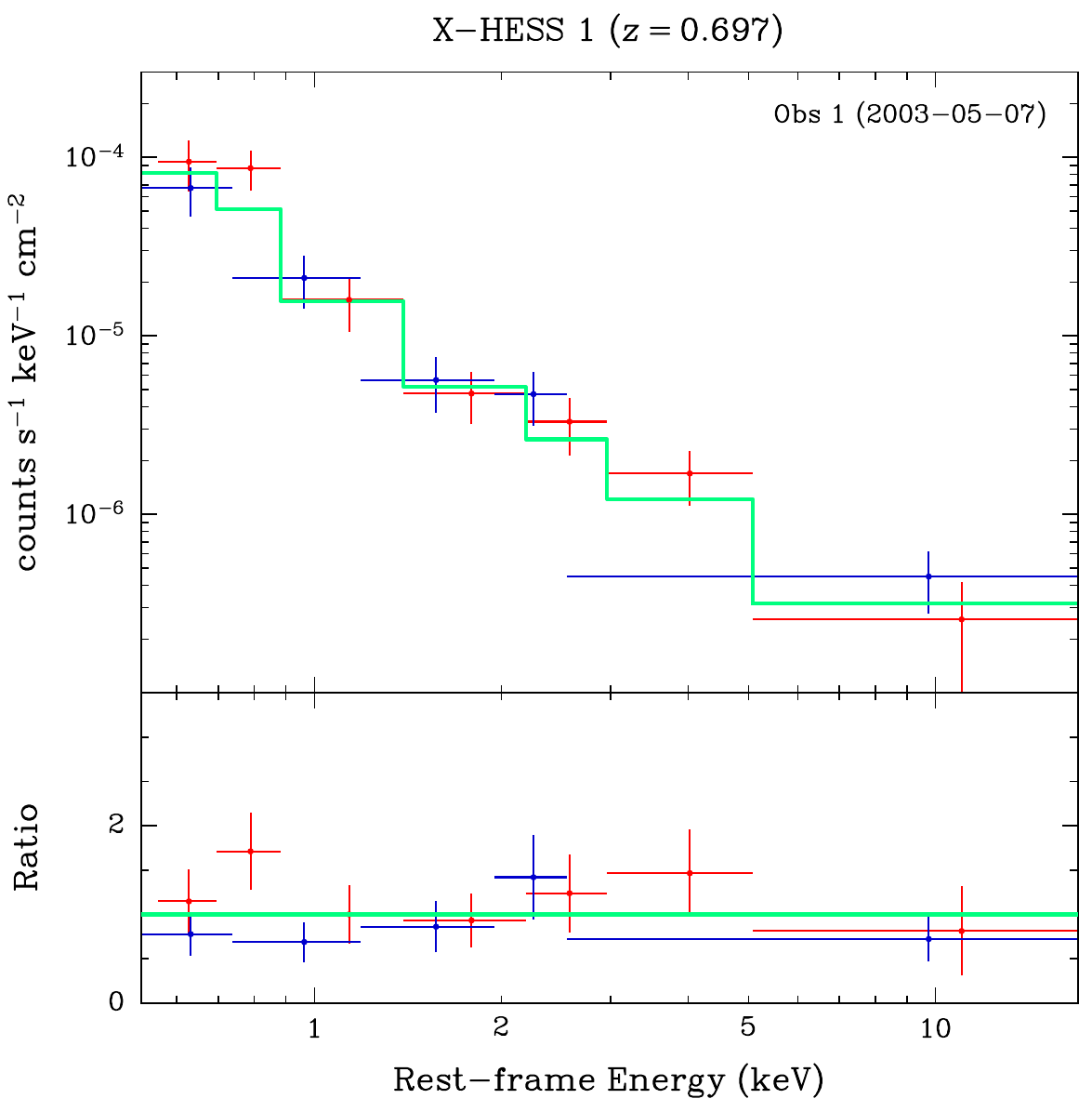}} &
     \subfloat{\includegraphics[width = 2.1in]{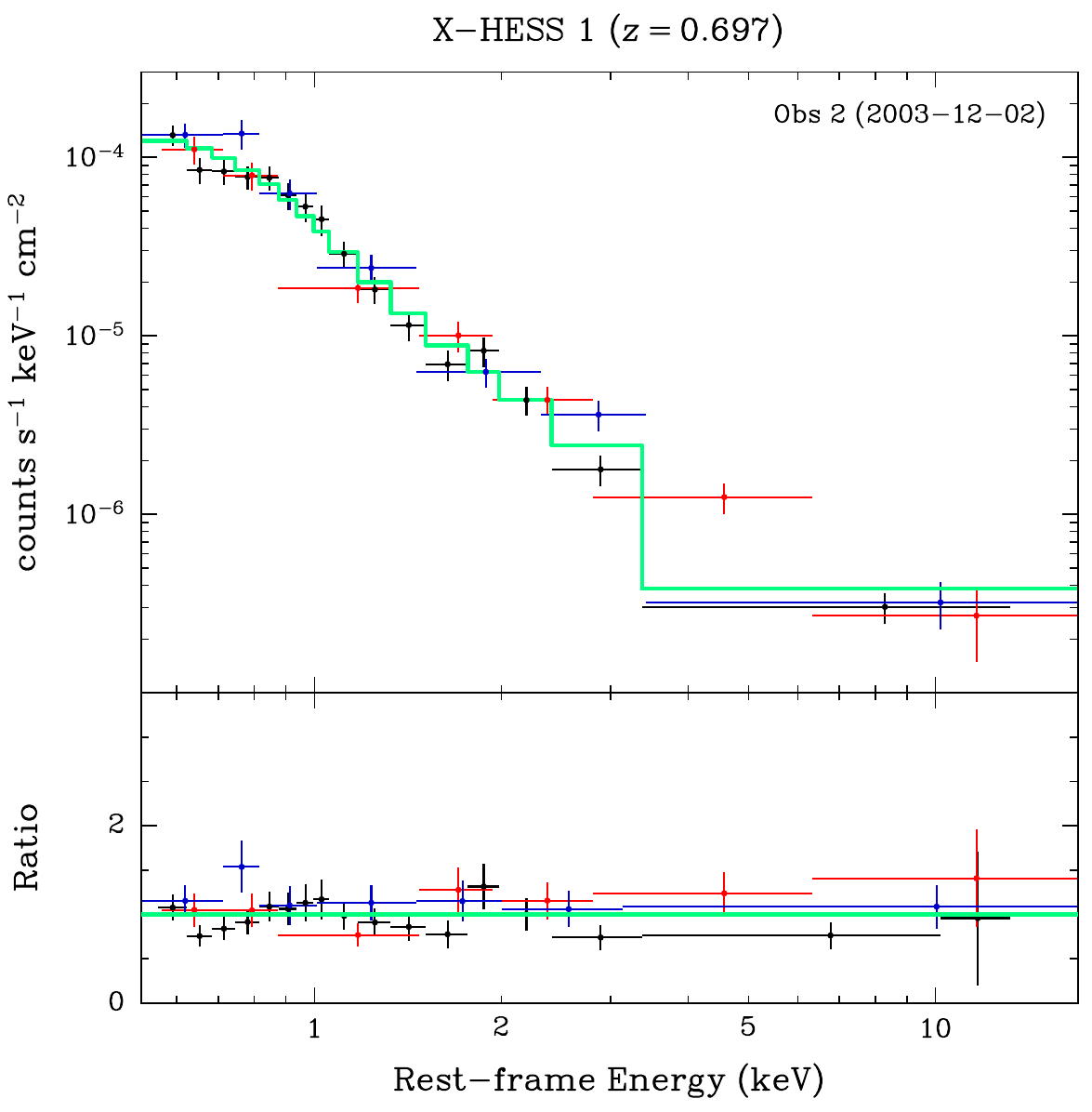}} &
     \subfloat{\includegraphics[width = 2.1in]{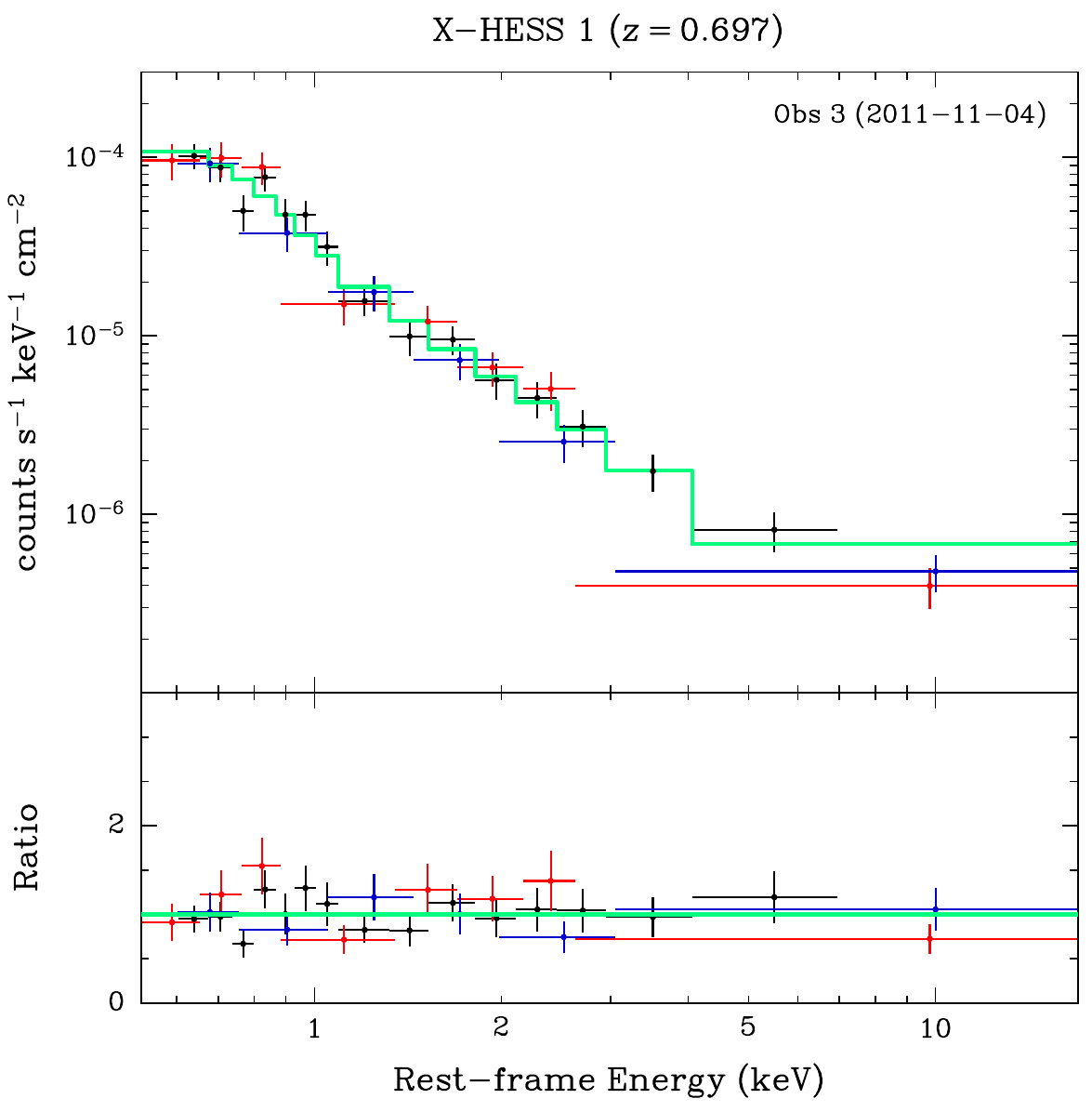}} \\
     \subfloat{\includegraphics[width = 2.1in]{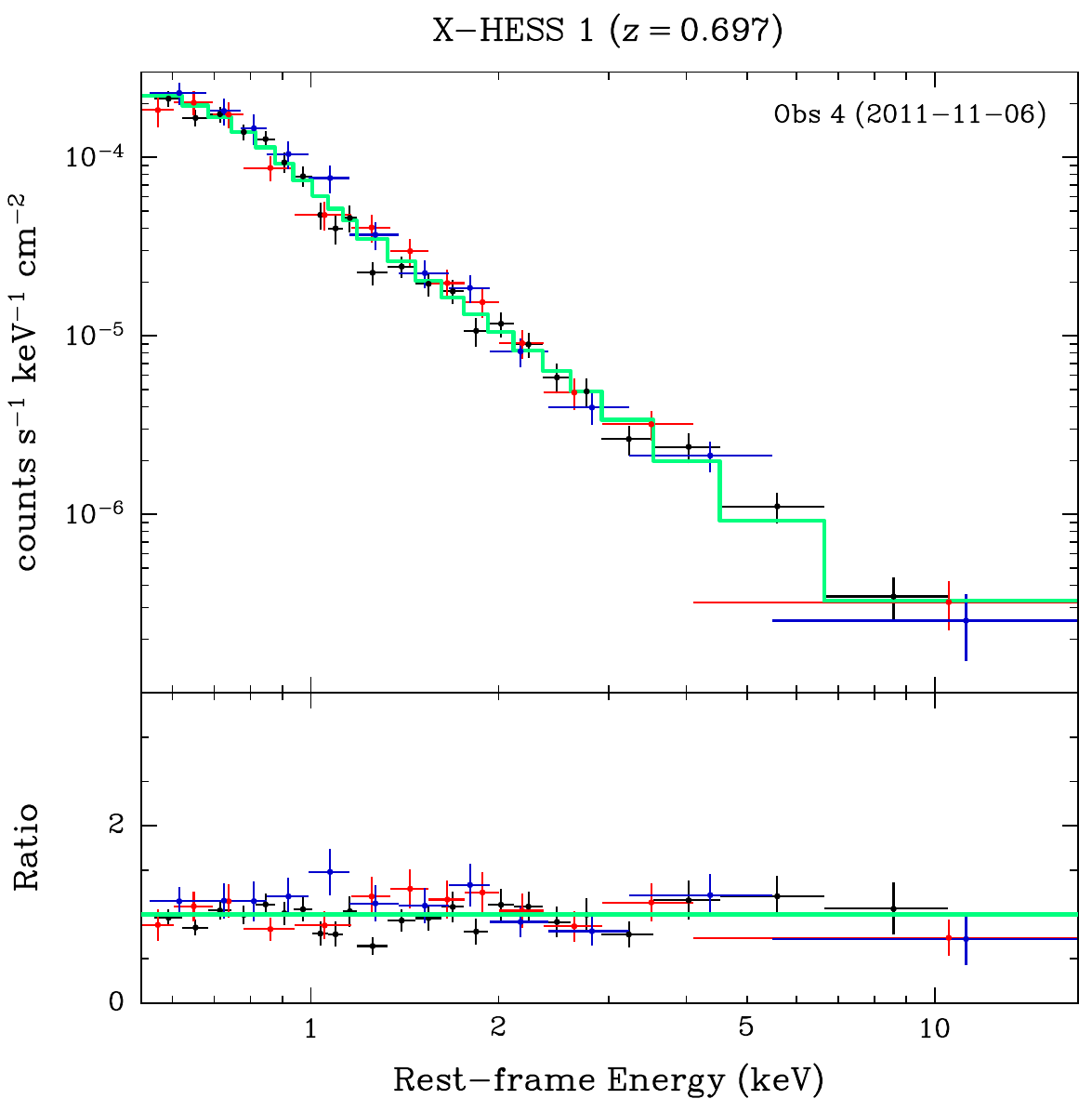}} &
     \subfloat{\includegraphics[width = 2.1in]{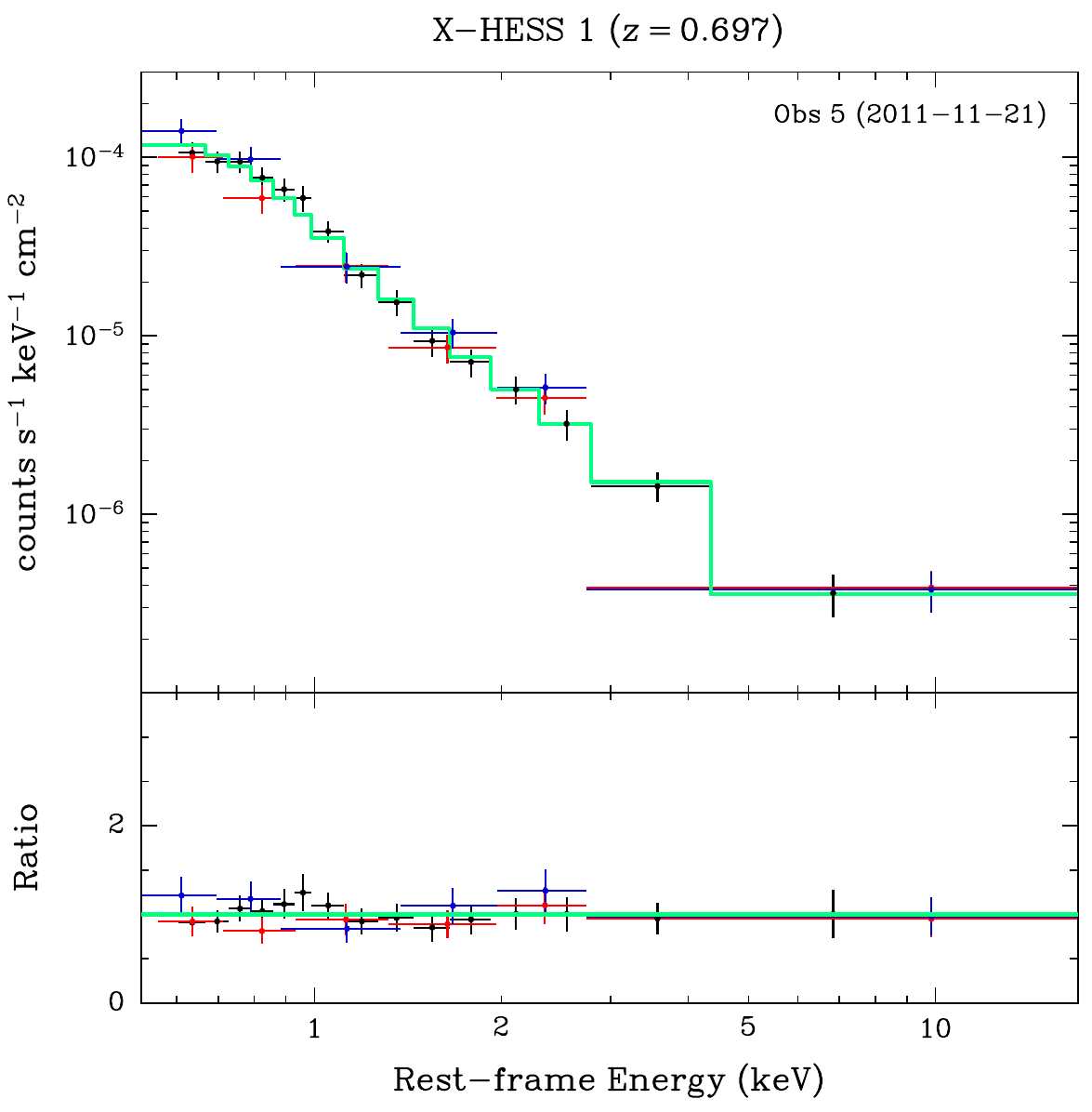}} &
     \subfloat{\includegraphics[width = 2.1in]{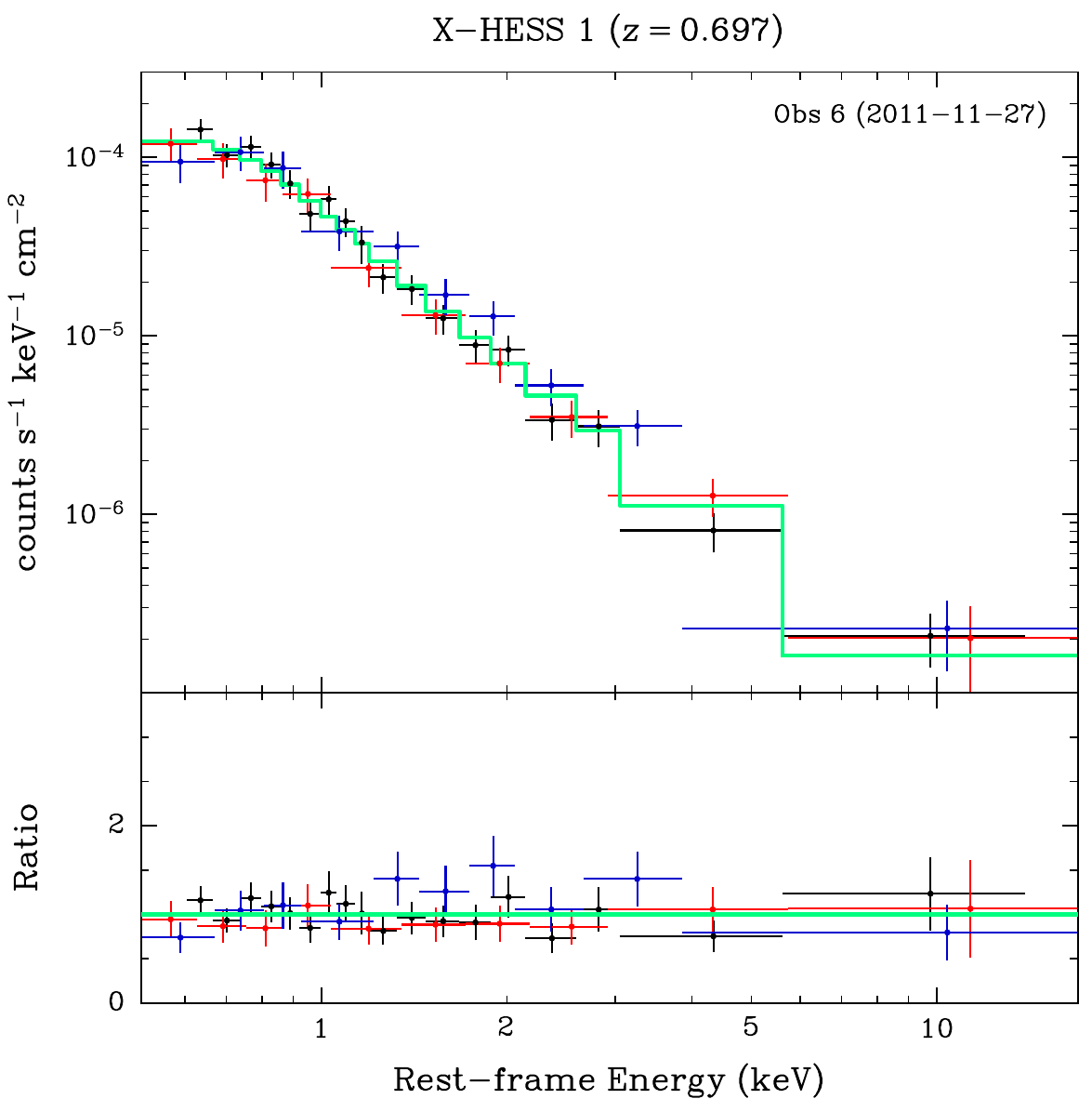}} \\
     \subfloat{\includegraphics[width = 2.1in]{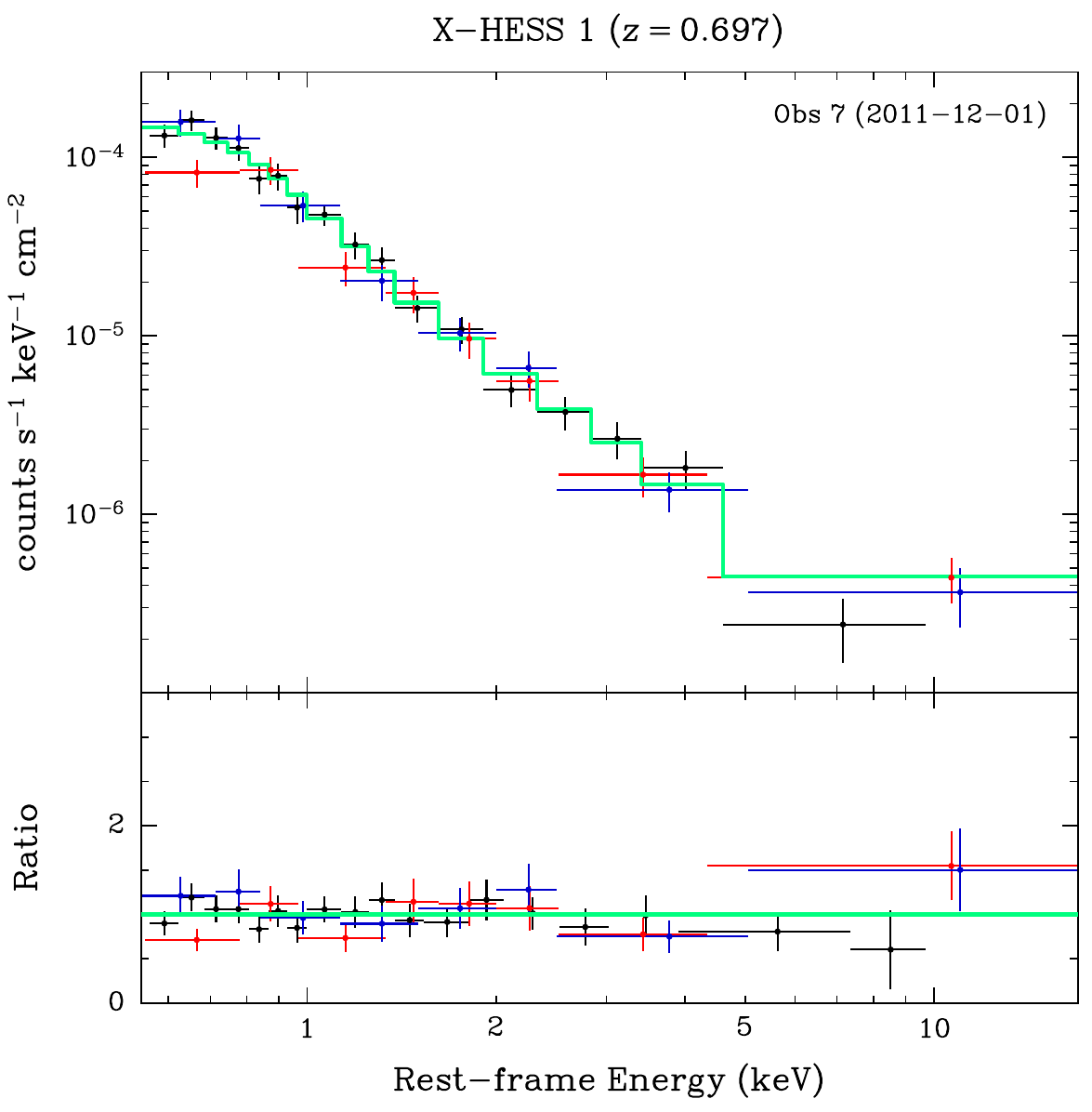}} &
     \subfloat{\includegraphics[width = 2.1in]{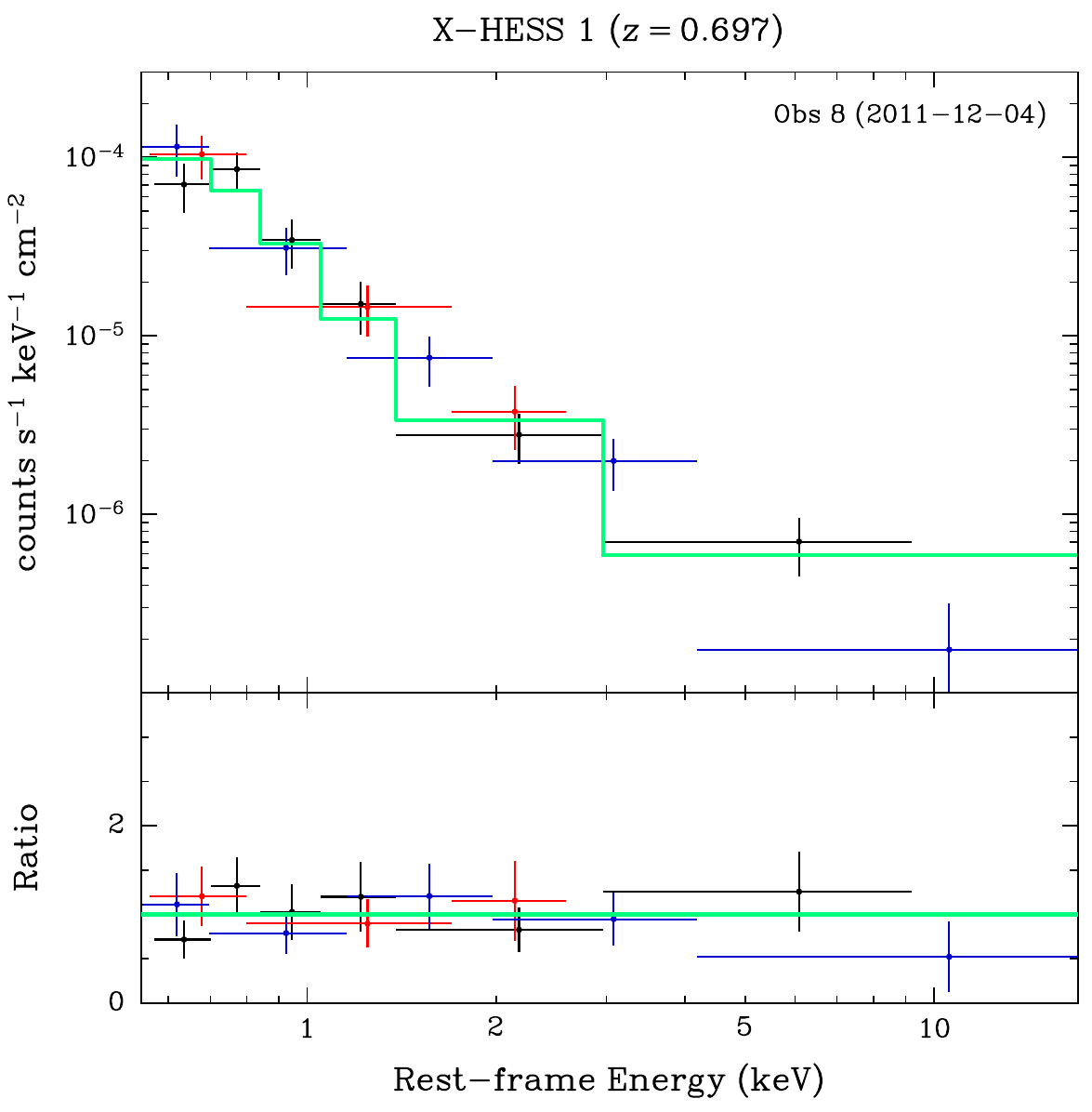}} &
     \subfloat{\includegraphics[width = 2.1in]{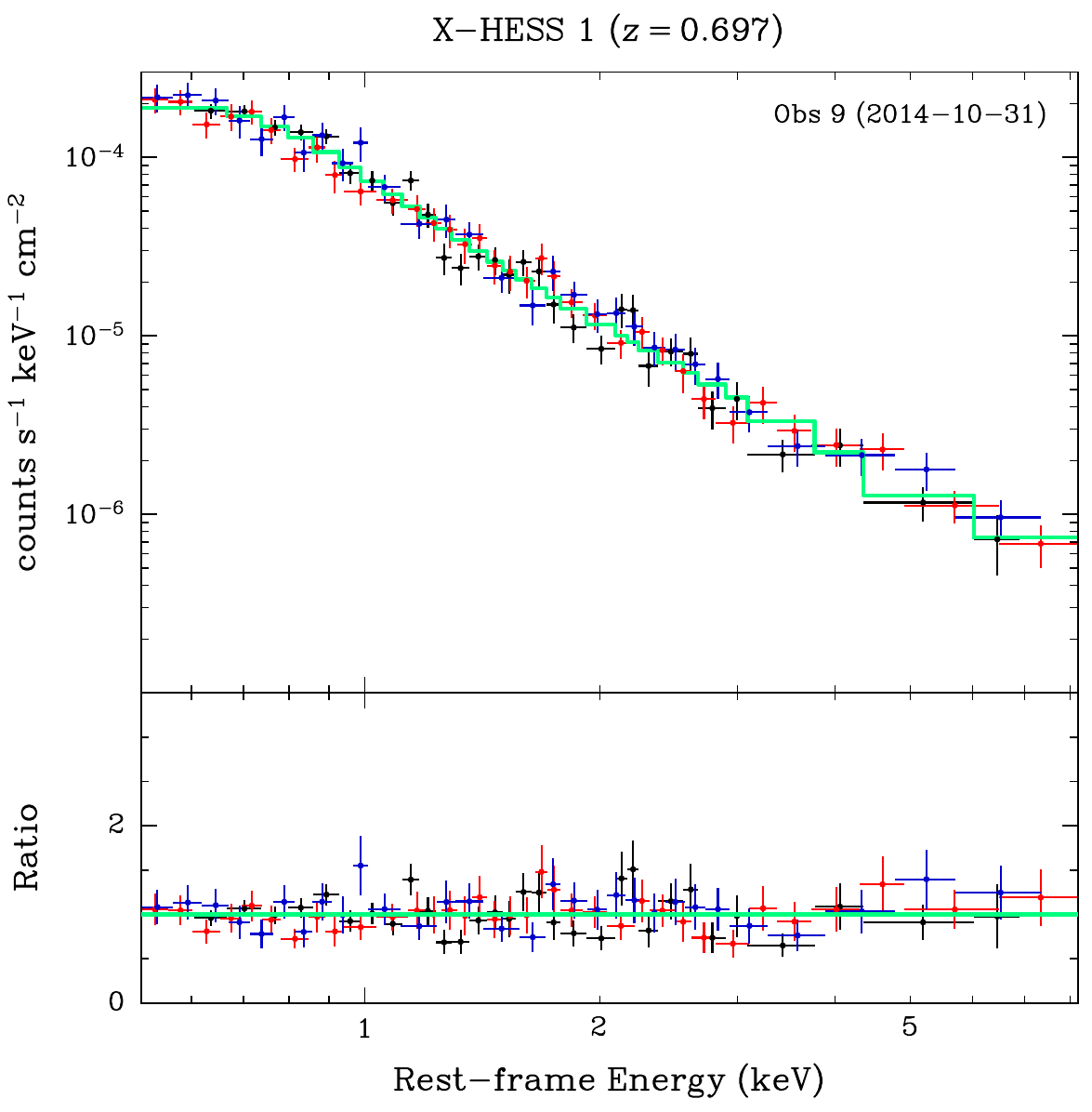}} \\
     \subfloat{\includegraphics[width = 2.1in]{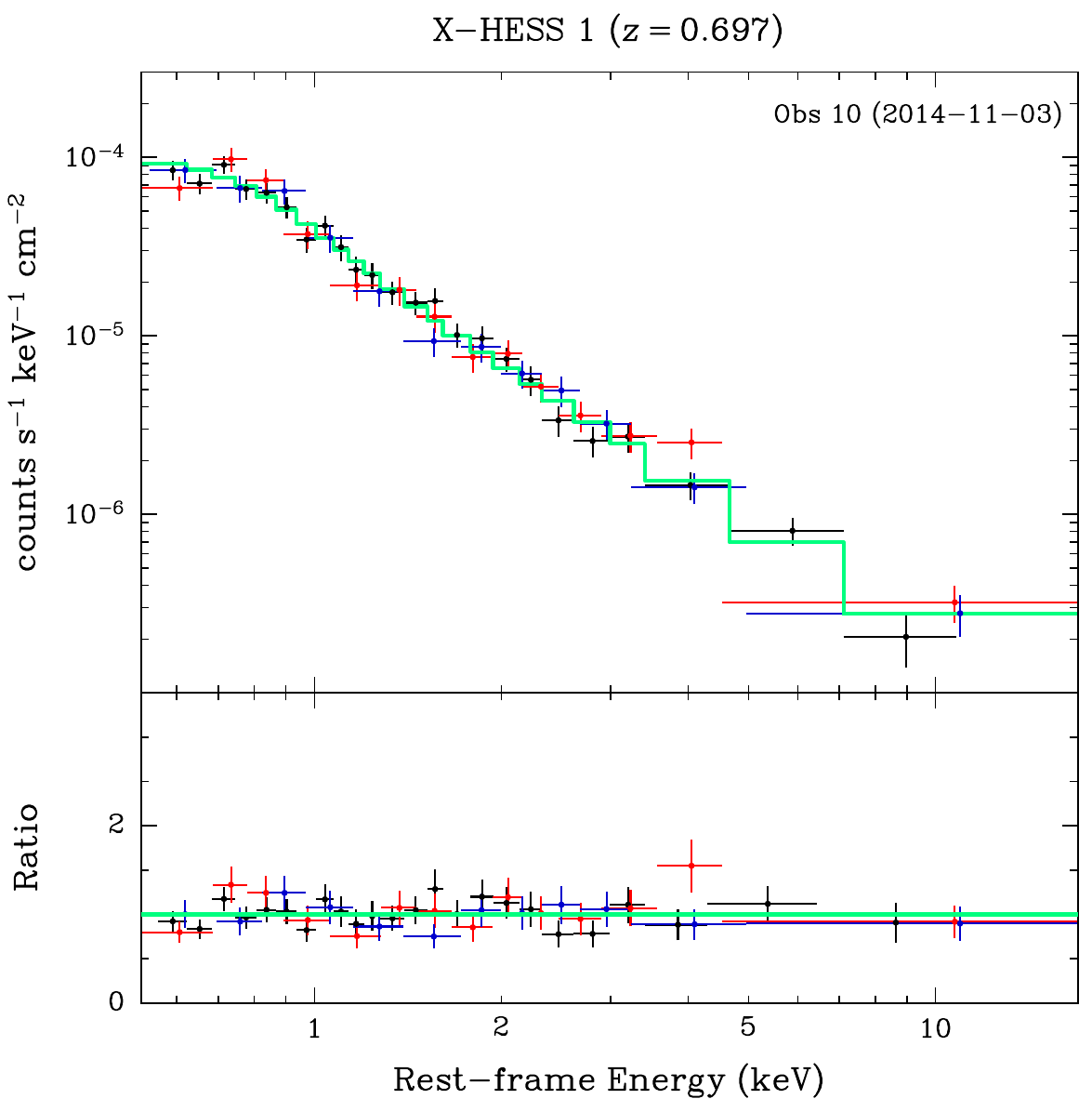}} &
     \subfloat{\includegraphics[width = 2.1in]{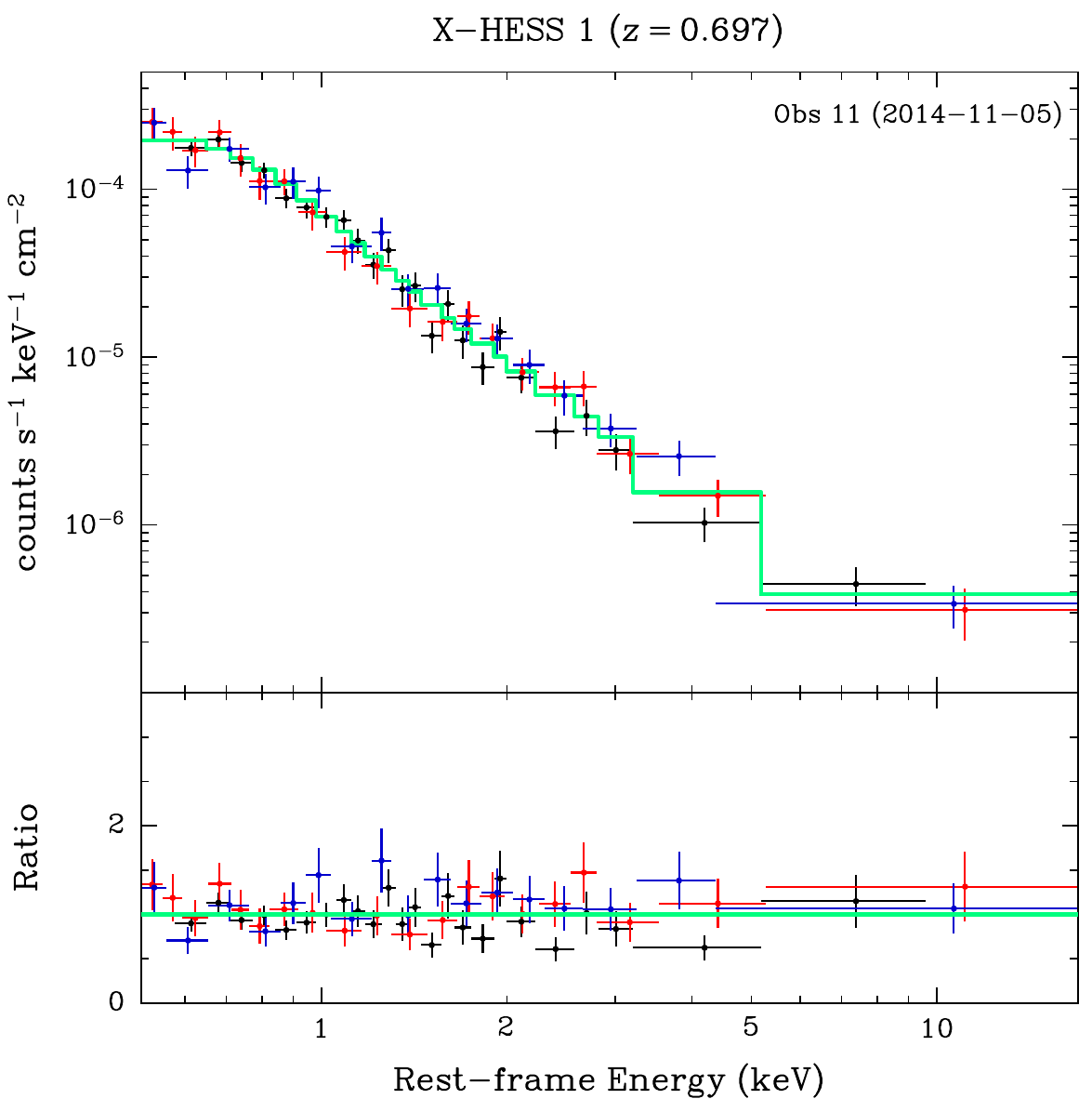}} &
     \subfloat{\includegraphics[width = 2.1in]{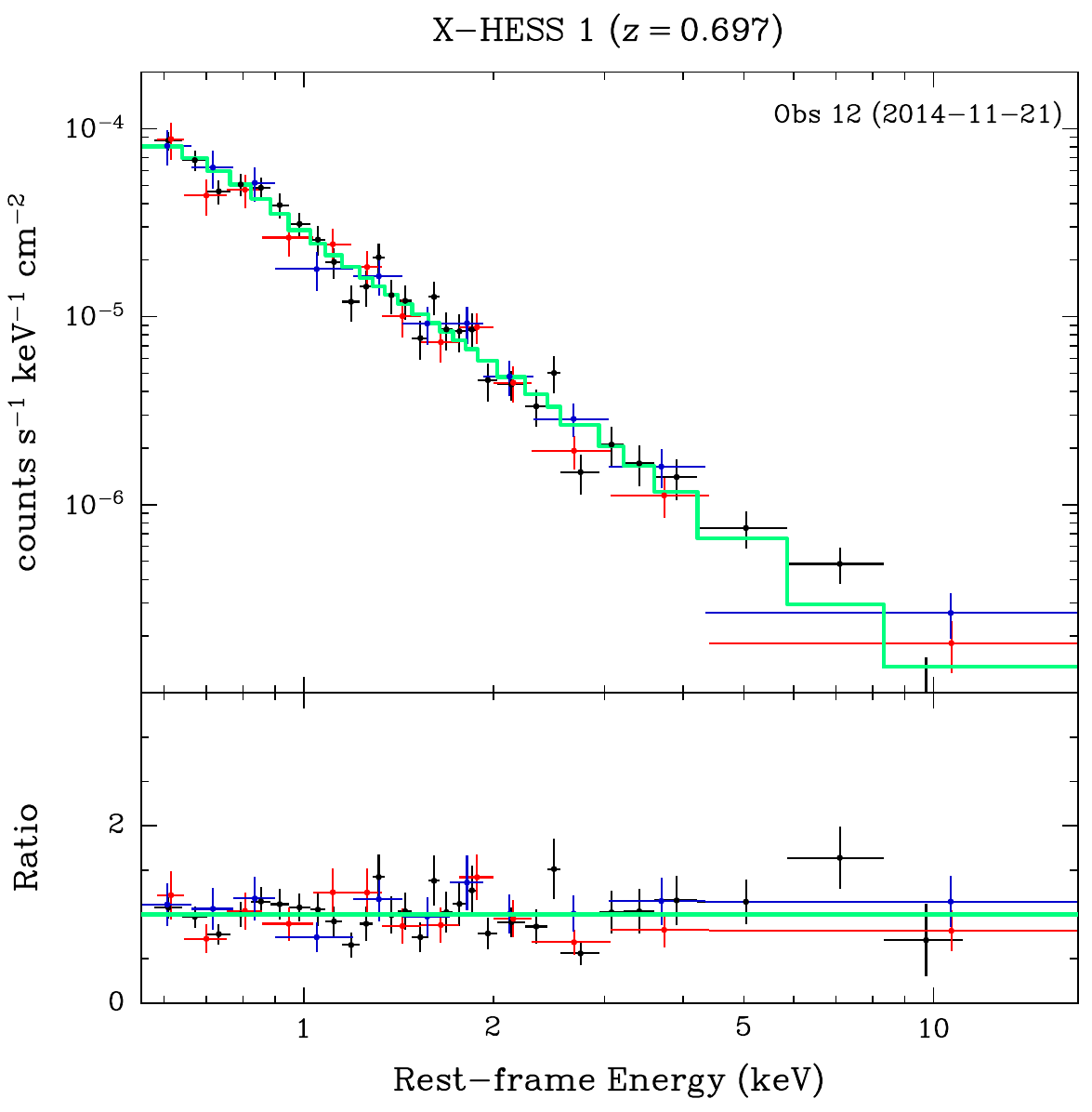}} \\
\end{tabular}

\begin{minipage}{\linewidth}
    \caption{\emph{XMM-Newton} spectra of the X-HESS AGN. Data from EPIC-pn, MOS1 and MOS2 are shown in black, red and blue, respectively. Data were divided by the response effective area for each particular channel. Best-fit model (solid green line) is shown for EPIC-pn only, for the sake of visual clarity. The bottom panel in each plot describes the ratio between the data and best-fit model. Spectra were rebinned with custom schemes for plotting purposes only.}\label{fig:xhess_spectra}
\end{minipage}

\end{figure}

\begin{figure}[h]
     \ContinuedFloat
     \centering
     \renewcommand{\arraystretch}{2}
     \begin{tabular}{ccc}
     \subfloat{\includegraphics[width = 2.1in]{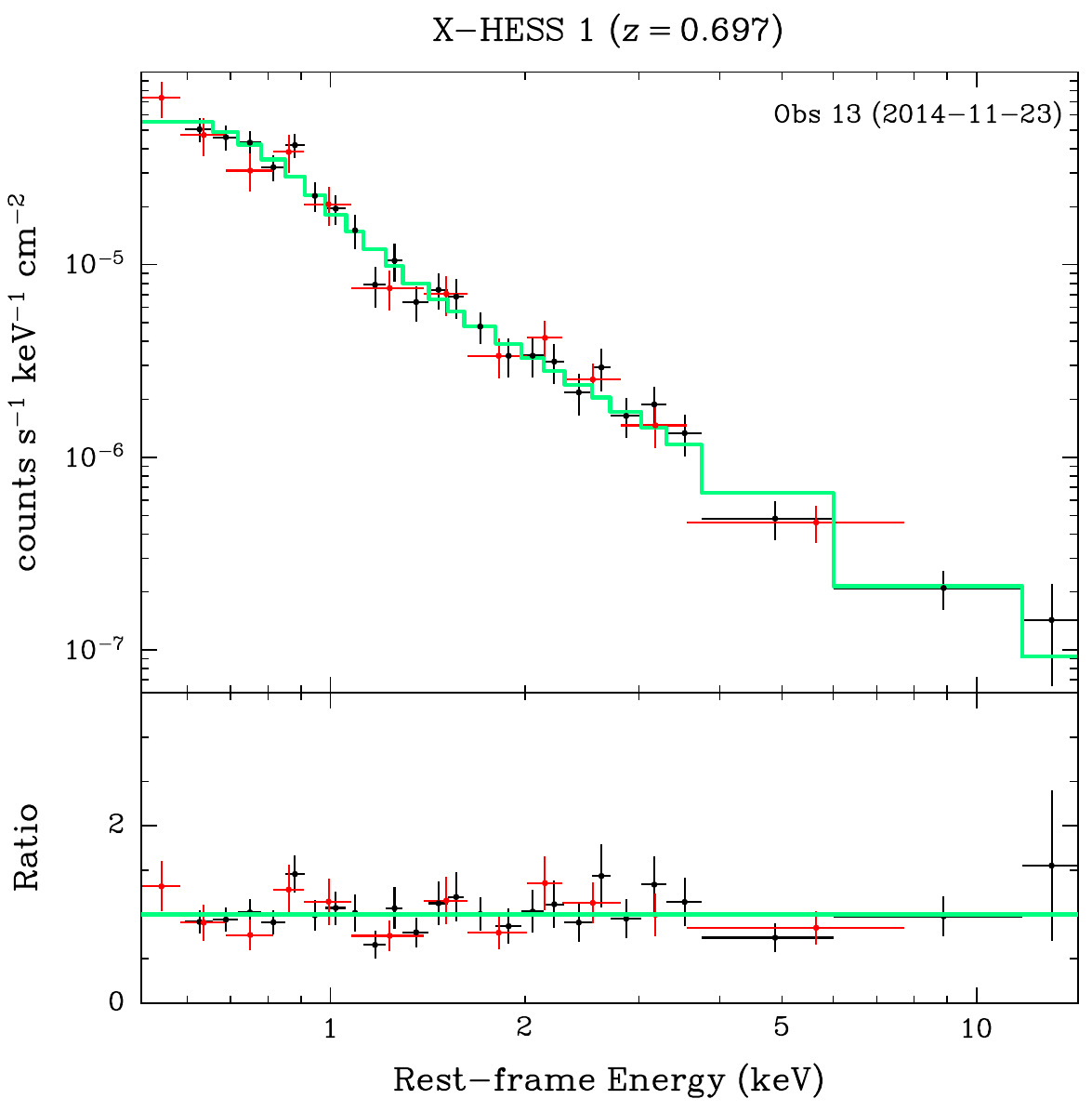}} &
     \subfloat{\includegraphics[width = 2.1in]{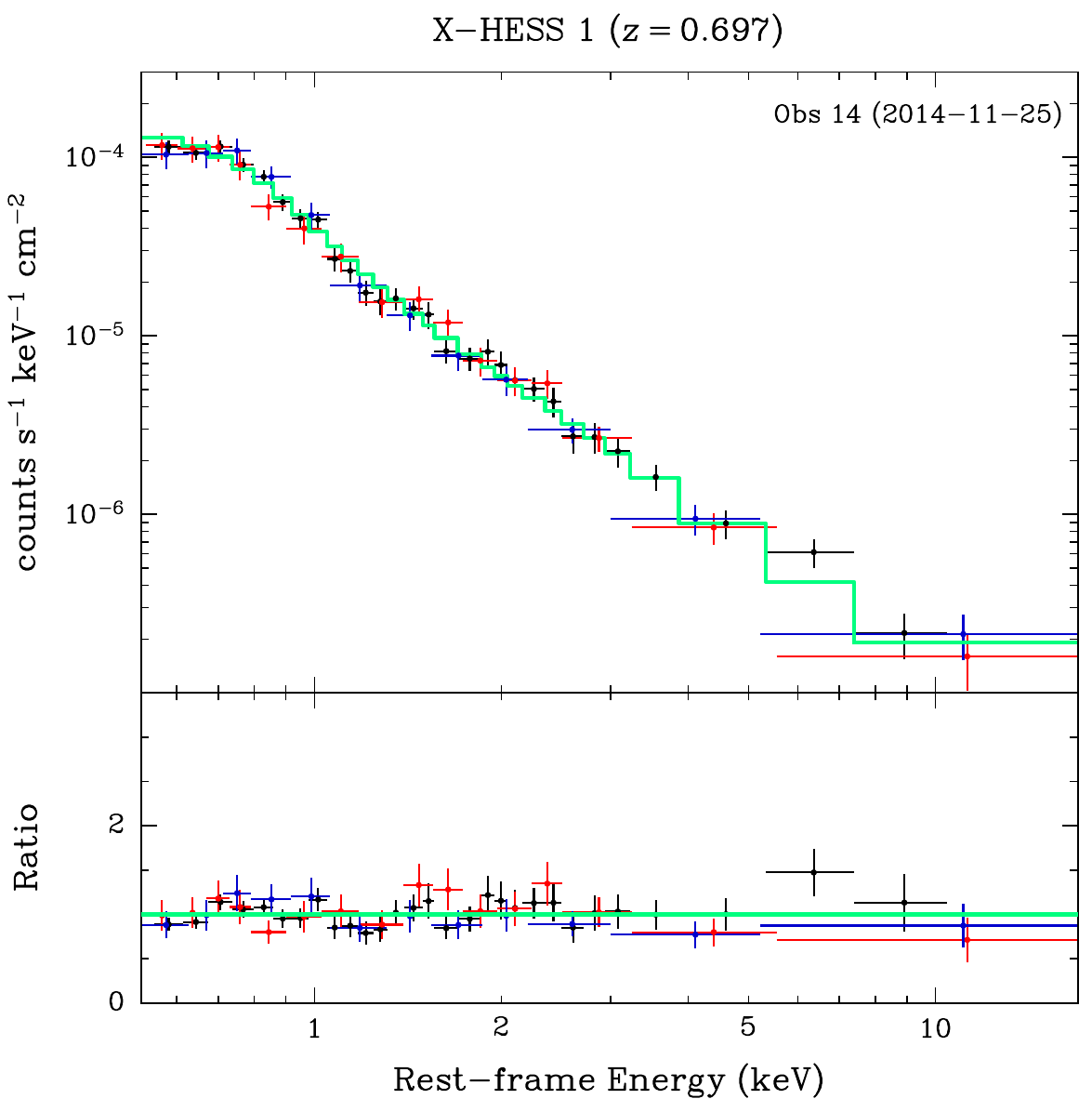}} &
     \subfloat{\includegraphics[width = 2.1in]{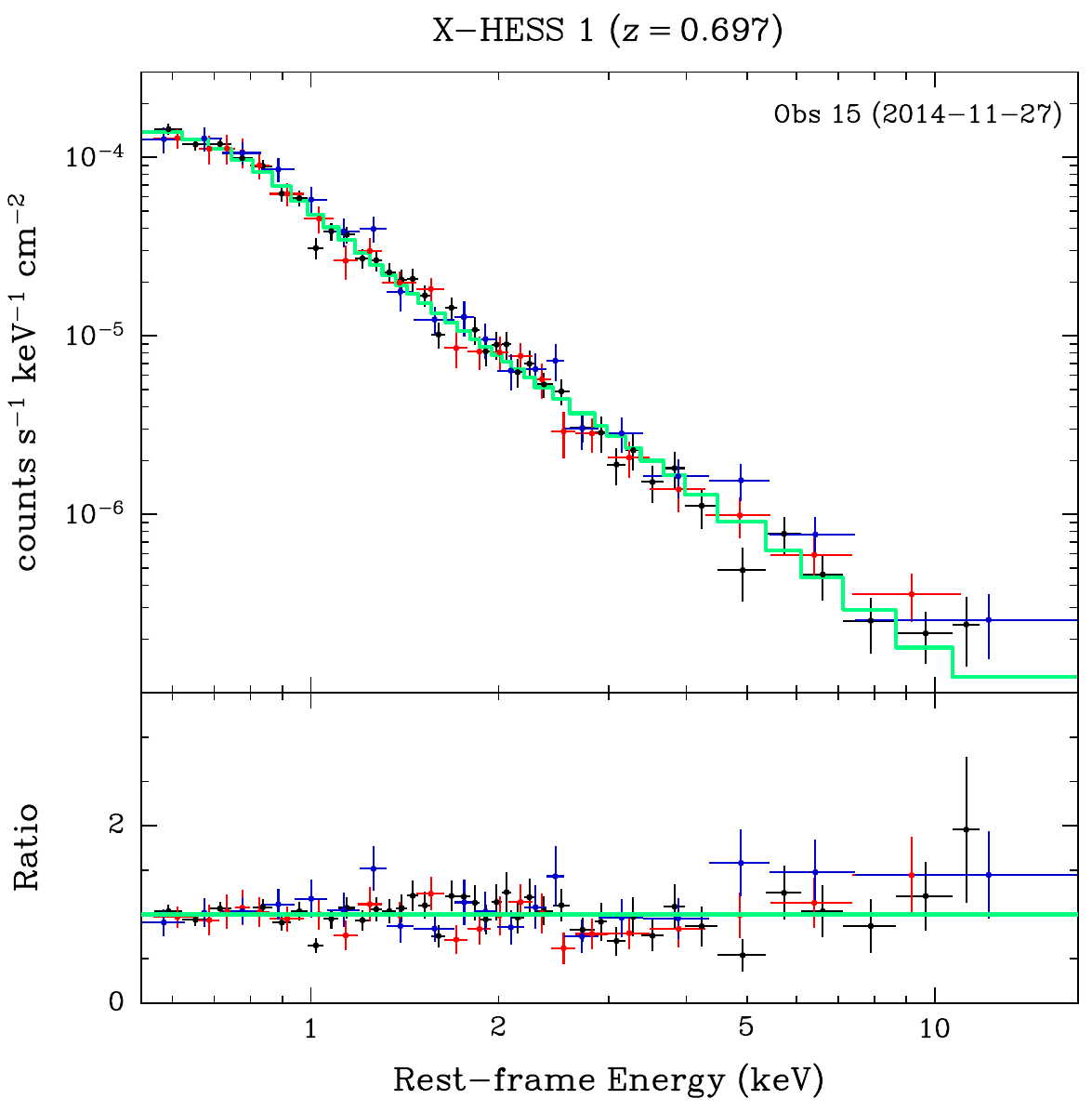}} \\
     \subfloat{\includegraphics[width = 2.1in]{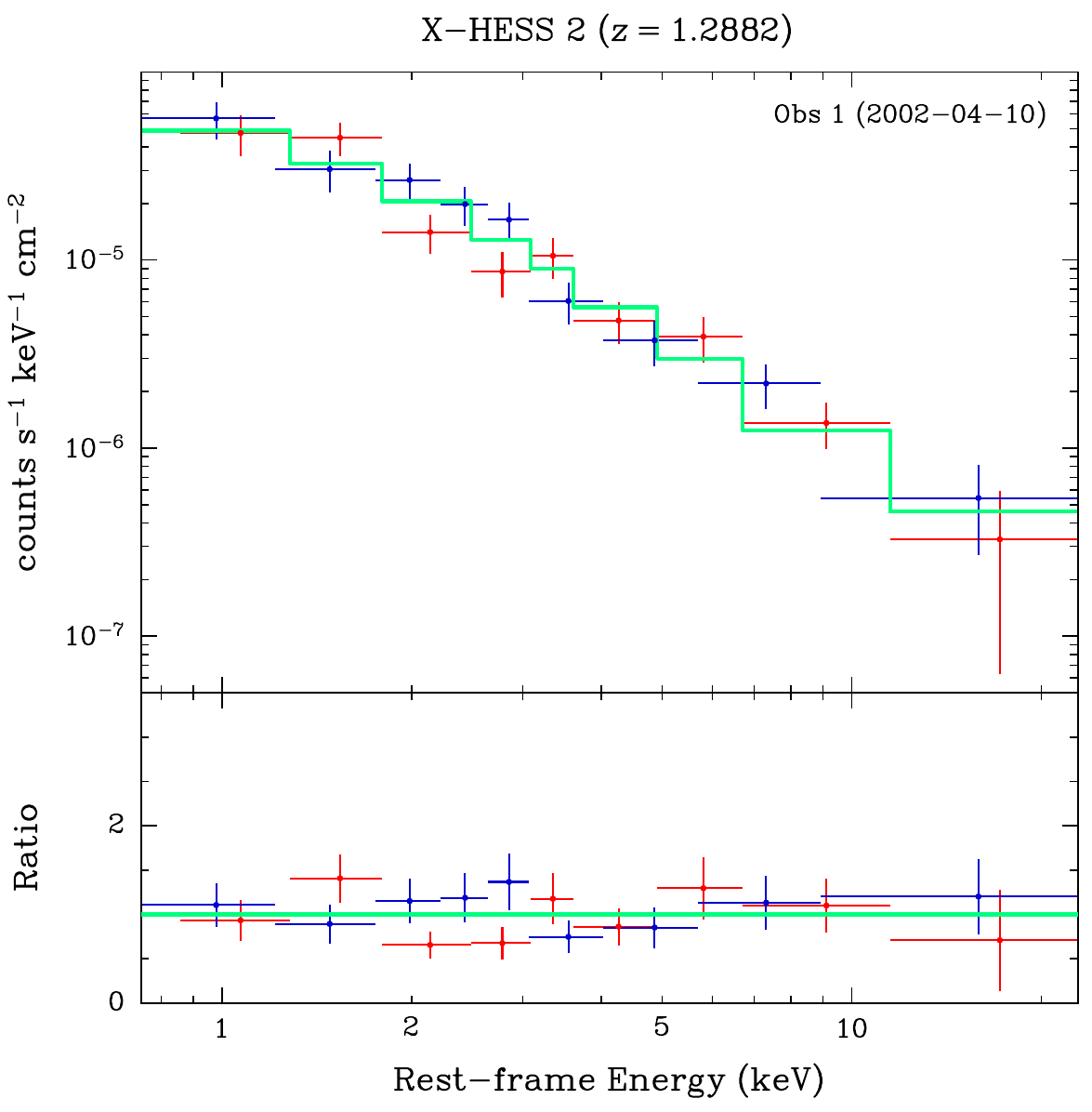}} &
     \subfloat{\includegraphics[width = 2.1in]{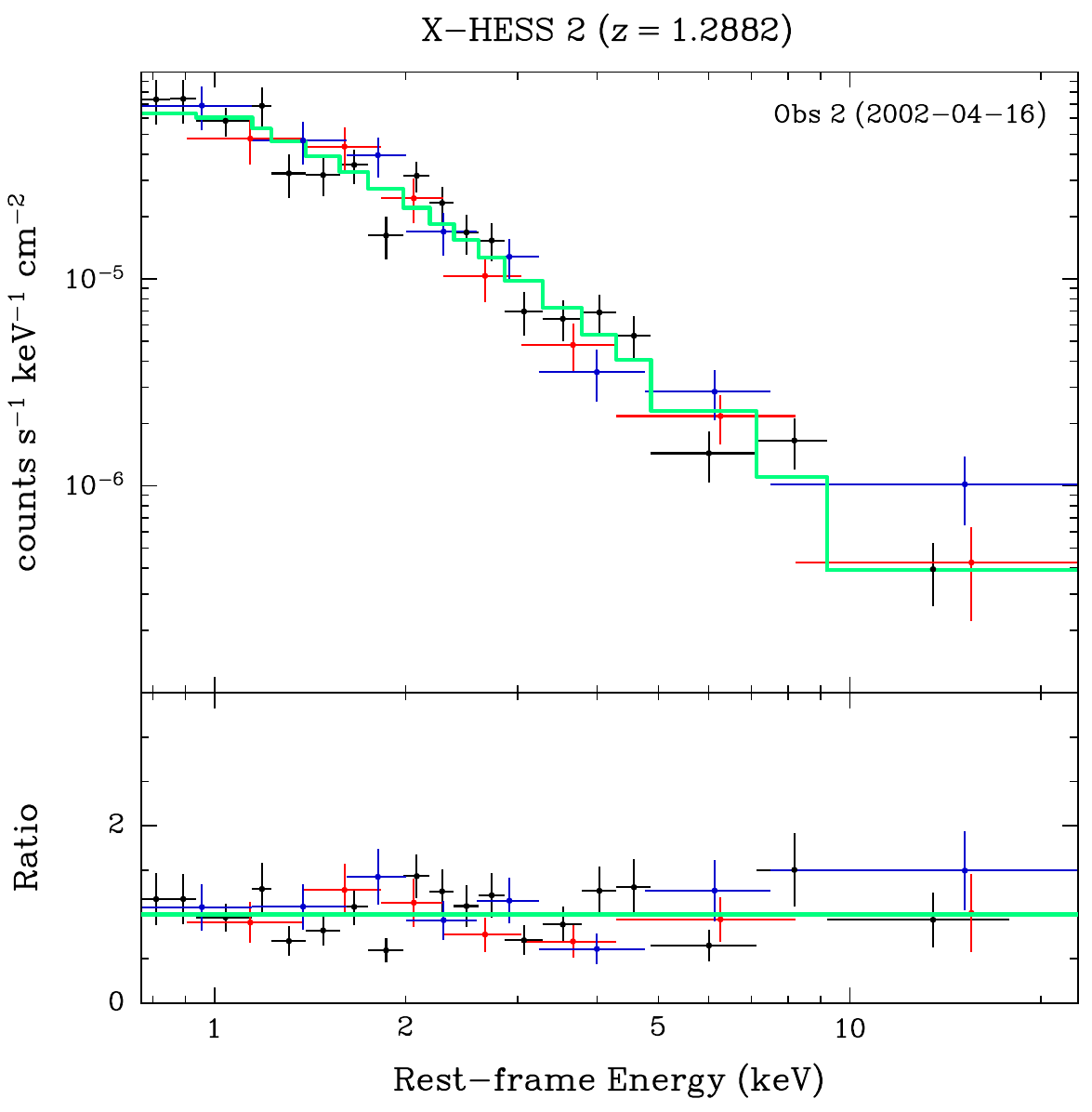}} &
     \subfloat{\includegraphics[width = 2.1in]{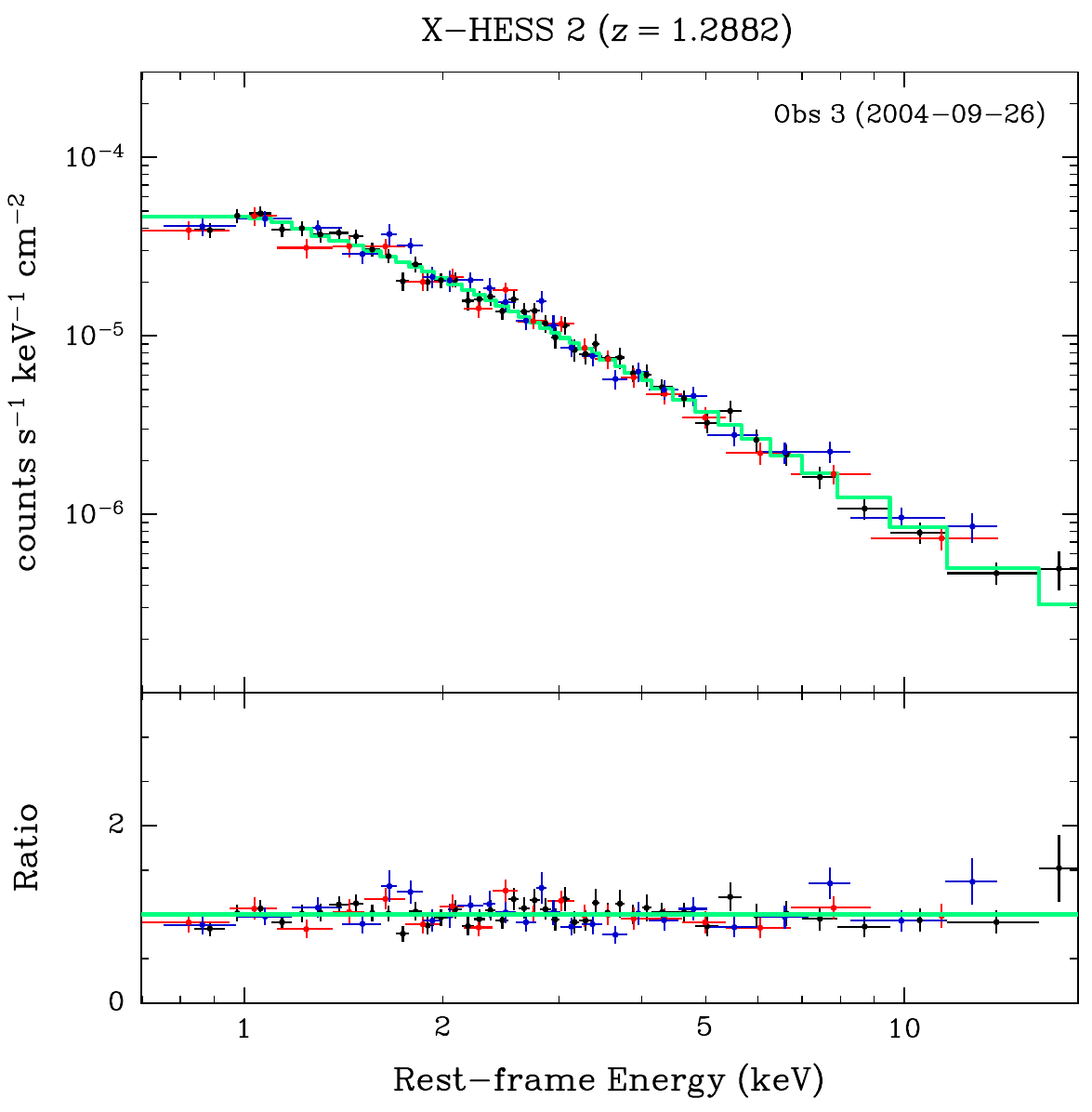}} \\
     \subfloat{\includegraphics[width = 2.1in]{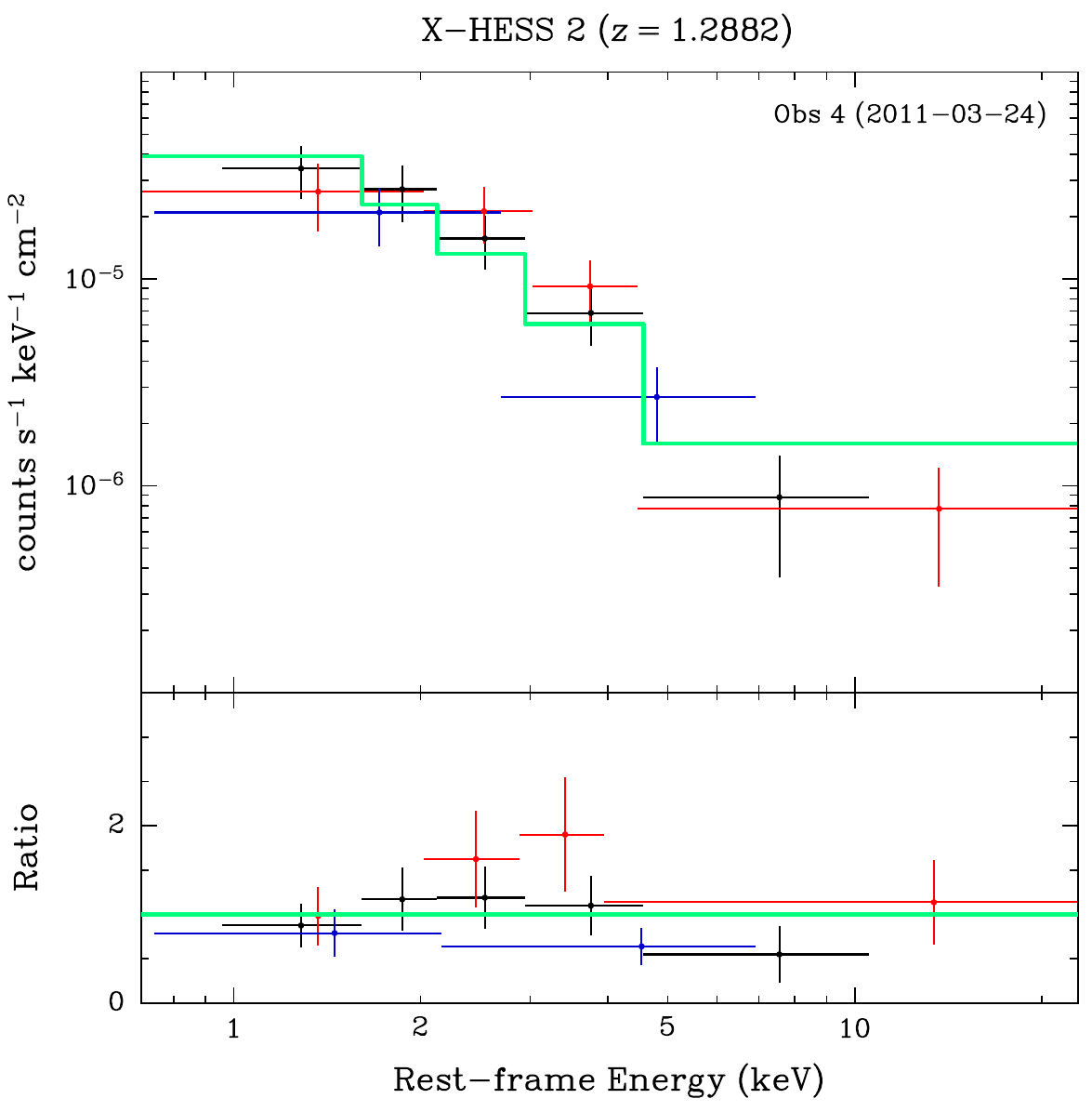}} &
     \subfloat{\includegraphics[width = 2.1in]{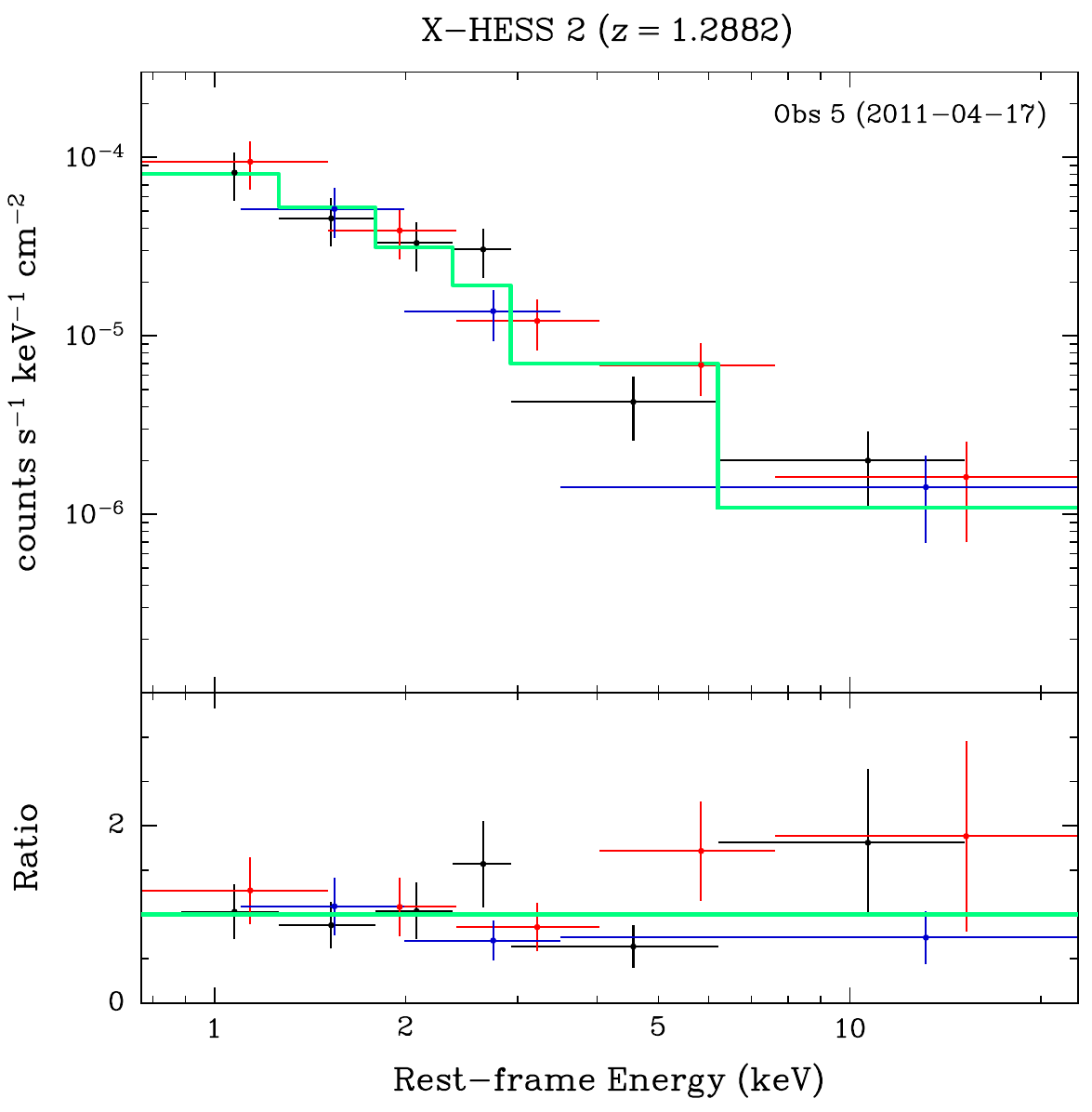}} &
     \subfloat{\includegraphics[width = 2.1in]{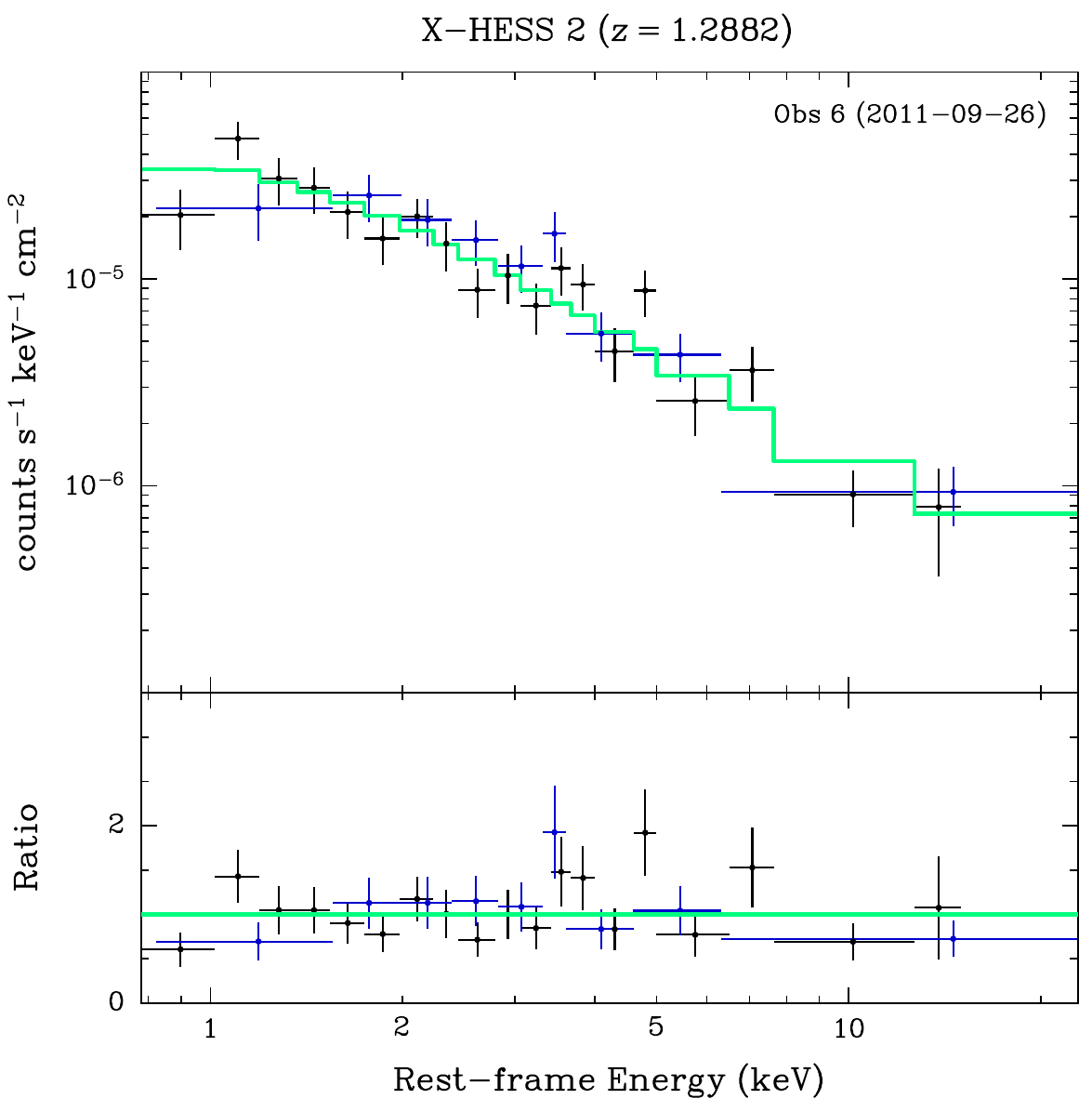}} \\
     \subfloat{\includegraphics[width = 2.1in]{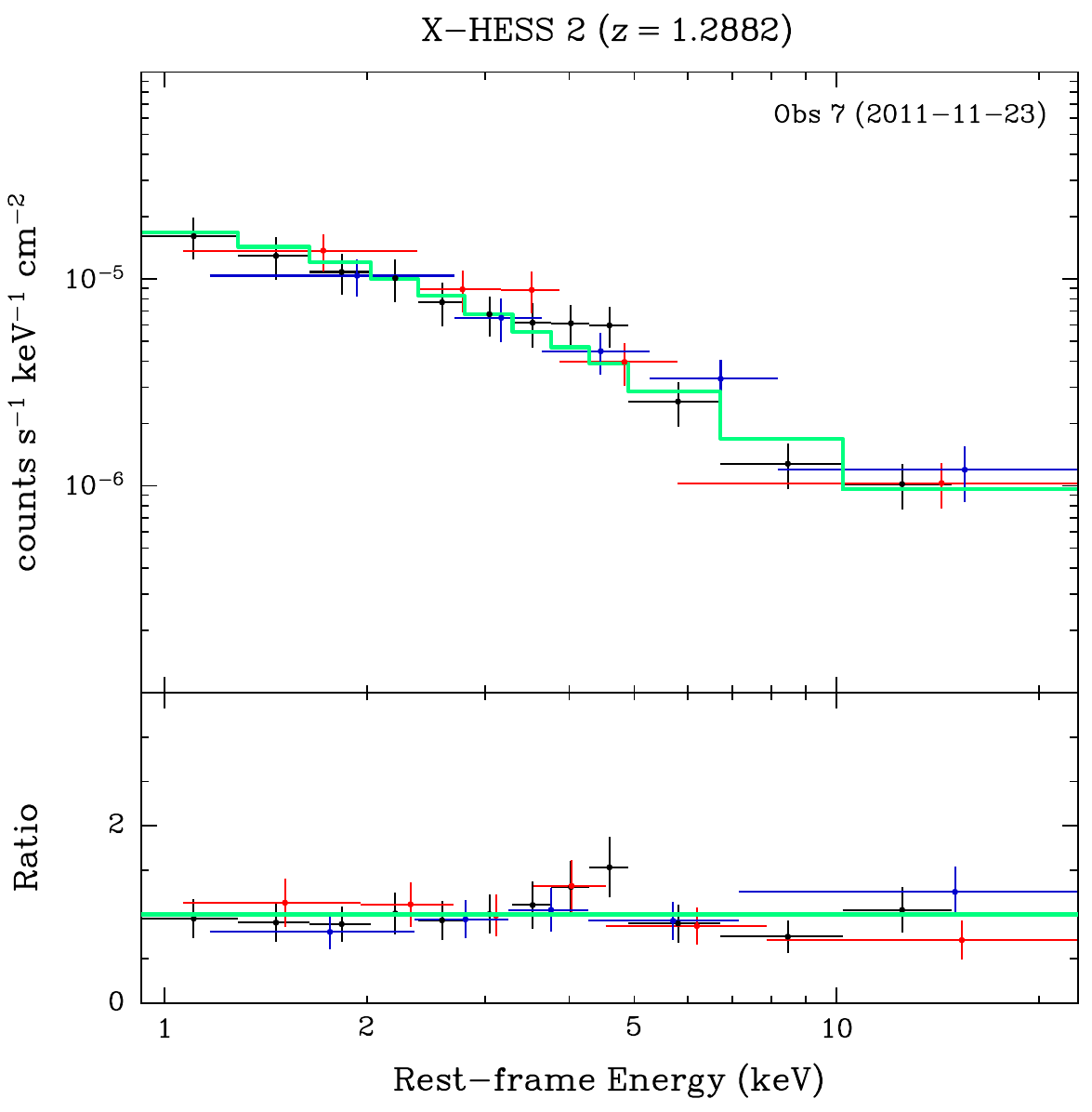}} &
     \subfloat{\includegraphics[width = 2.1in]{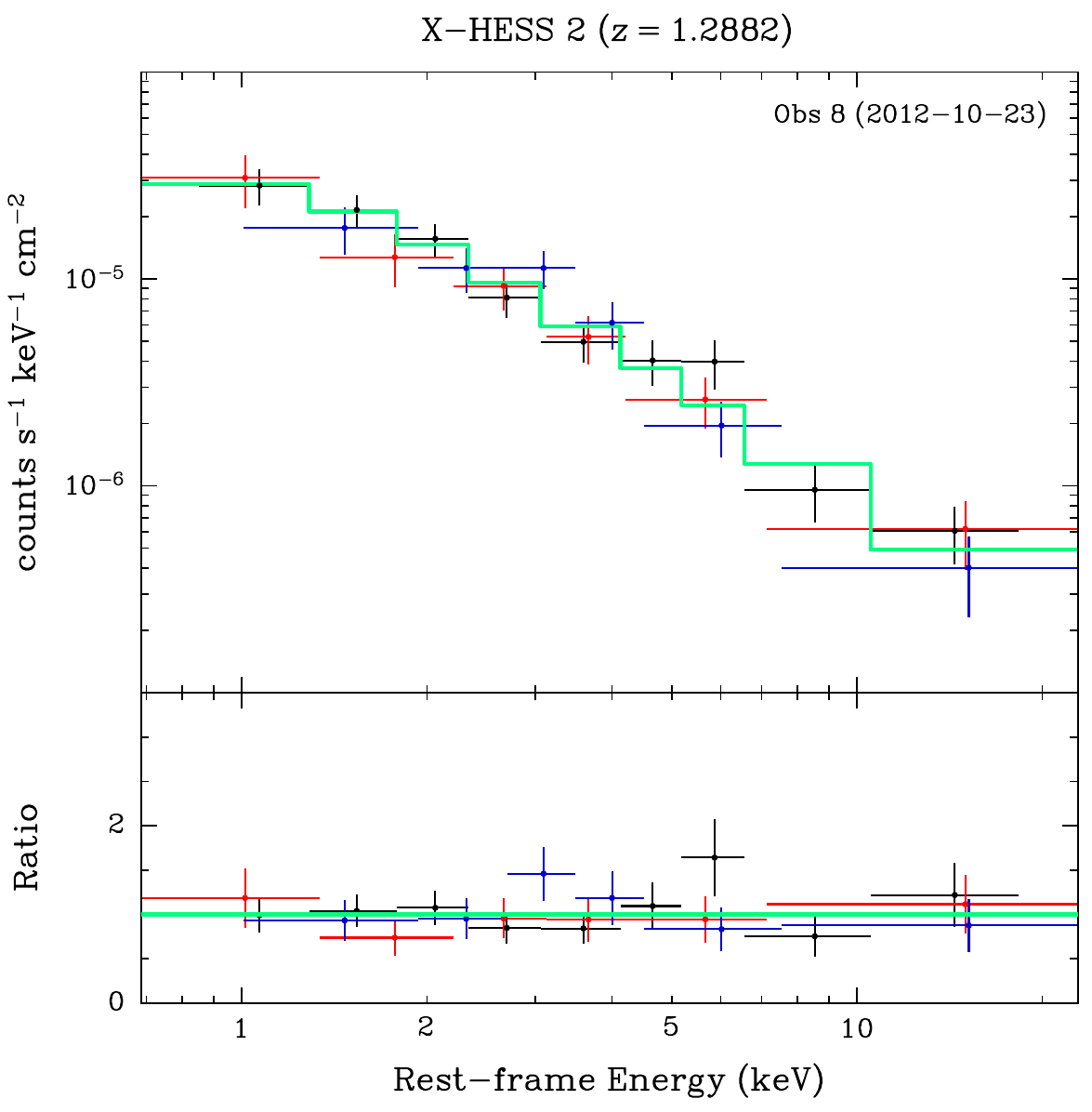}} &
     \subfloat{\includegraphics[width = 2.1in]{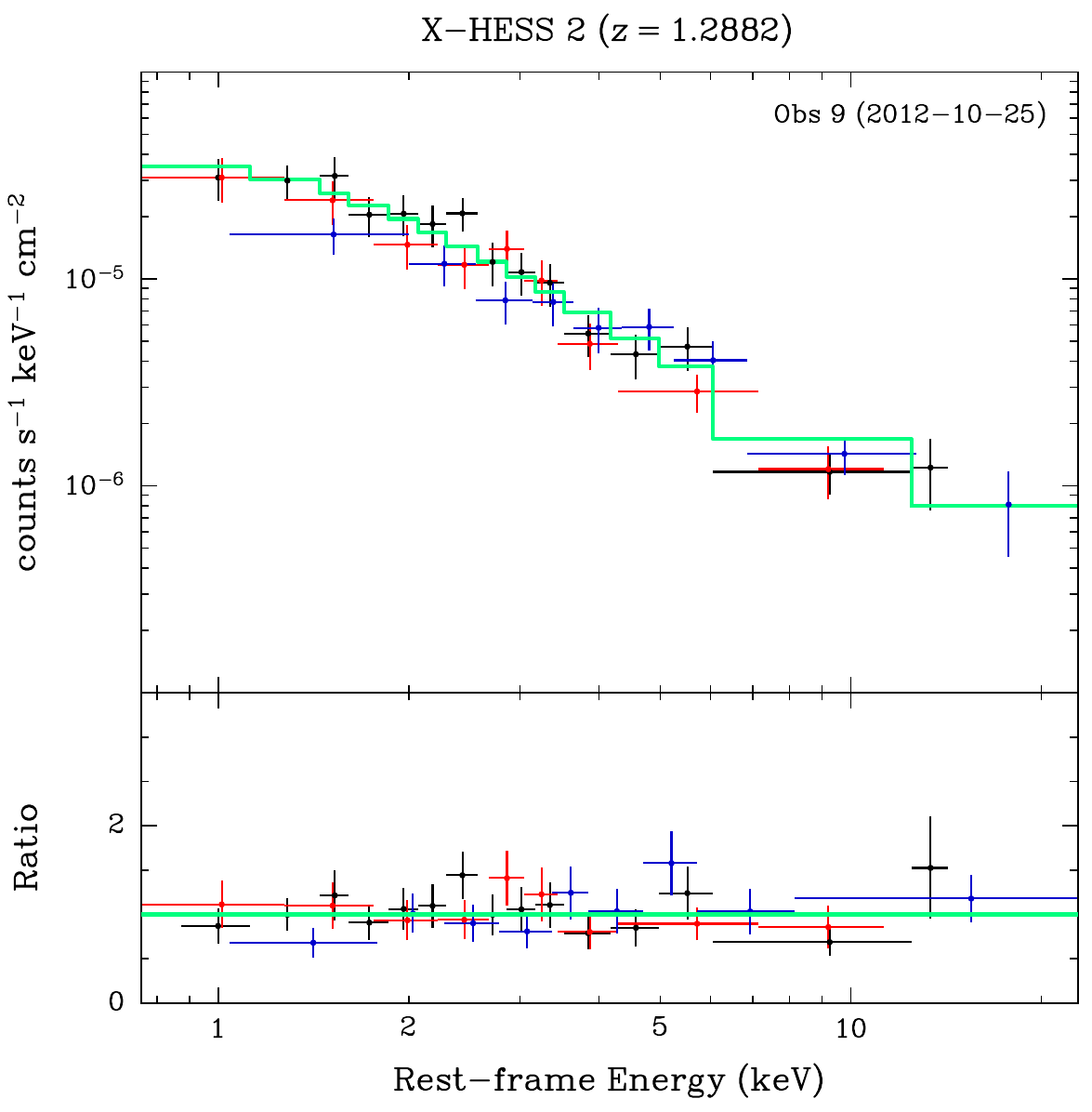}} \\
     
\end{tabular}
\begin{minipage}{1.2\linewidth}
\centering 
{Continuation of Fig. \ref{fig:xhess_spectra}.}
\end{minipage}

\end{figure}

\begin{figure}[h]
     \ContinuedFloat
     \centering
     \renewcommand{\arraystretch}{2}
     \begin{tabular}{ccc}
     \subfloat{\includegraphics[width = 2.1in]{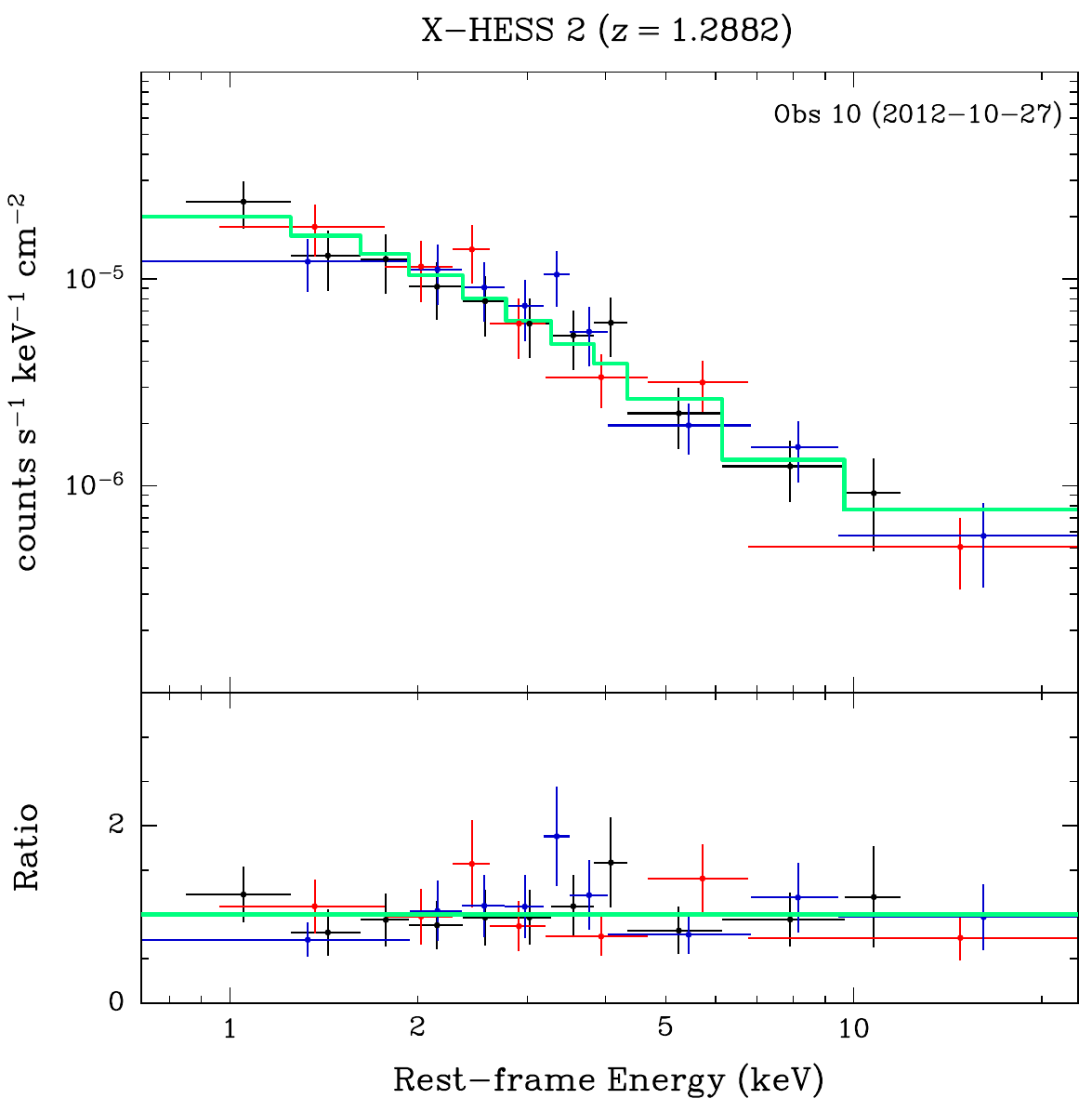}} &
     \subfloat{\includegraphics[width = 2.1in]{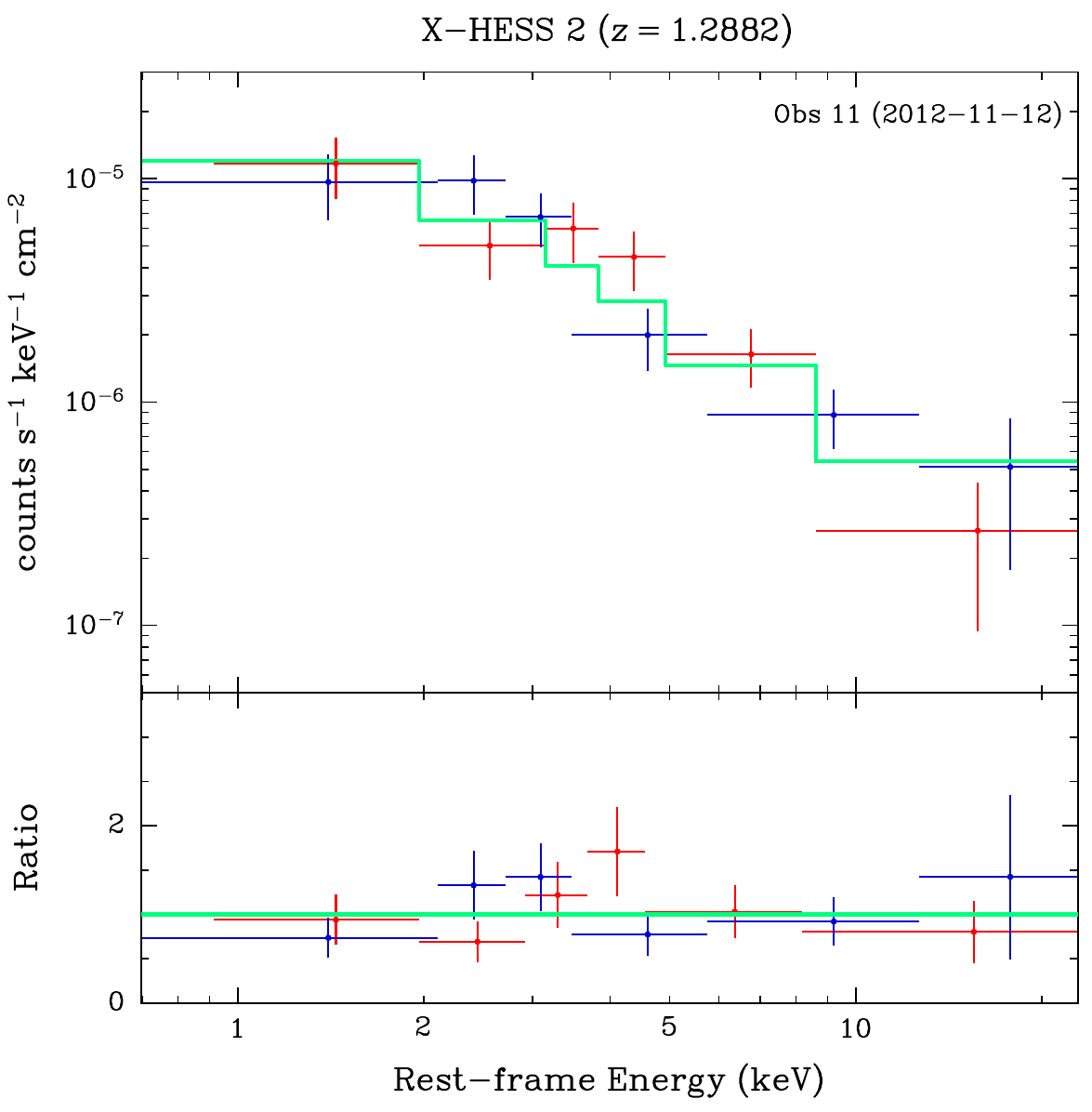}} &
     \subfloat{\includegraphics[width = 2.1in]{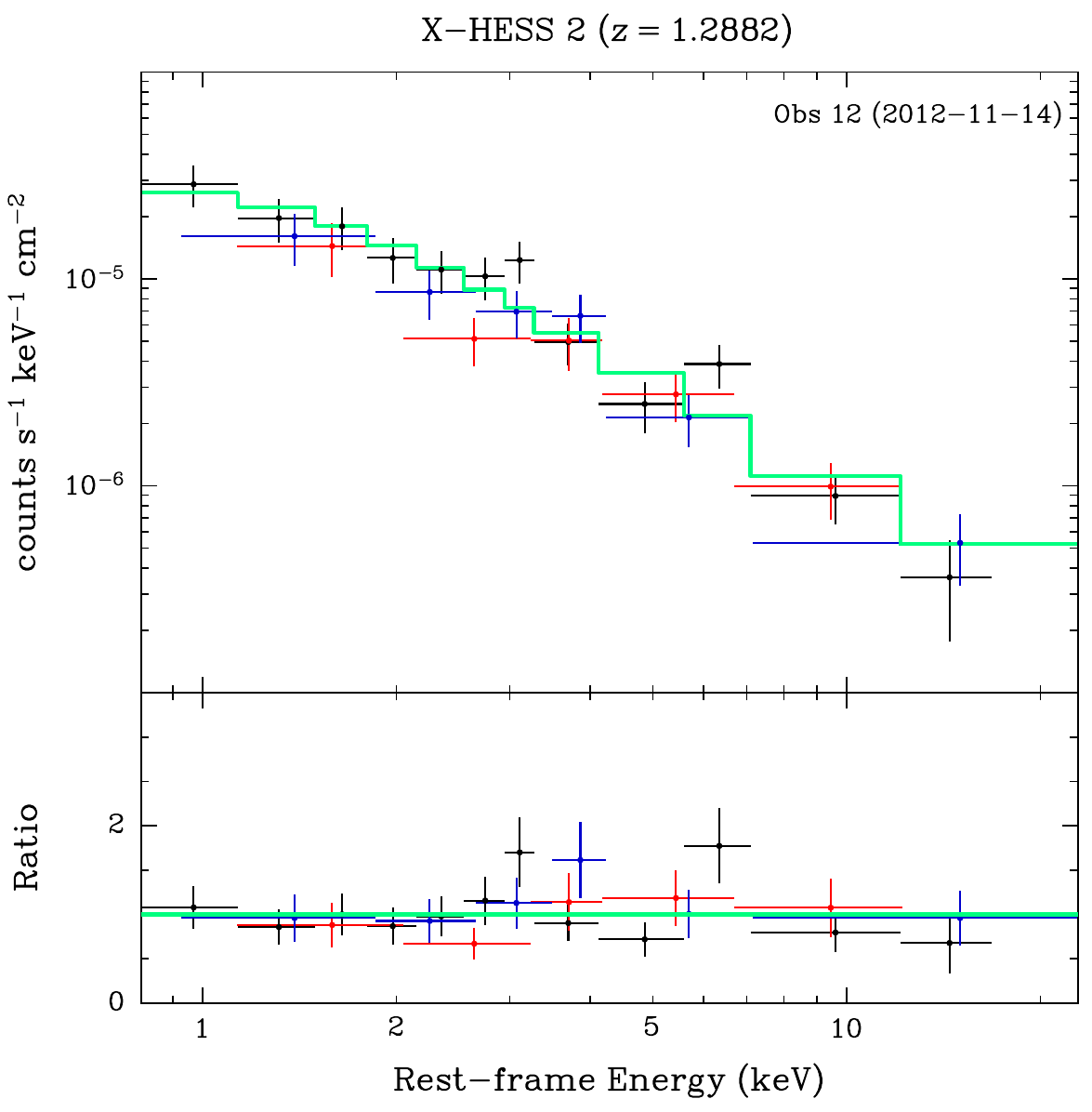}} \\
     \subfloat{\includegraphics[width = 2.1in]{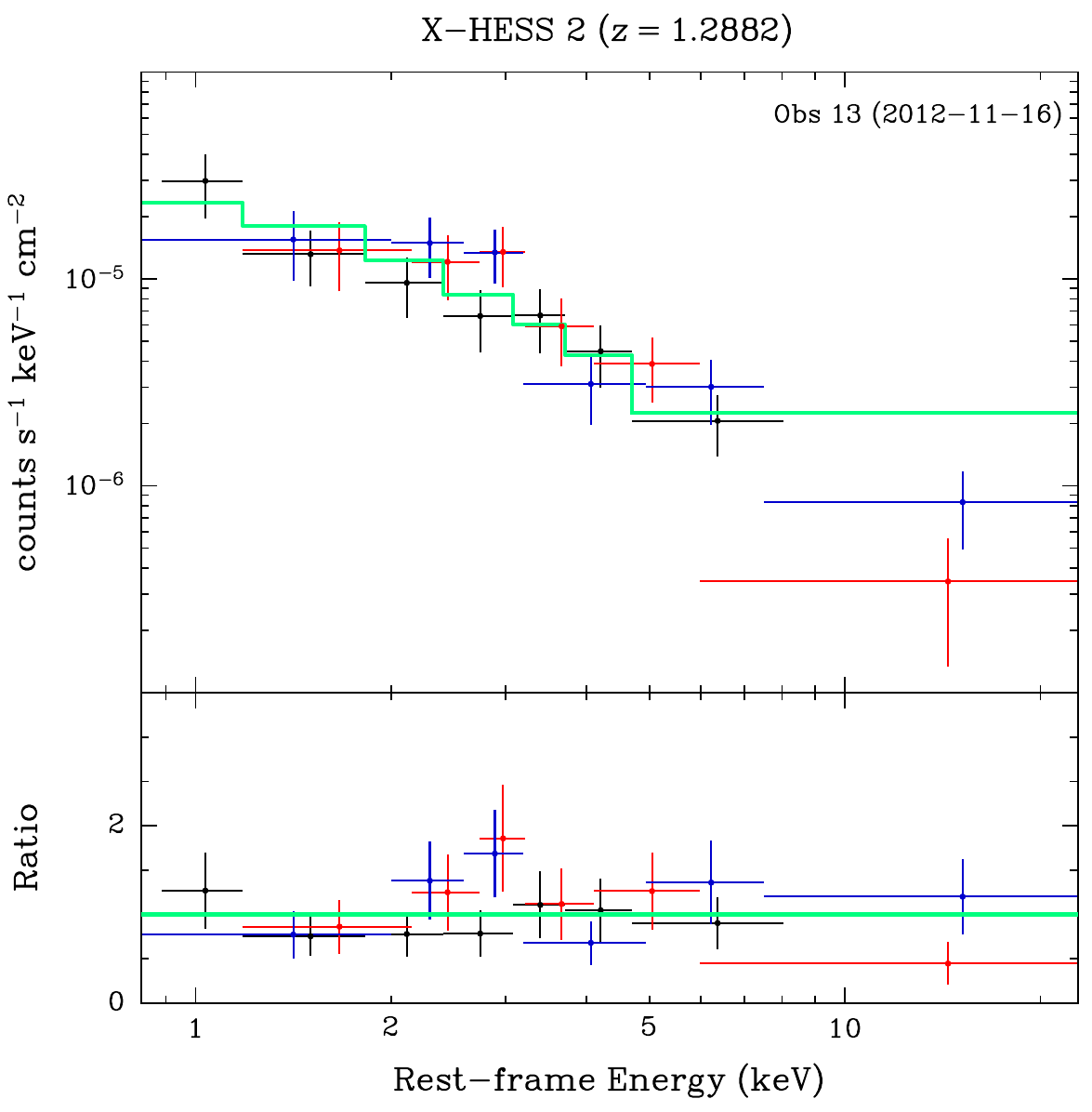}} &
     \subfloat{\includegraphics[width = 2.1in]{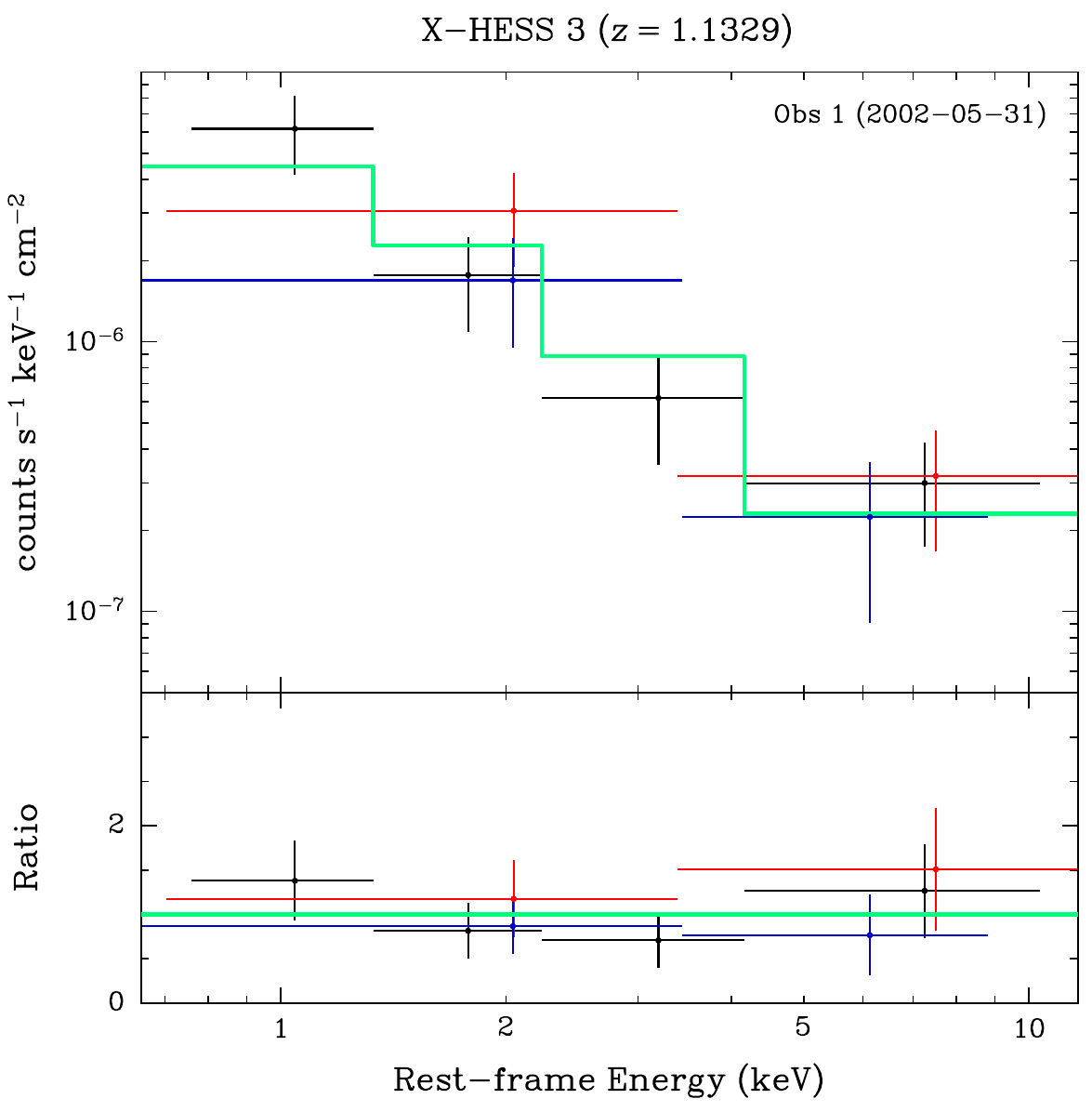}} &
     \subfloat{\includegraphics[width = 2.1in]{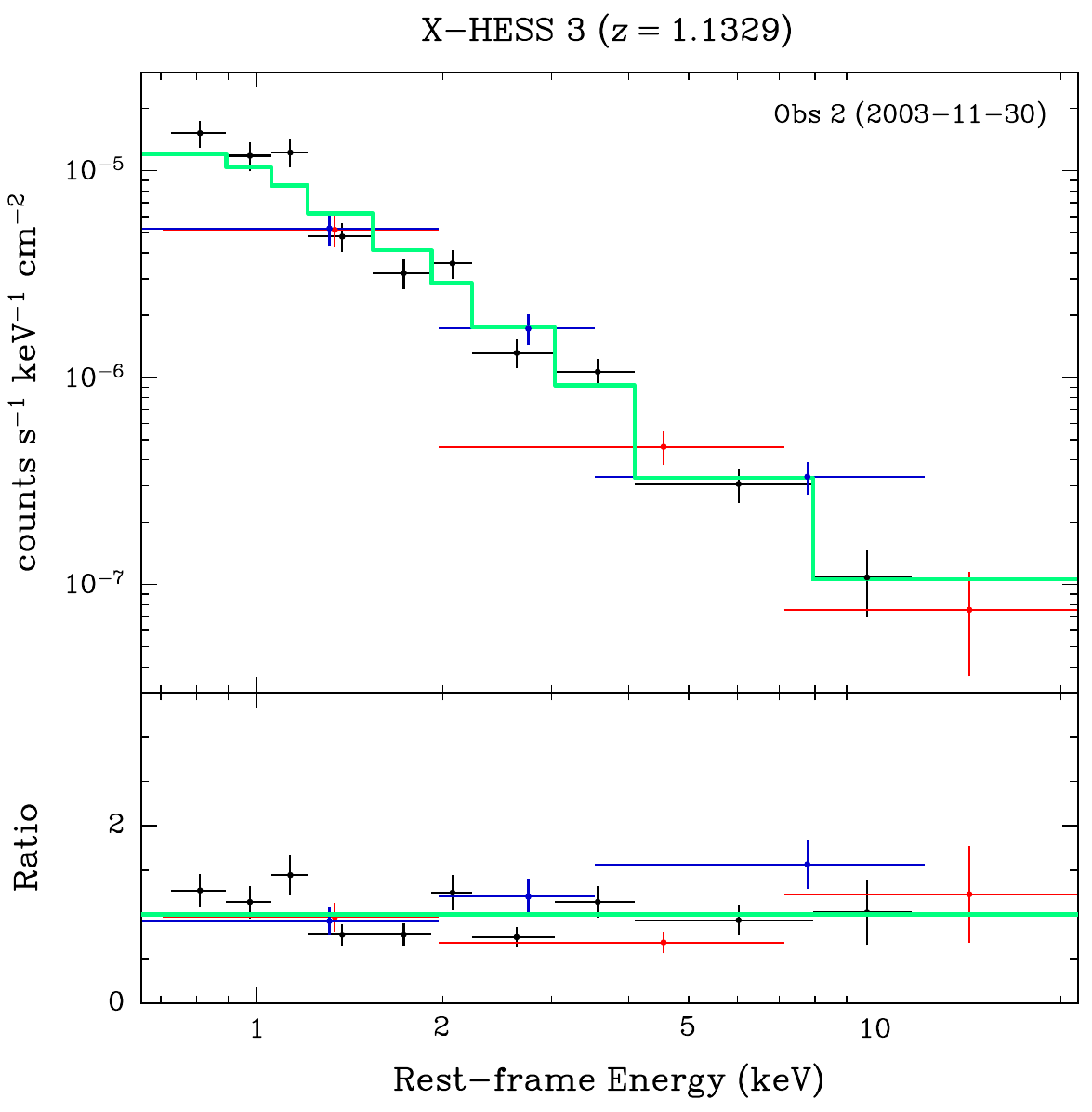}} \\
     \subfloat{\includegraphics[width = 2.1in]{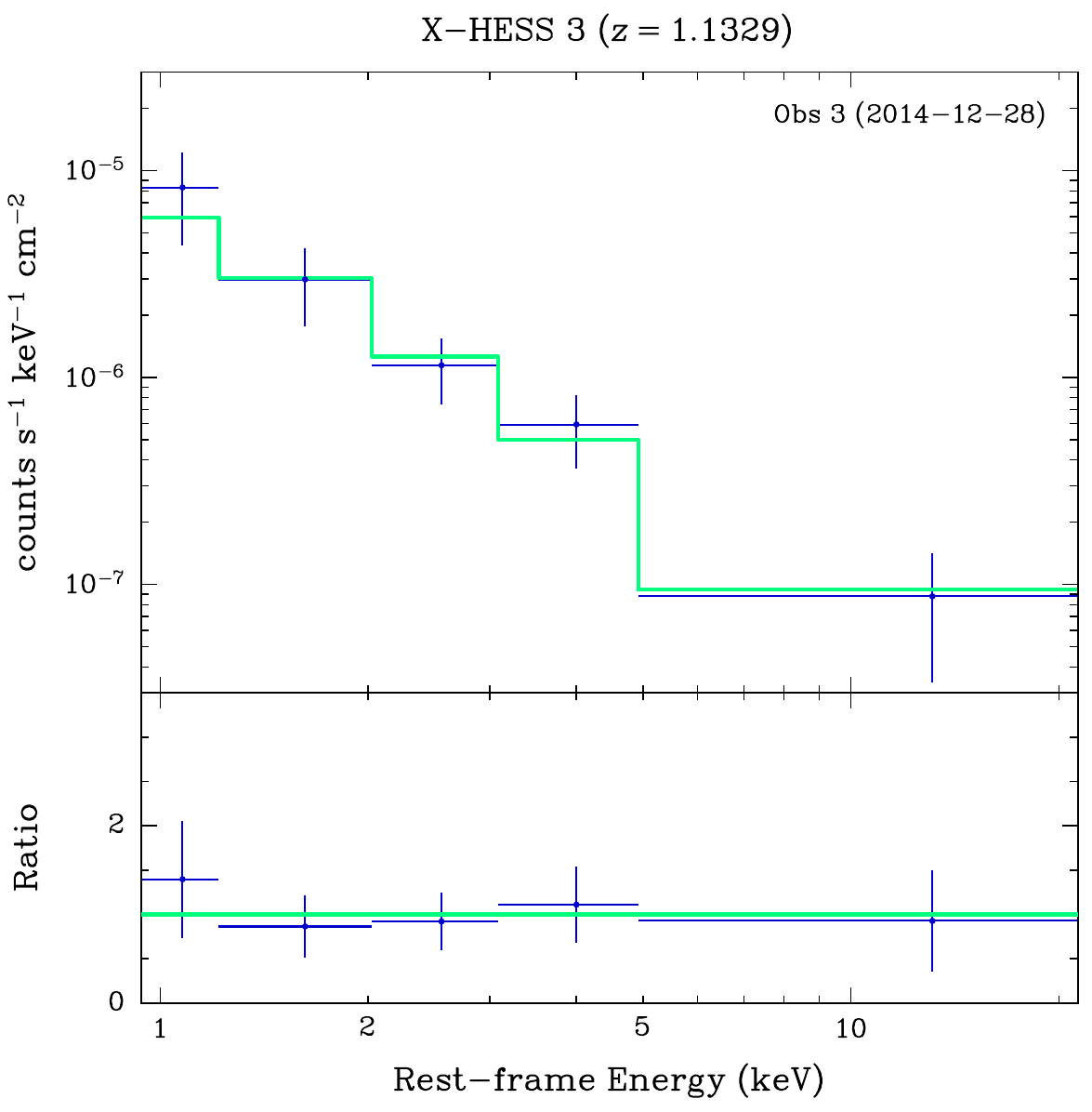}} &
     \subfloat{\includegraphics[width = 2.1in]{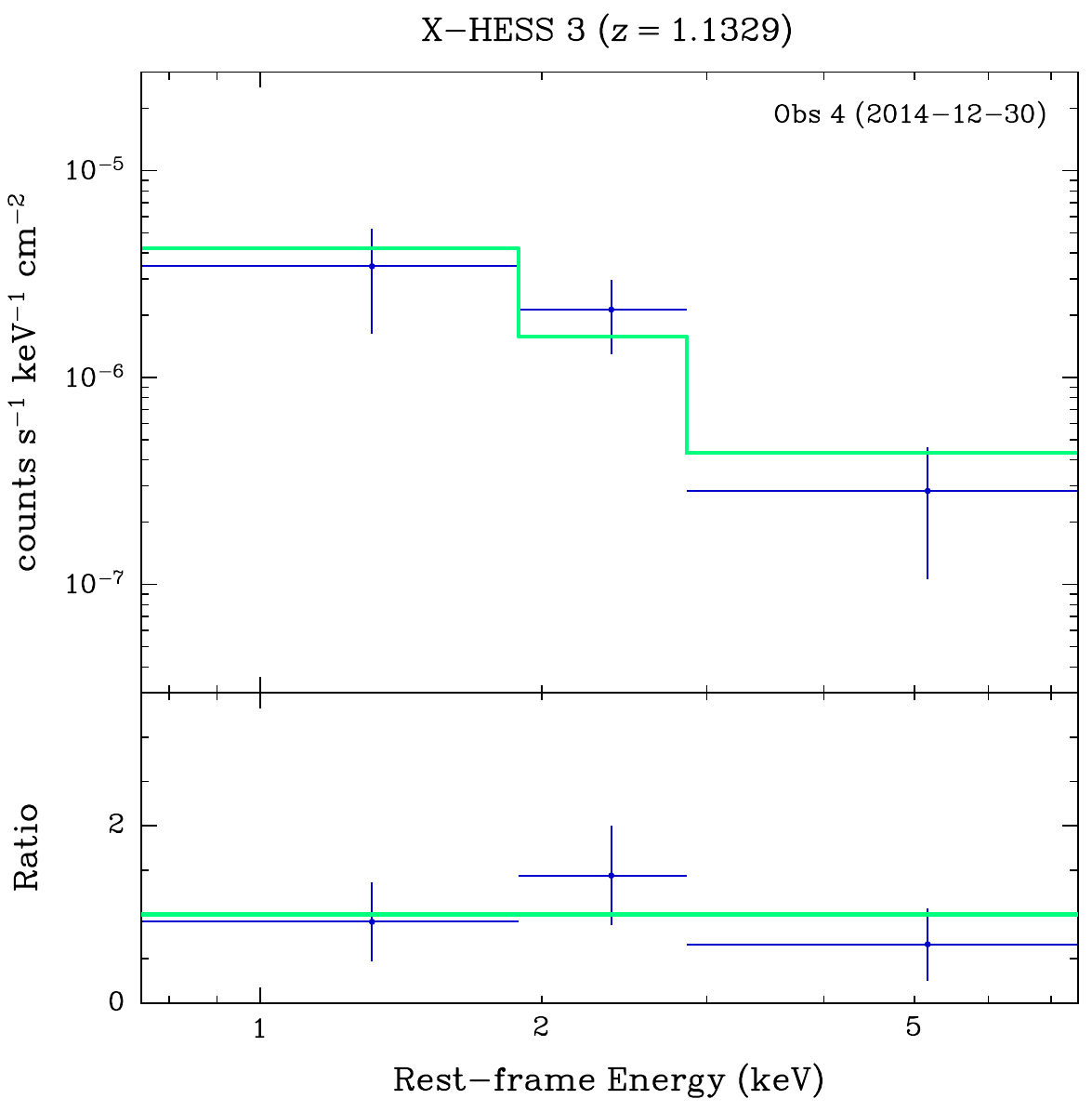}} &
     \subfloat{\includegraphics[width = 2.1in]{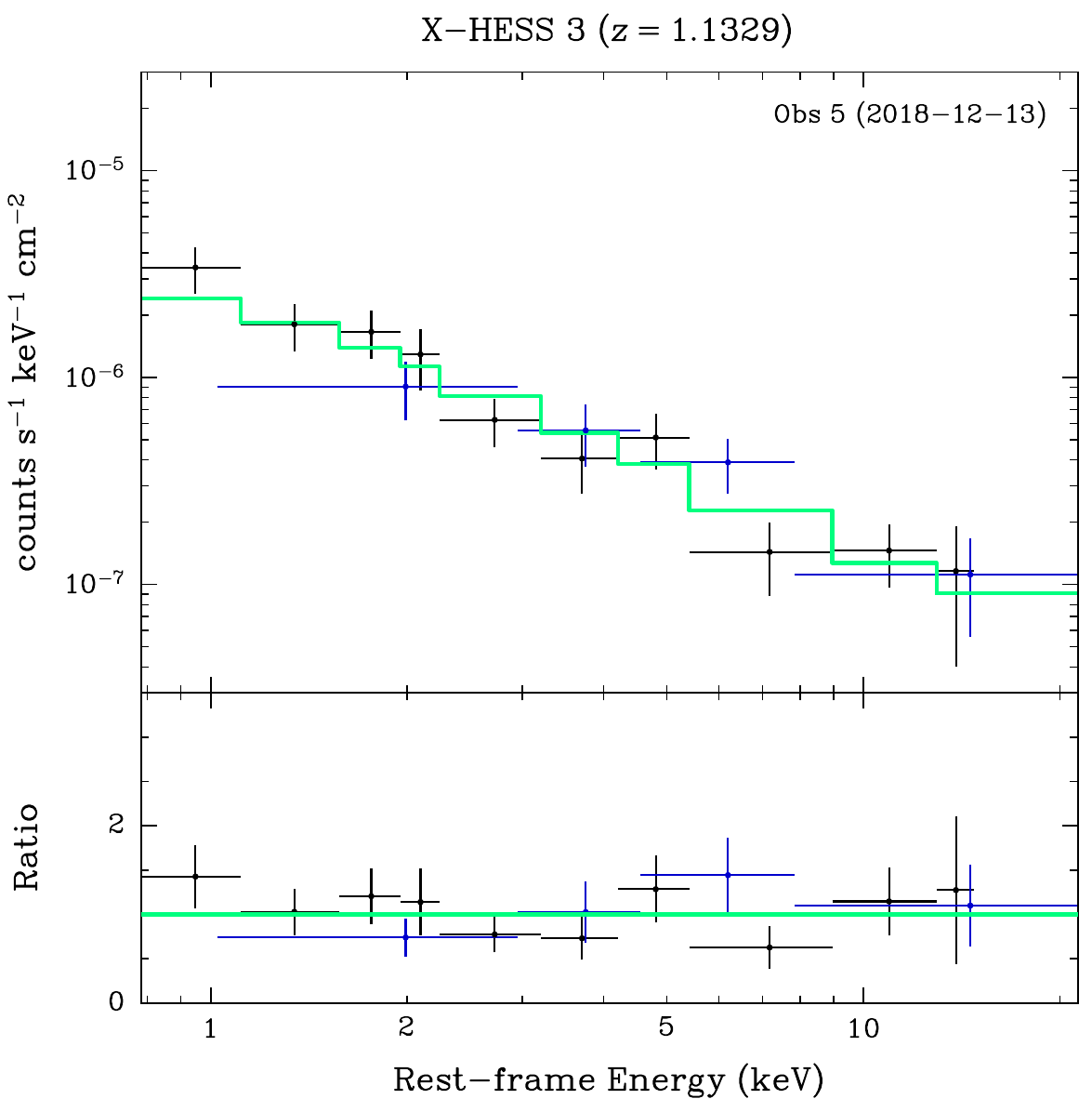}} \\
     \subfloat{\includegraphics[width = 2.1in]{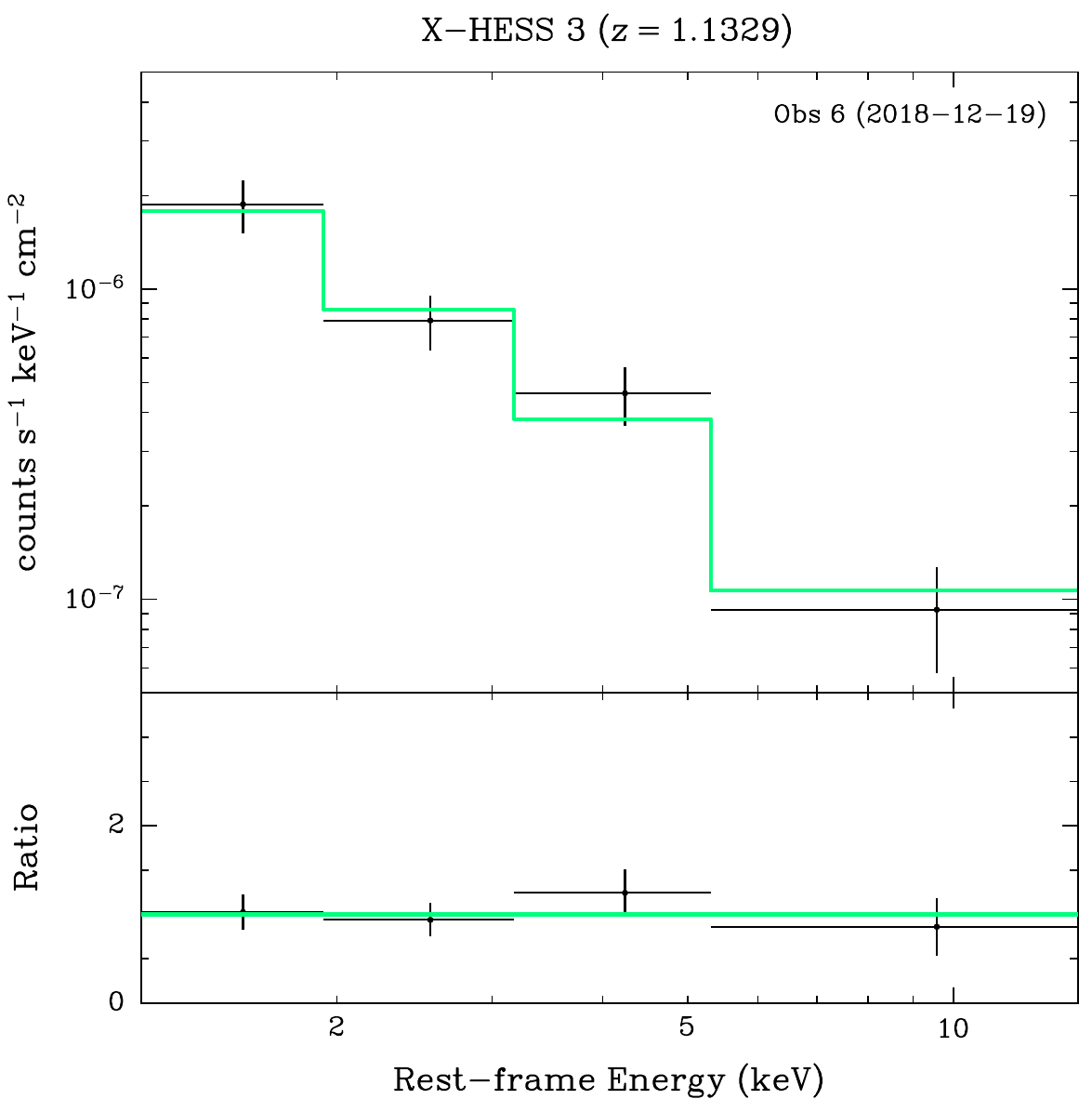}} &
     \subfloat{\includegraphics[width = 2.1in]{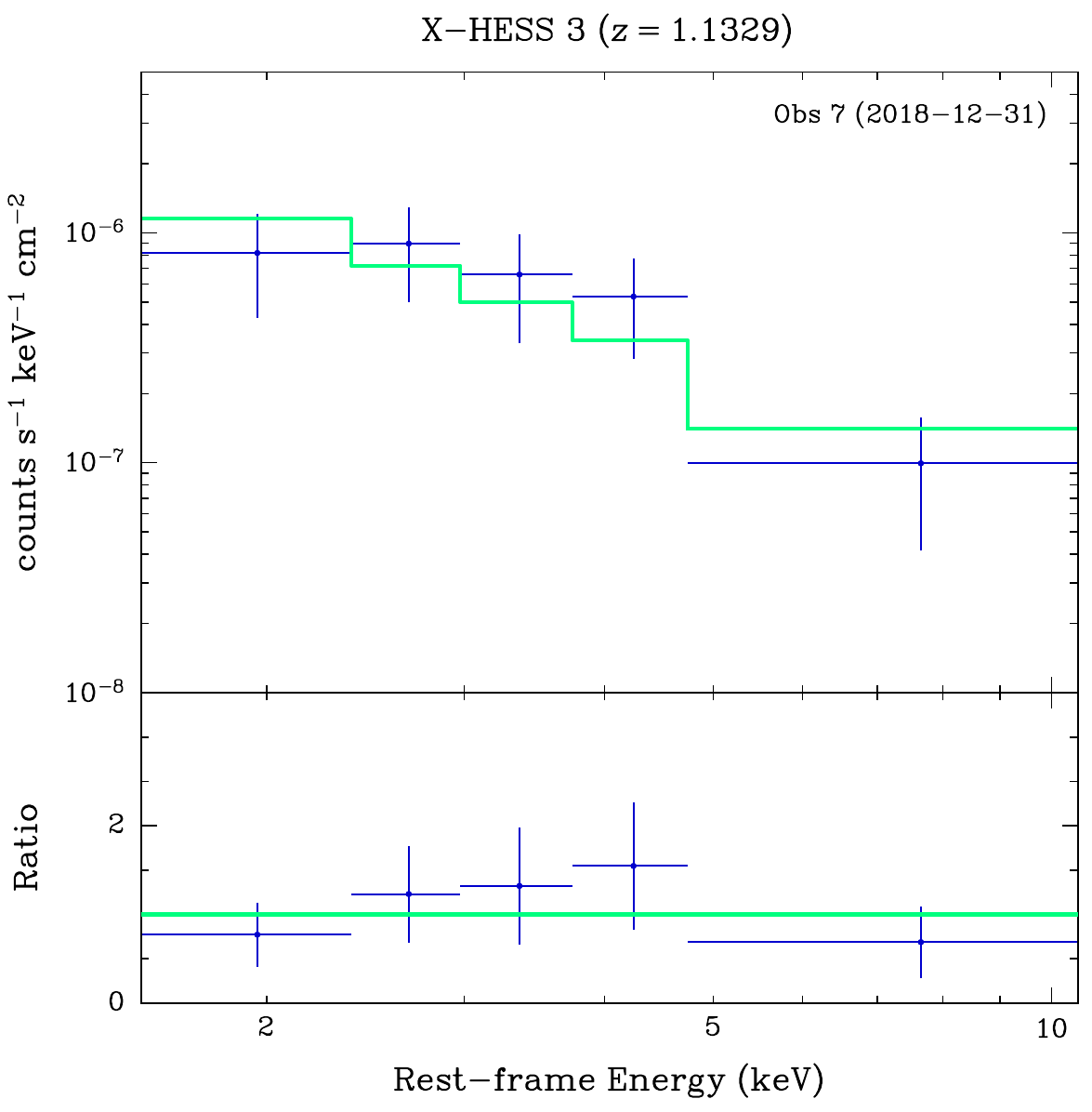}} &
     \subfloat{\includegraphics[width = 2.1in]{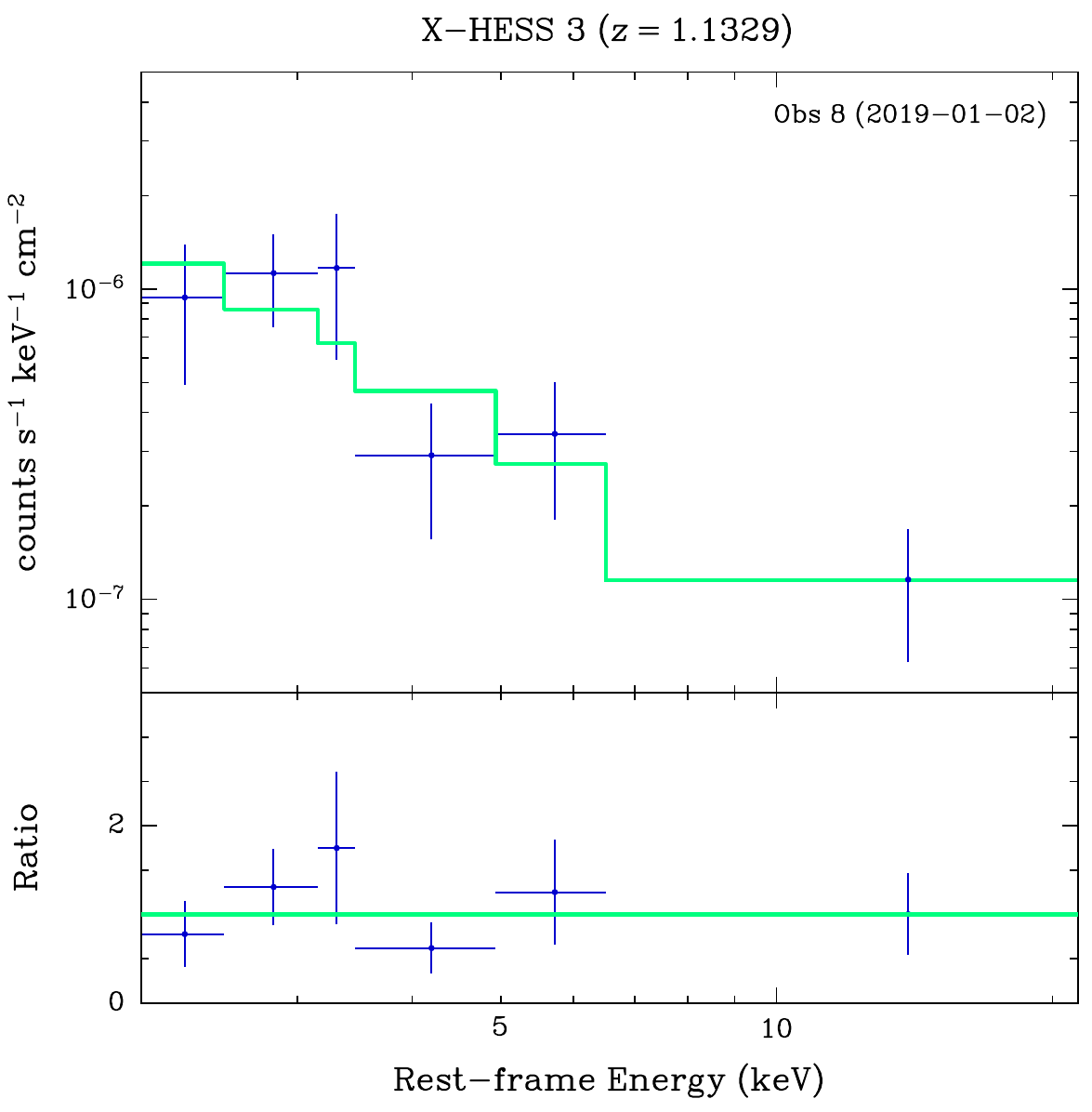}} \\
     \end{tabular}
\begin{minipage}{1.2\linewidth}
    \centering 
    {Continuation of Fig. \ref{fig:xhess_spectra}.}
\end{minipage}
\end{figure}

\begin{figure}[h]
     \ContinuedFloat
     \centering
     \renewcommand{\arraystretch}{2}
     \begin{tabular}{ccc}
     \subfloat{\includegraphics[width = 2.1in]{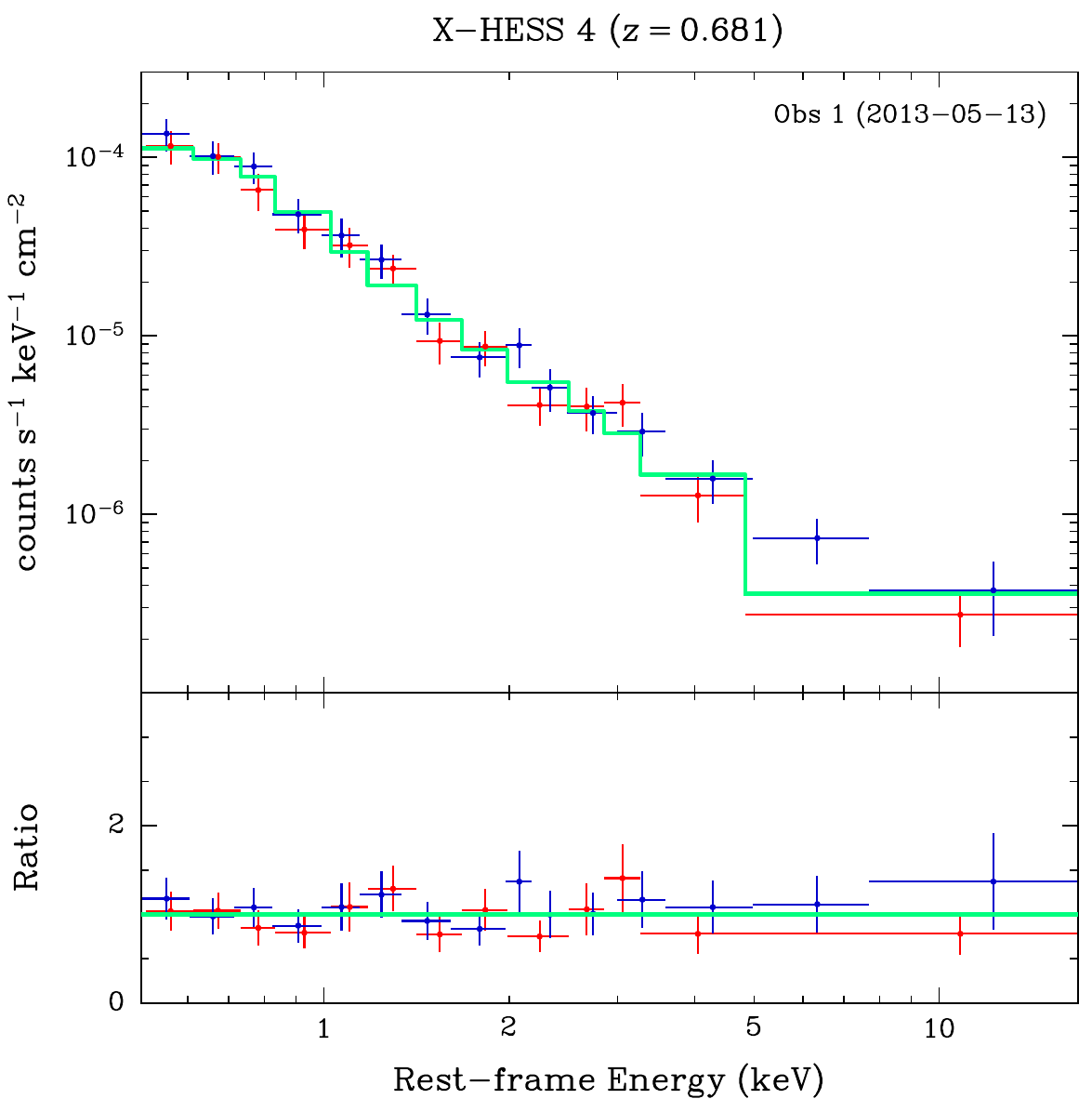}} &
     \subfloat{\includegraphics[width = 2.1in]{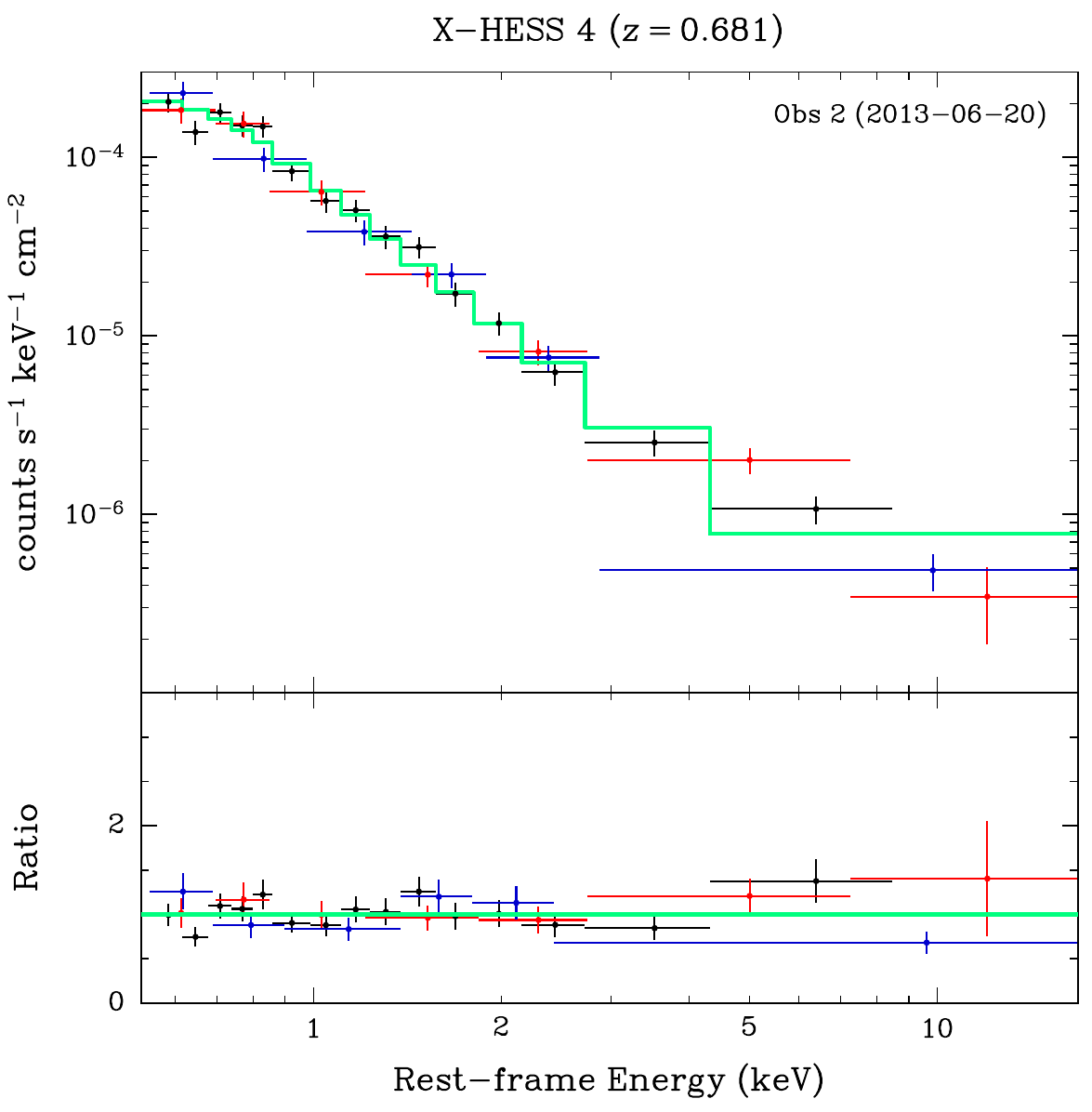}} &
     \subfloat{\includegraphics[width = 2.1in]{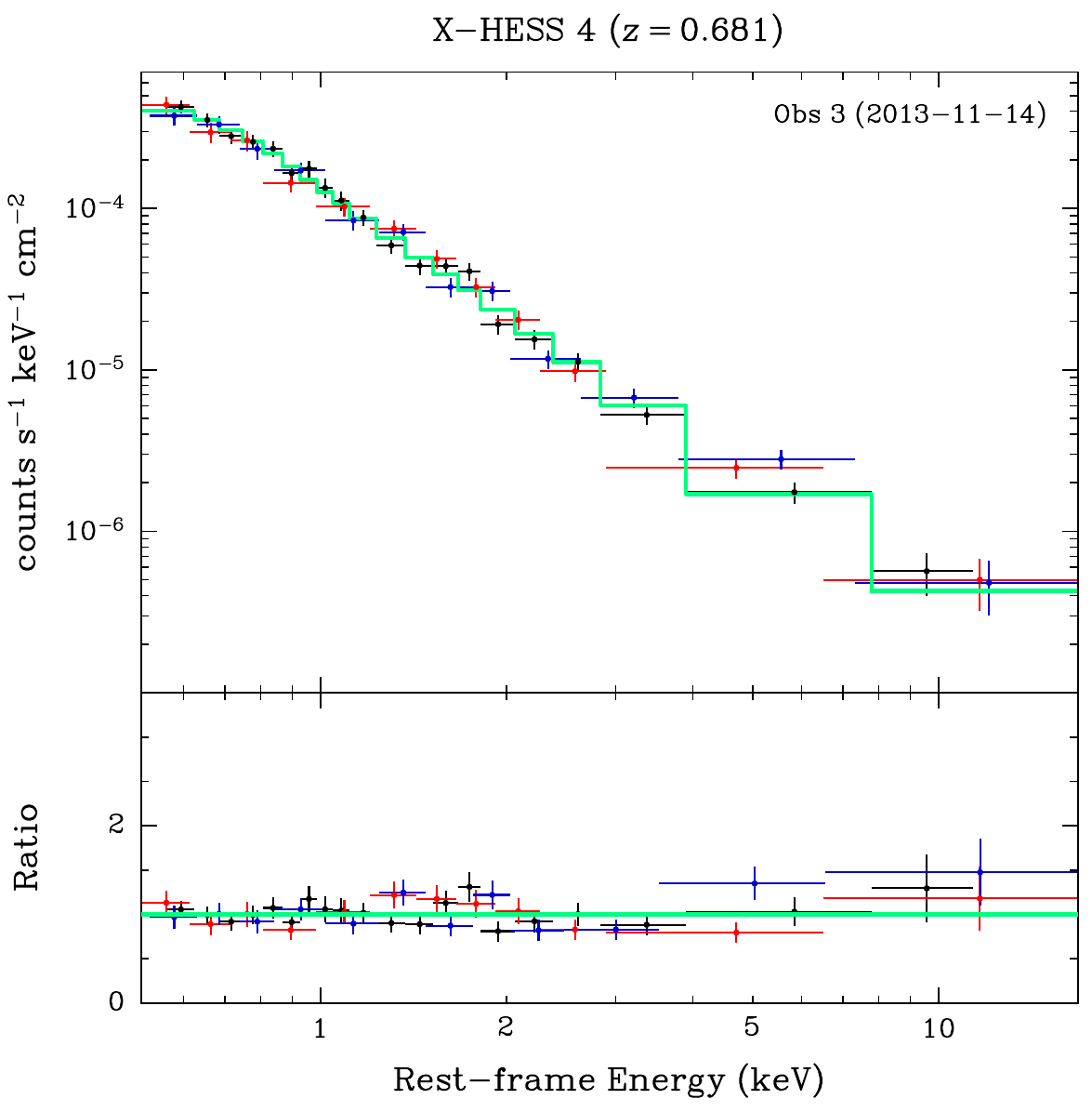}} \\
     \subfloat{\includegraphics[width = 2.1in]{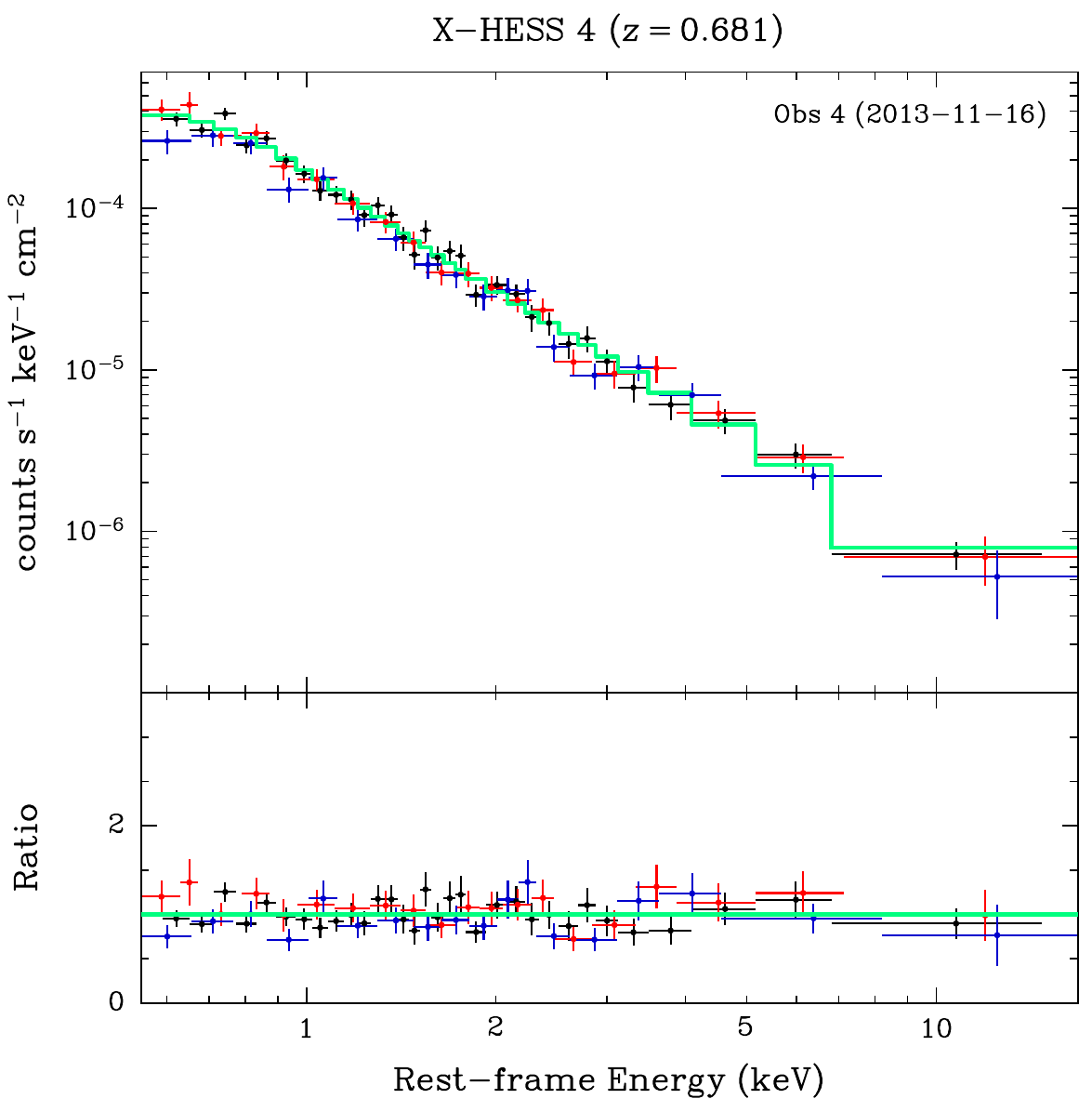}} &
     \subfloat{\includegraphics[width = 2.1in]{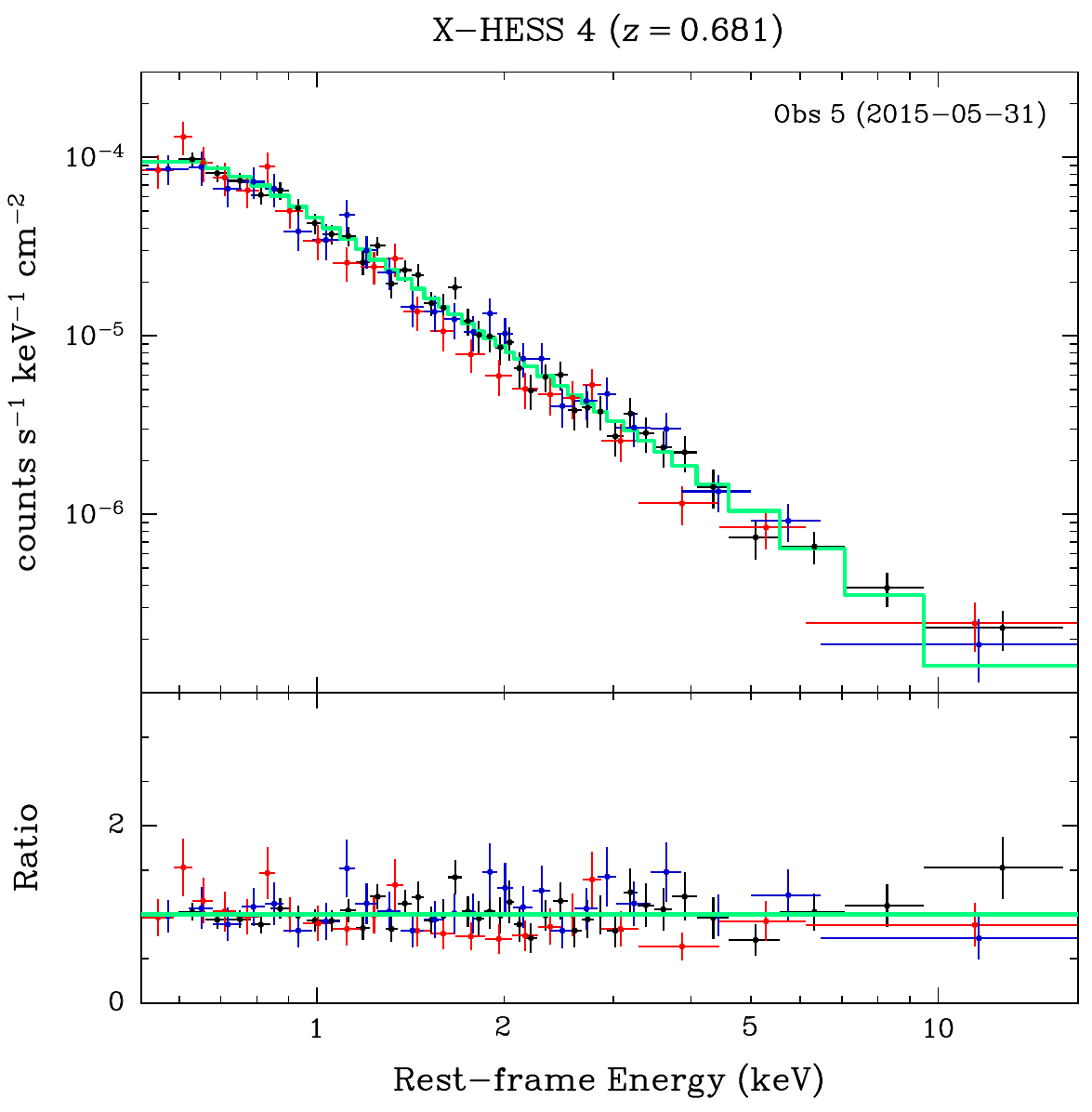}} &
     \subfloat{\includegraphics[width = 2.1in]{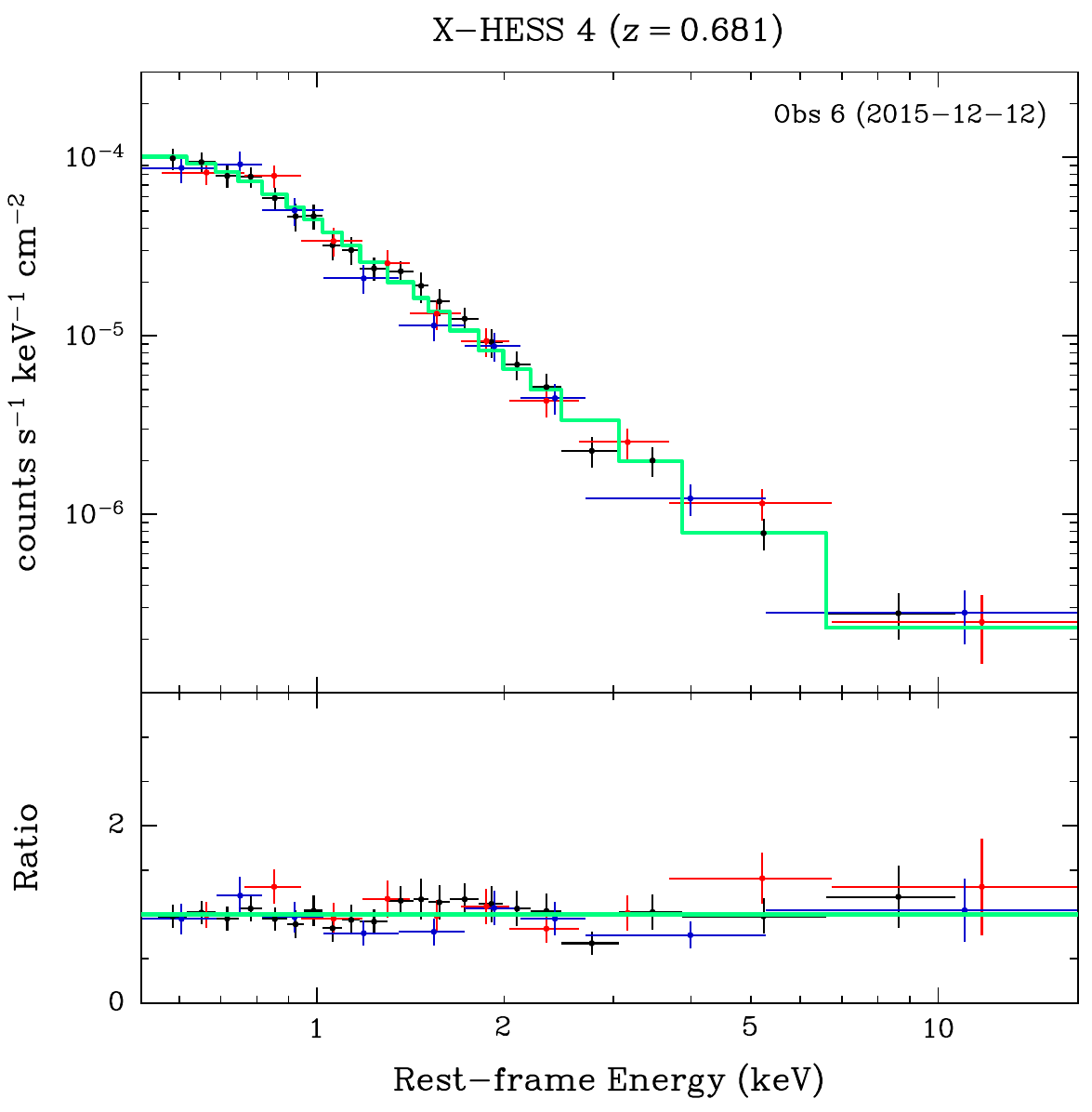}} \\
     \subfloat{\includegraphics[width = 2.1in]{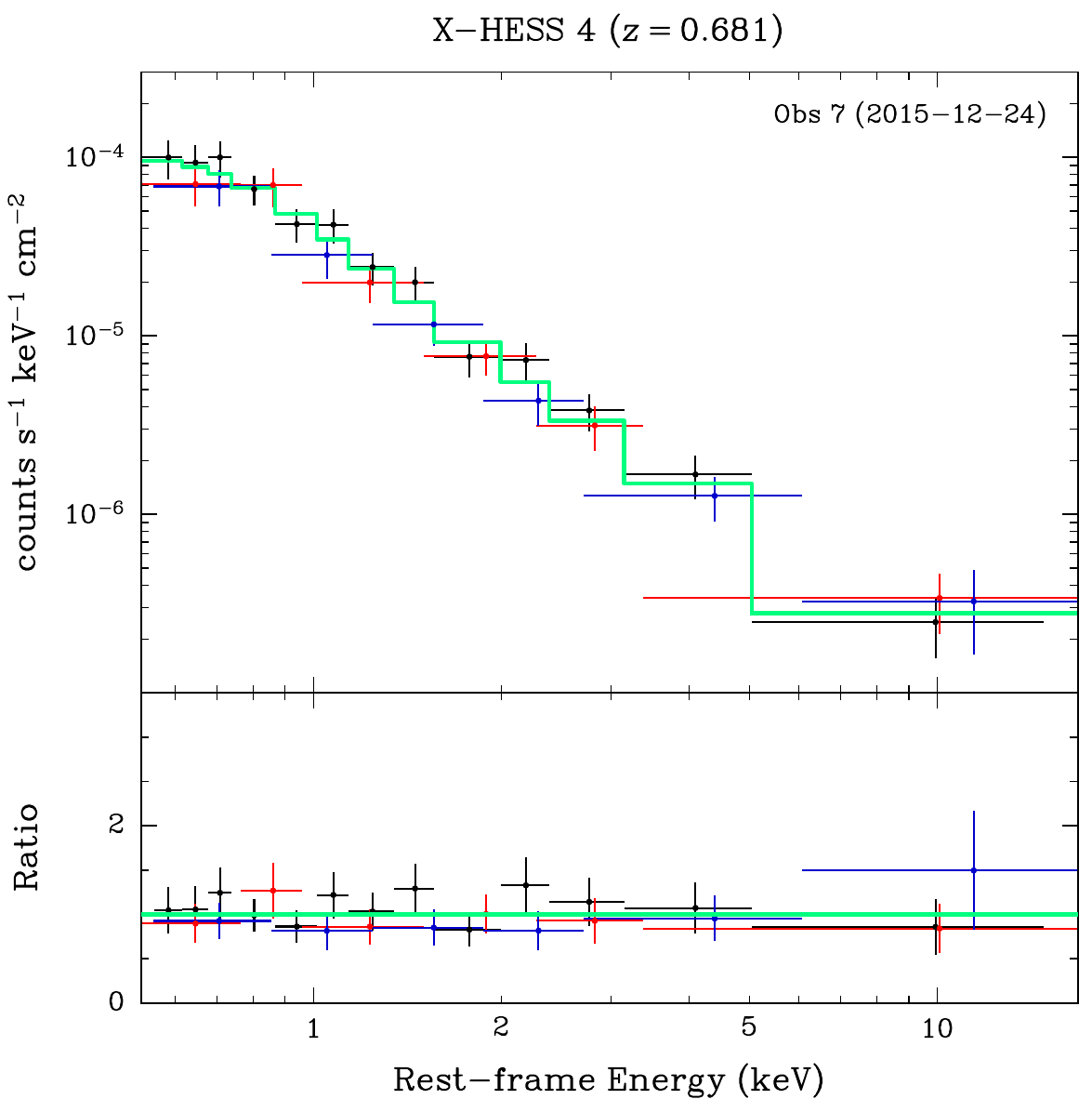}} &
     \subfloat{\includegraphics[width = 2.1in]{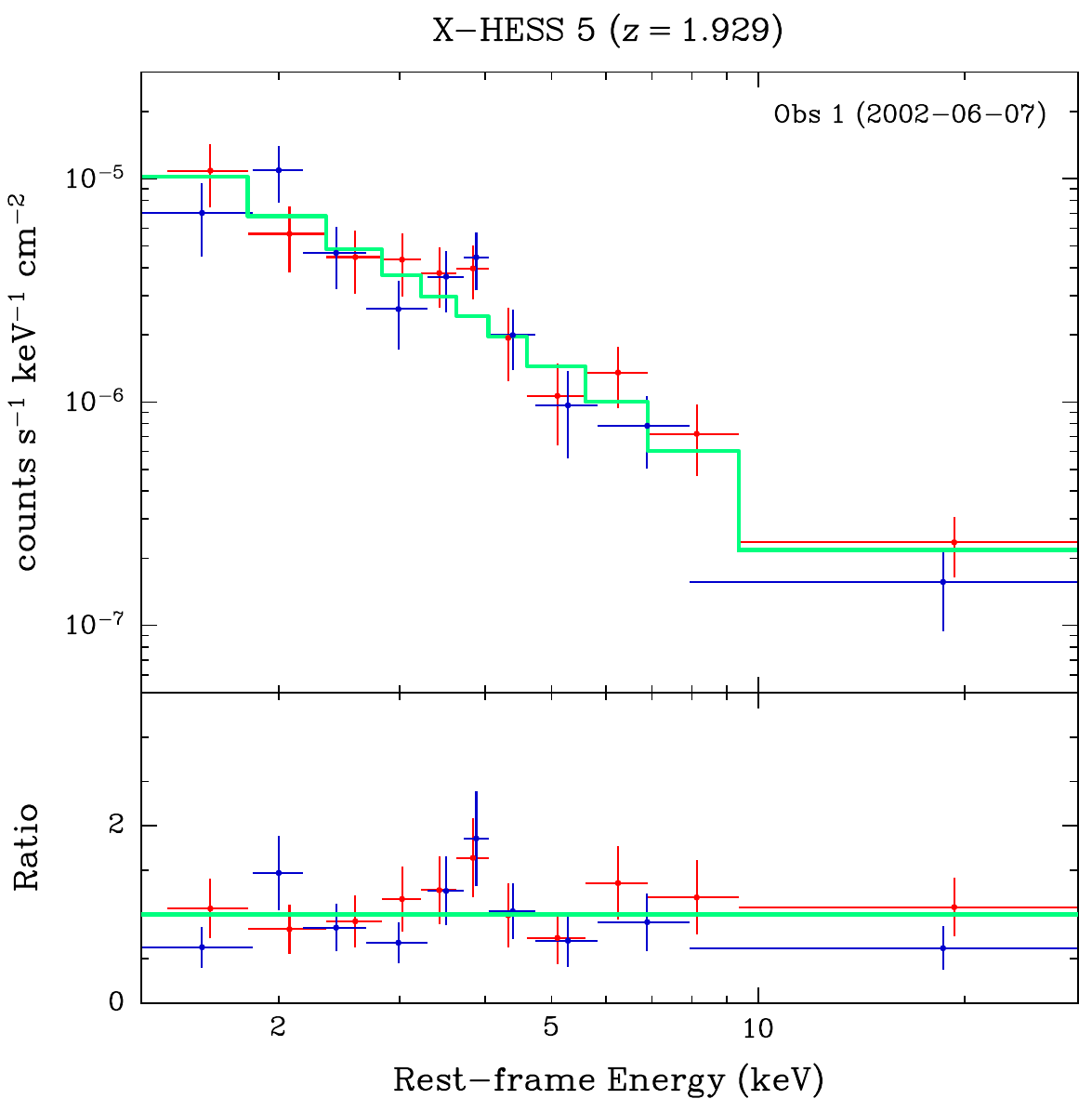}} &
     \subfloat{\includegraphics[width = 2.1in]{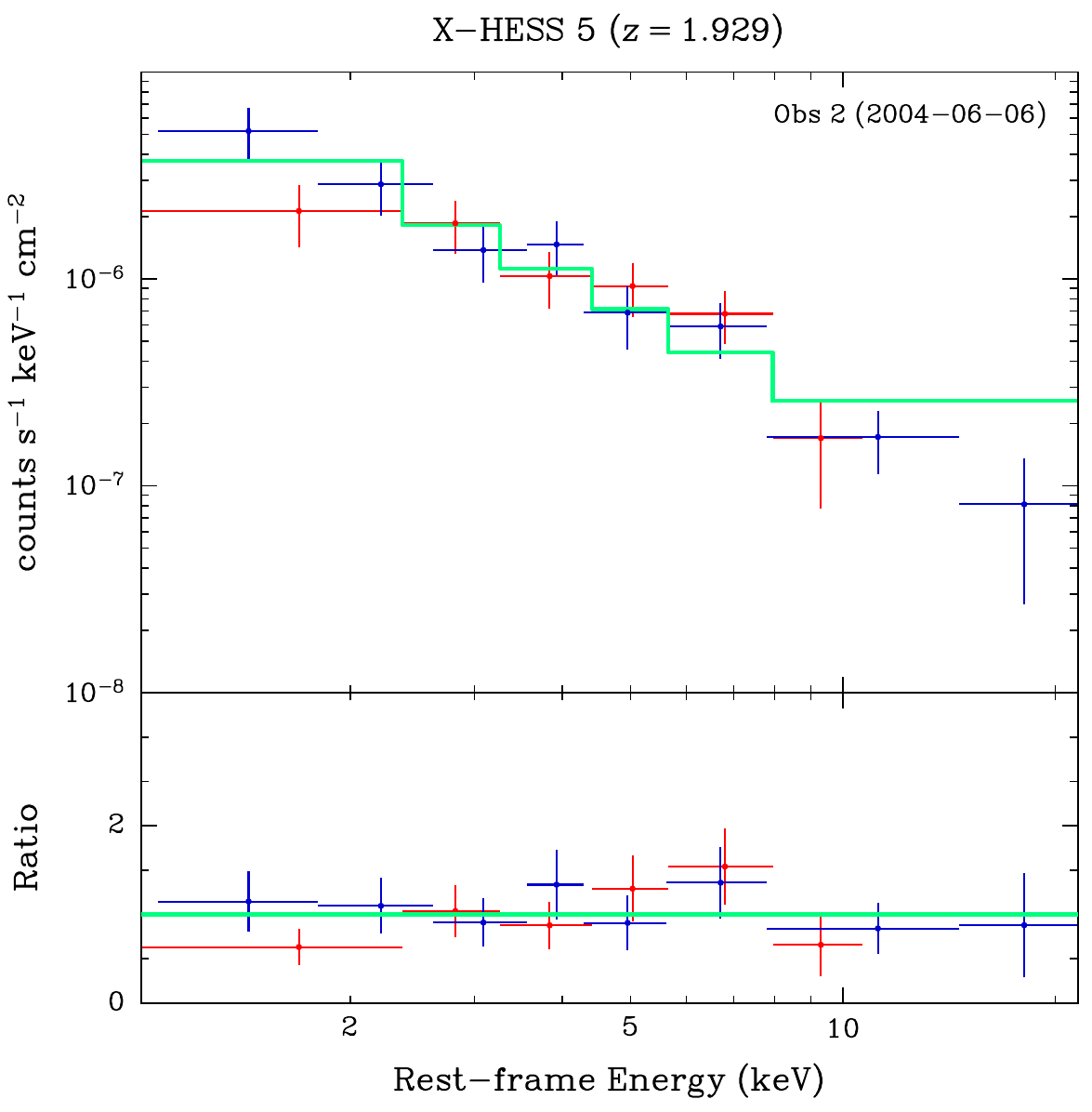}} \\
     \subfloat{\includegraphics[width = 2.1in]{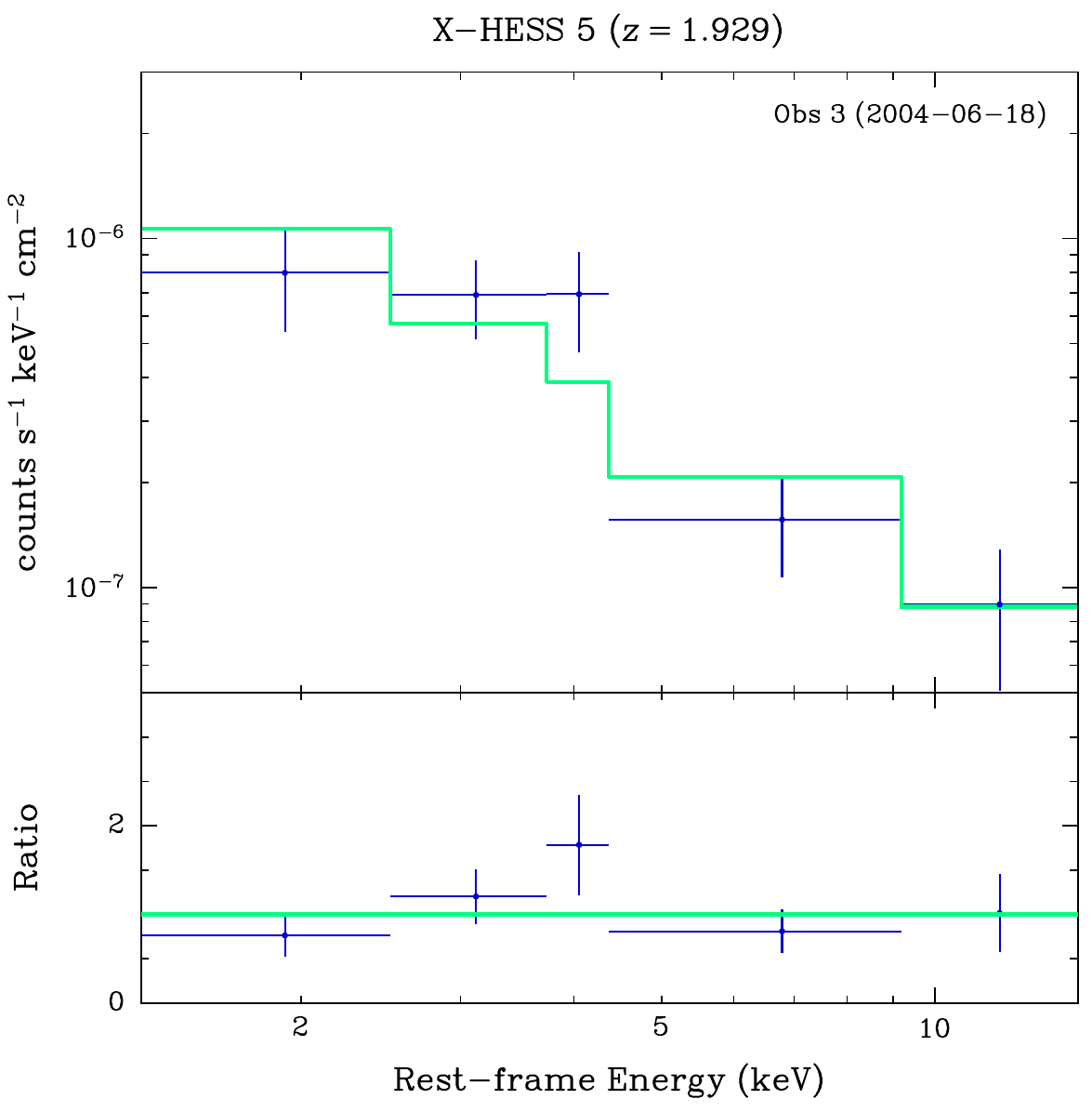}} &
     \subfloat{\includegraphics[width = 2.1in]{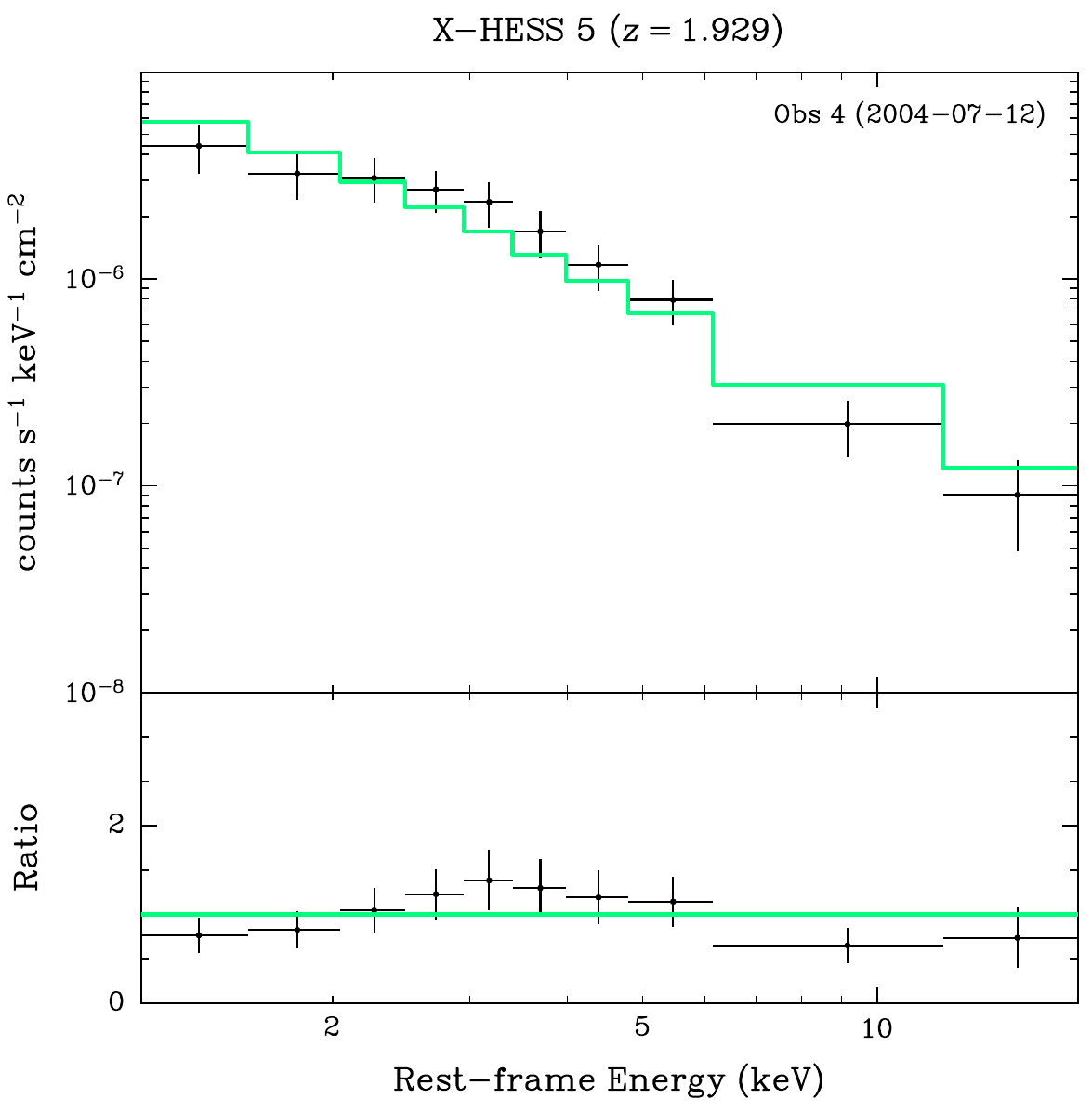}} &
     \subfloat{\includegraphics[width = 2.1in]{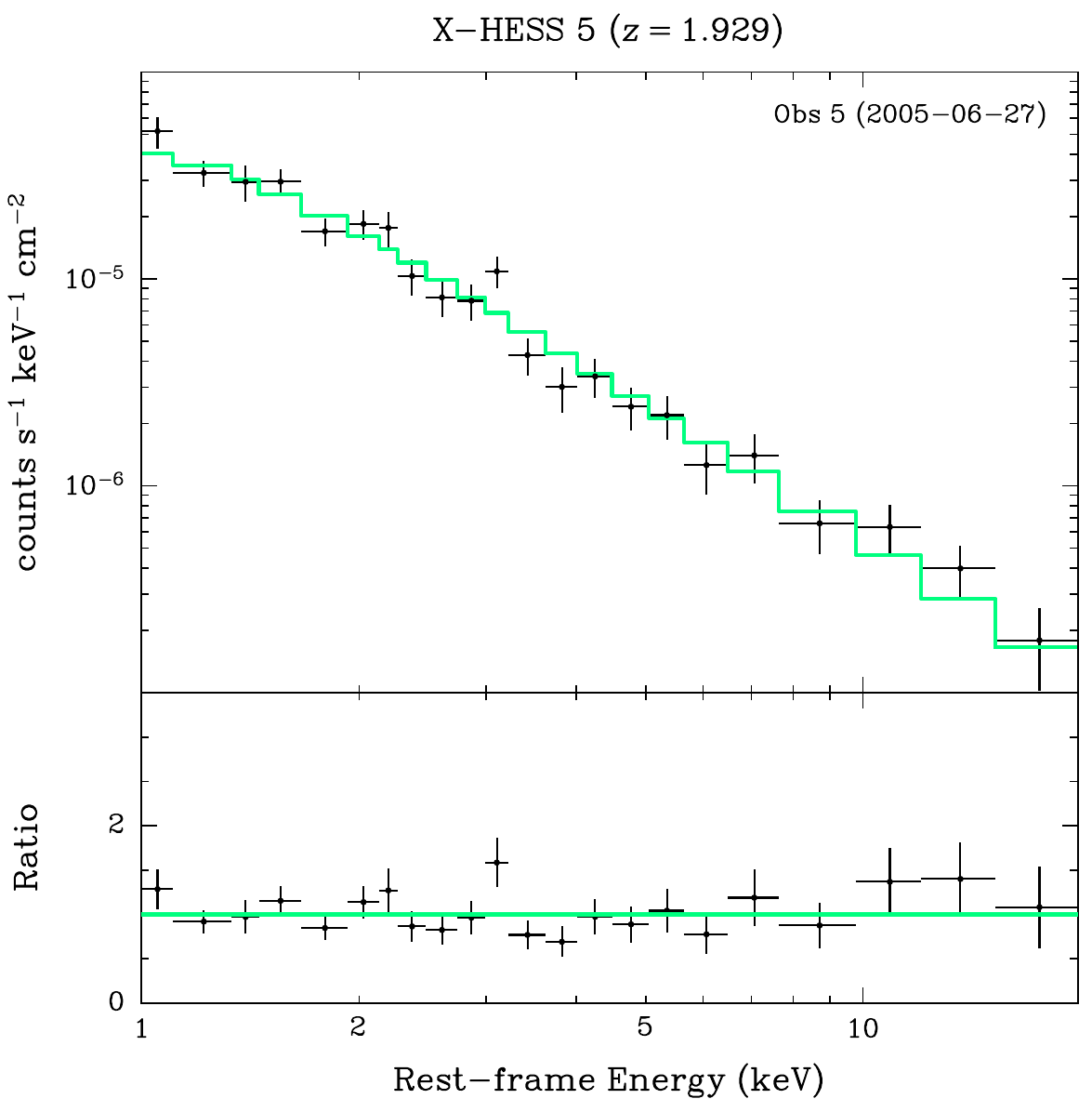}} \\
     
\end{tabular}
\begin{minipage}{1.2\linewidth}
    \centering 
    {Continuation of Fig. \ref{fig:xhess_spectra}.}
\end{minipage}

\end{figure}

\begin{figure}[h]
     \ContinuedFloat
     \centering
     \renewcommand{\arraystretch}{2}
     \begin{tabular}{ccc}
     \subfloat{\includegraphics[width = 2.1in]{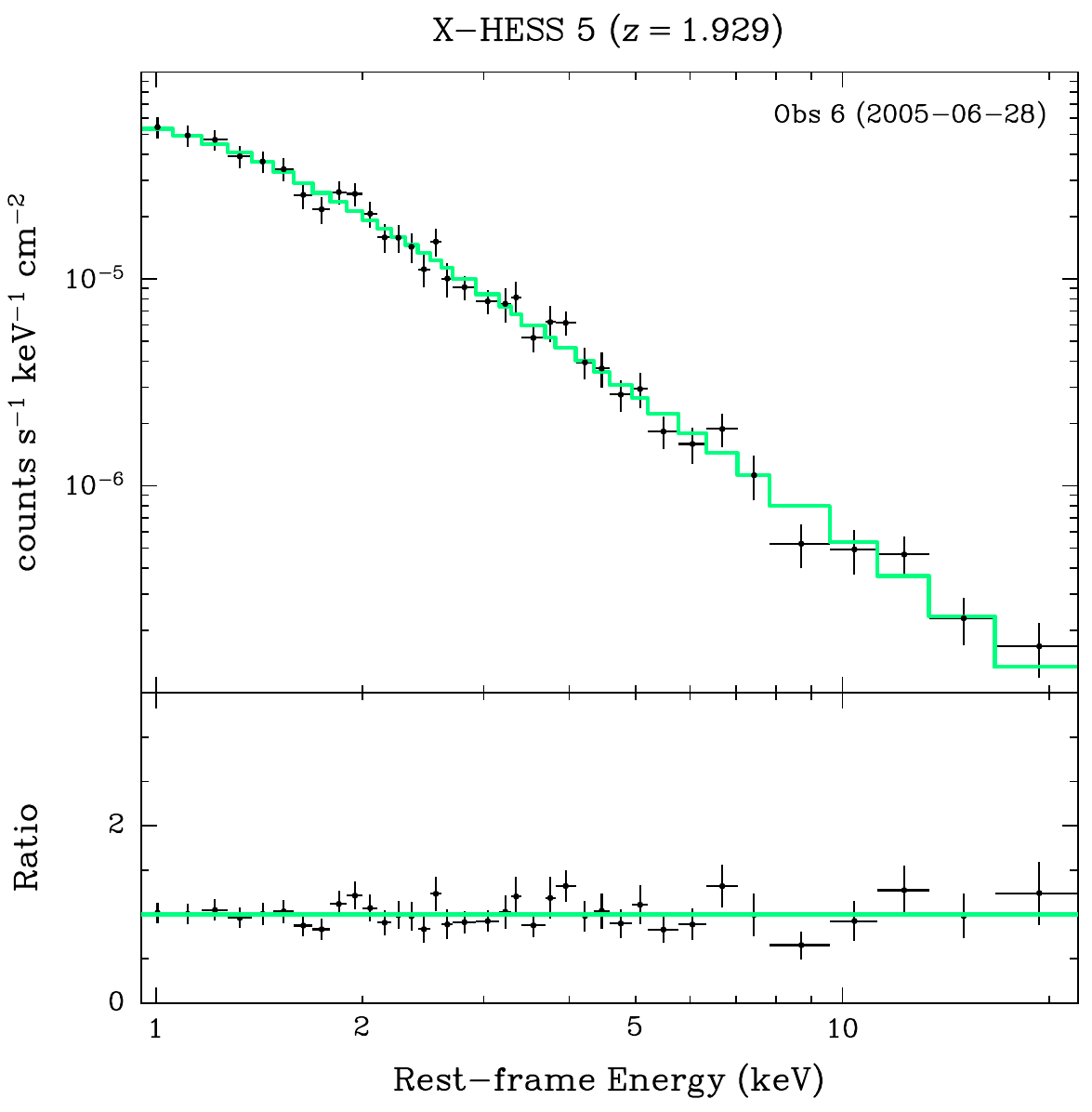}} &
     \subfloat{\includegraphics[width = 2.1in]{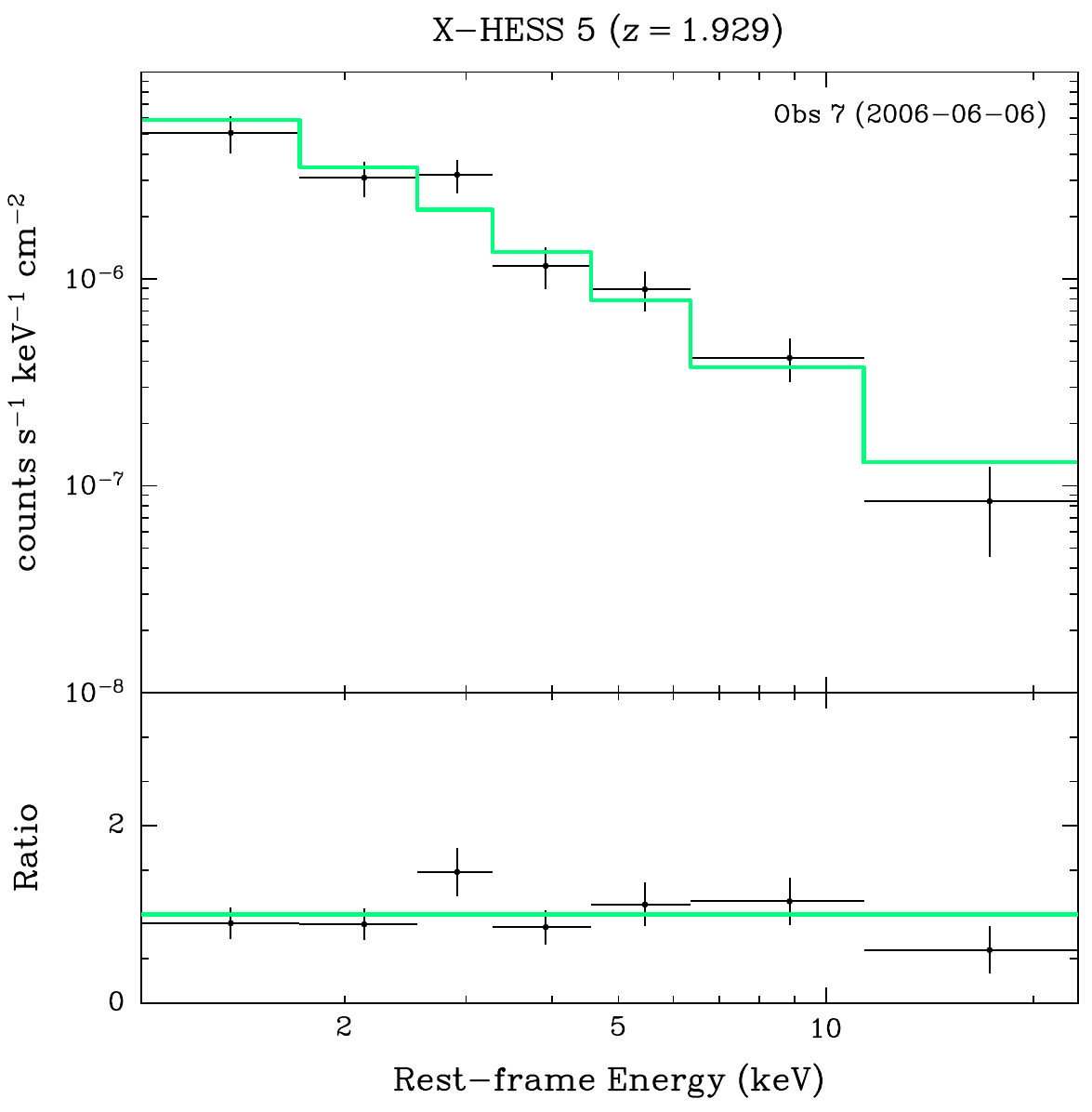}} &
     \subfloat{\includegraphics[width = 2.1in]{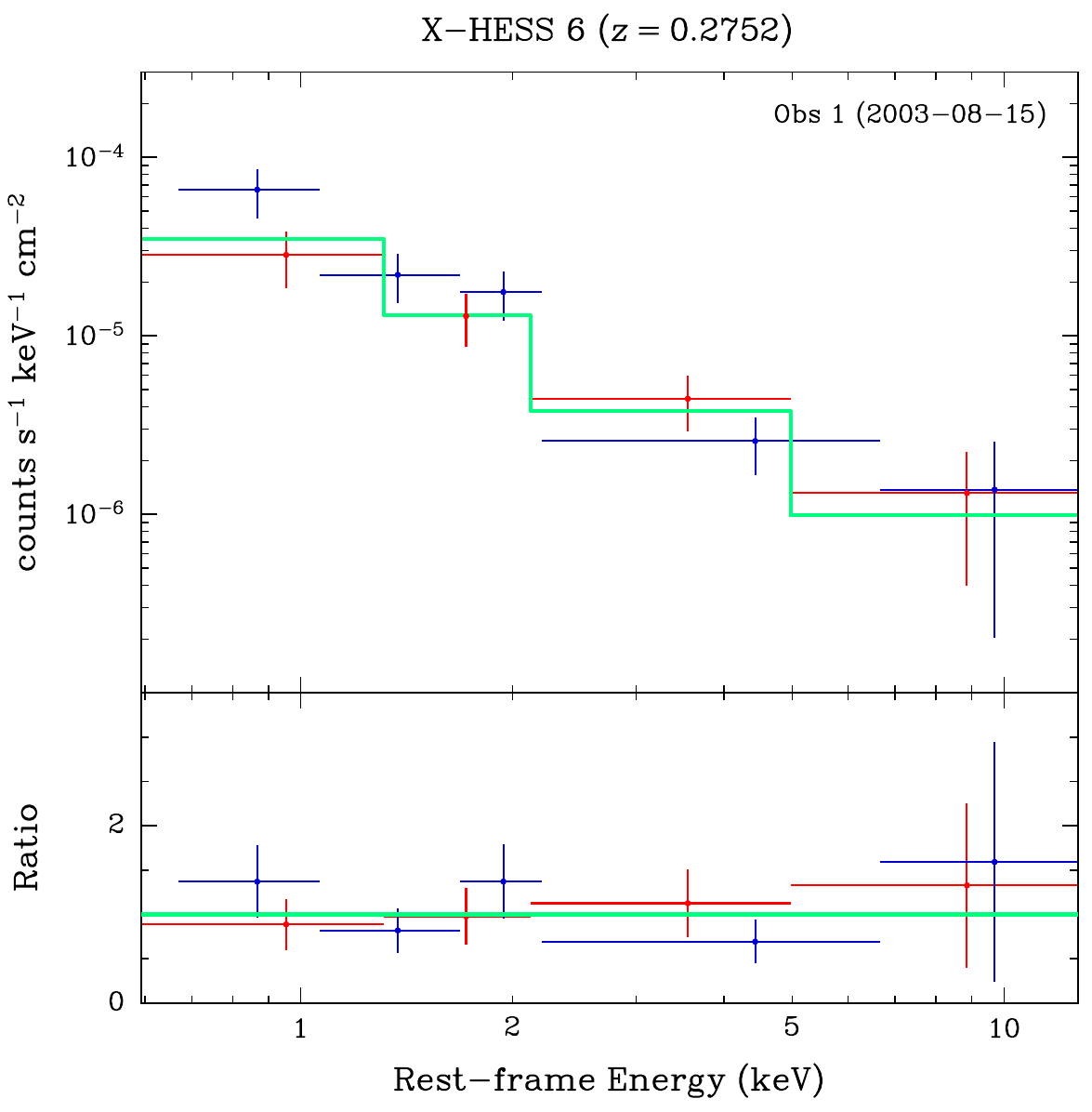}} \\
     \subfloat{\includegraphics[width = 2.1in]{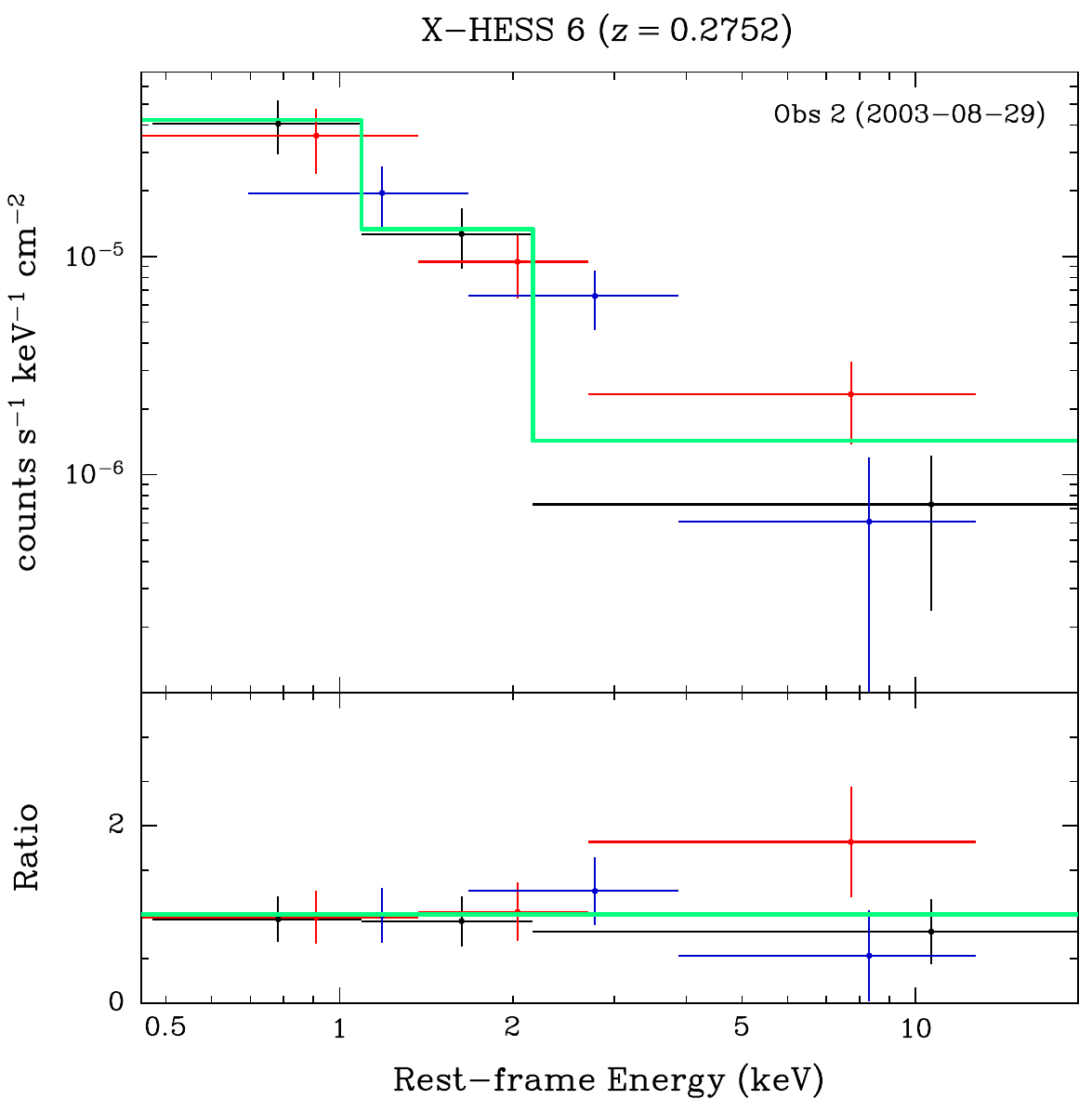}} &
     \subfloat{\includegraphics[width = 2.1in]{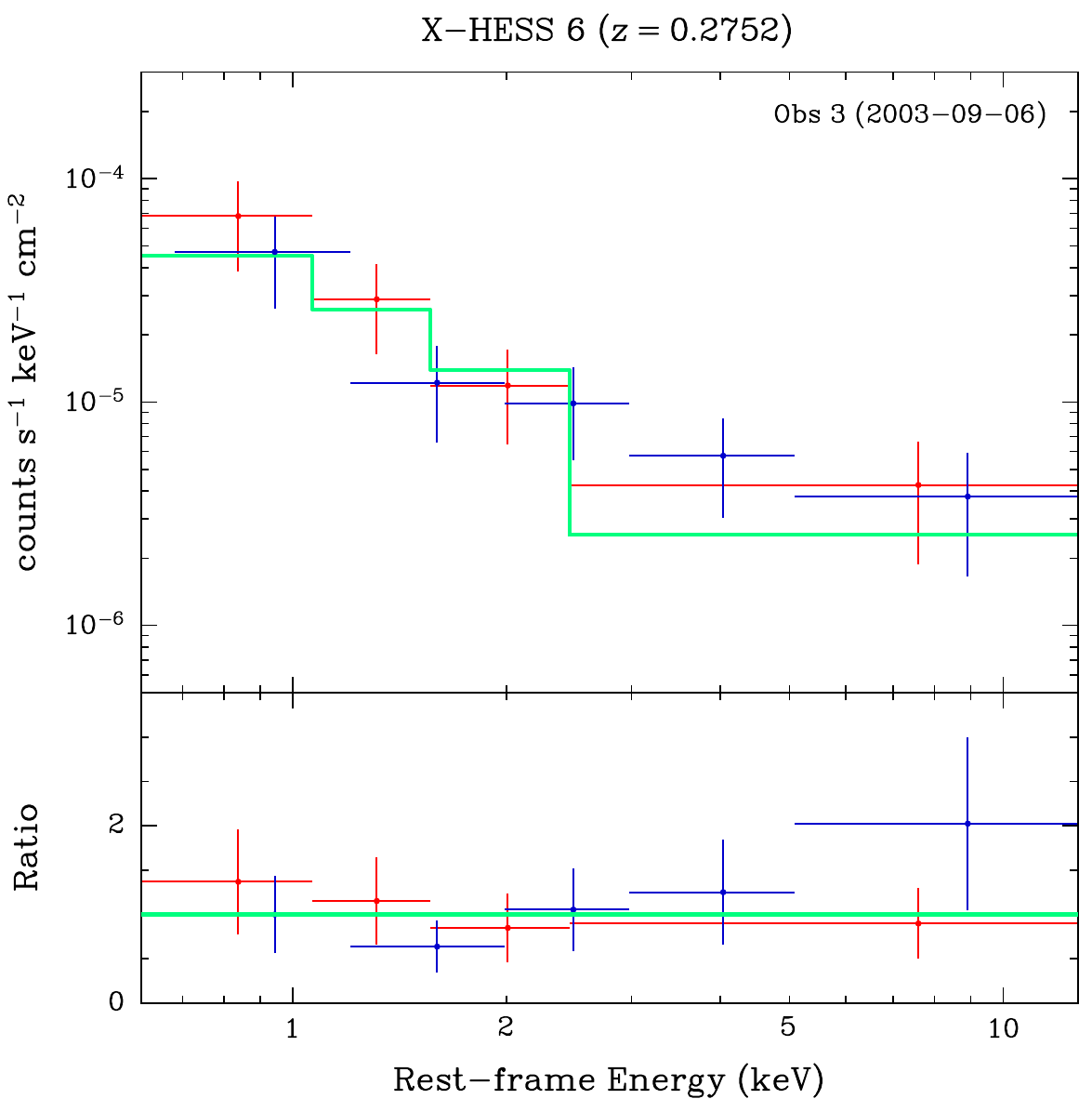}} &
     \subfloat{\includegraphics[width = 2.1in]{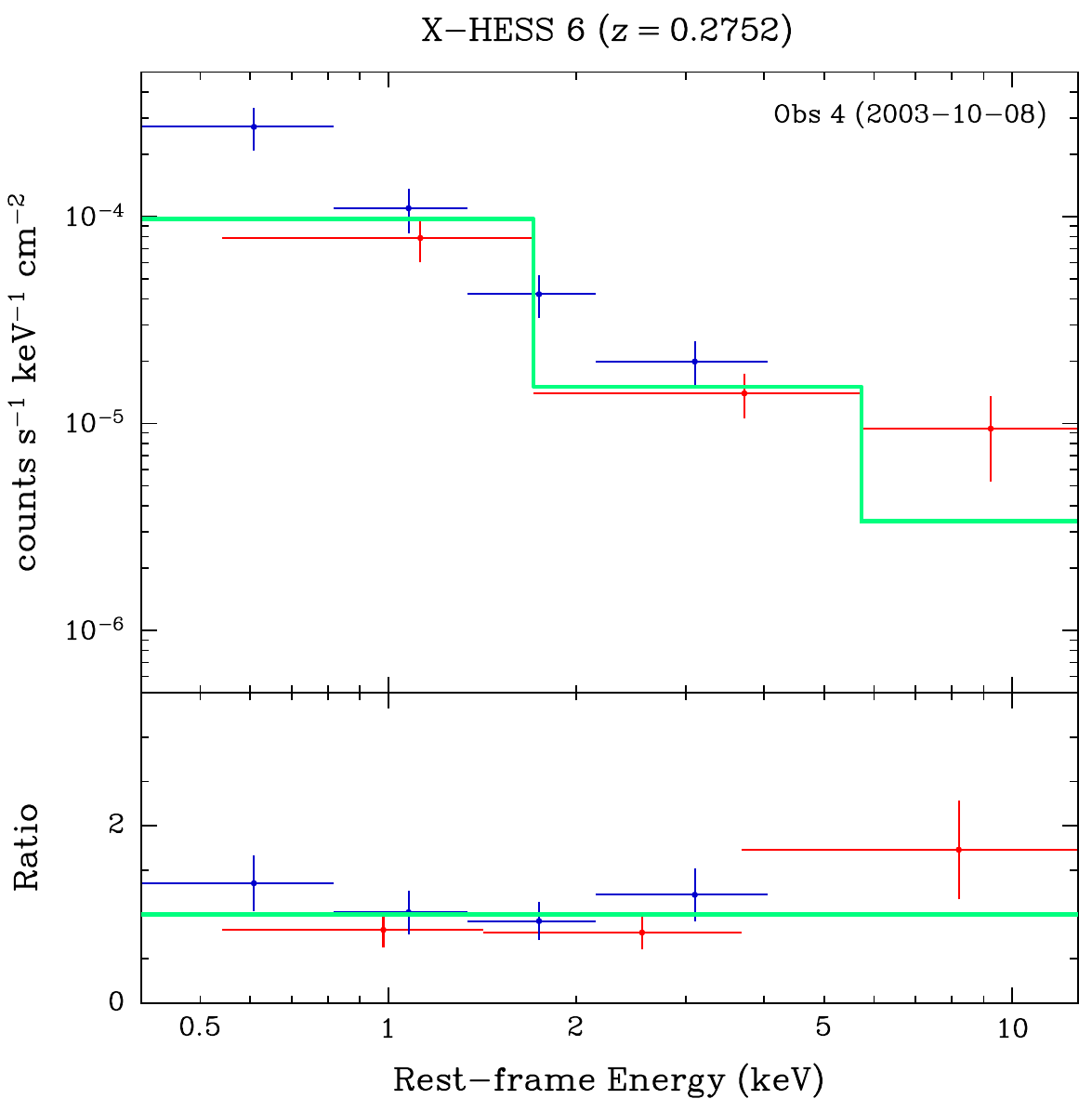}} \\
     \subfloat{\includegraphics[width = 2.1in]{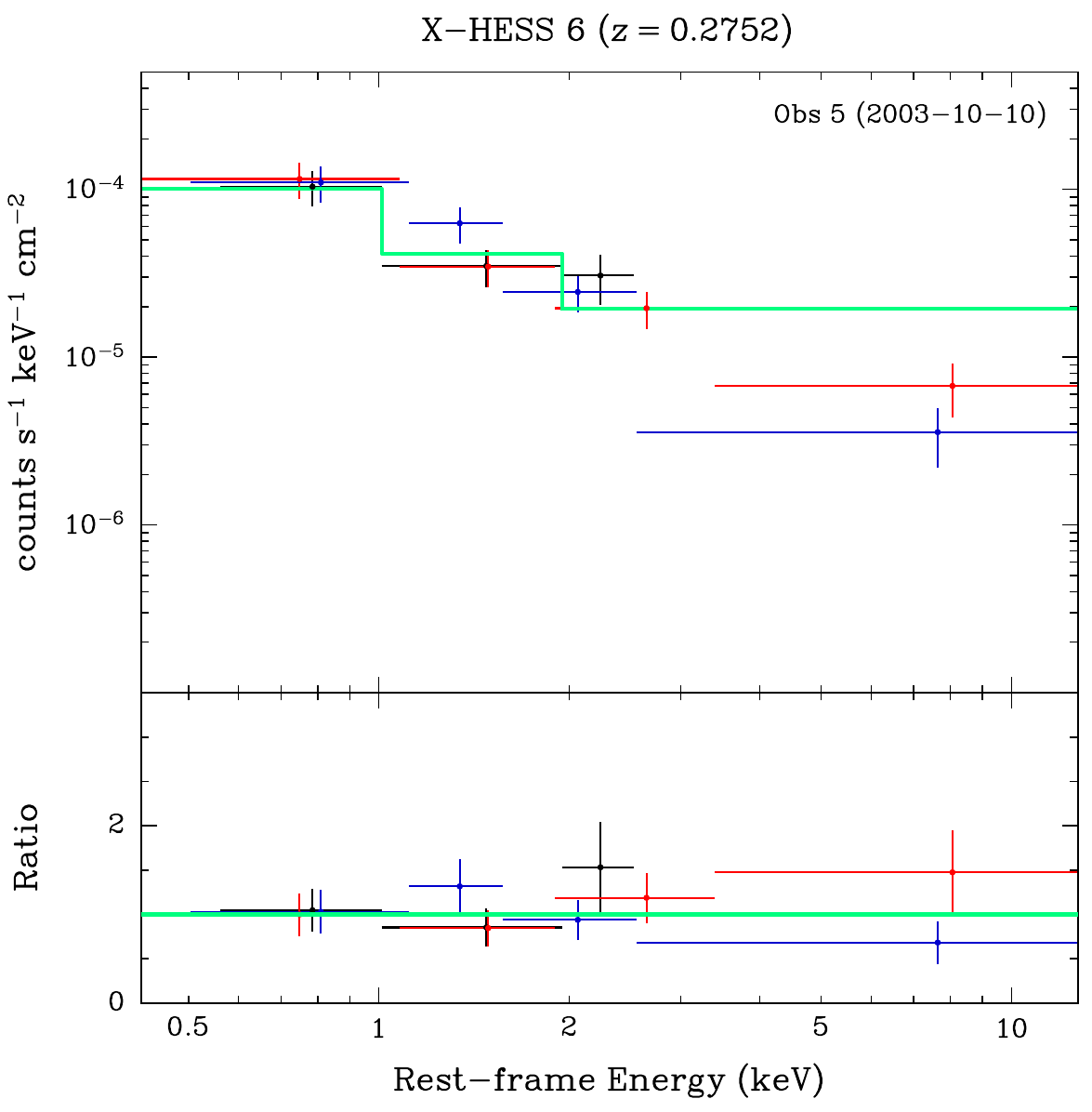}} &
     \subfloat{\includegraphics[width = 2.1in]{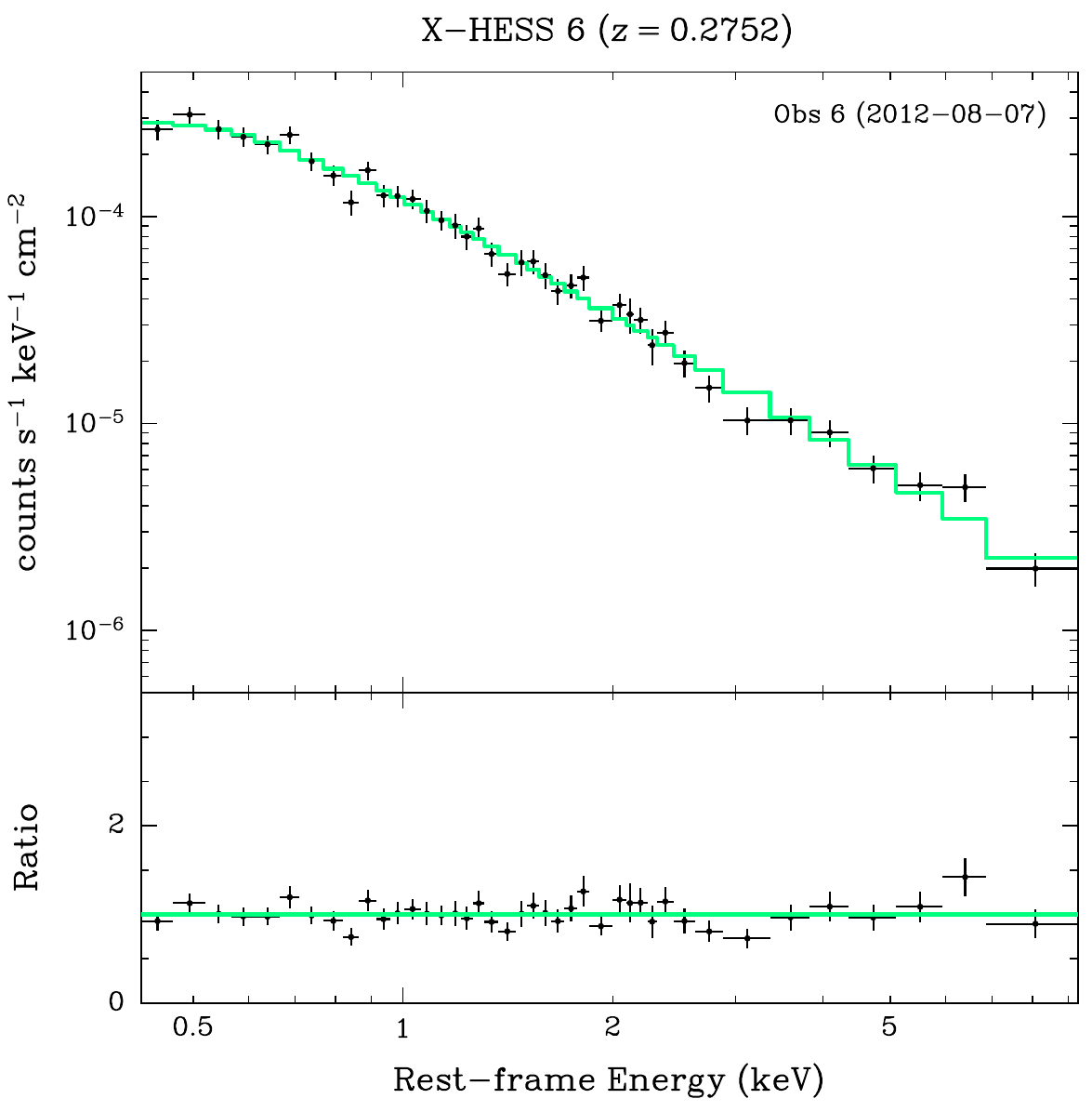}} &
     \subfloat{\includegraphics[width = 2.1in]{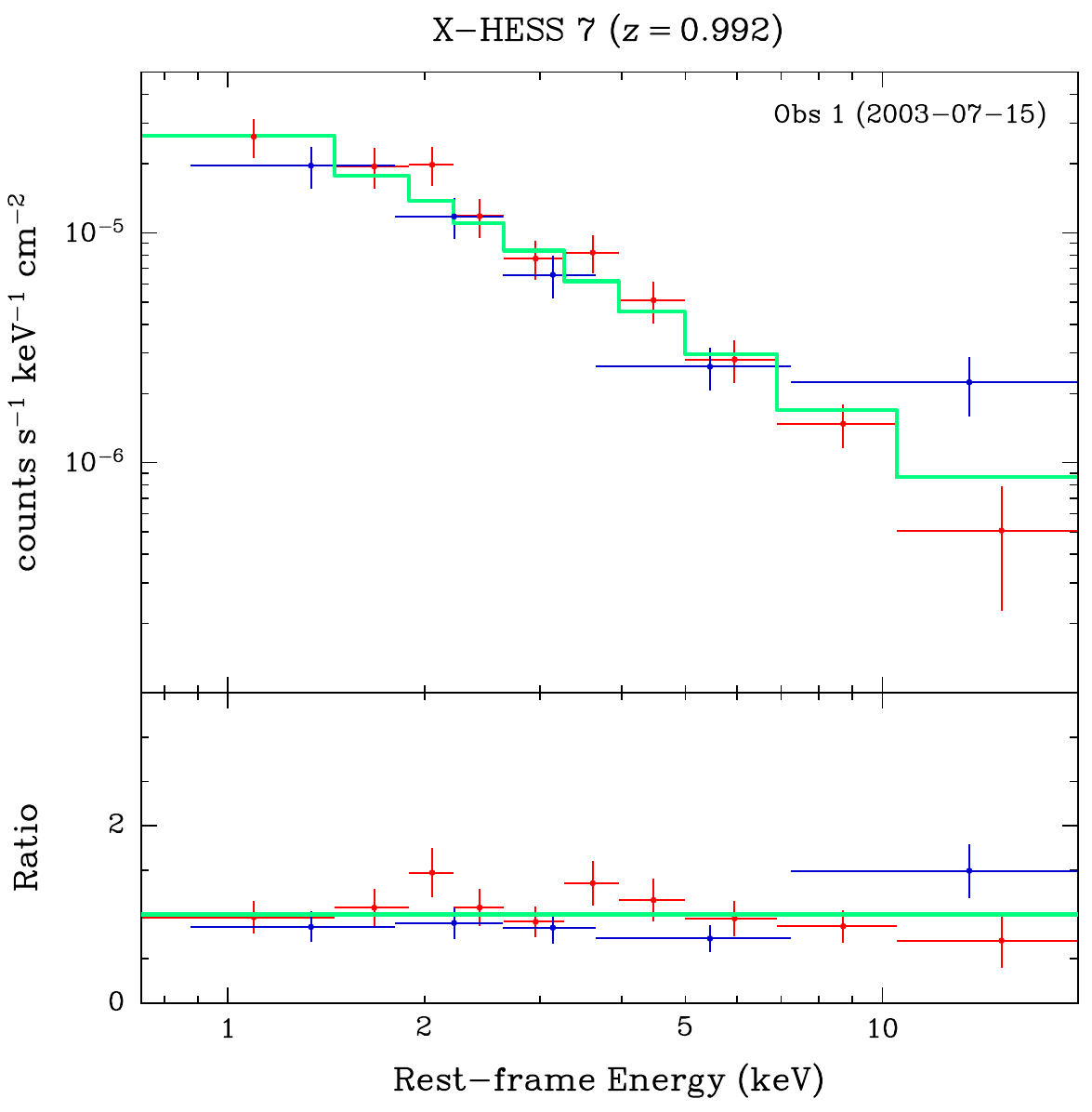}} \\
     \subfloat{\includegraphics[width = 2.1in]{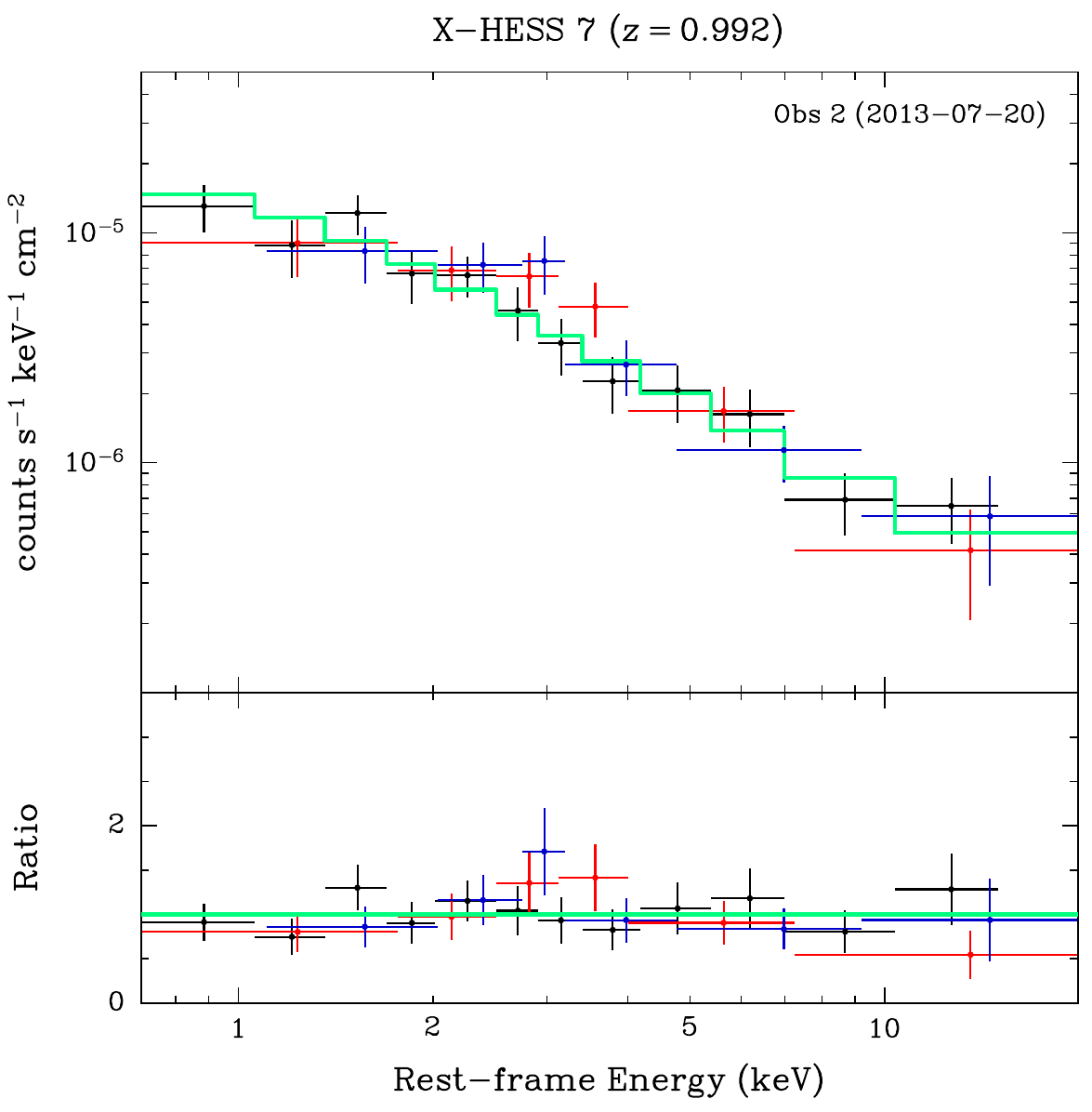}} &
     \subfloat{\includegraphics[width = 2.1in]{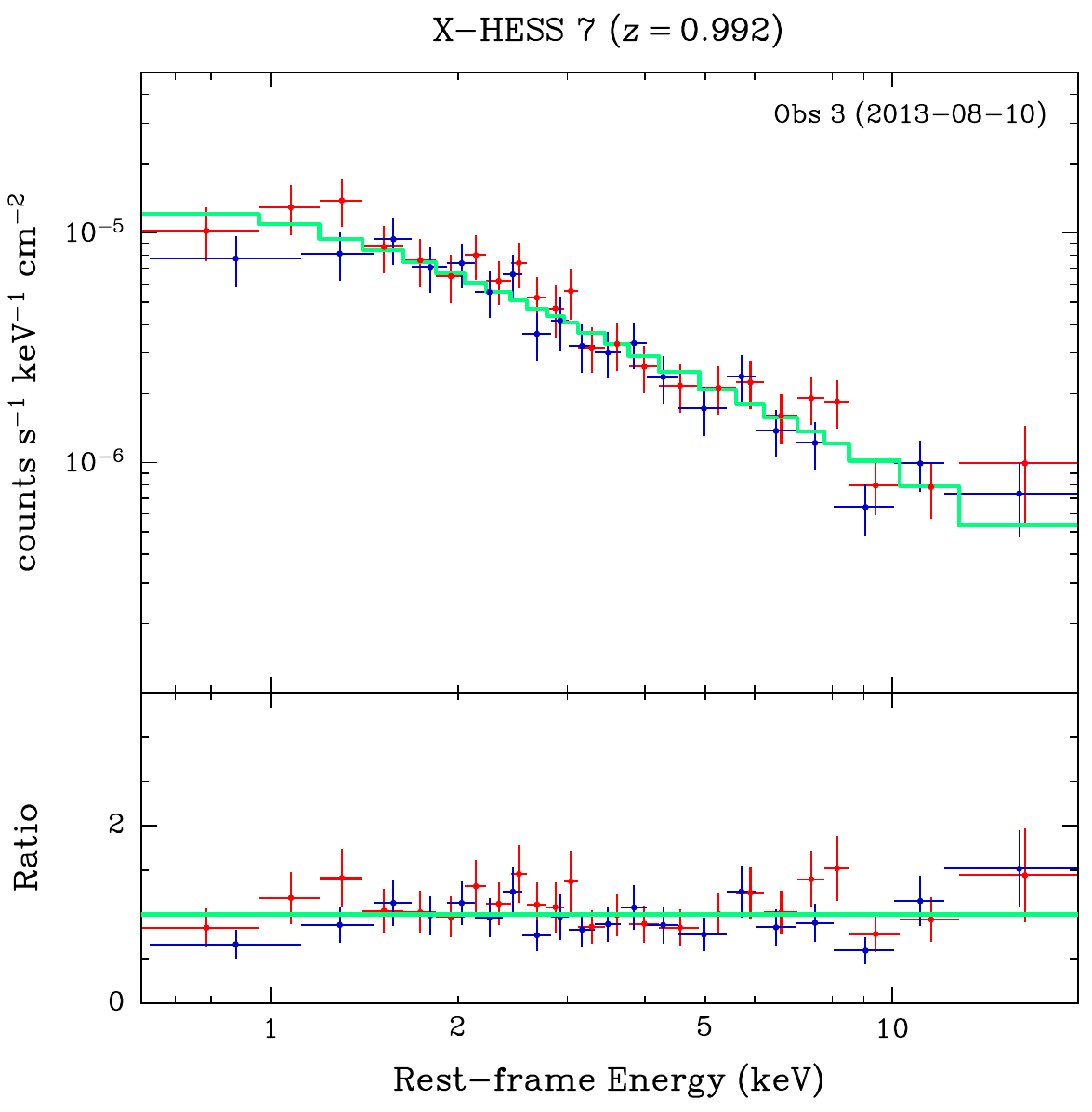}} &
     \subfloat{\includegraphics[width = 2.1in]{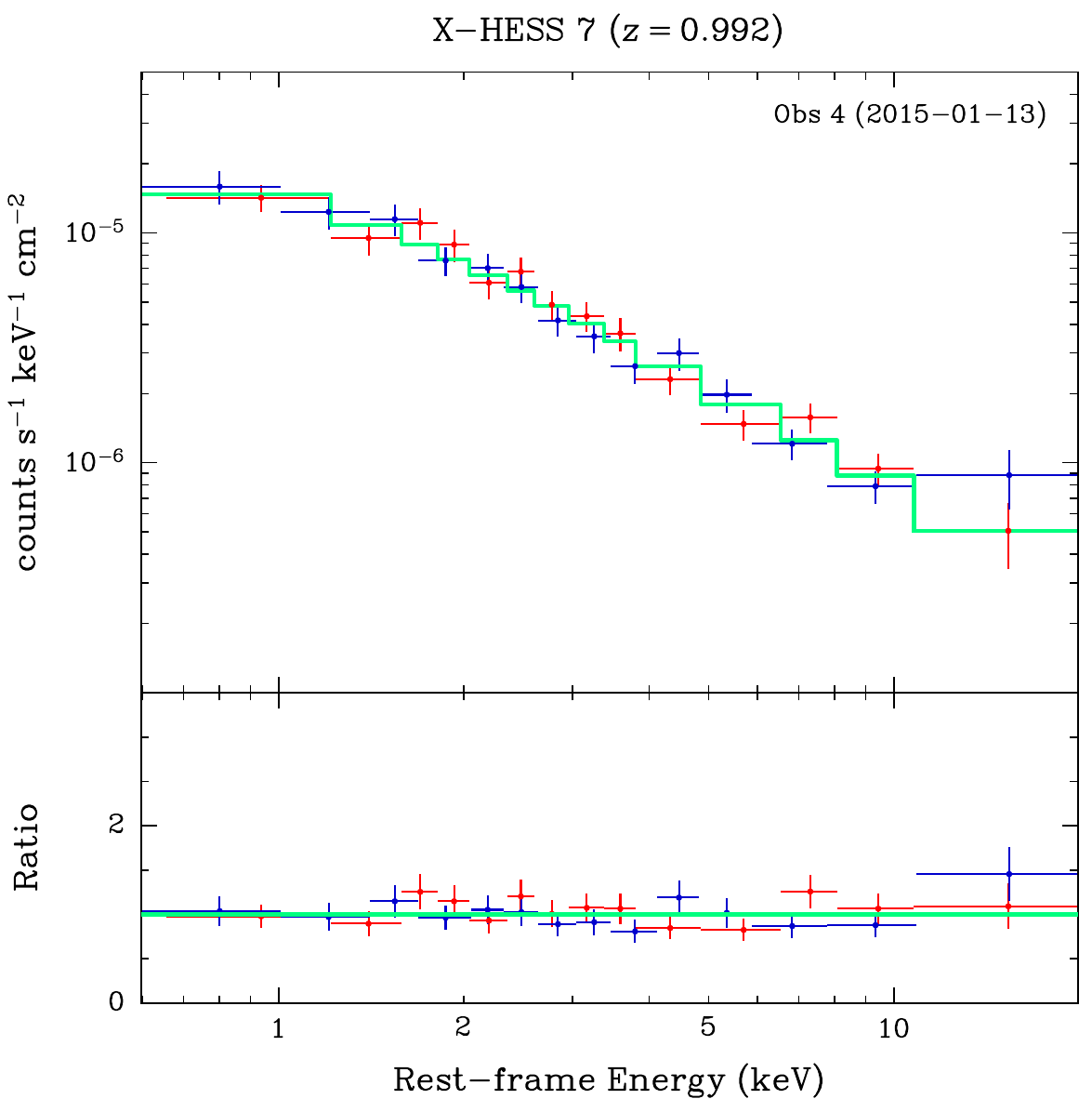}} \\
     \end{tabular}
\begin{minipage}{1.2\linewidth}
     \centering
     {Continuation of Fig. \ref{fig:xhess_spectra}.}
\end{minipage}

\end{figure}

\begin{figure}[h]
     \ContinuedFloat
     \centering
     \renewcommand{\arraystretch}{2}
     \begin{tabular}{ccc}
     \subfloat{\includegraphics[width = 2.1in]{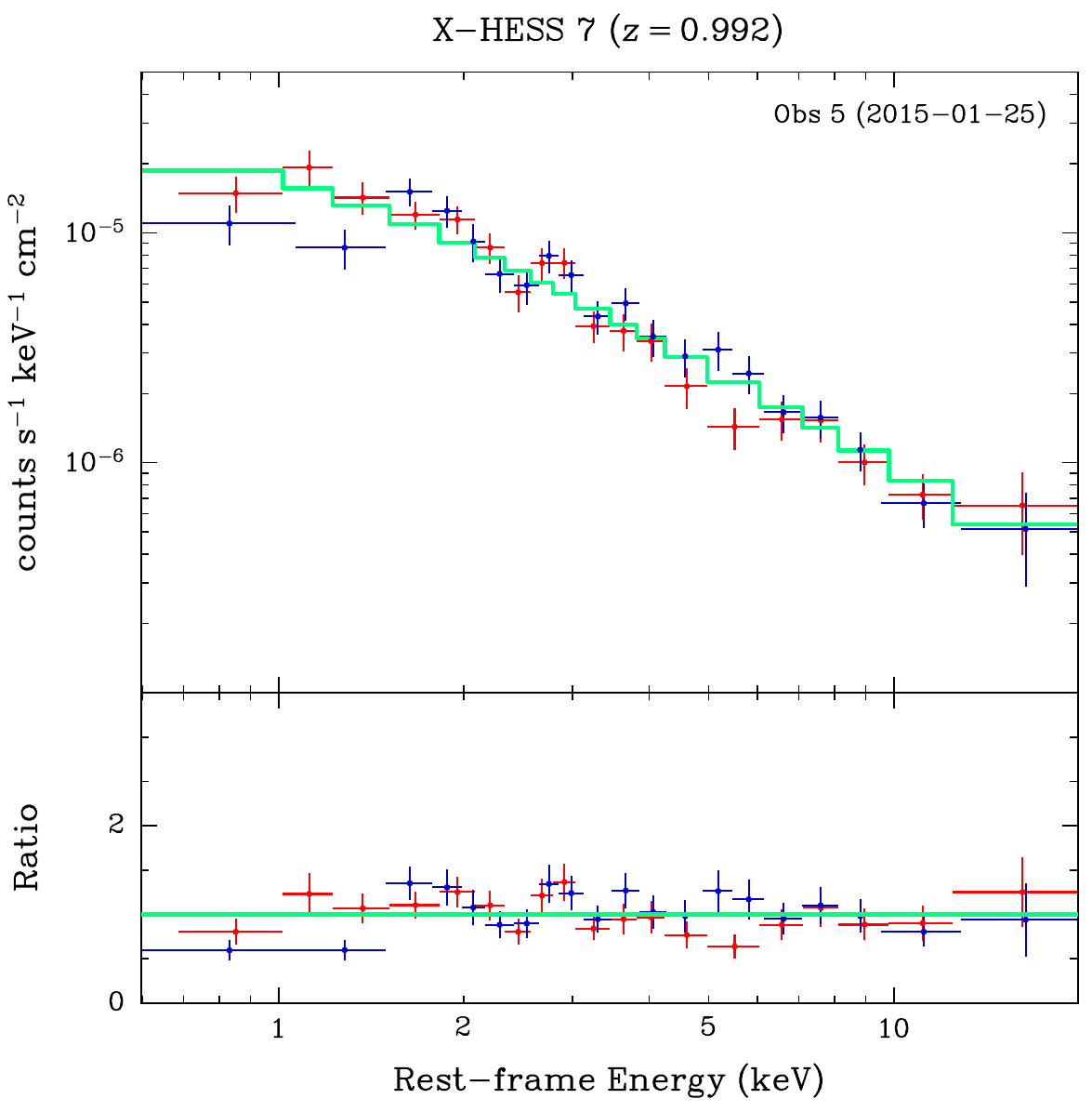}} &
     \subfloat{\includegraphics[width = 2.1in]{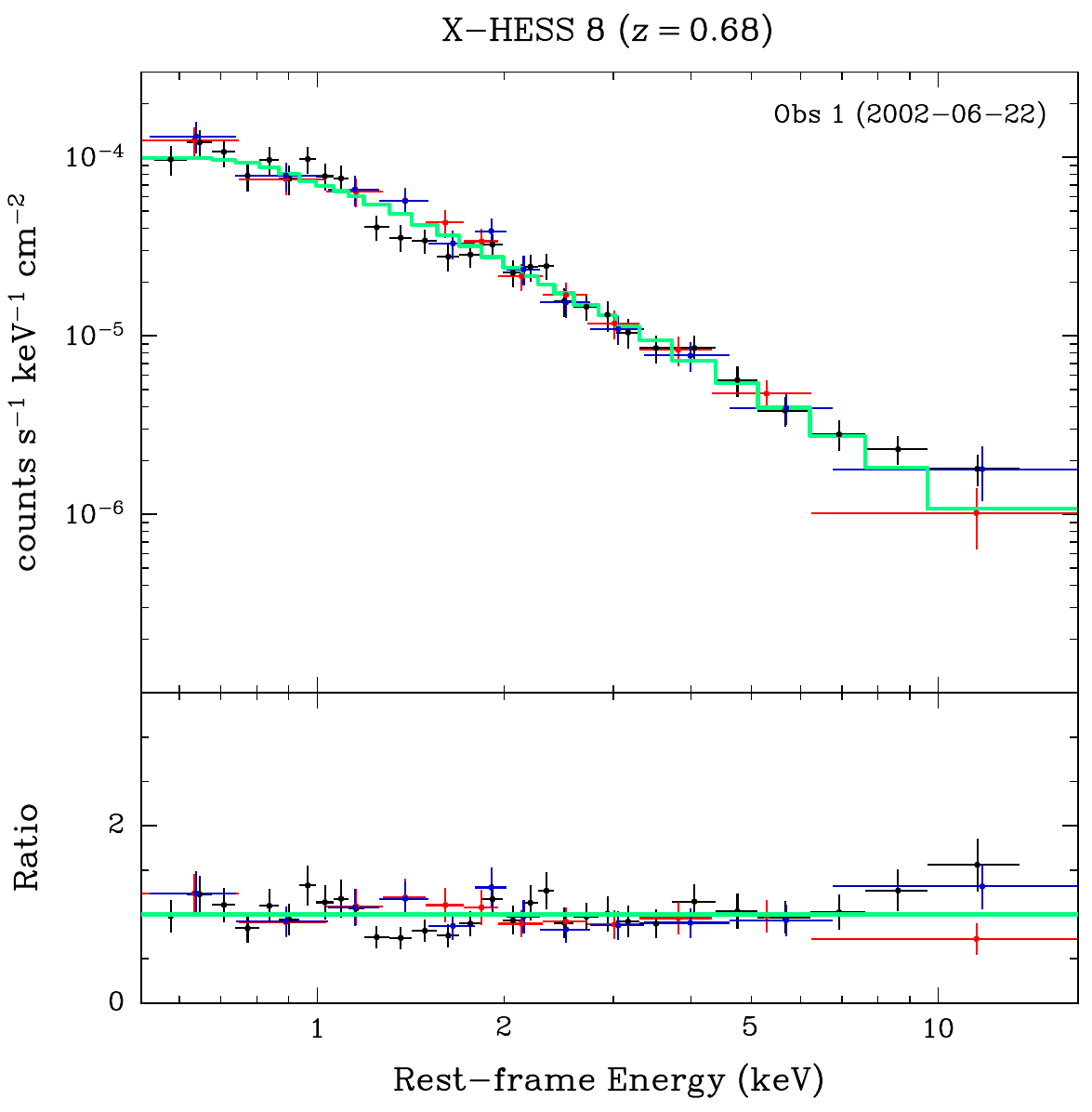}} &
     \subfloat{\includegraphics[width = 2.1in]{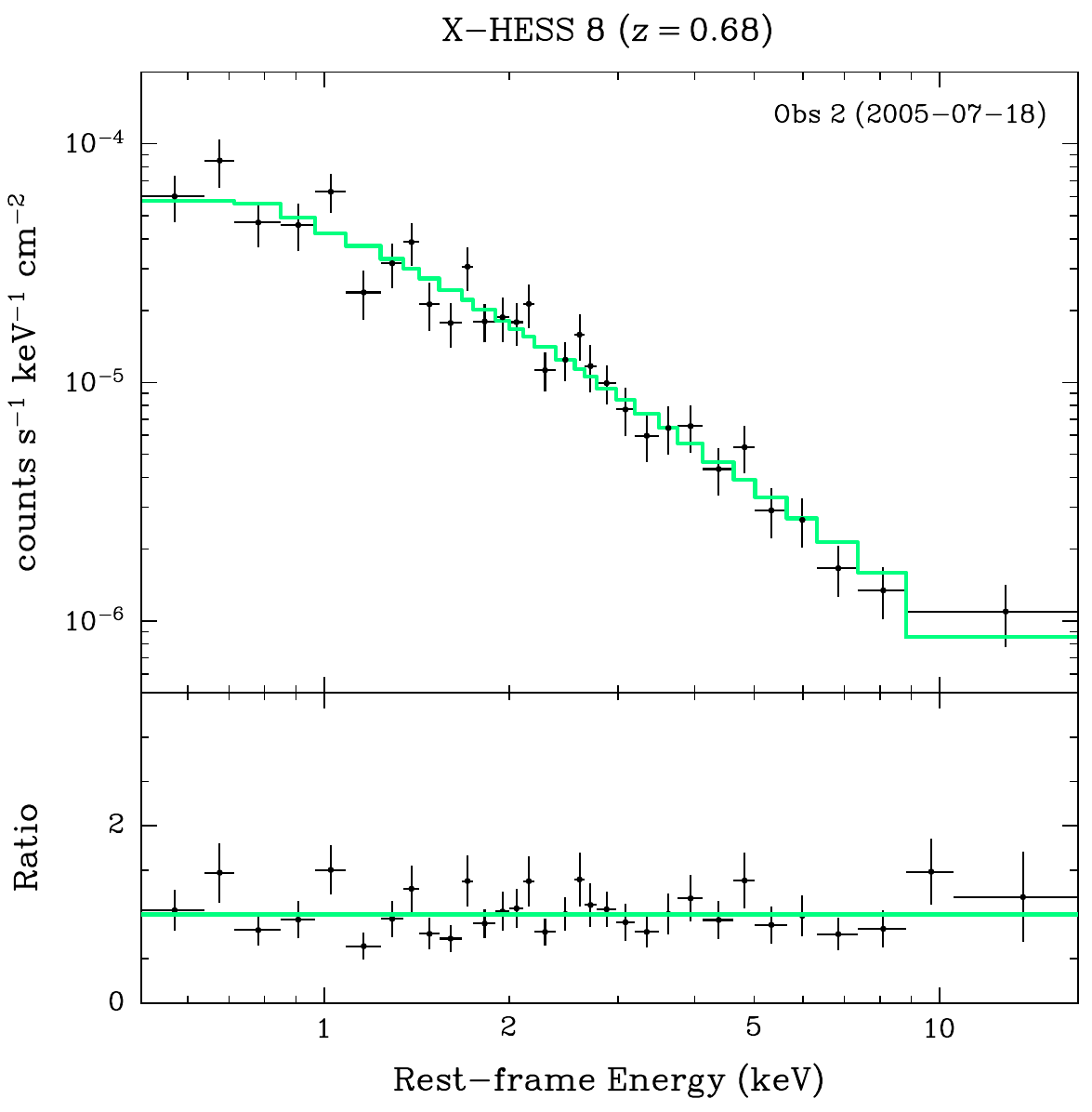}} \\
     \subfloat{\includegraphics[width = 2.1in]{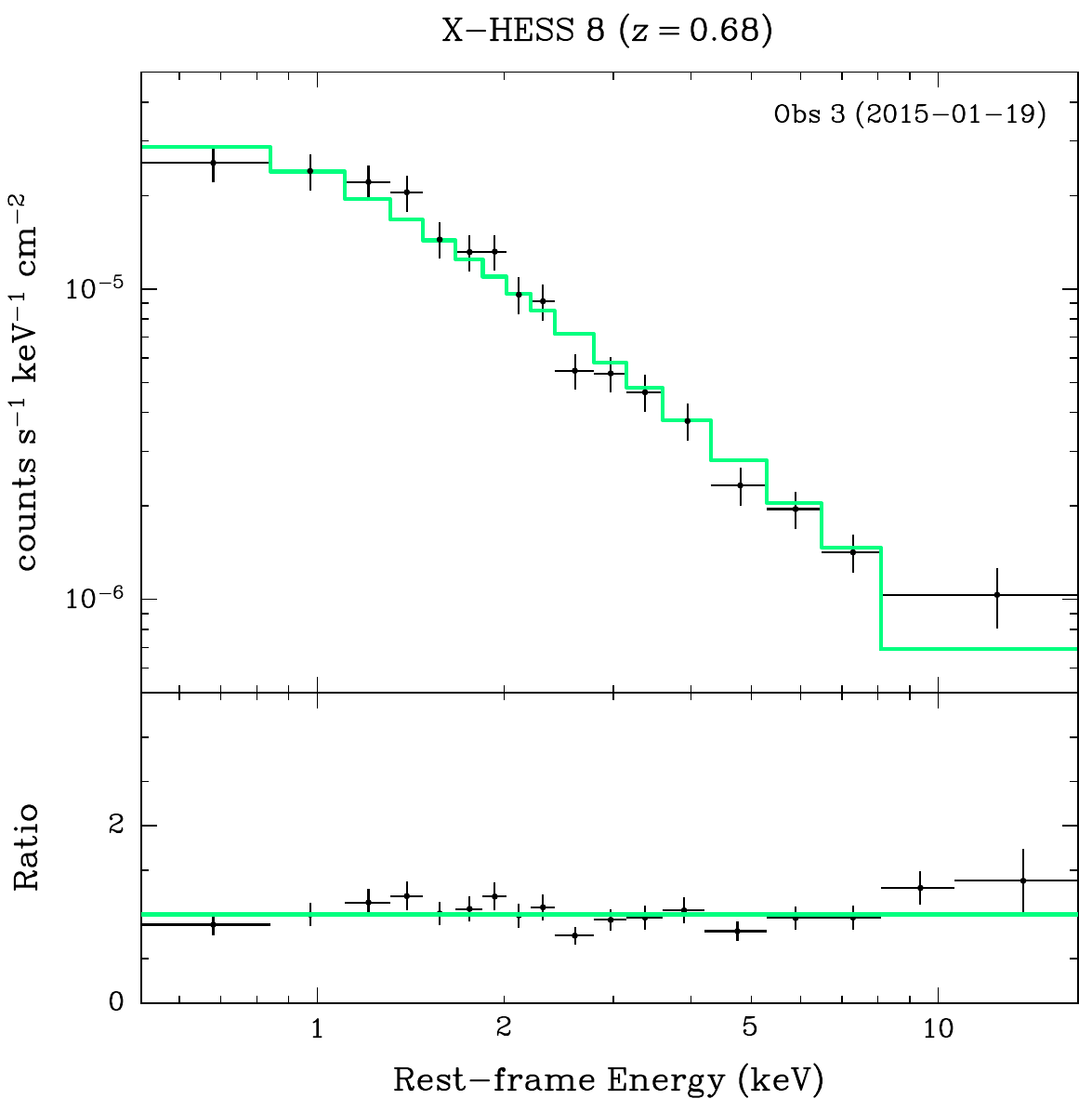}} &
     \subfloat{\includegraphics[width = 2.1in]{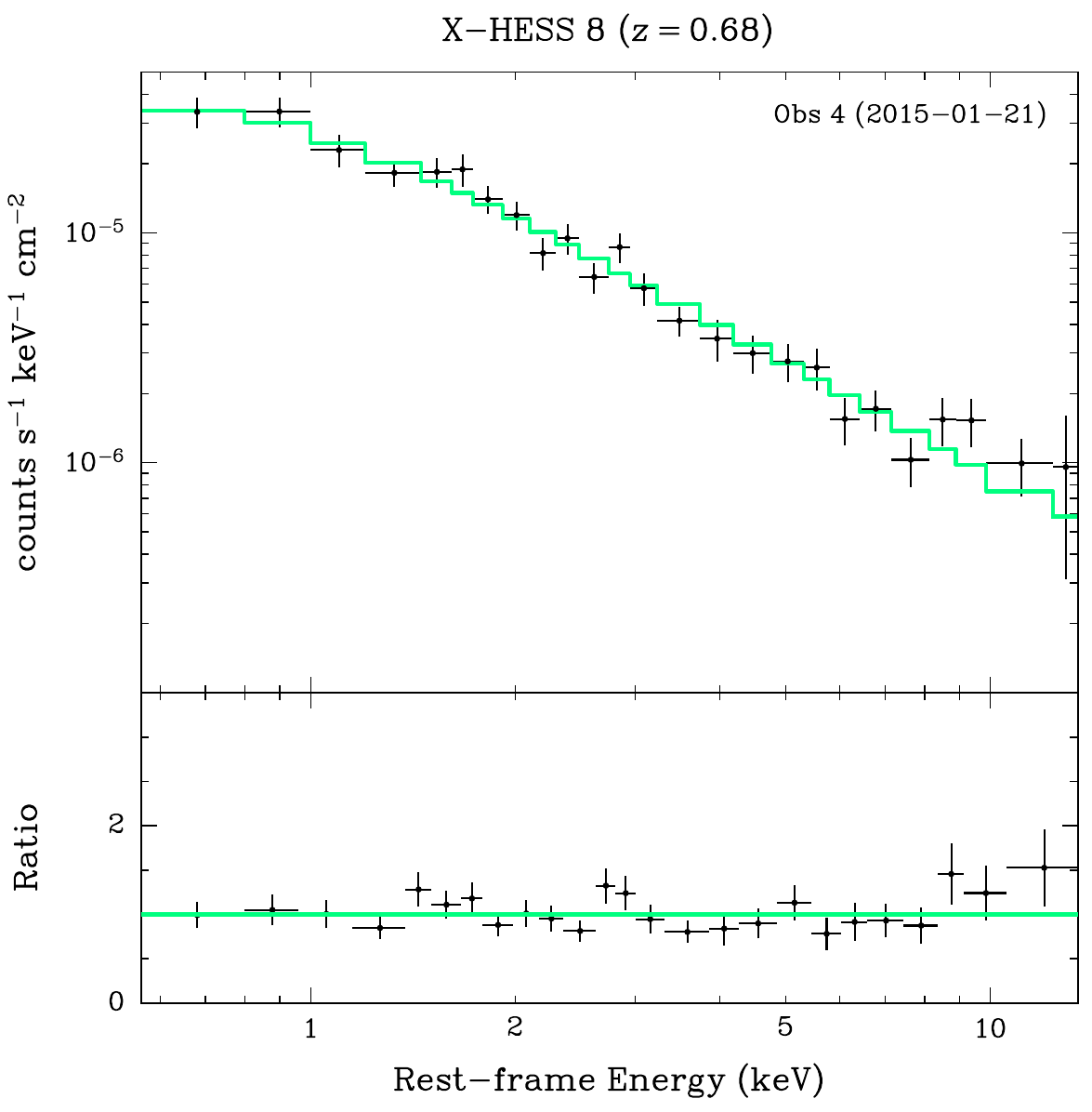}} &
     \subfloat{\includegraphics[width = 2.1in]{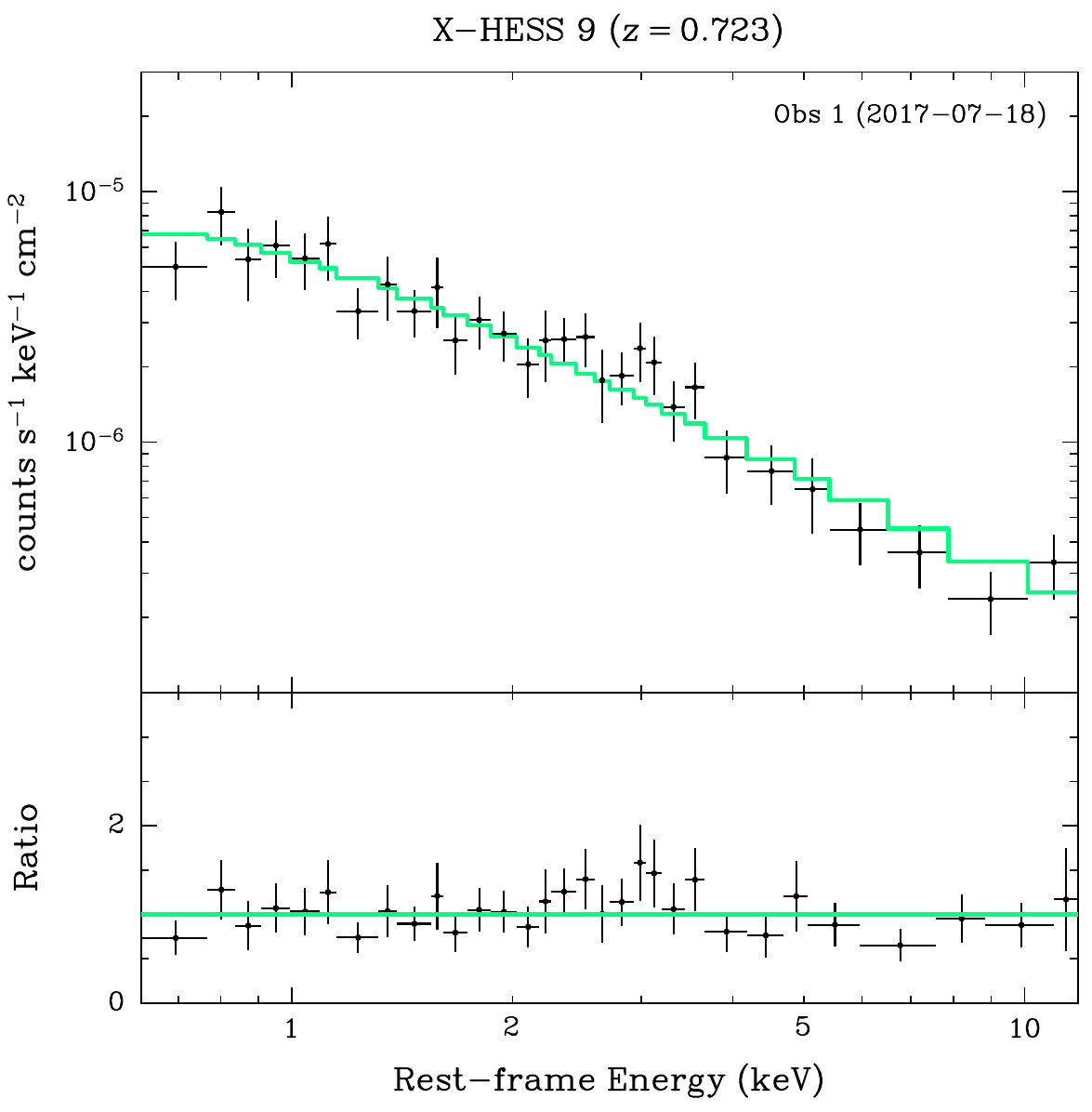}} \\
     \subfloat{\includegraphics[width = 2.1in]{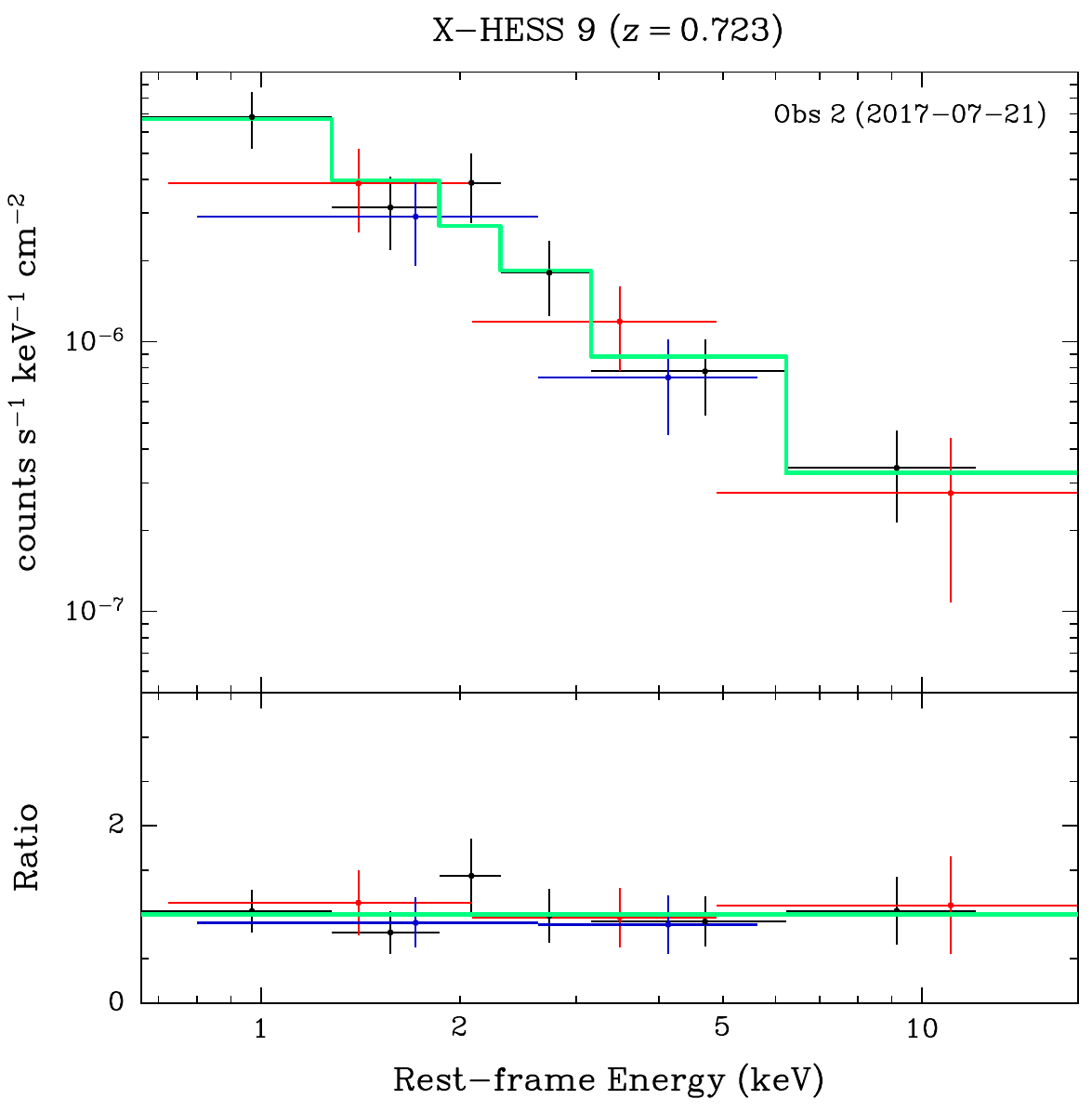}} &
     \subfloat{\includegraphics[width = 2.1in]{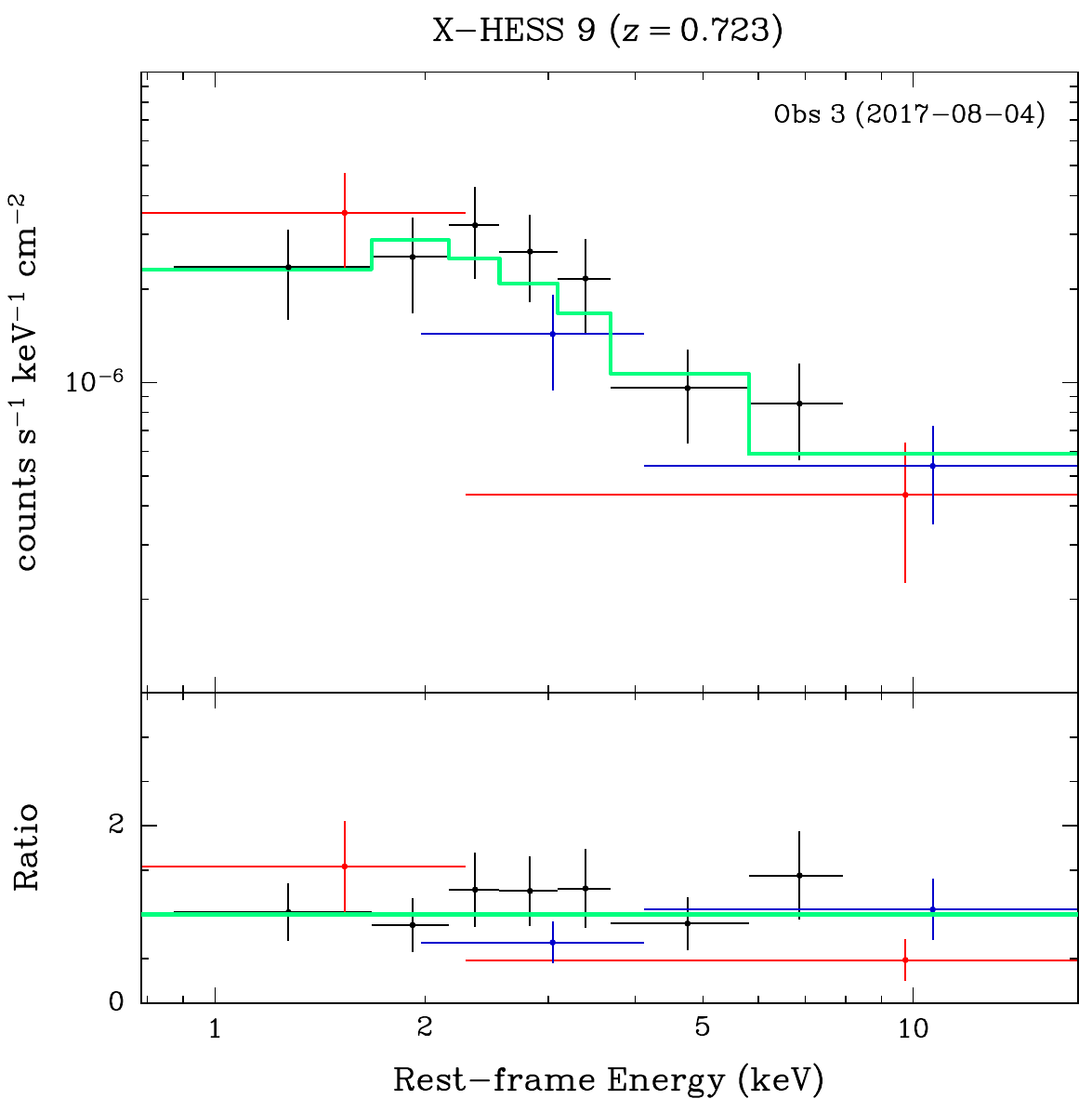}} &
     \subfloat{\includegraphics[width = 2.1in]{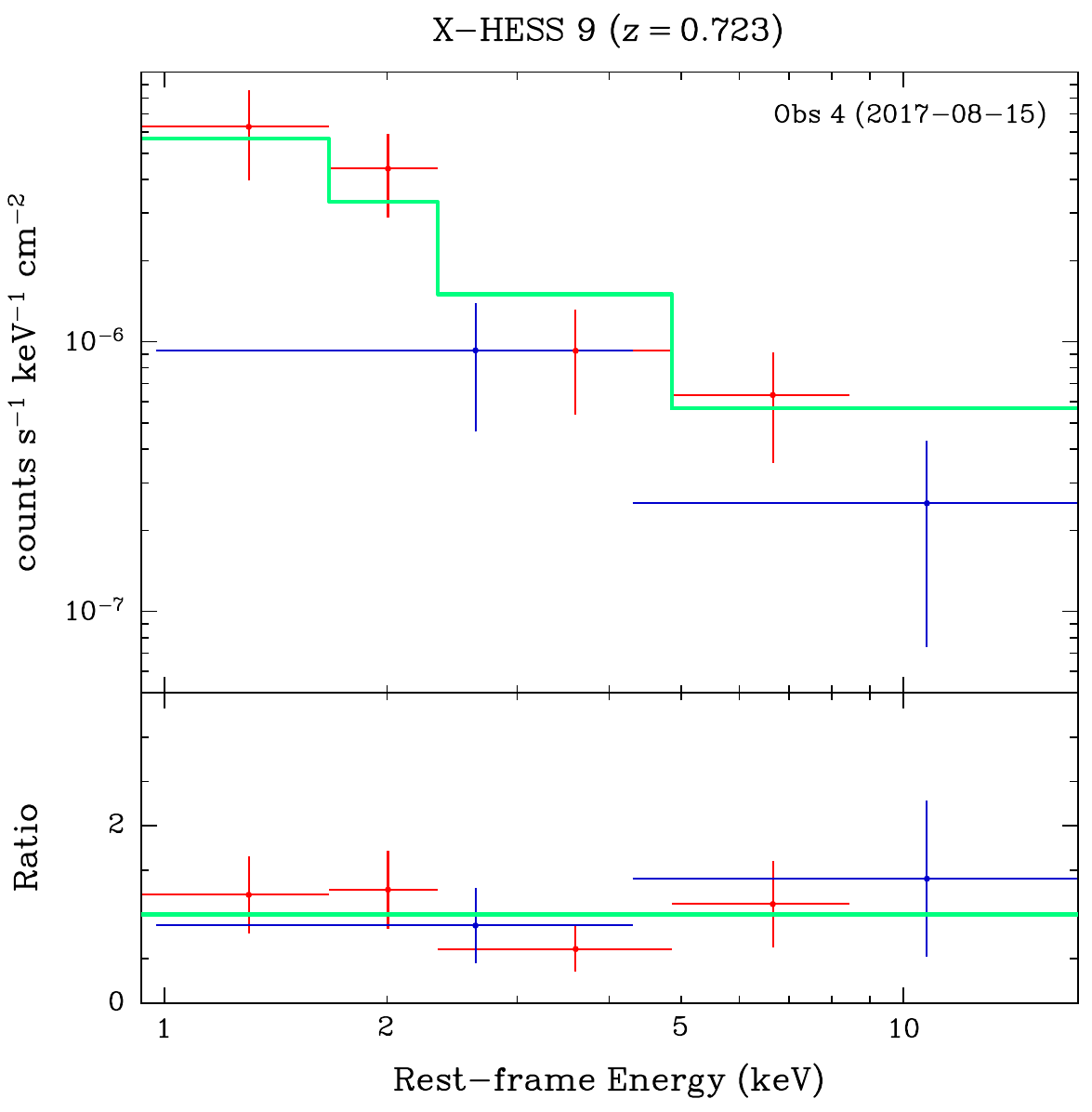}} \\
     \subfloat{\includegraphics[width = 2.1in]{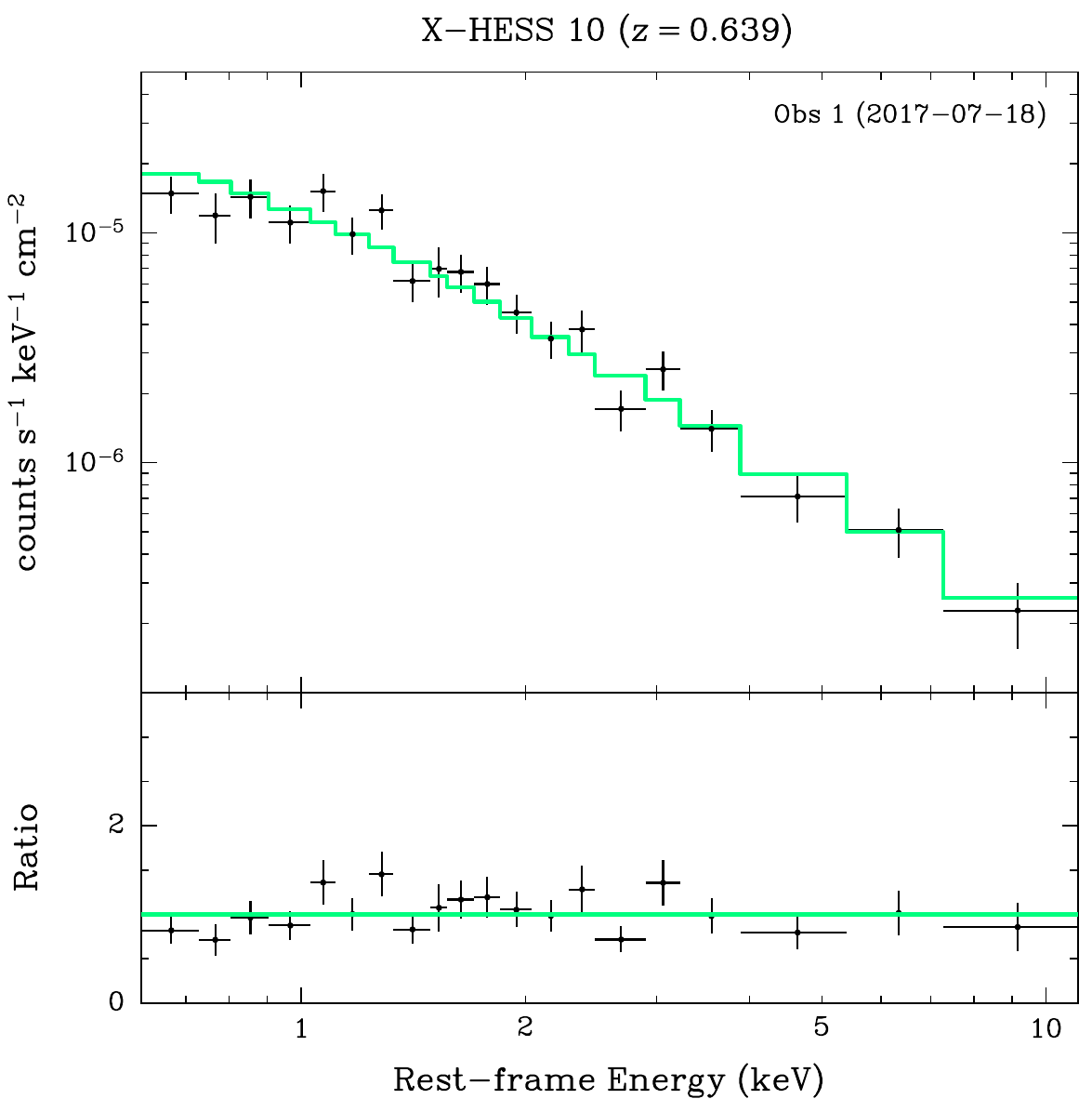}} &
     \subfloat{\includegraphics[width = 2.1in]{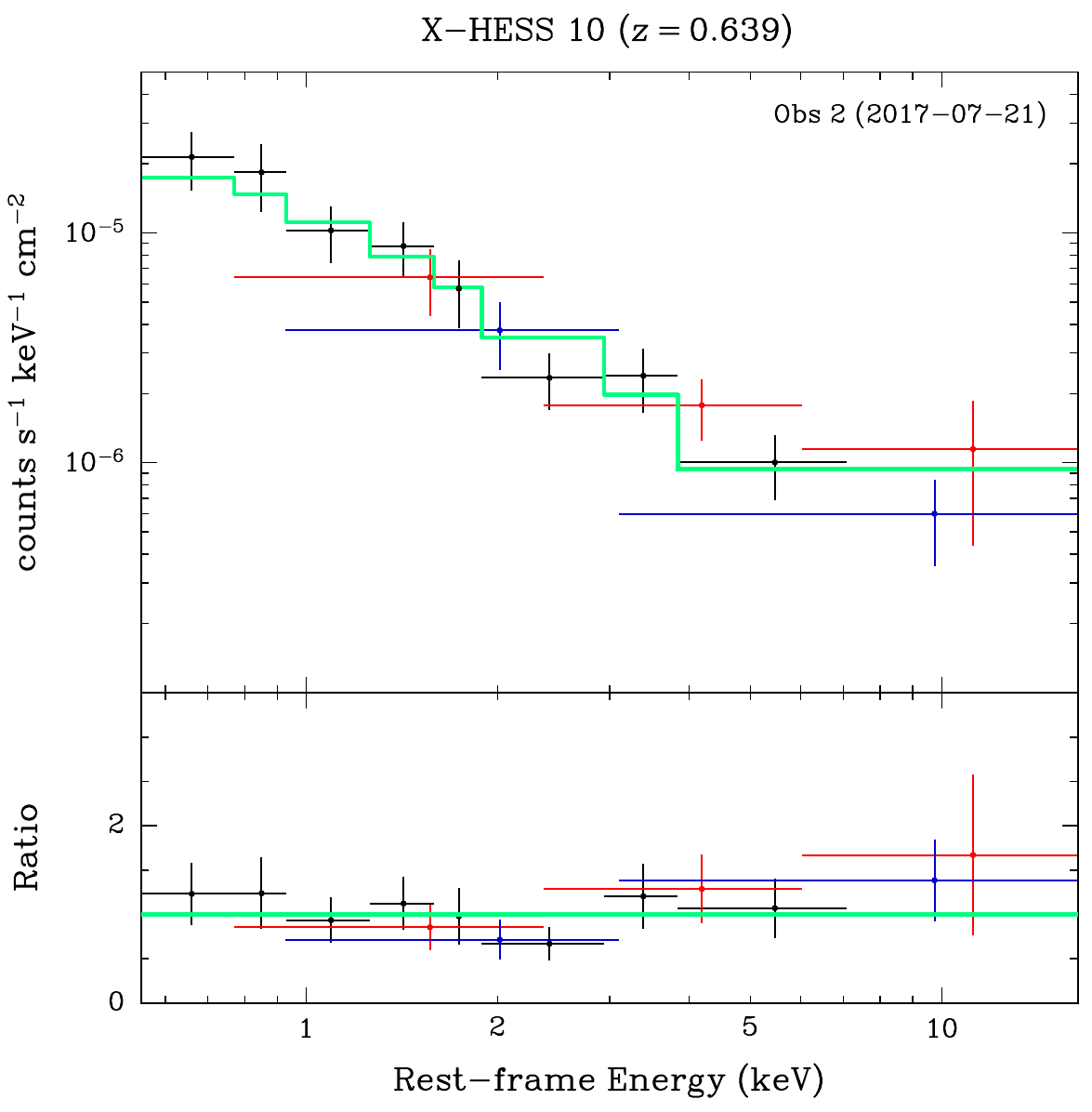}} &
     \subfloat{\includegraphics[width = 2.1in]{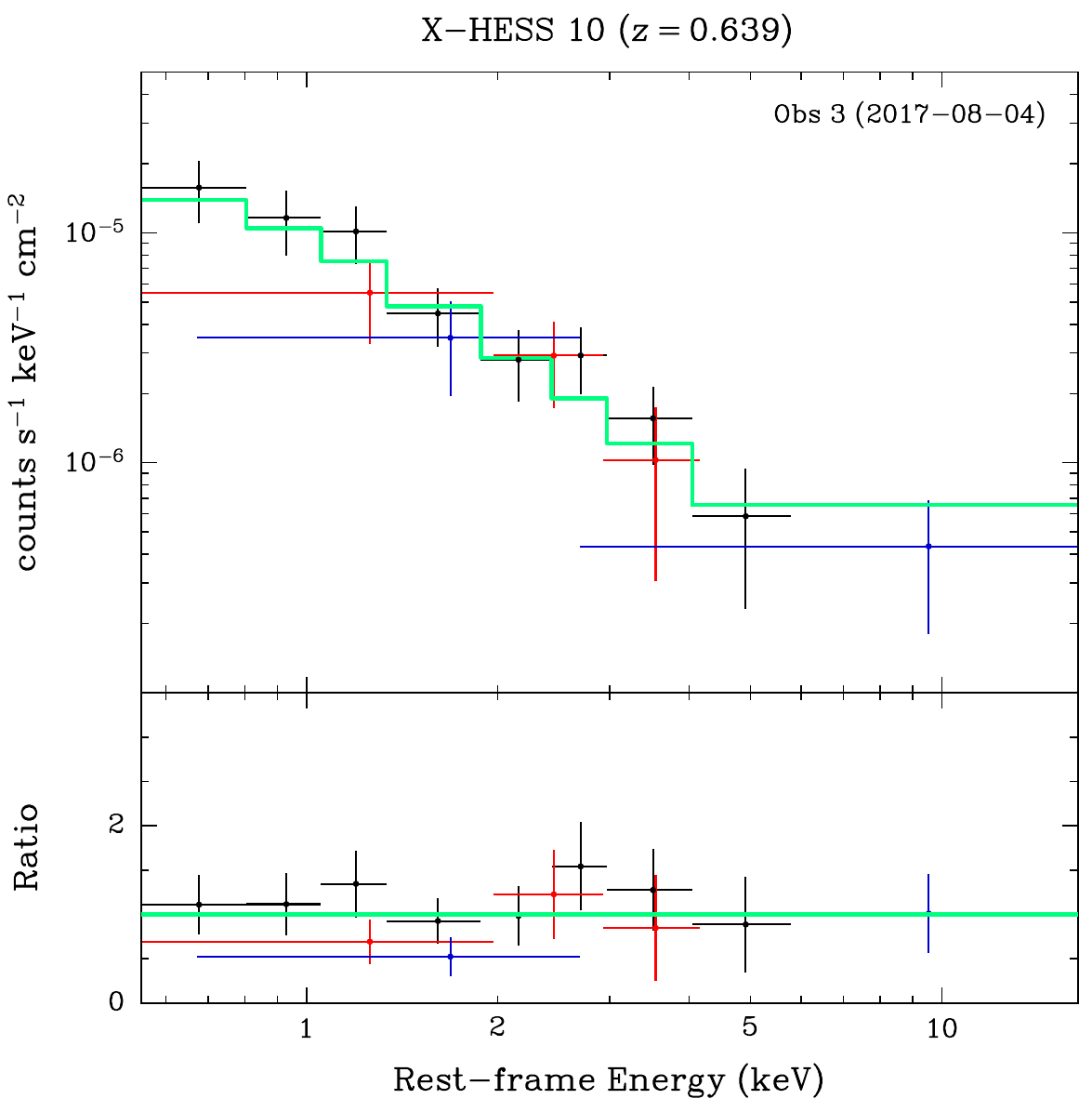}} \\     
     \end{tabular}
\begin{minipage}{1.2\linewidth}
    \centering 
    {Continuation of Fig. \ref{fig:xhess_spectra}.}
\end{minipage}

\end{figure}

\begin{figure}[h]
     \ContinuedFloat
     \centering
     \renewcommand{\arraystretch}{2}
     \begin{tabular}{ccc}
     \subfloat{\includegraphics[width = 2.1in]{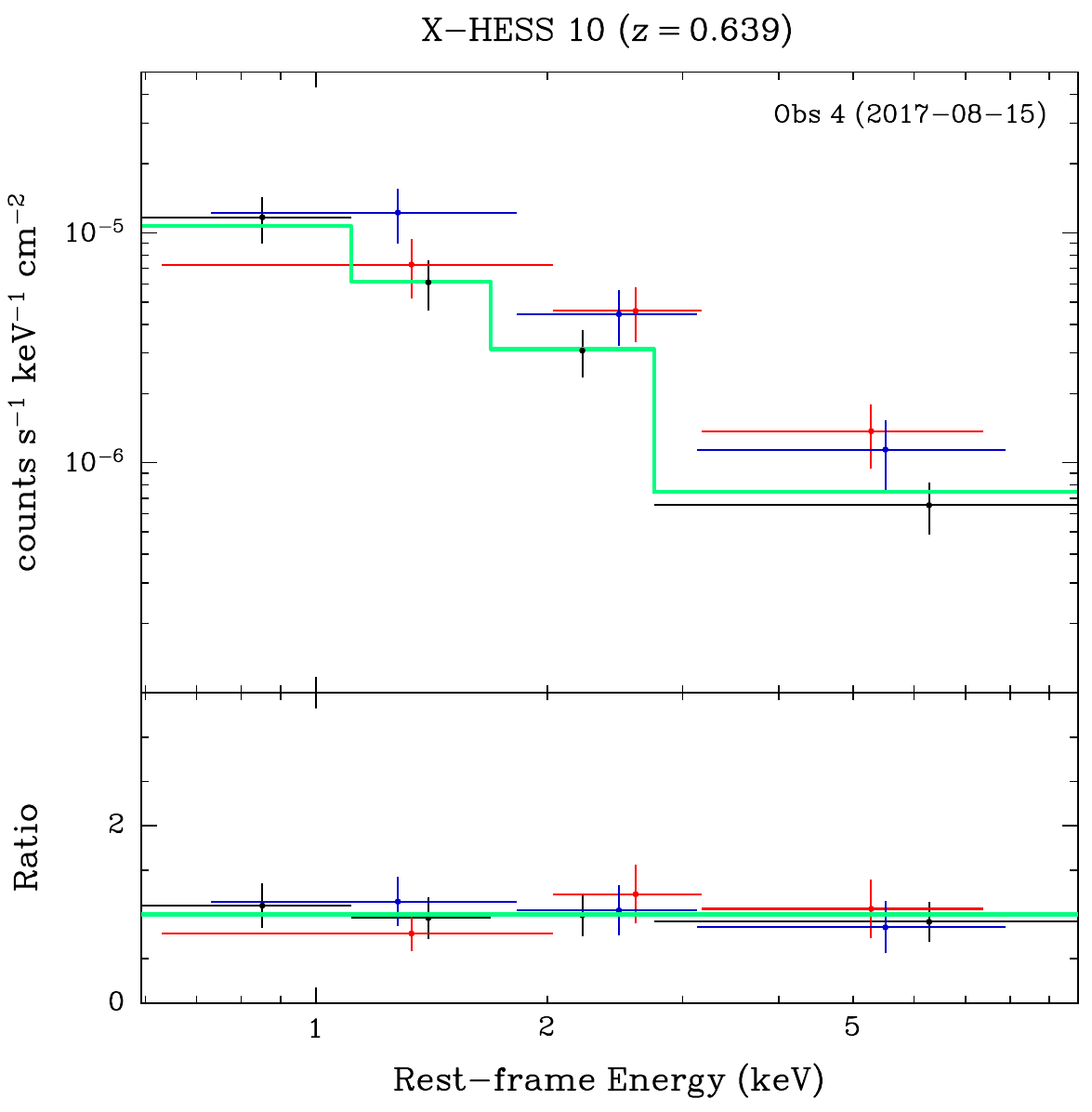}} &
     \subfloat{\includegraphics[width = 2.1in]{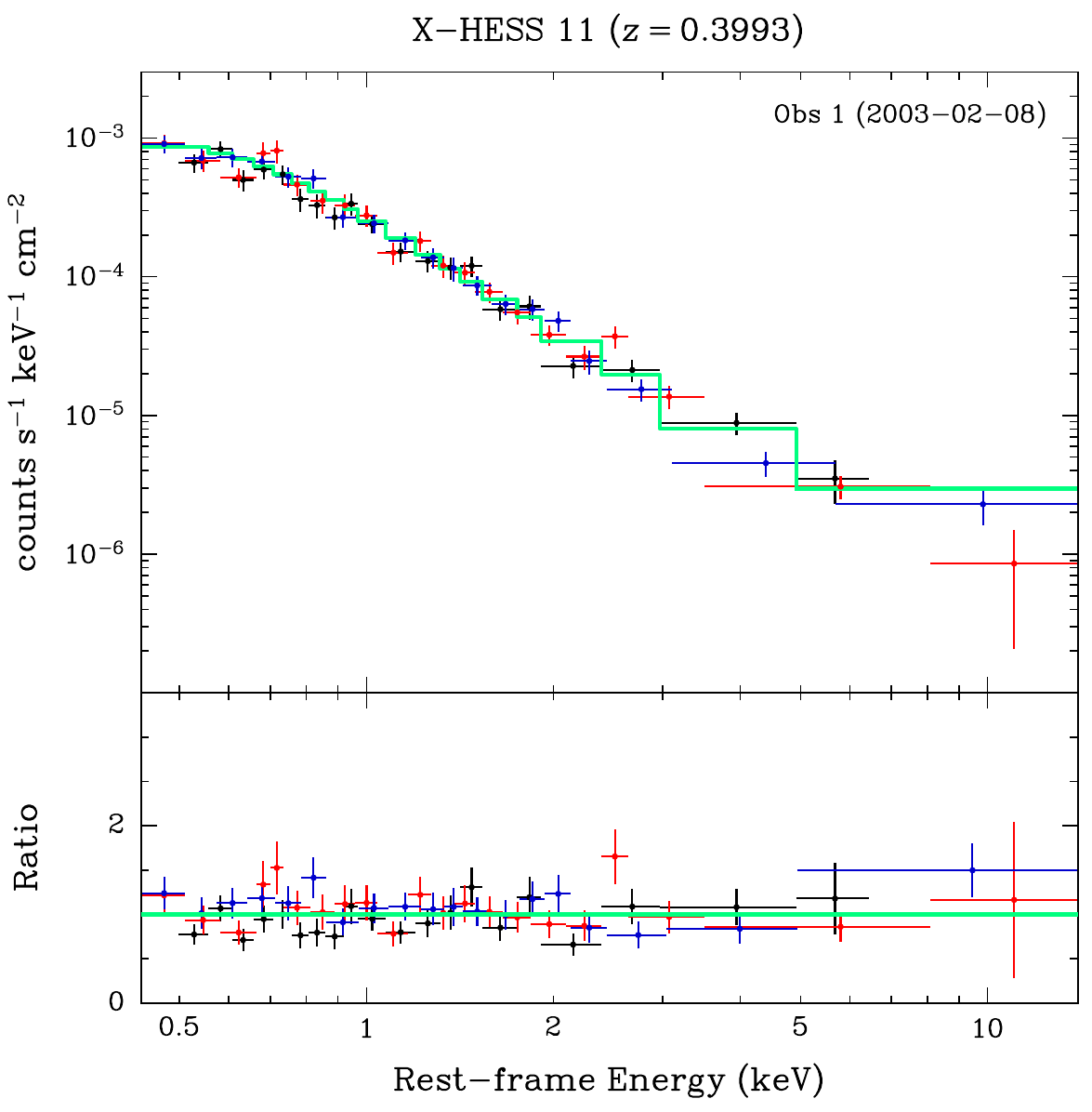}} &
     \subfloat{\includegraphics[width = 2.1in]{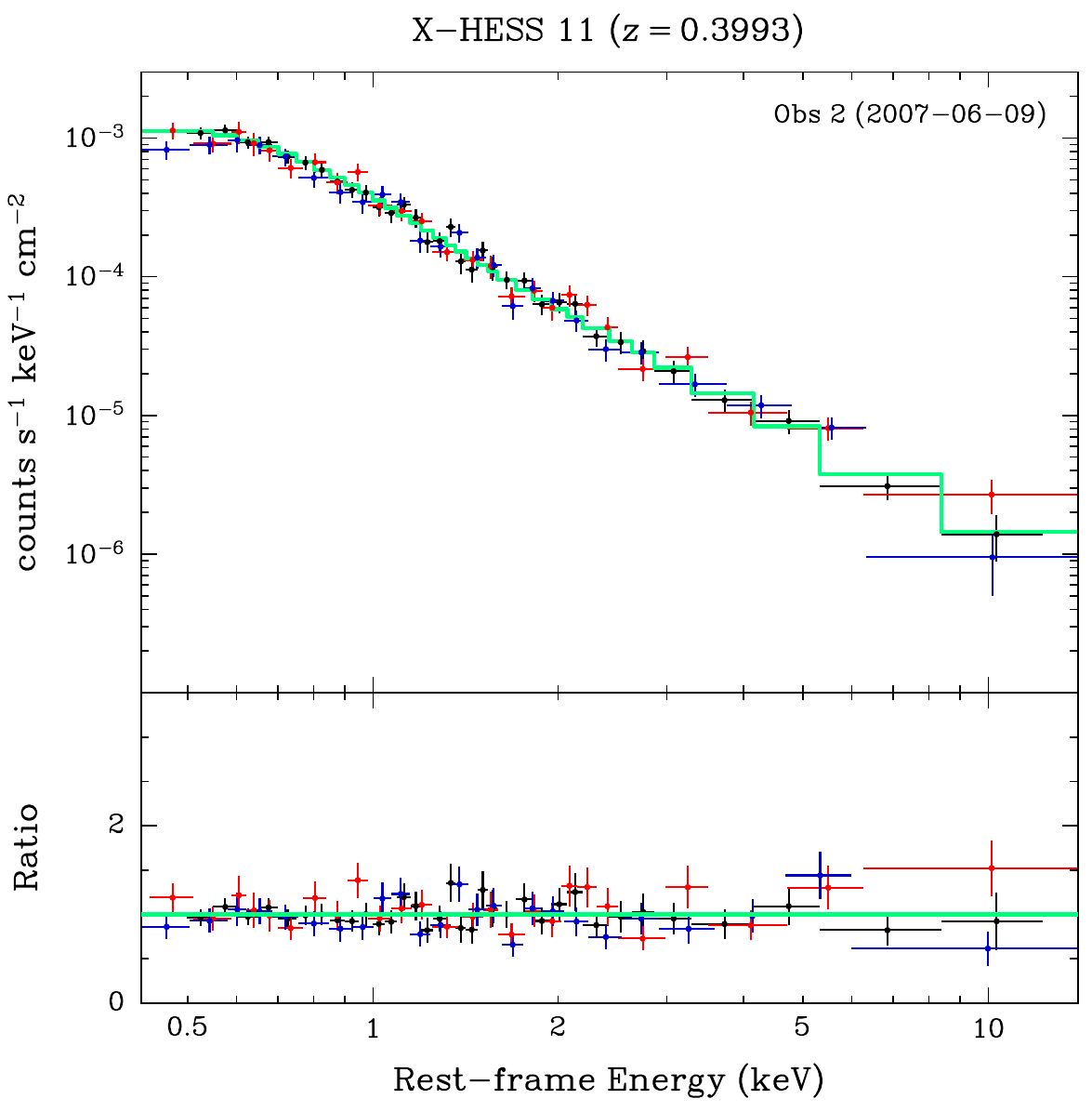}} \\
     \subfloat{\includegraphics[width = 2.1in]{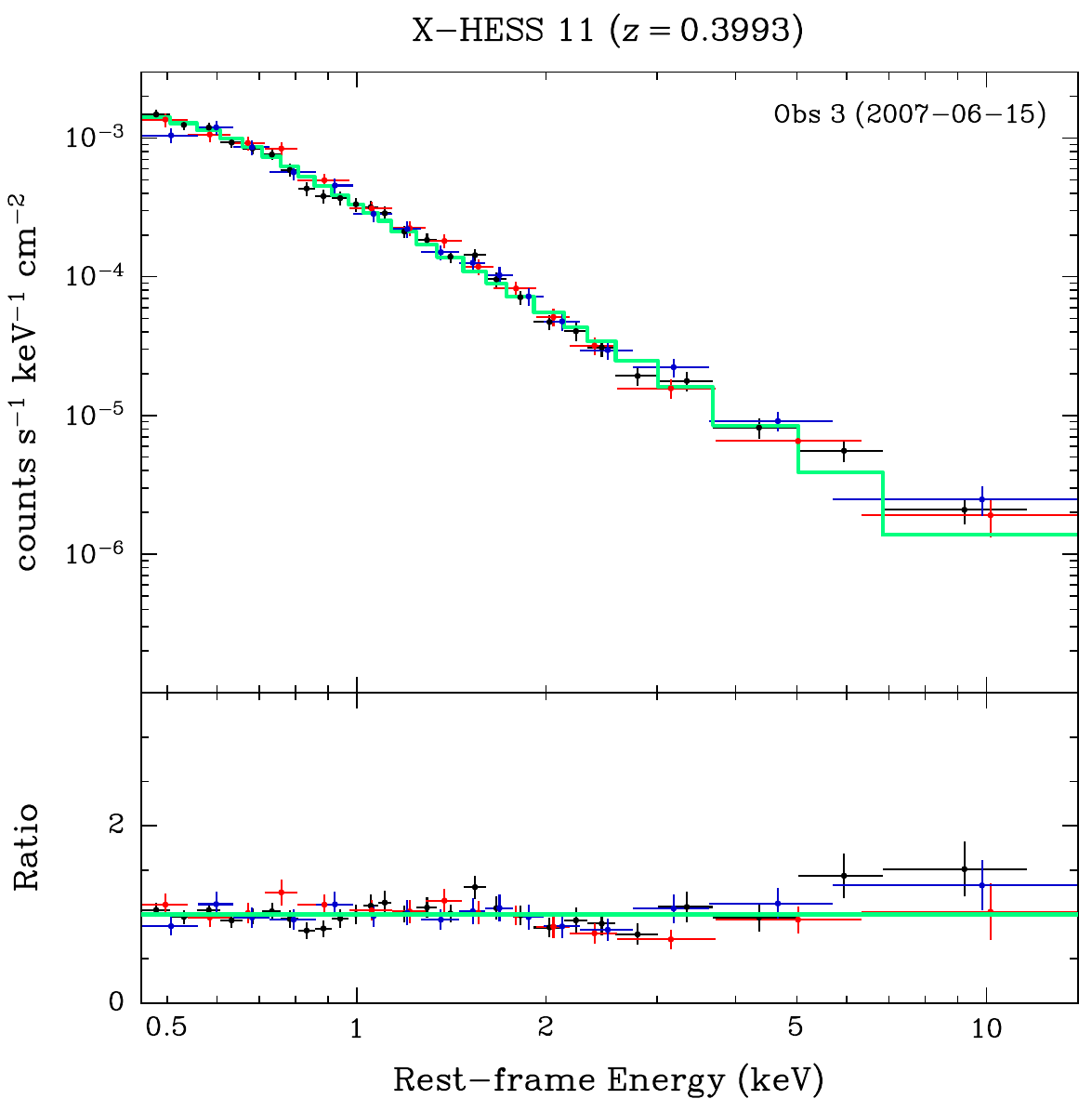}} &
     \subfloat{\includegraphics[width = 2.1in]{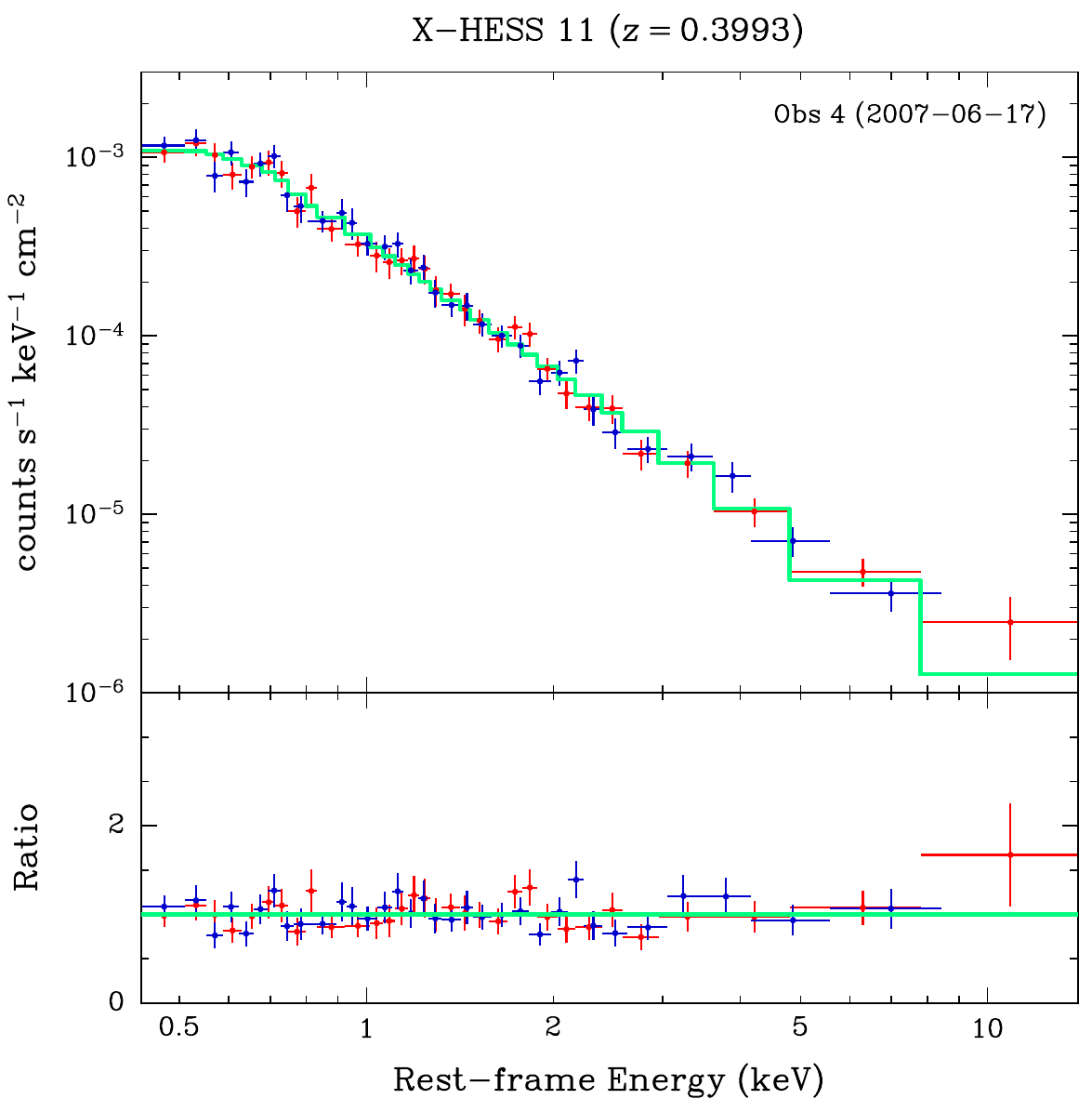}} &
     \subfloat{\includegraphics[width = 2.1in]{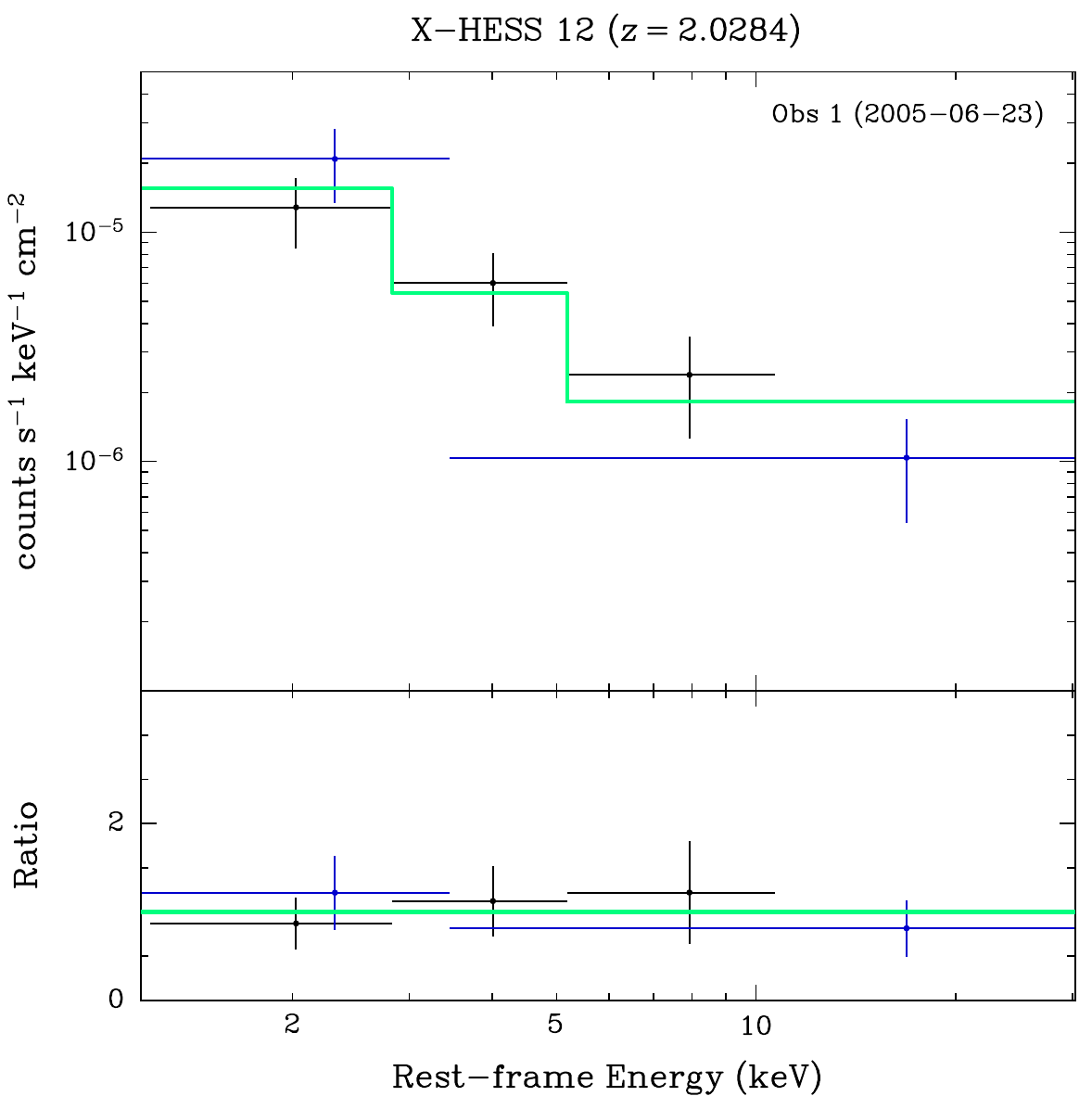}} \\
     \subfloat{\includegraphics[width = 2.1in]{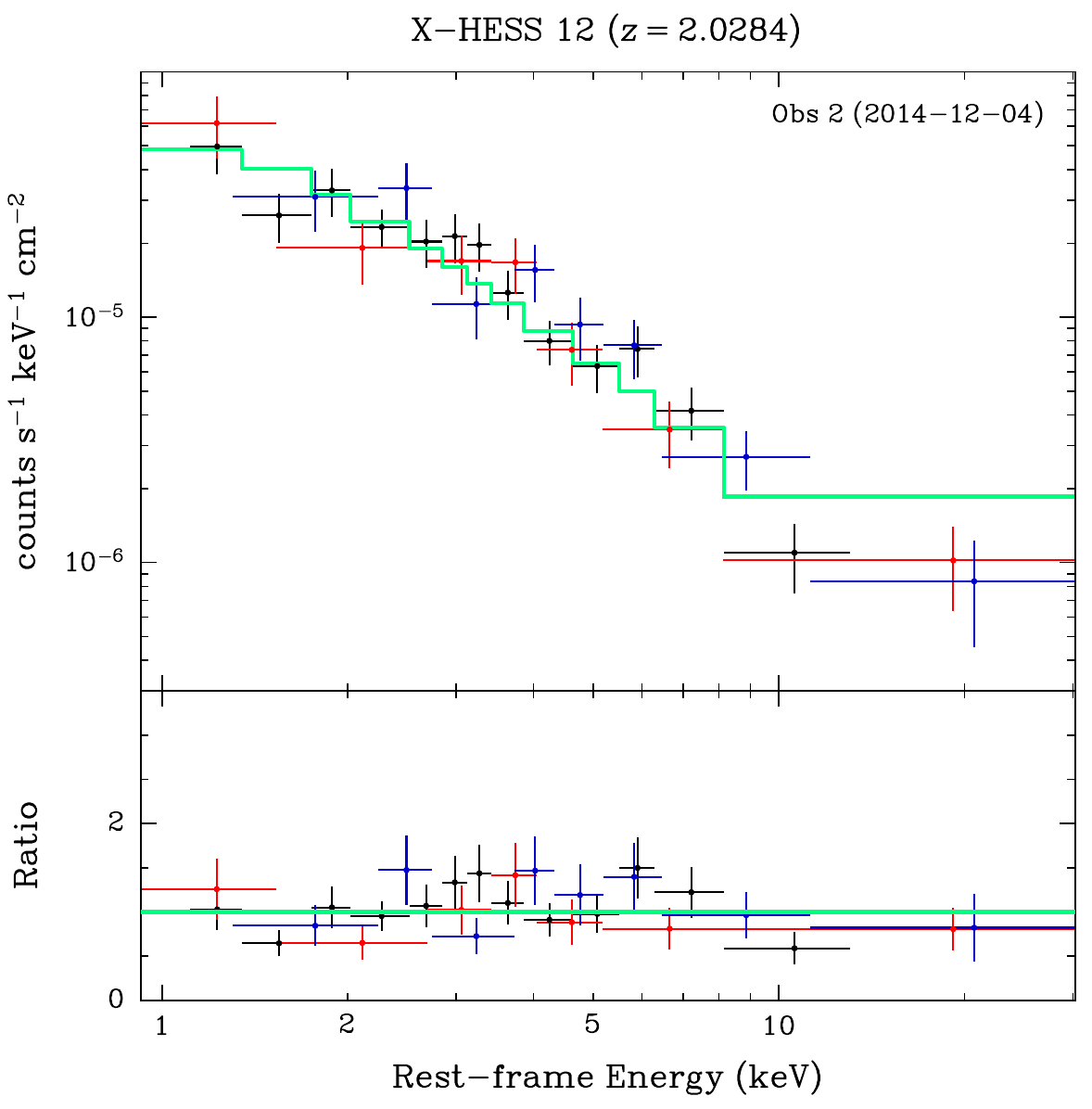}} &
     \subfloat{\includegraphics[width = 2.1in]{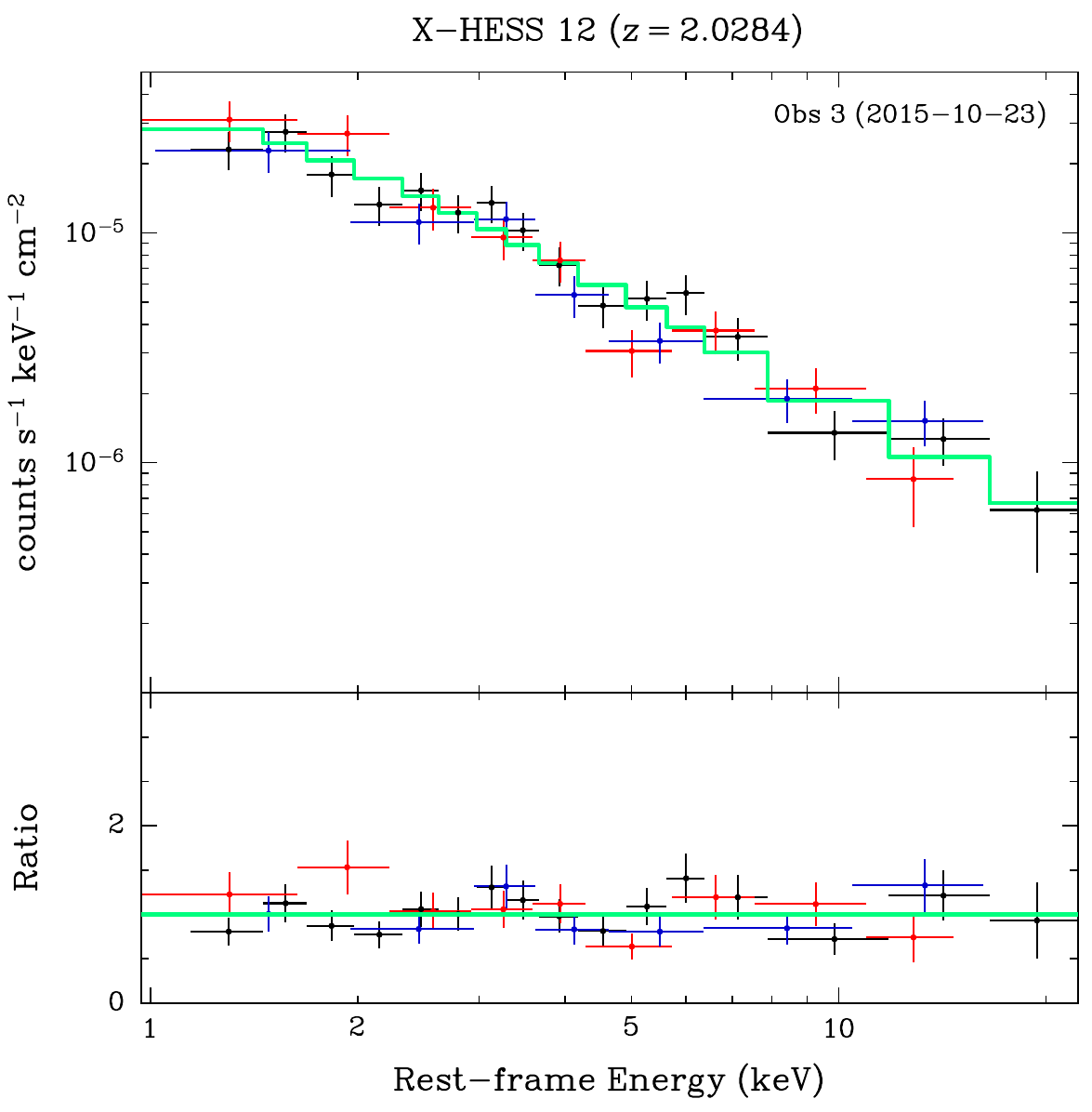}} &
     \subfloat{\includegraphics[width = 2.1in]{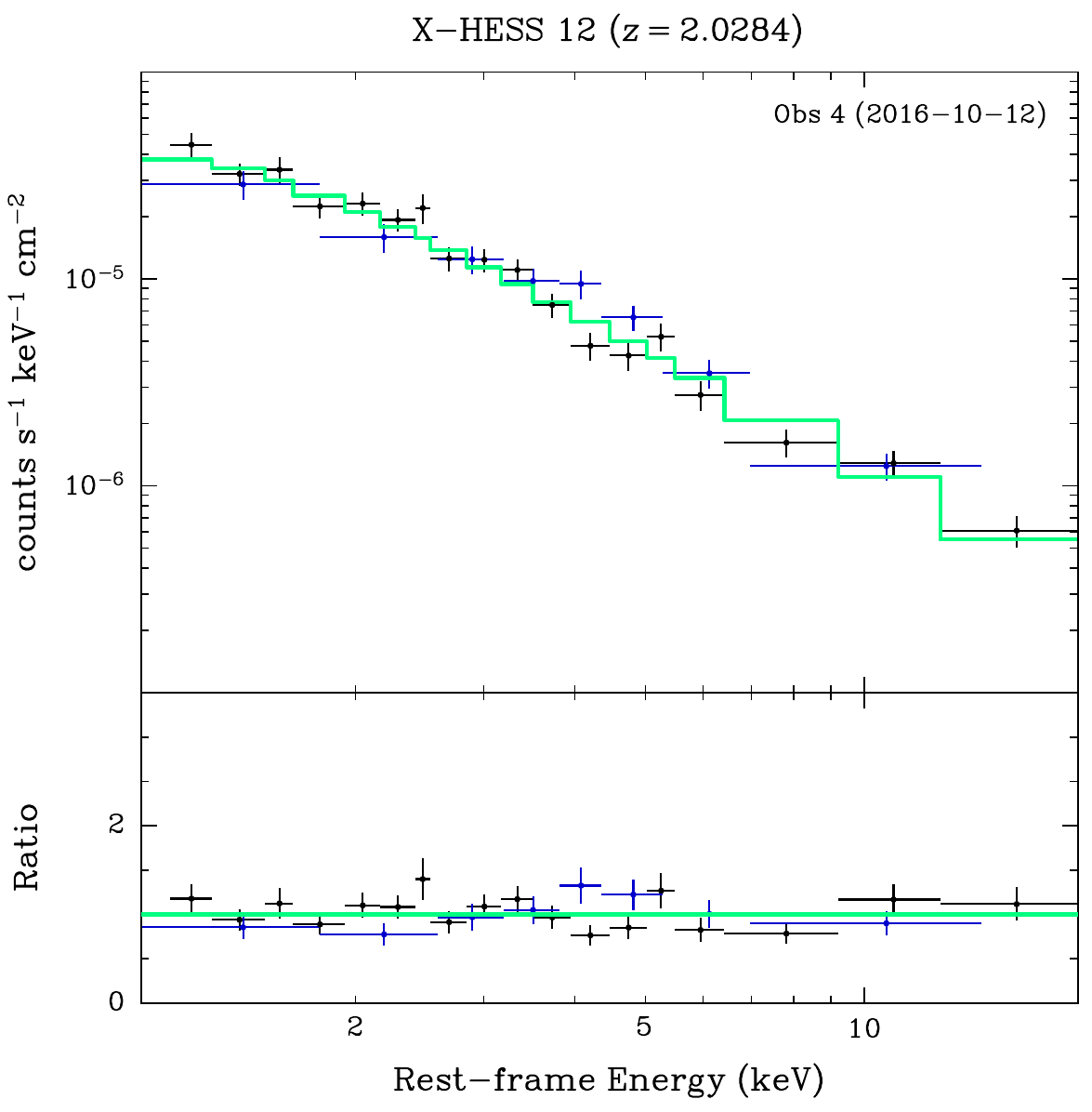}} \\
     \subfloat{\includegraphics[width = 2.1in]{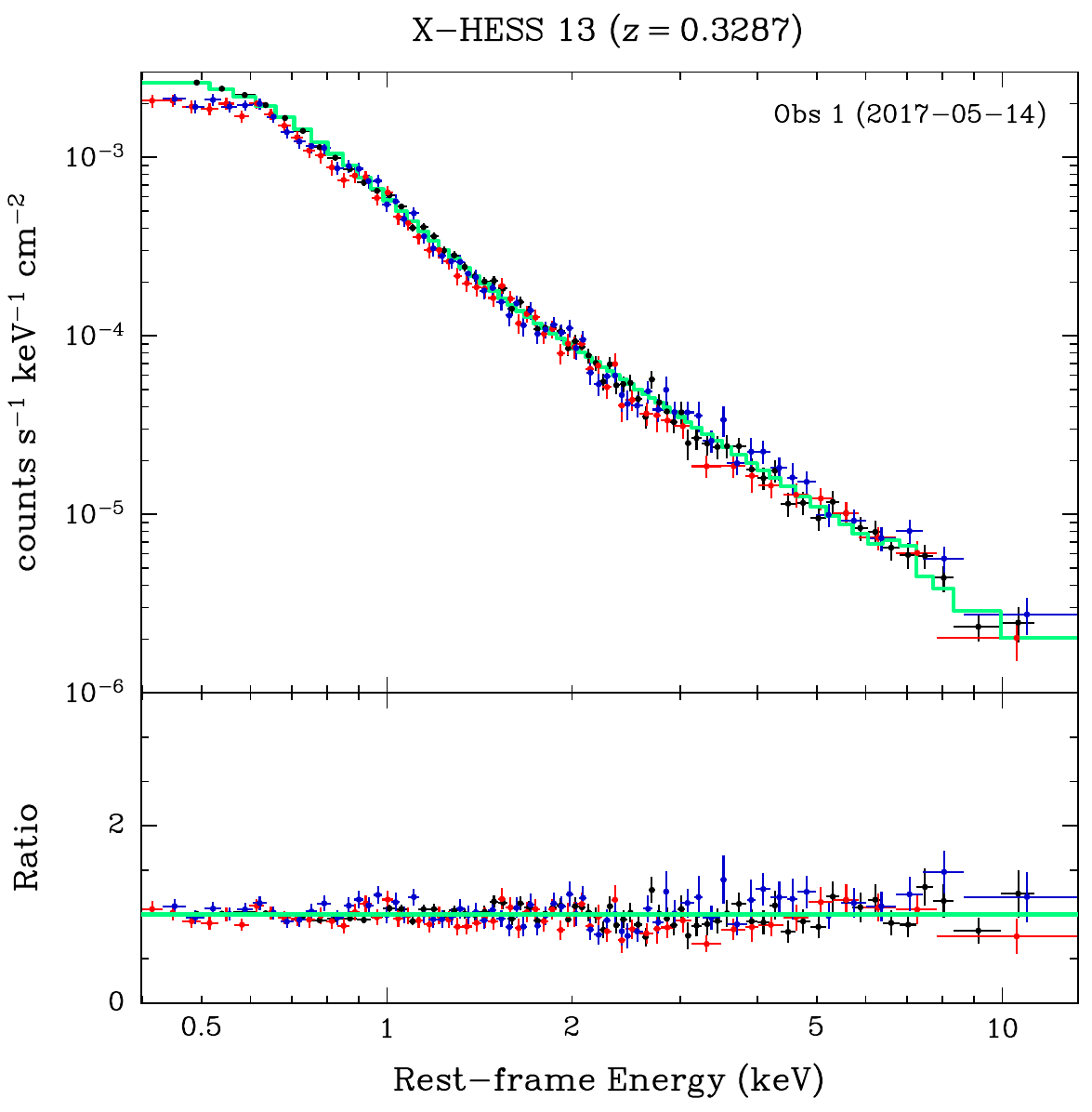}} &
     \subfloat{\includegraphics[width = 2.1in]{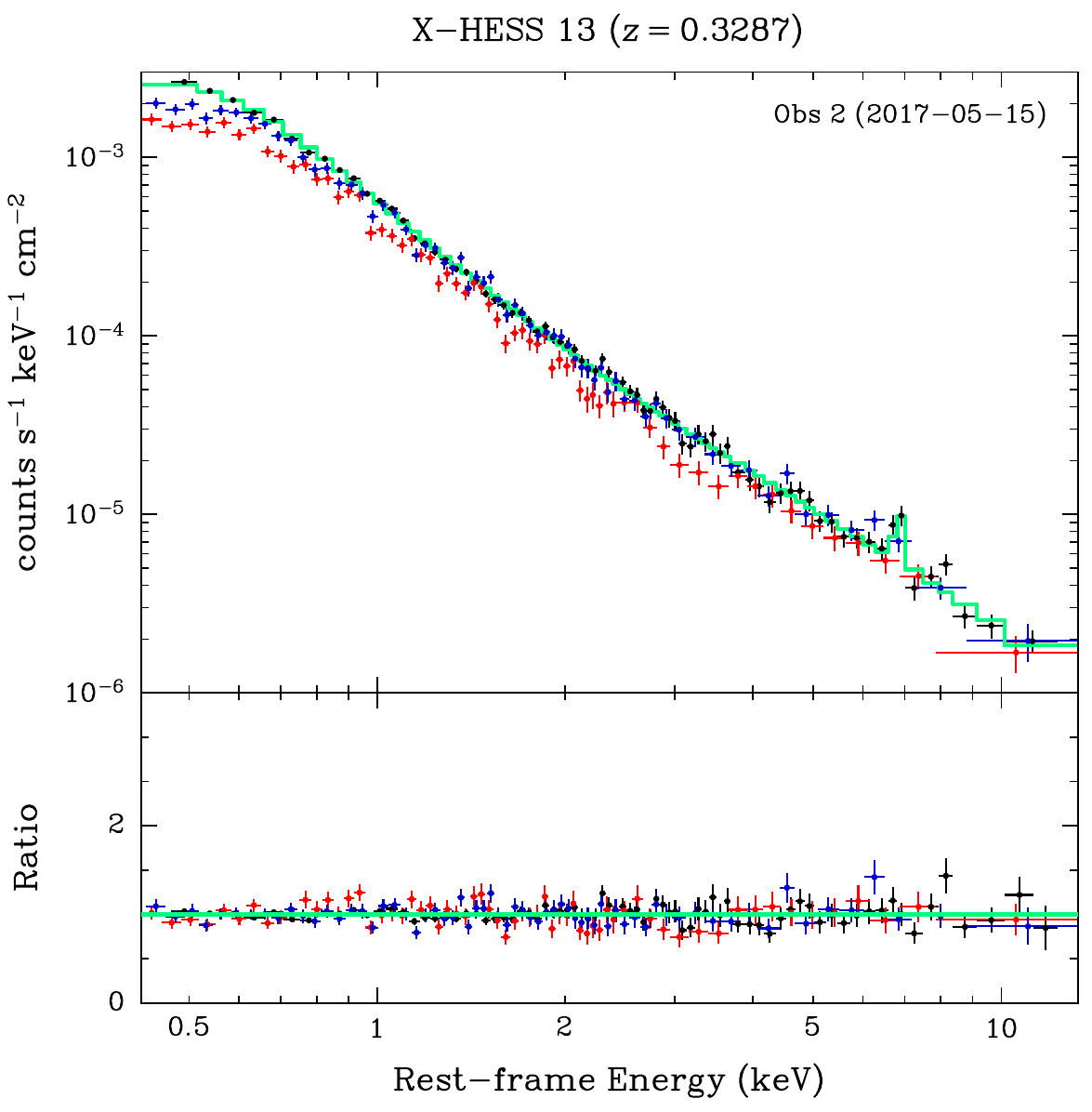}} &
     \subfloat{\includegraphics[width = 2.1in]{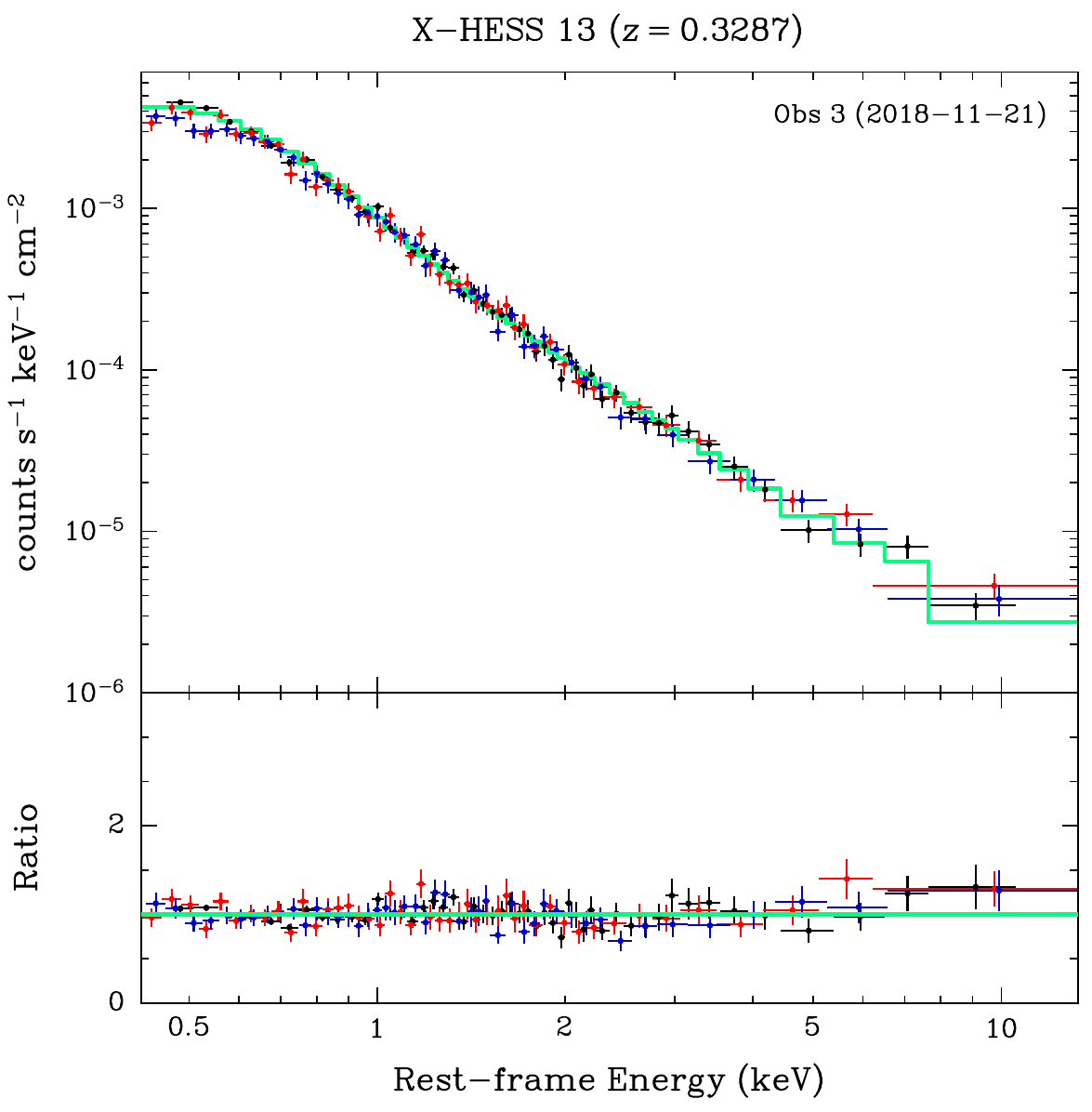}} \\
     \end{tabular}
\begin{minipage}{1.2\linewidth}
    \centering 
    {Continuation of Fig. \ref{fig:xhess_spectra}.}
\end{minipage}

\end{figure}

\begin{figure}[h]
     \ContinuedFloat
     \centering
     \renewcommand{\arraystretch}{2}
     \begin{tabular}{ccc}
     \subfloat{\includegraphics[width = 2.1in]{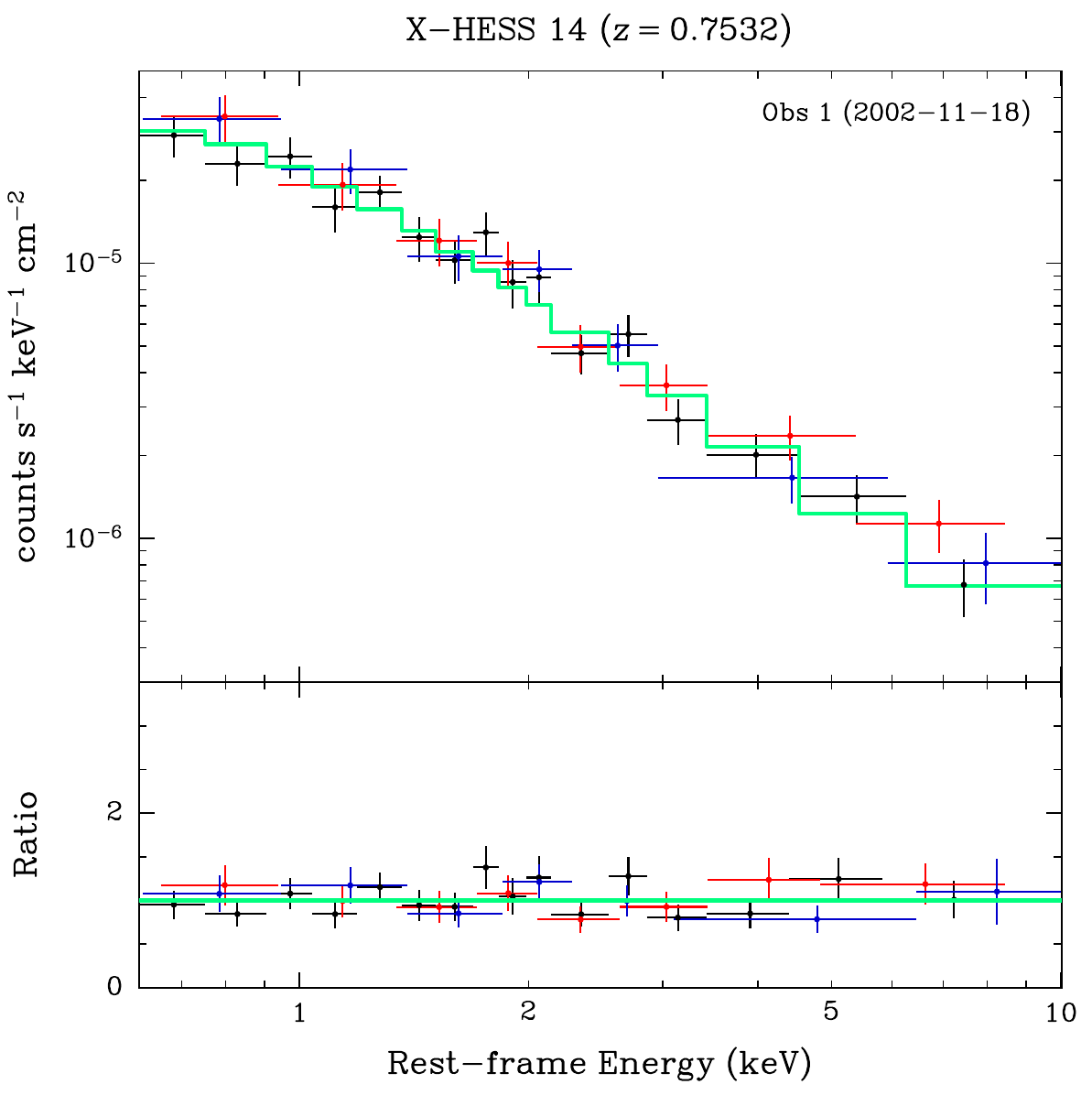}} &
     \subfloat{\includegraphics[width = 2.1in]{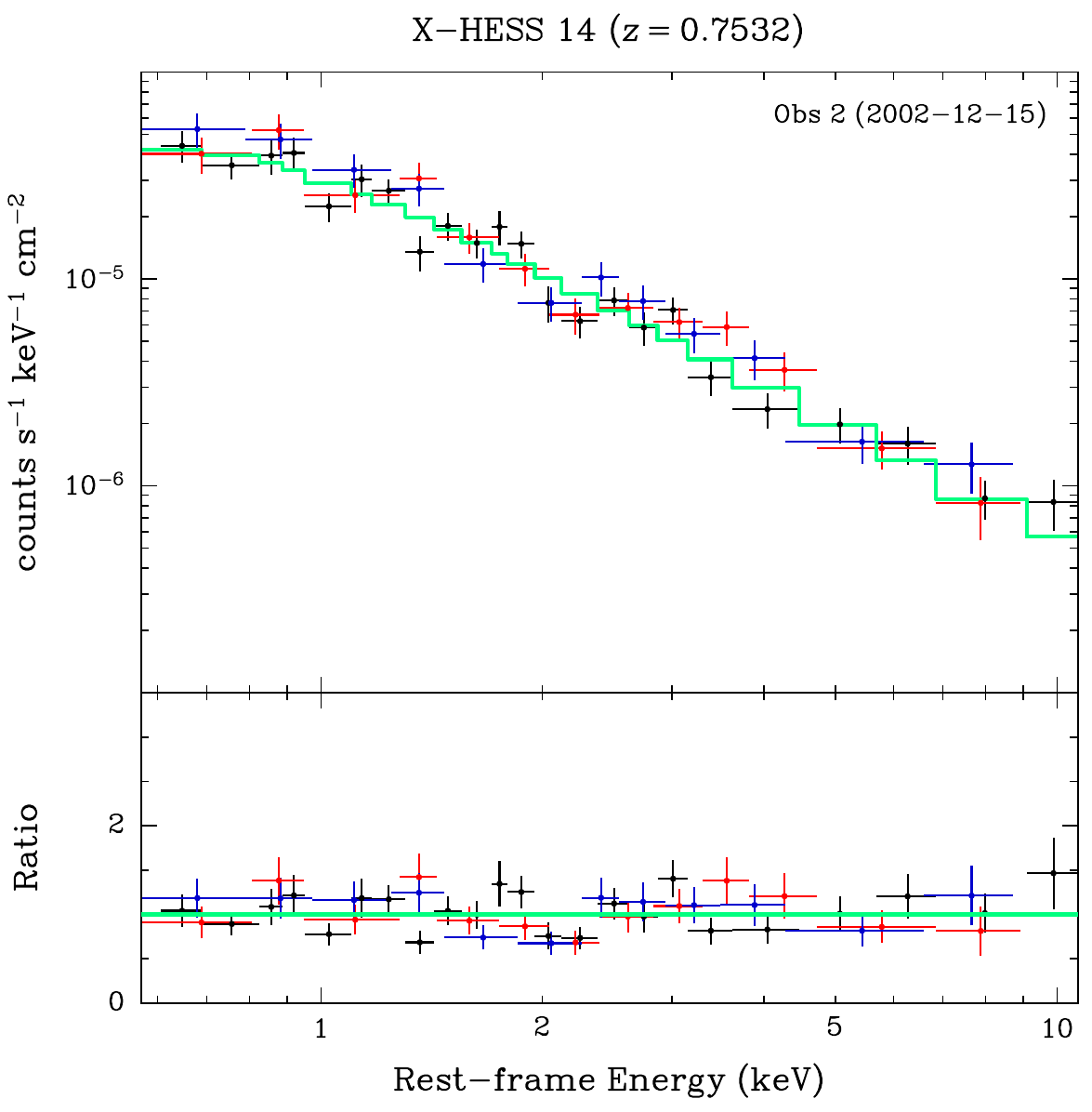}} &
     \subfloat{\includegraphics[width = 2.1in]{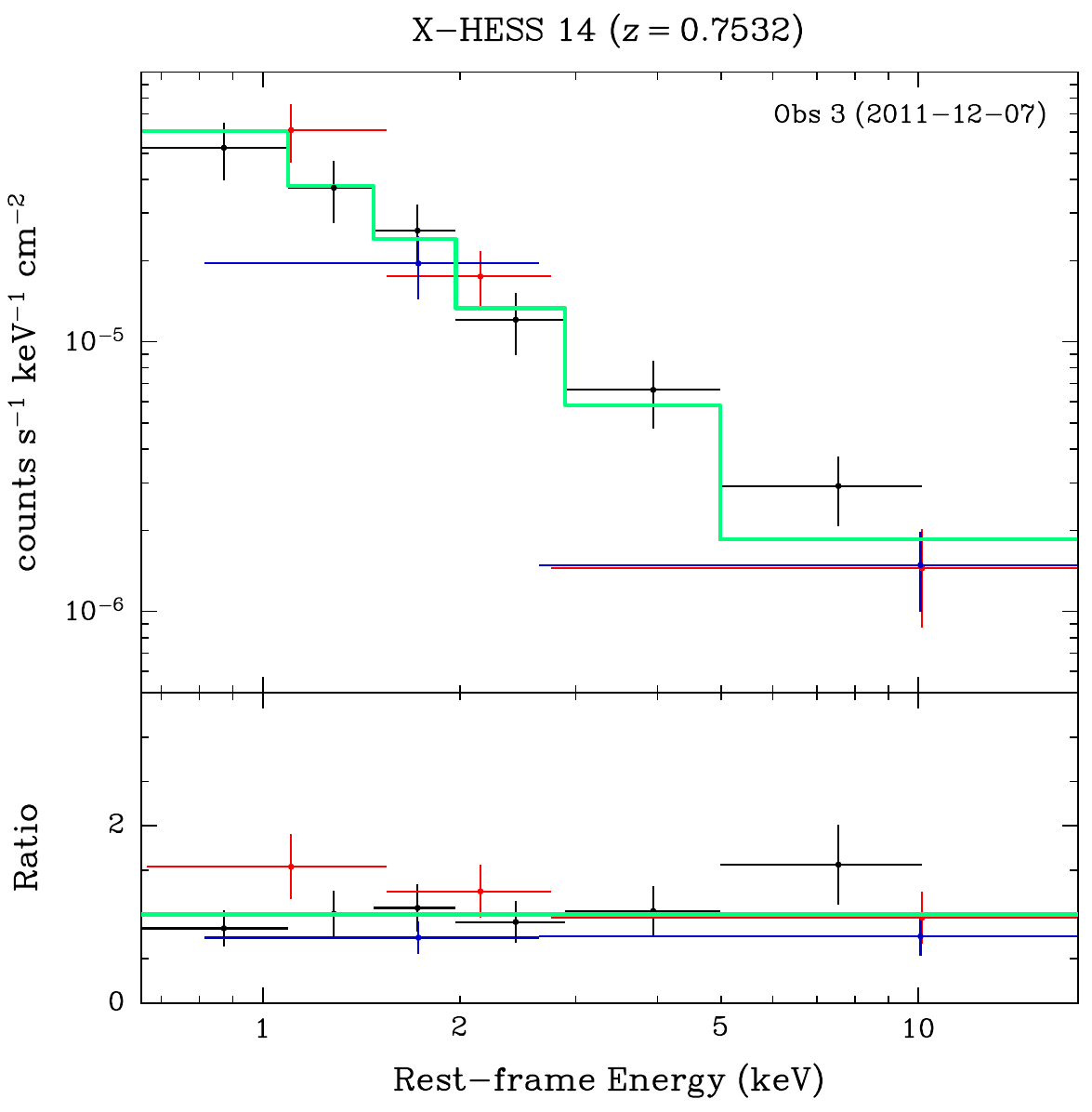}} \\
     \subfloat{\includegraphics[width = 2.1in]{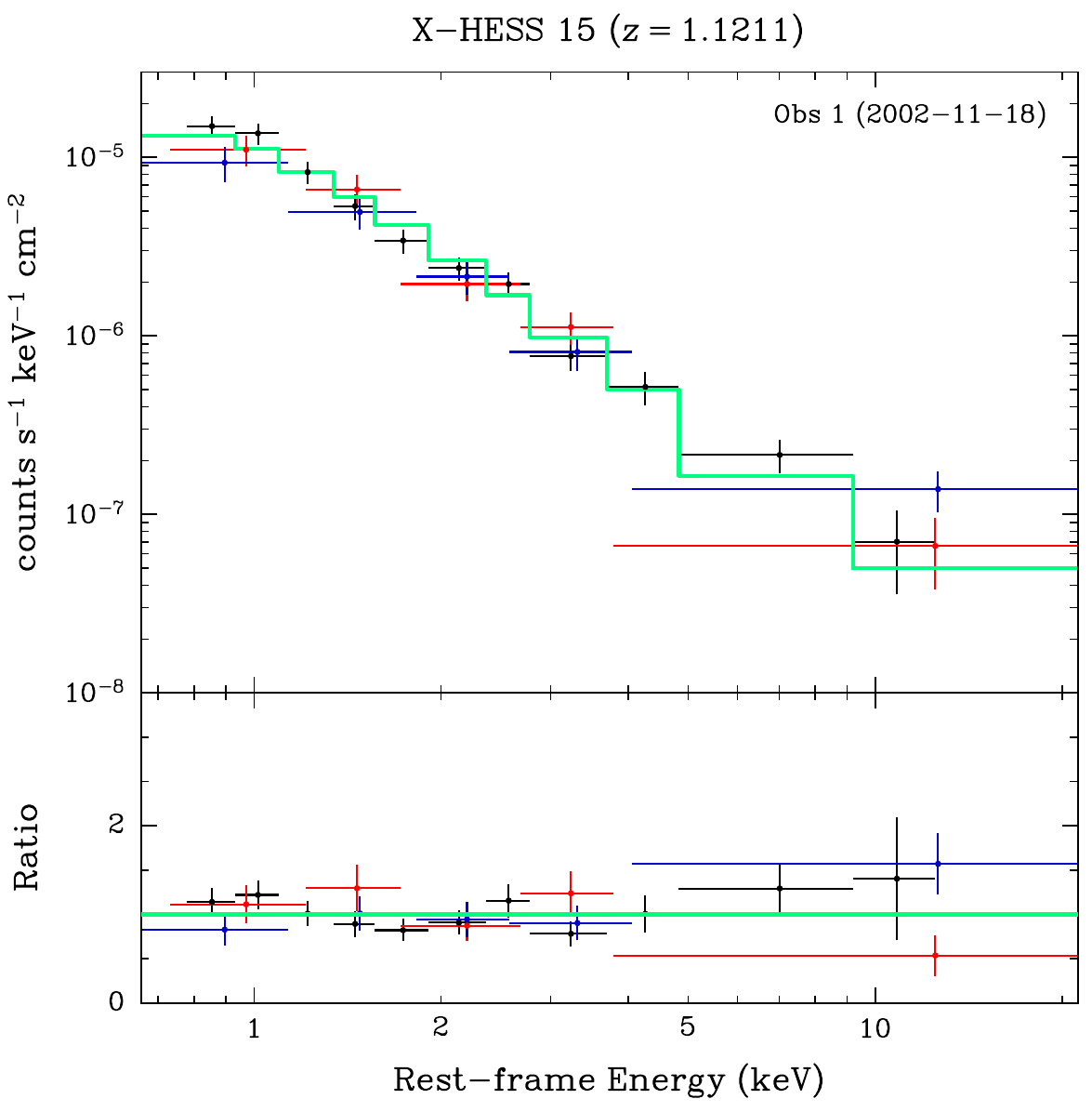}} &
     \subfloat{\includegraphics[width = 2.1in]{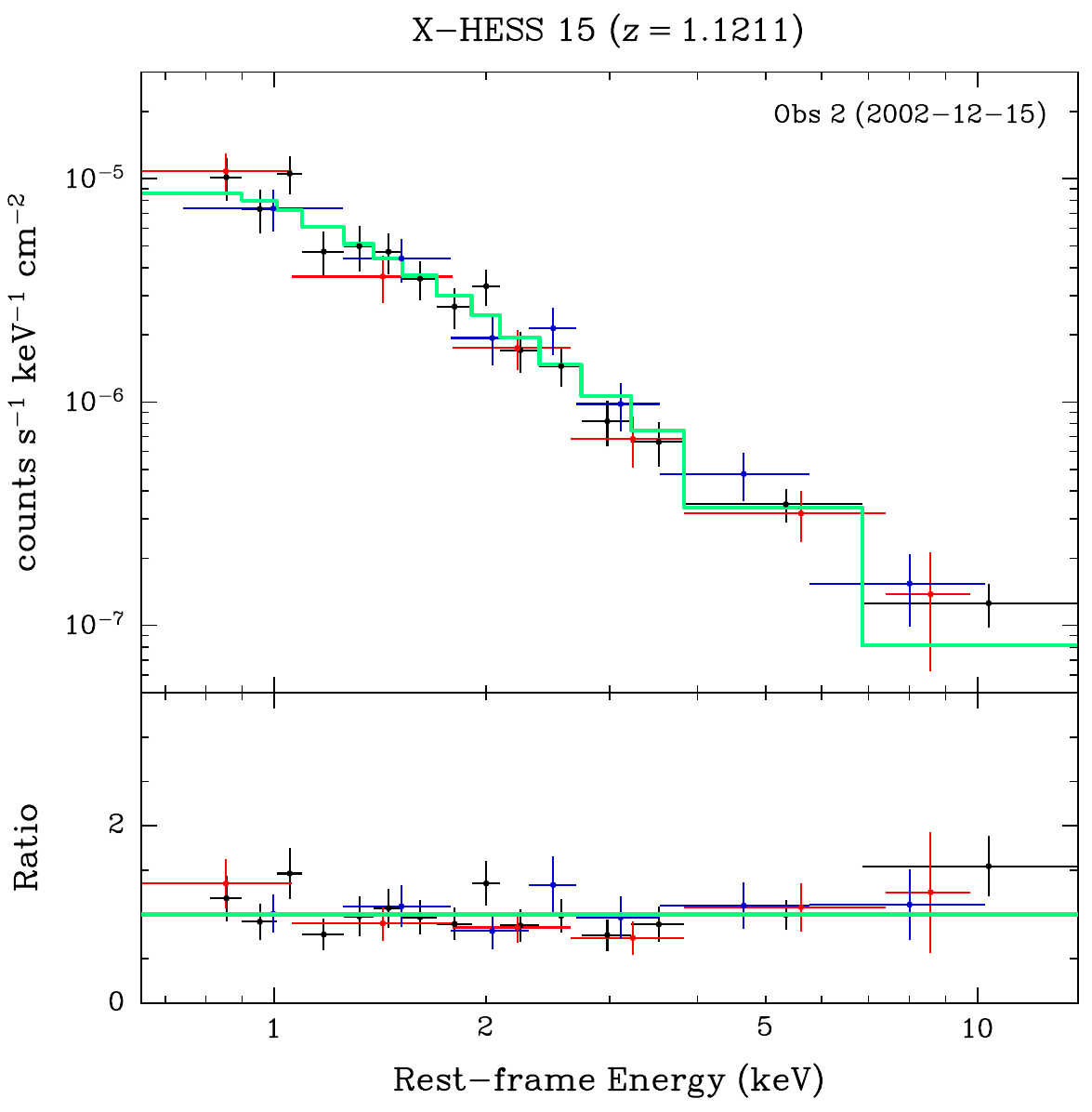}} &
     \subfloat{\includegraphics[width = 2.1in]{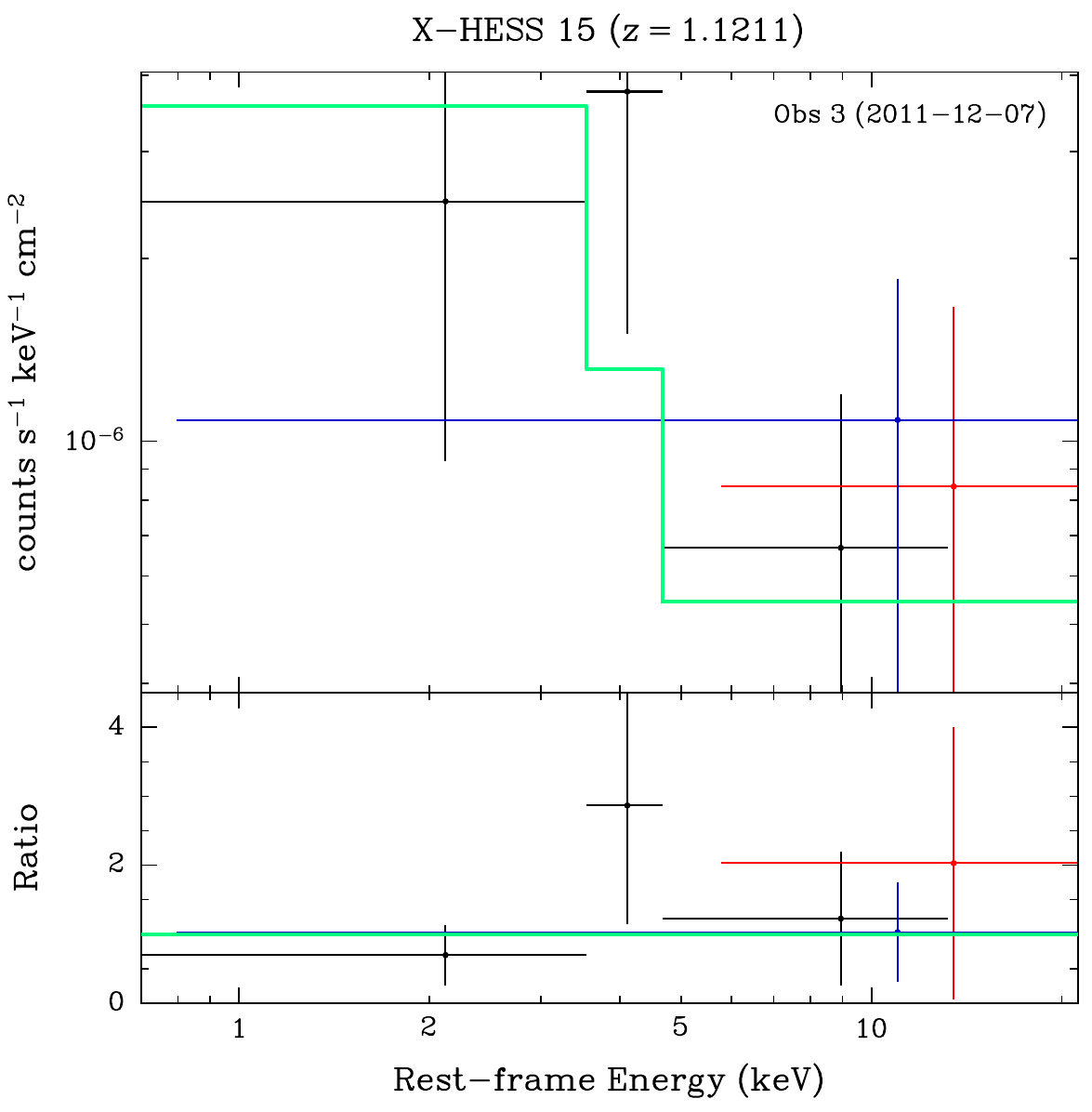}} \\
     \subfloat{\includegraphics[width = 2.1in]{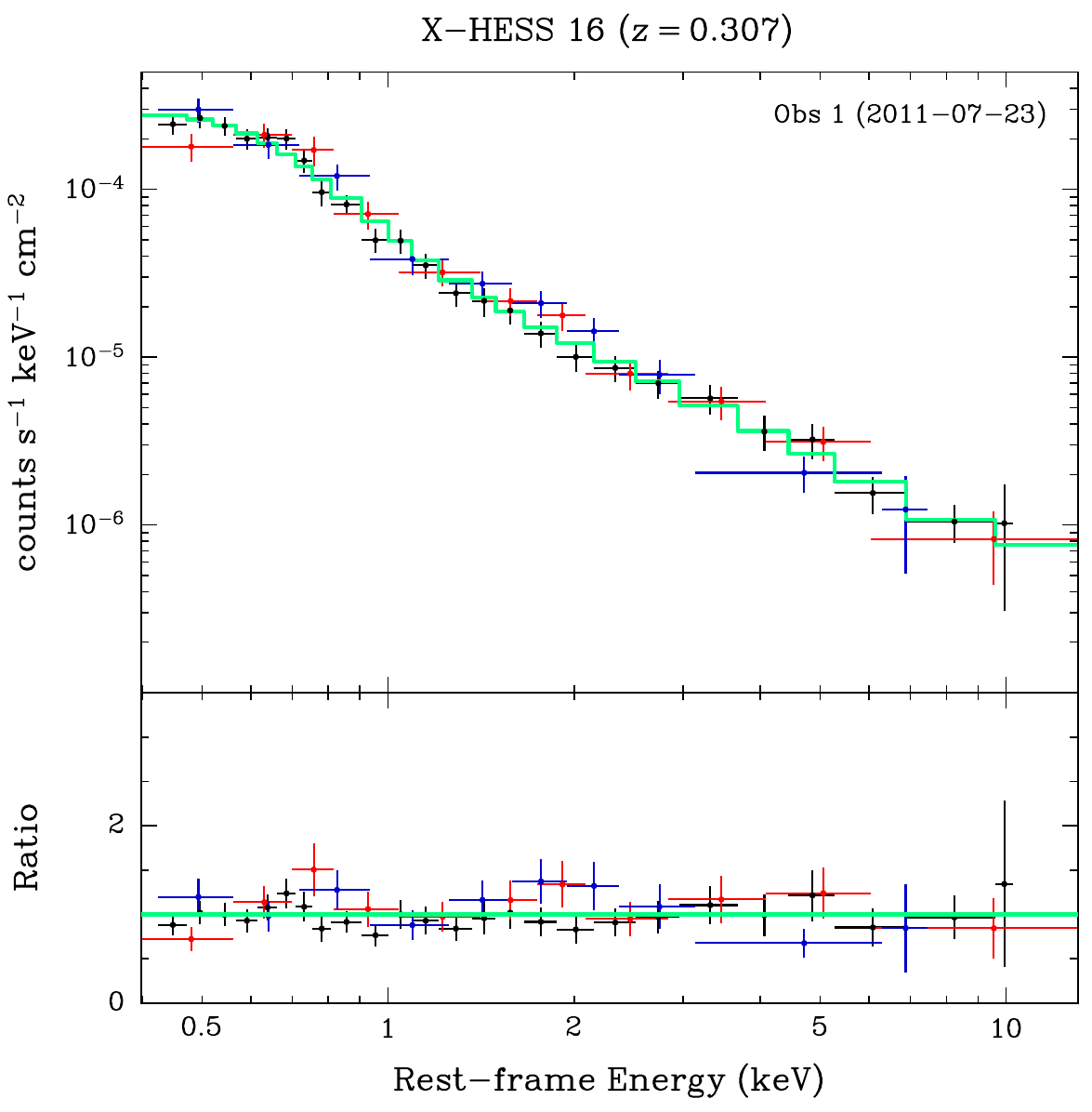}} &
     \subfloat{\includegraphics[width = 2.1in]{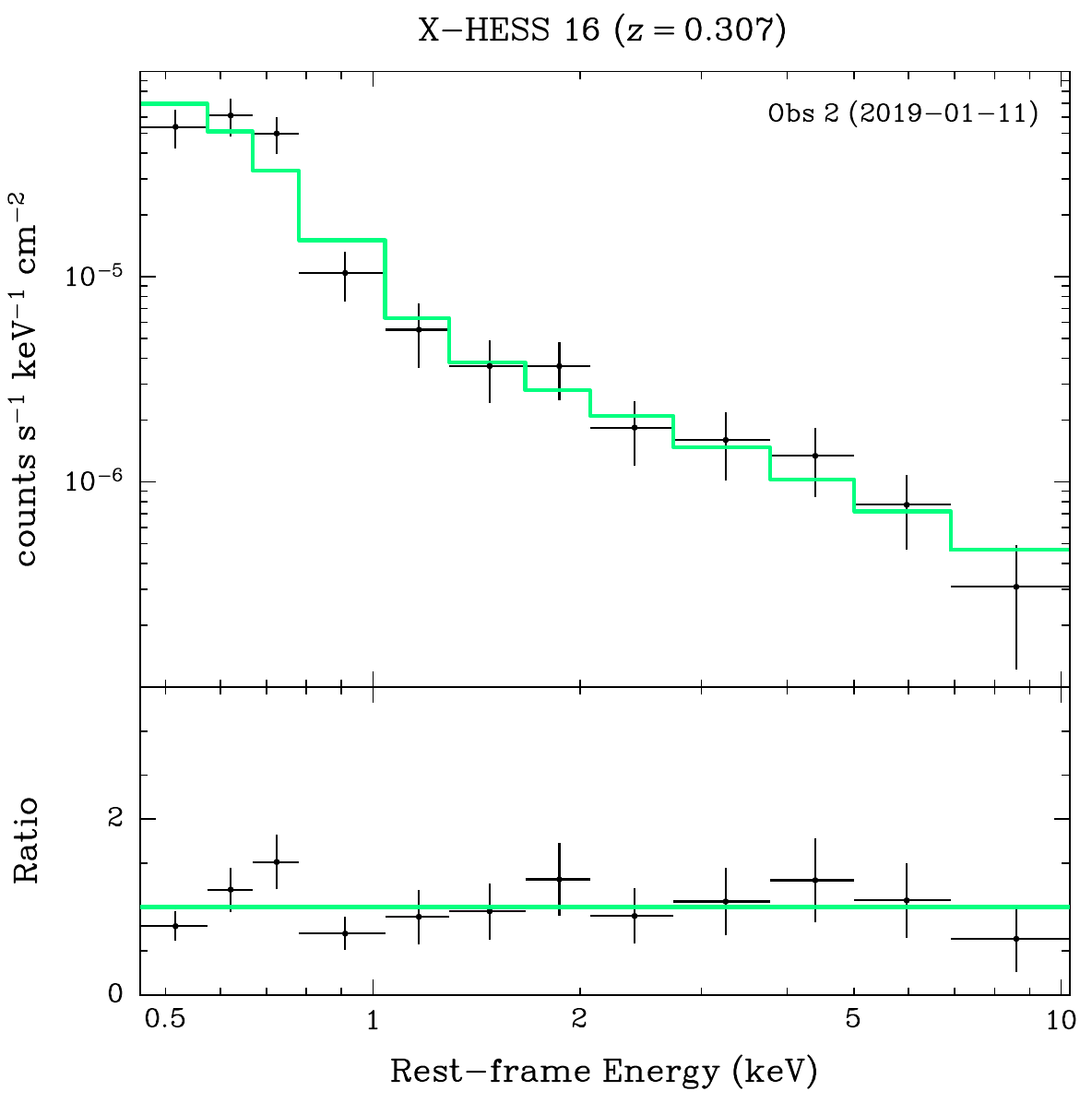}} &
     \subfloat{\includegraphics[width = 2.1in]{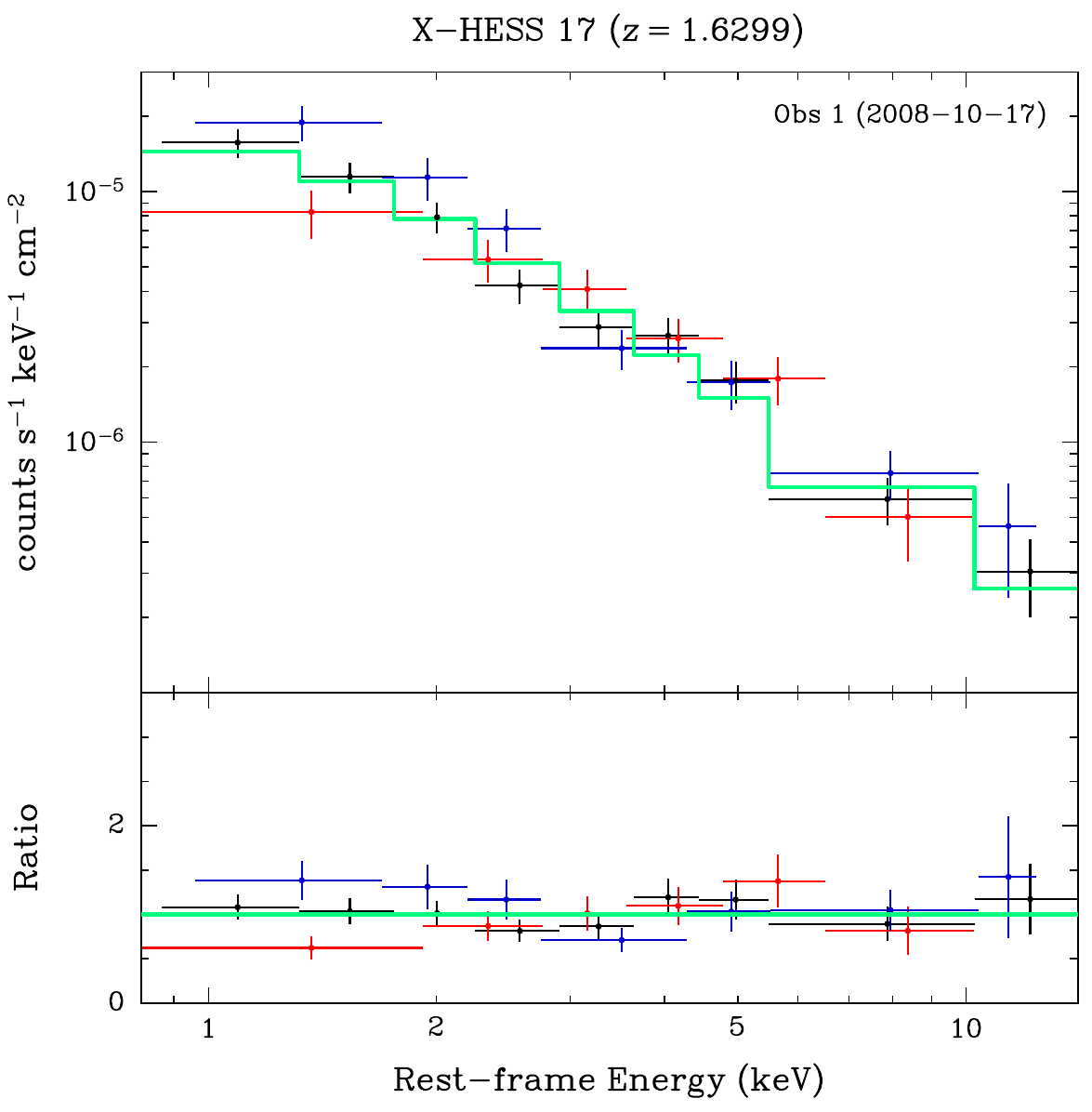}} \\
     \subfloat{\includegraphics[width = 2.1in]{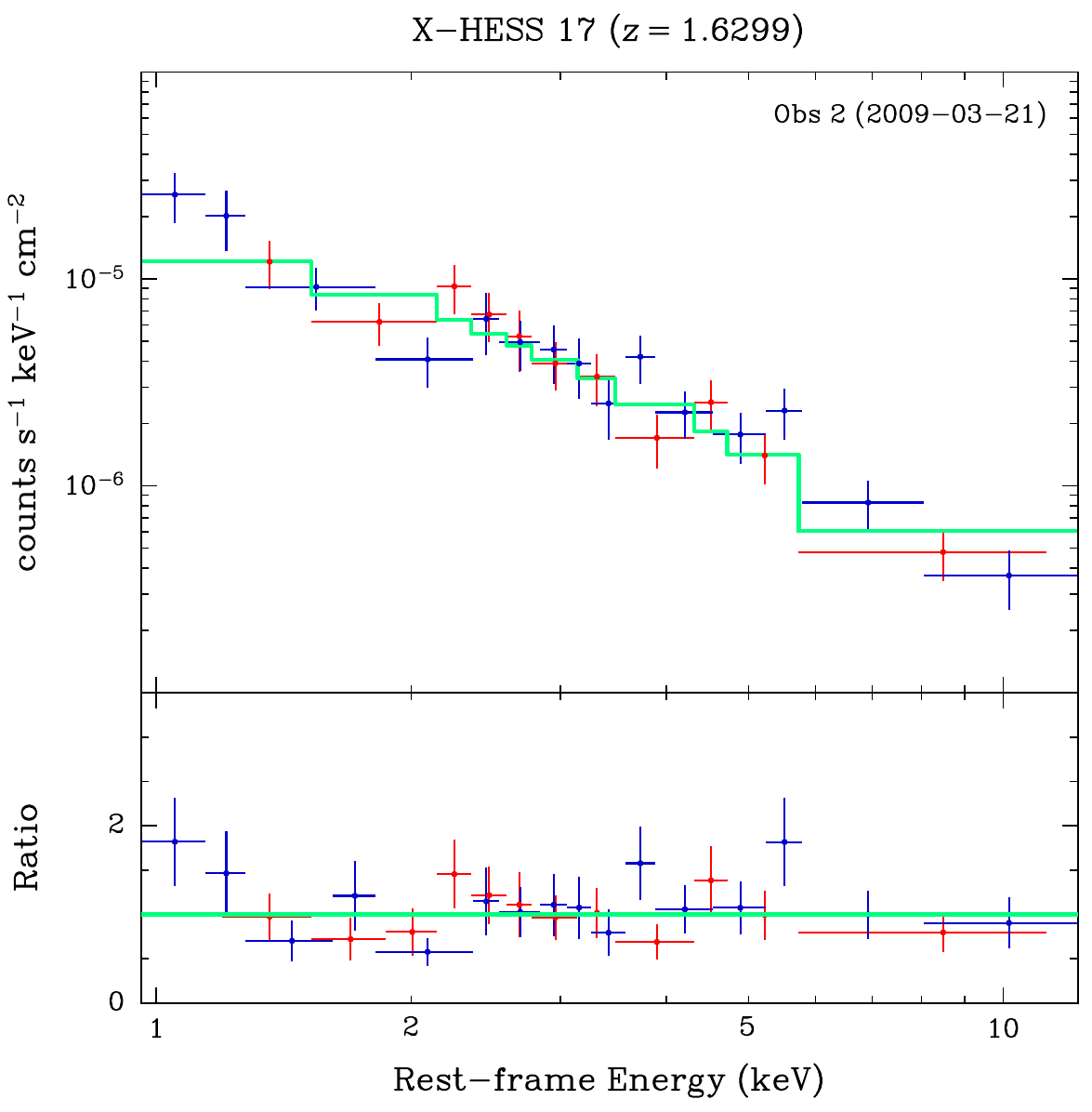}} &
     \subfloat{\includegraphics[width = 2.1in]{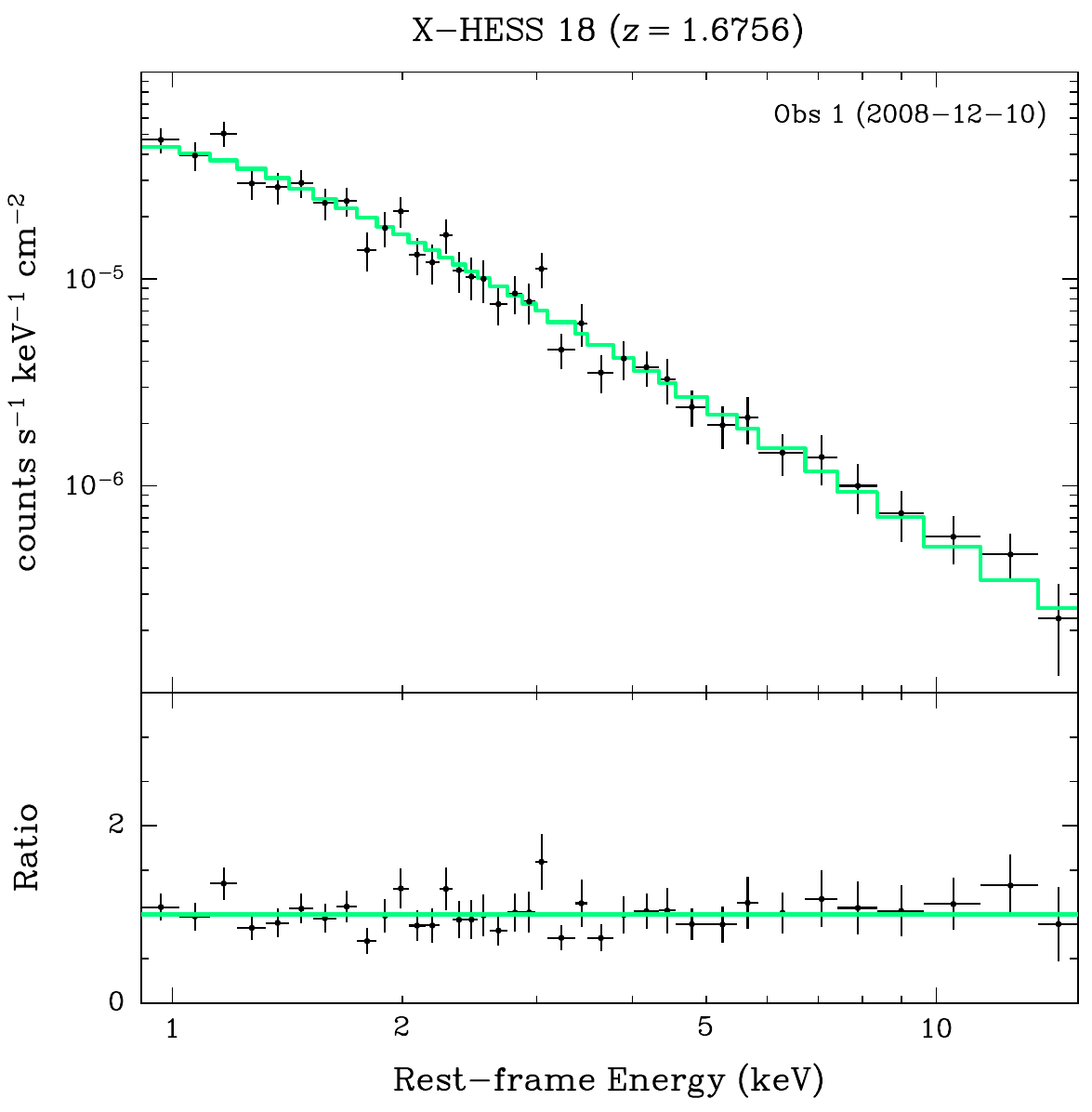}} &
     \subfloat{\includegraphics[width = 2.1in]{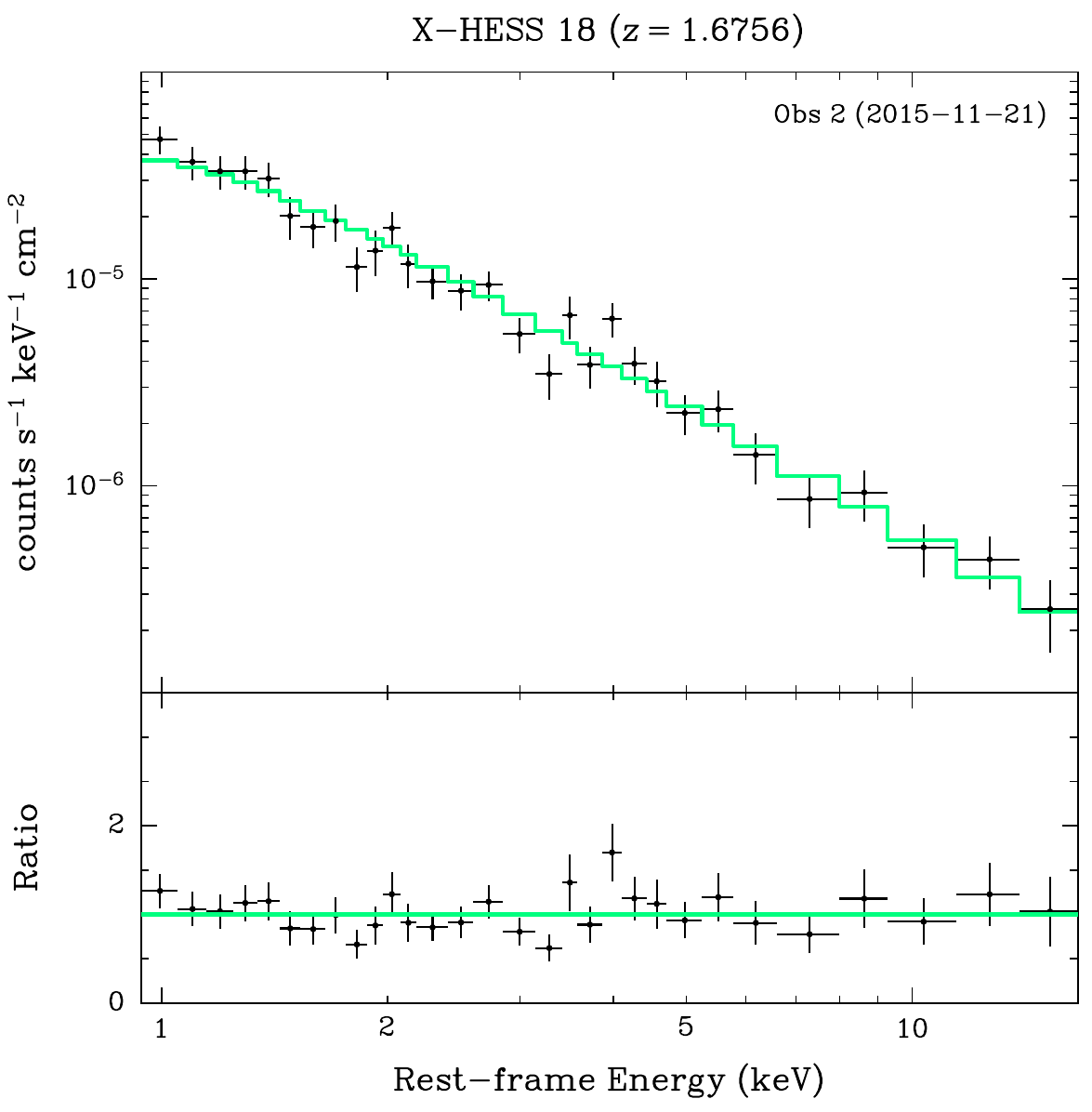}} \\
     \end{tabular}
\begin{minipage}{1.2\linewidth}
    \centering 
    {Continuation of Fig. \ref{fig:xhess_spectra}.}
\end{minipage}

\end{figure}

\begin{figure}[h]
     \ContinuedFloat
     \centering
     \renewcommand{\arraystretch}{2}
     \begin{tabular}{ccc}
     \subfloat{\includegraphics[width = 2.1in]{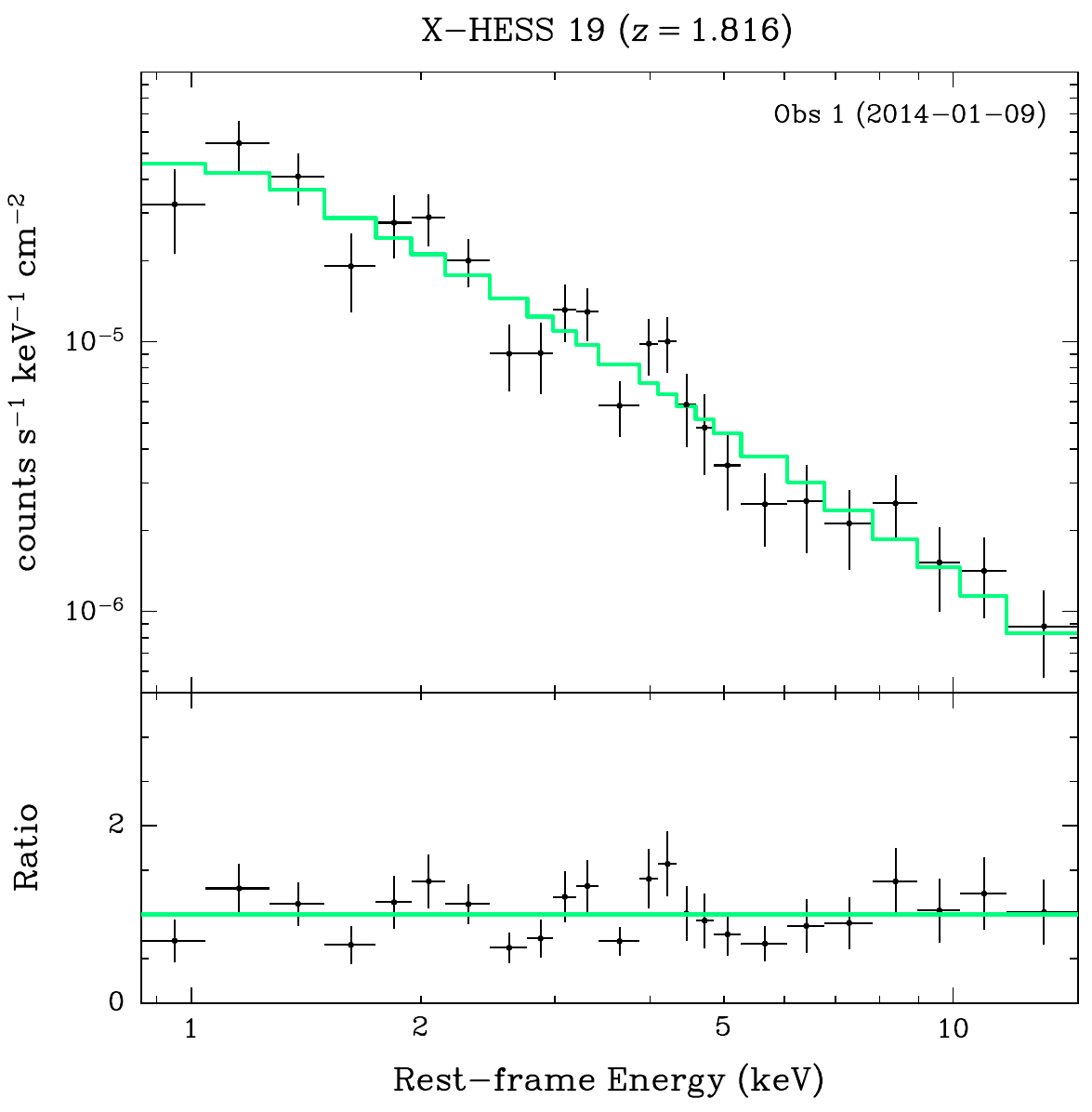}} &
     \subfloat{\includegraphics[width = 2.1in]{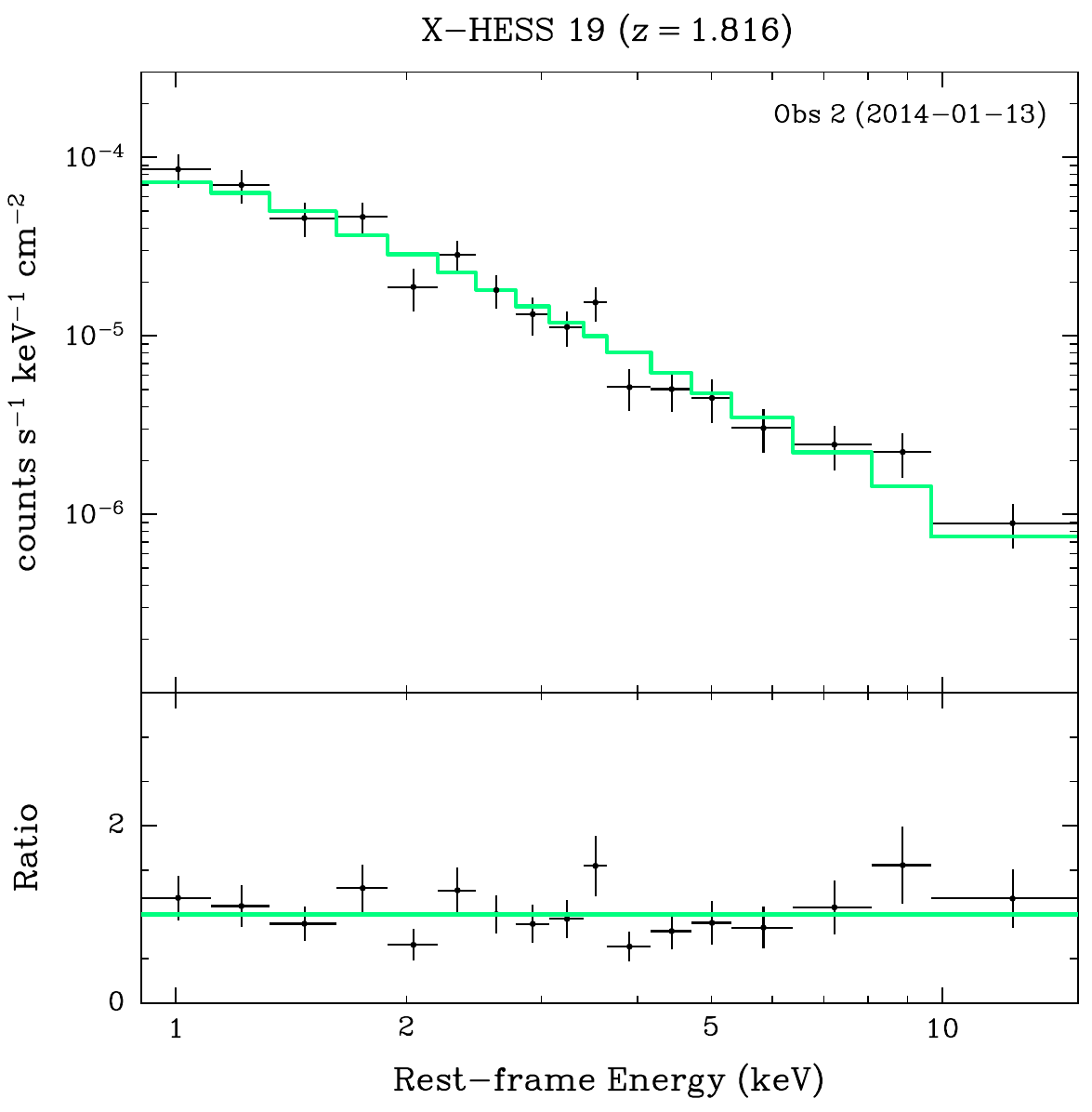}} &
     \subfloat{\includegraphics[width = 2.1in]{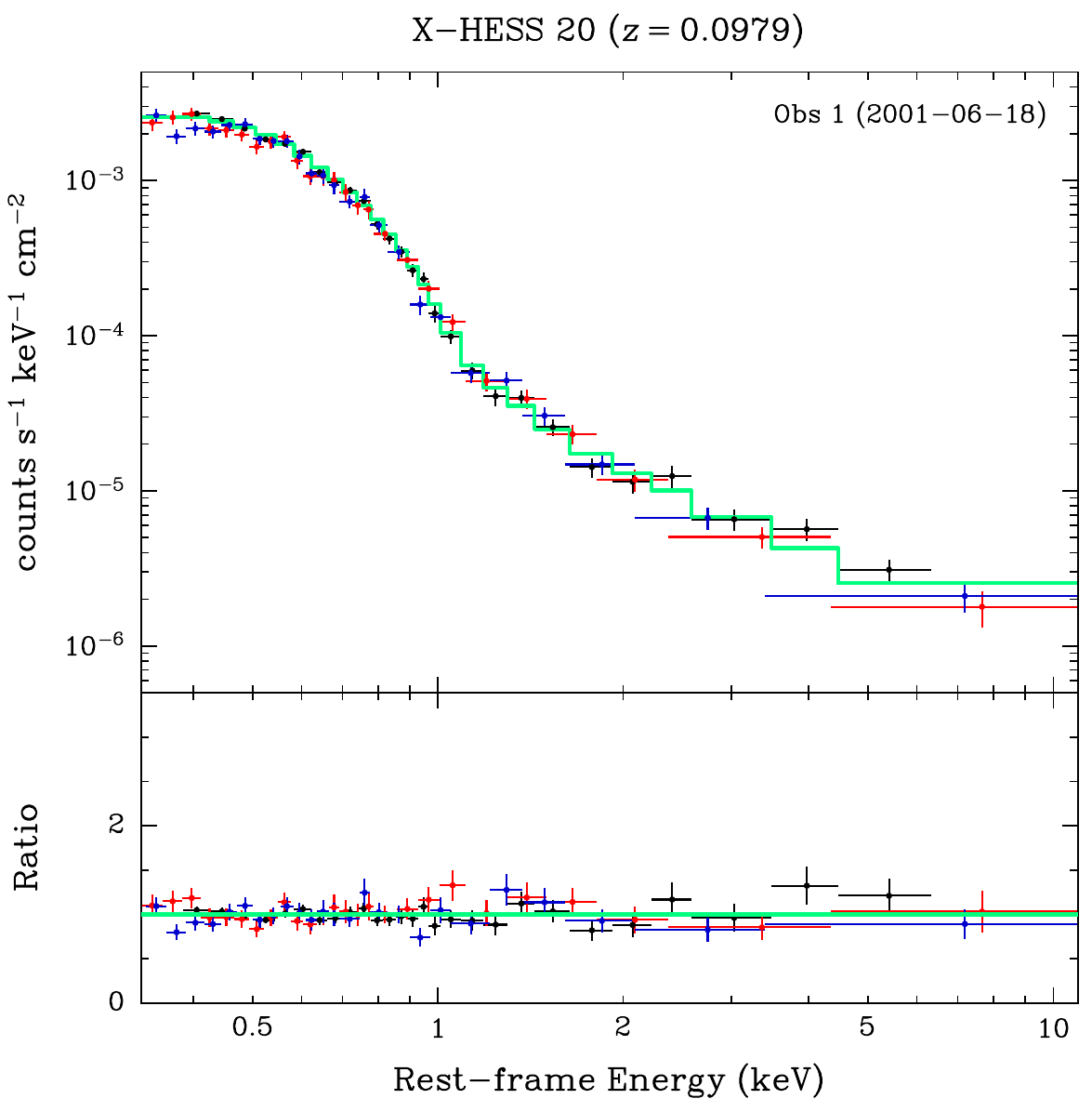}} \\
     \subfloat{\includegraphics[width = 2.1in]{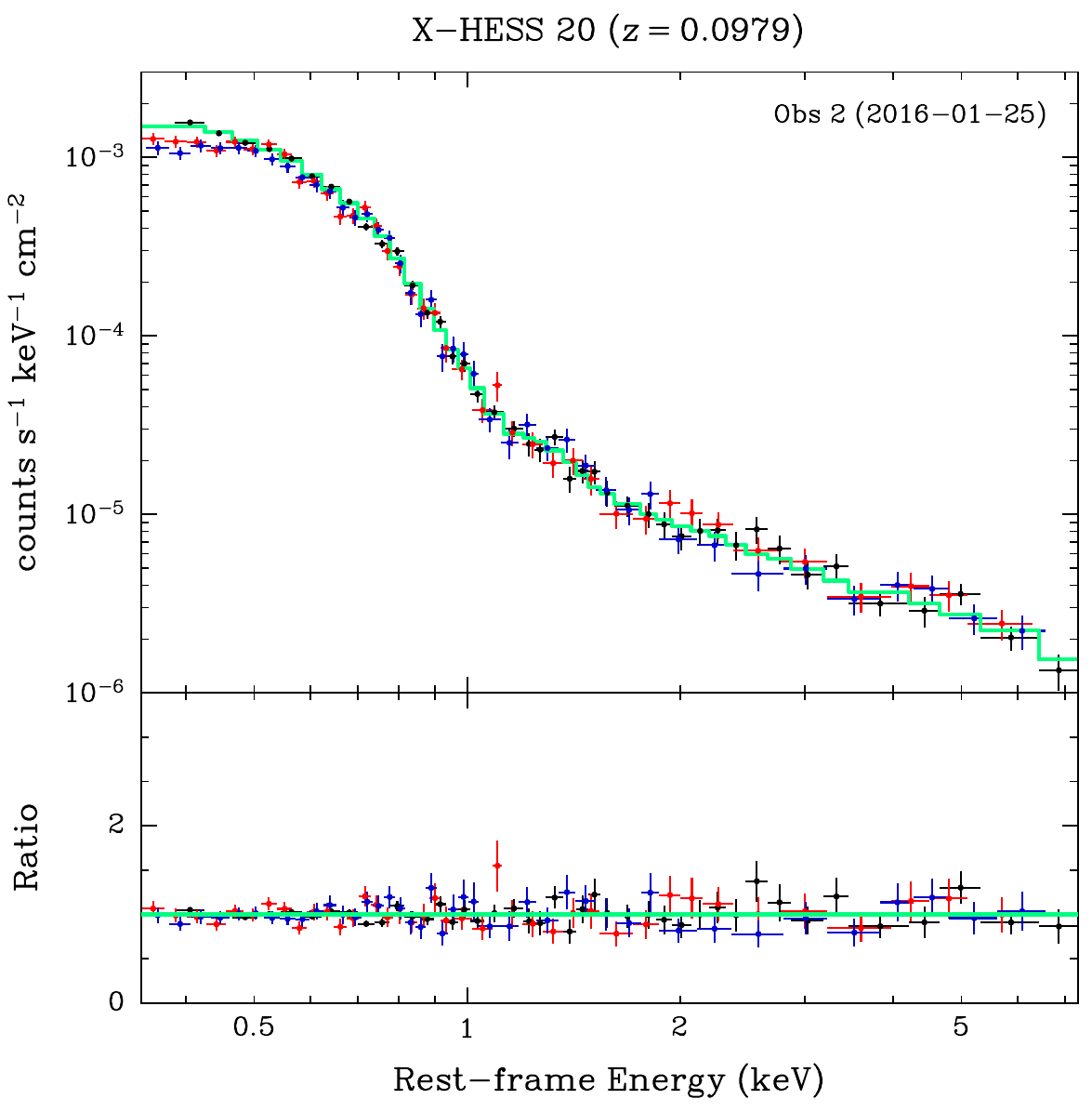}} &
     \subfloat{\includegraphics[width = 2.1in]{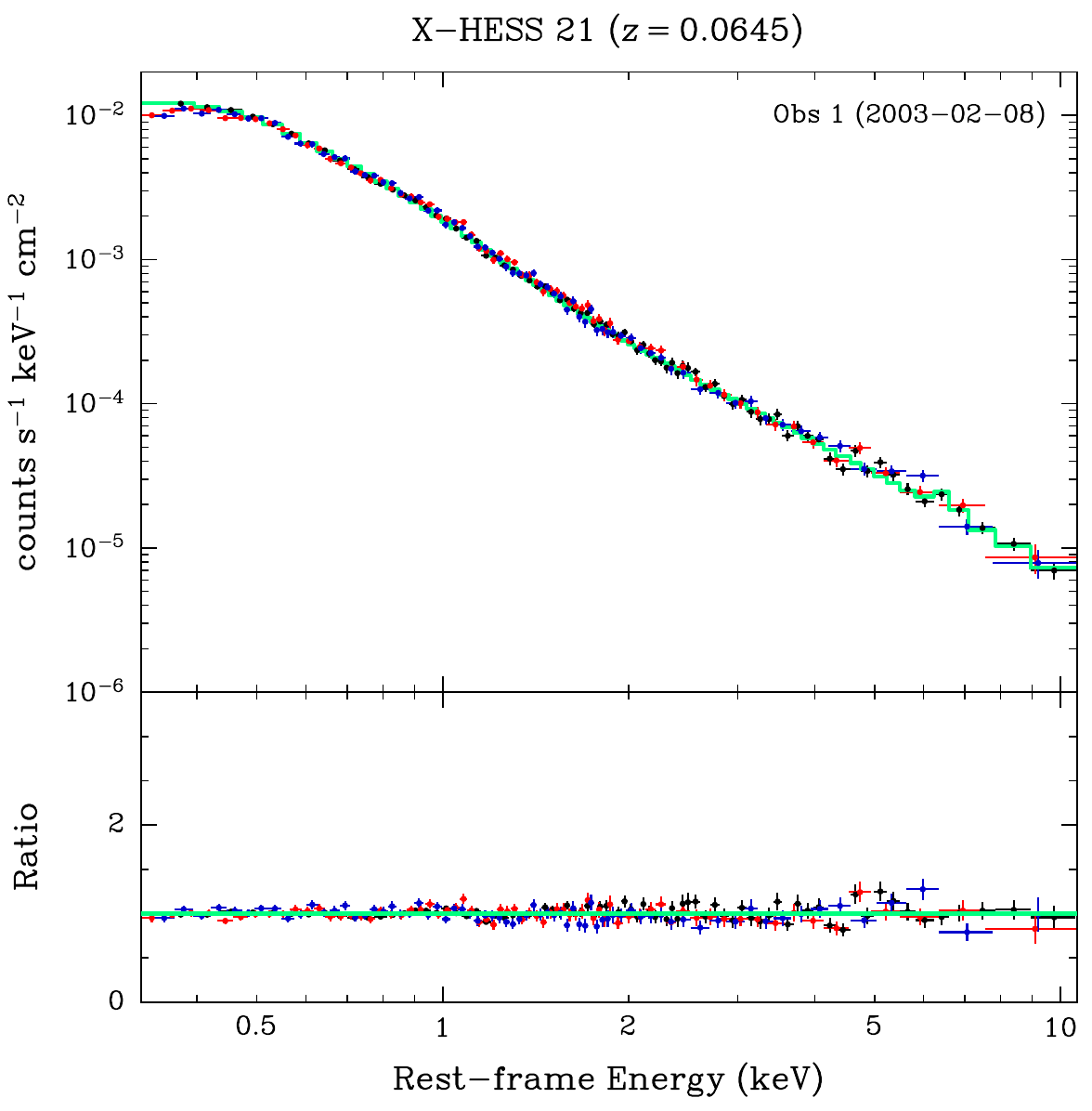}} &
     \subfloat{\includegraphics[width = 2.1in]{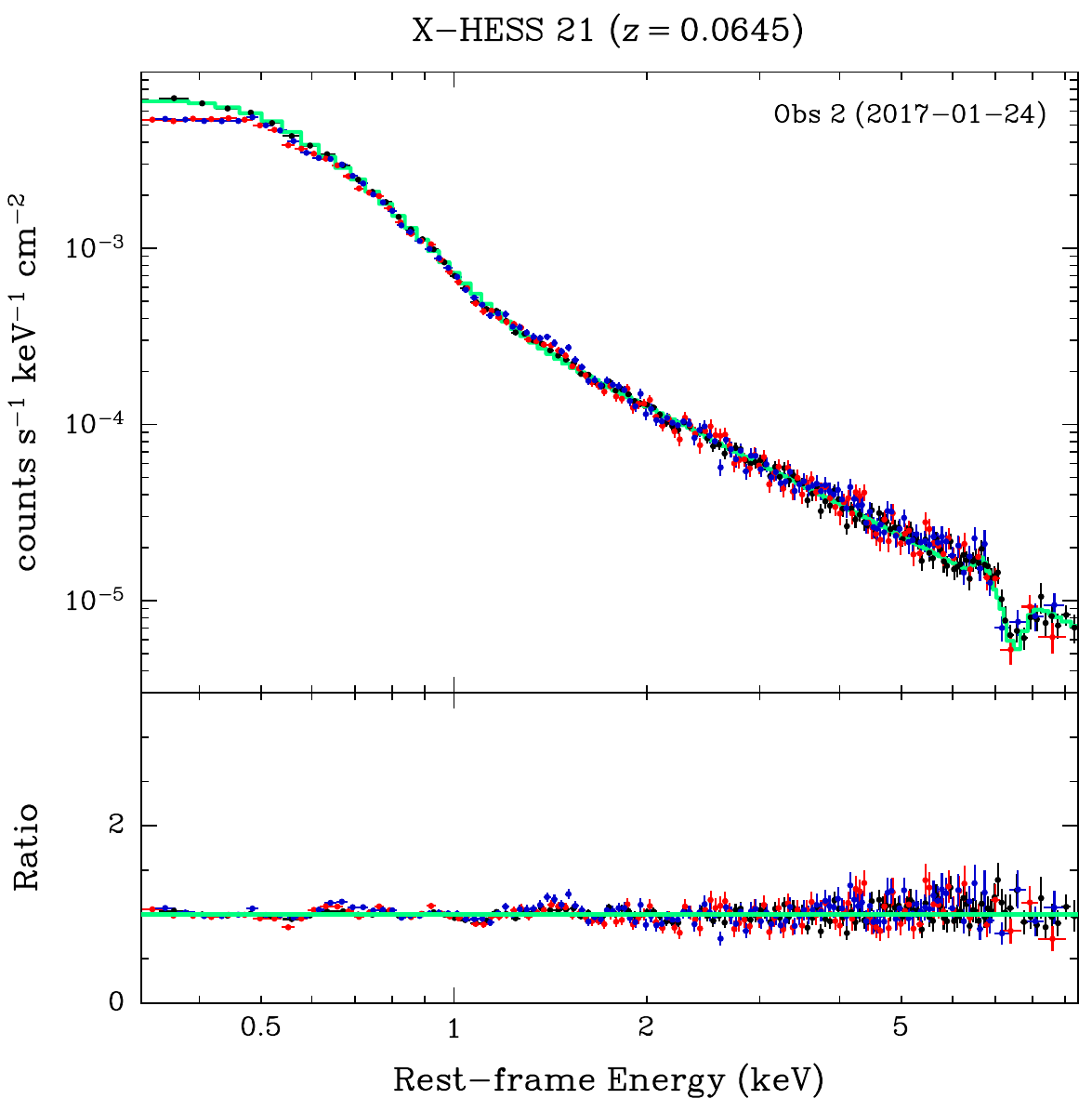}} \\
     \subfloat{\includegraphics[width = 2.1in]{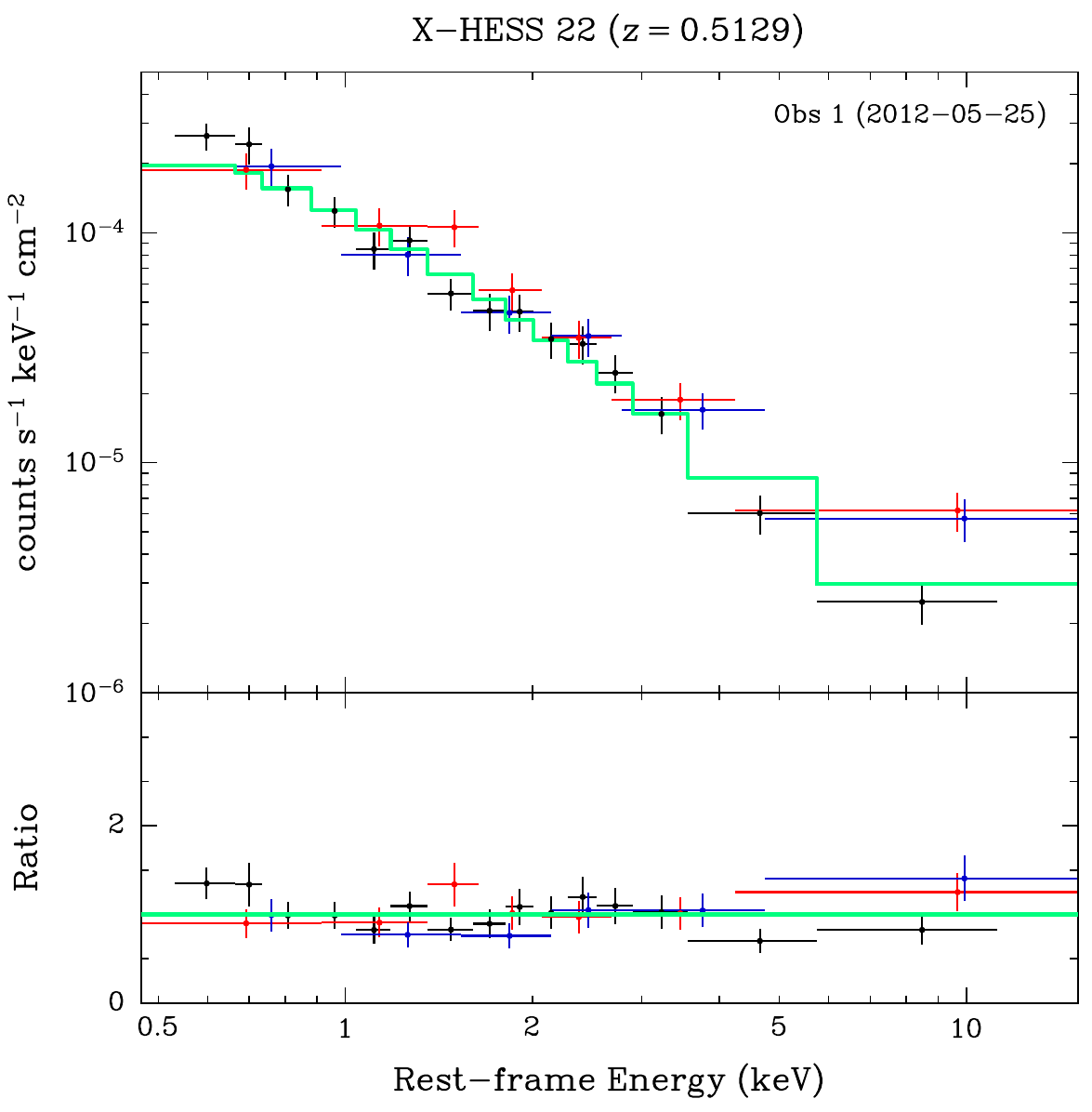}} &
     \subfloat{\includegraphics[width = 2.1in]{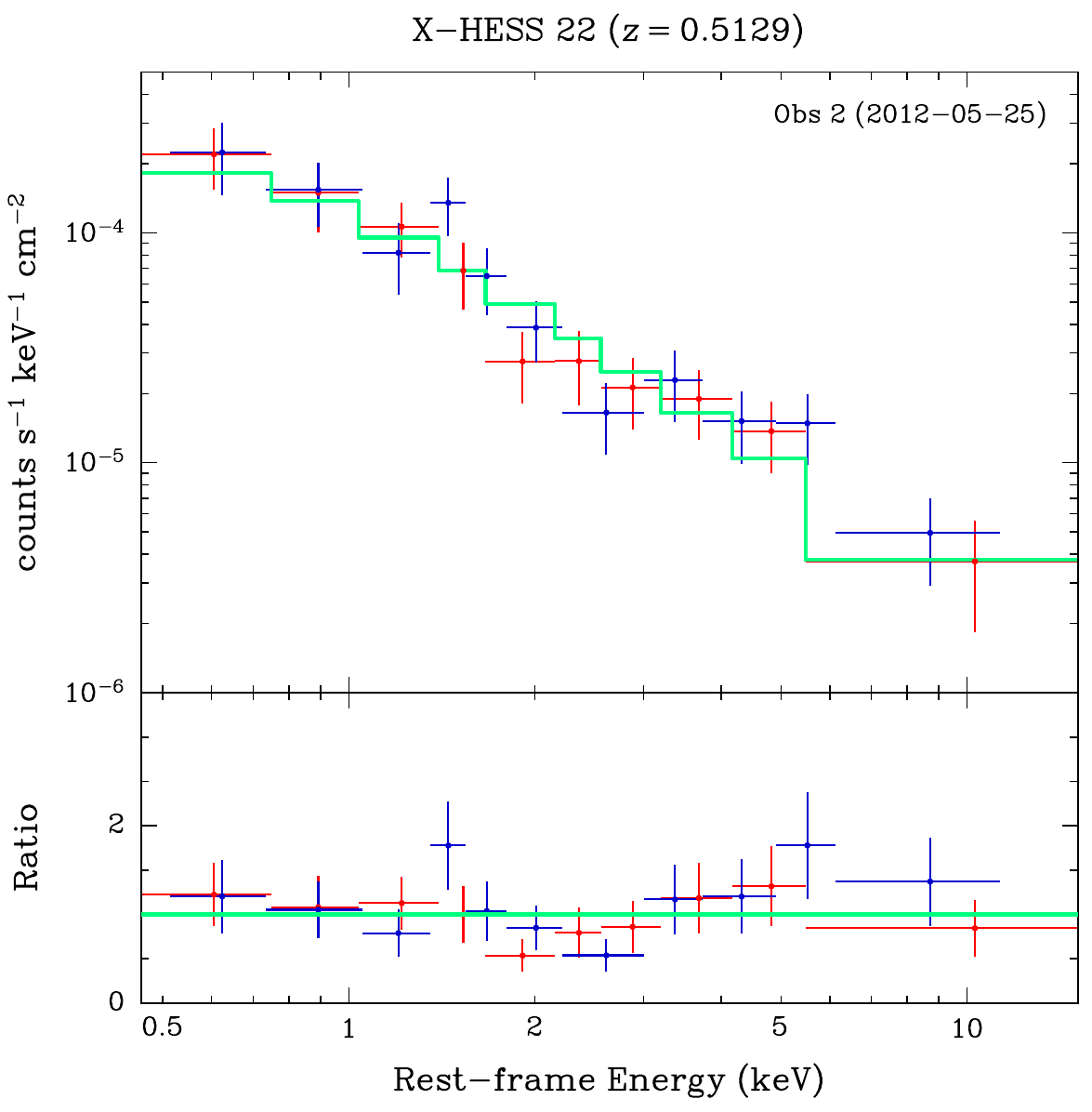}} &
     \subfloat{\includegraphics[width = 2.1in]{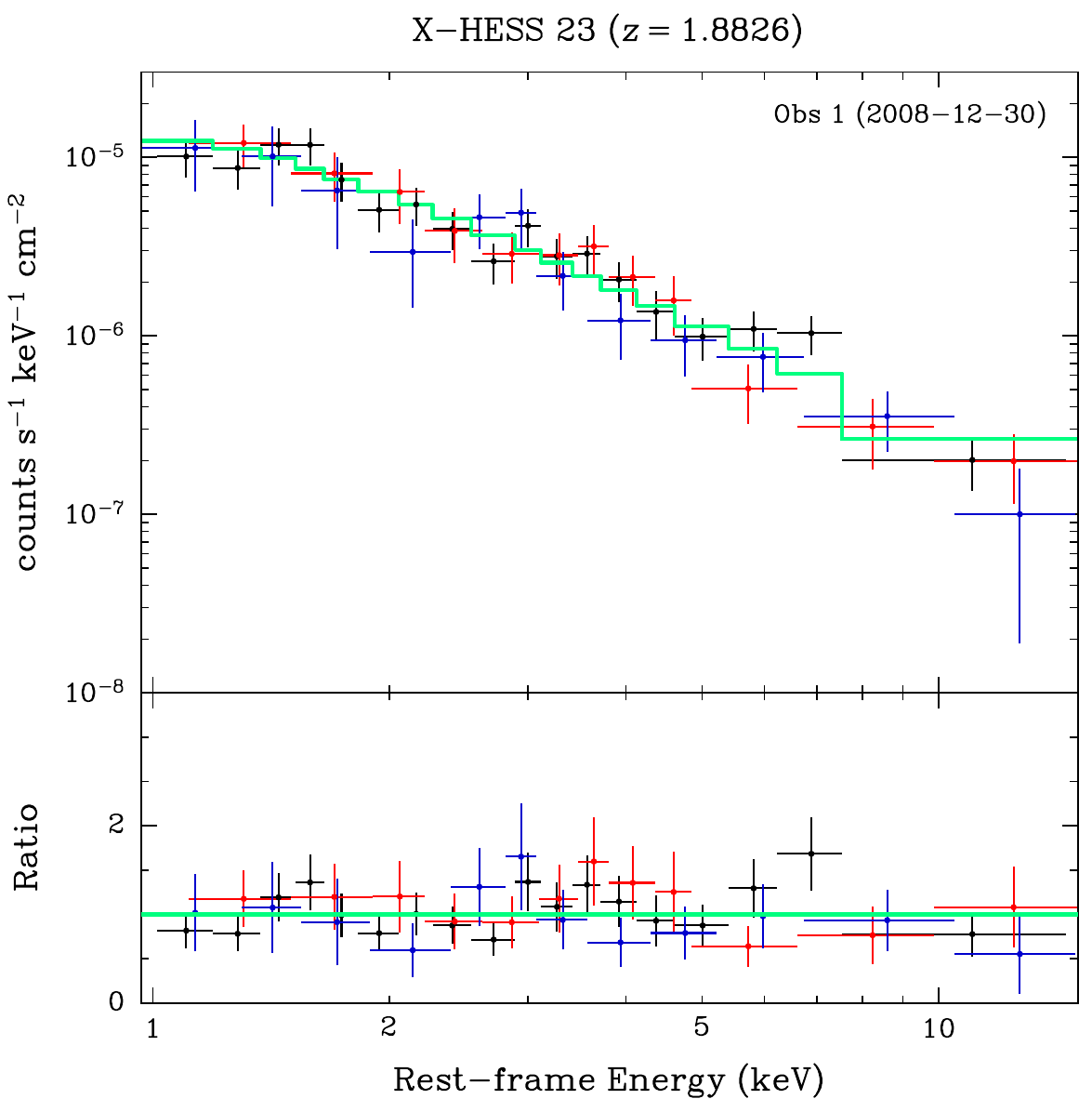}} \\
     \subfloat{\includegraphics[width = 2.1in]{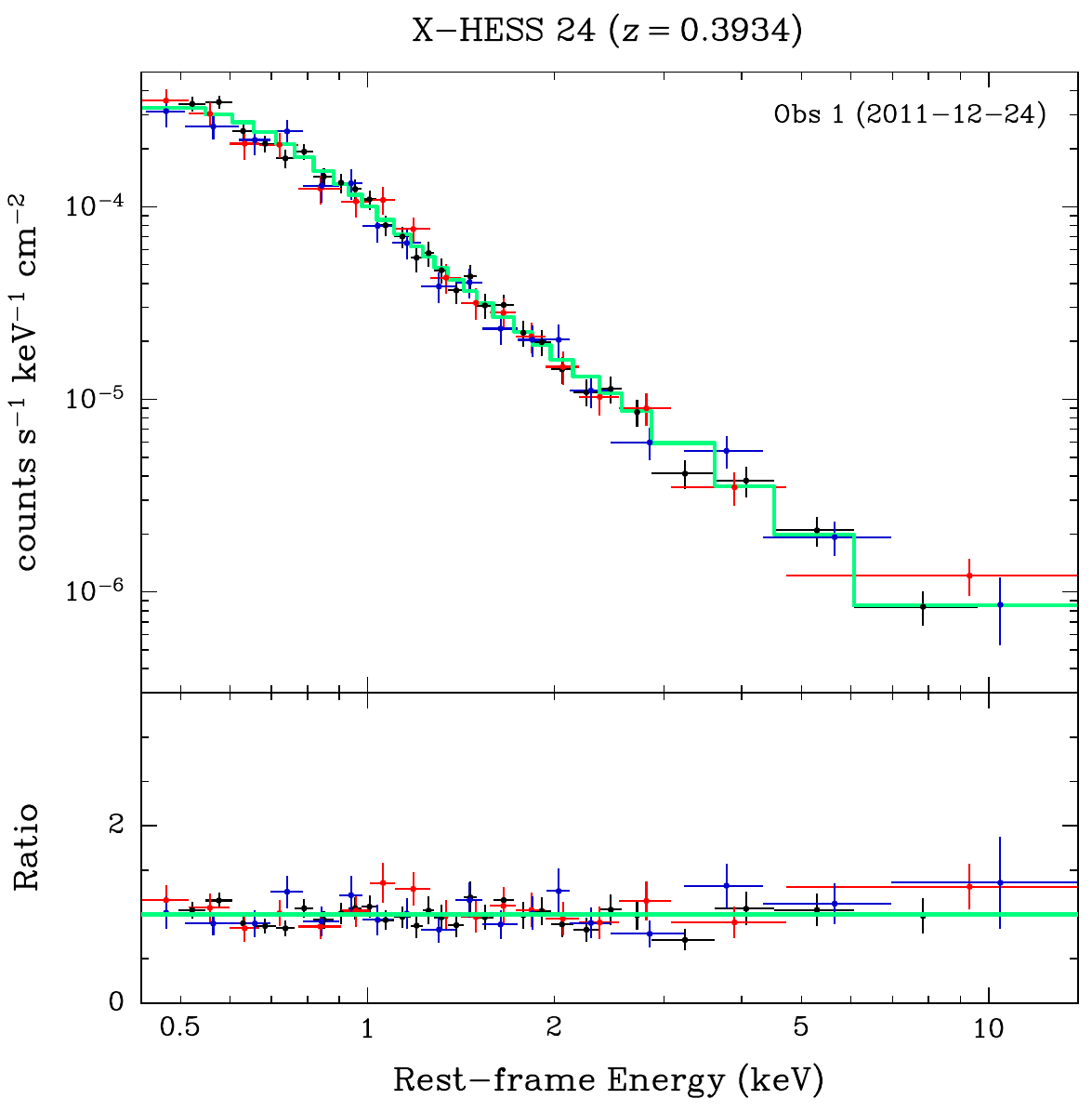}} &
     \subfloat{\includegraphics[width = 2.1in]{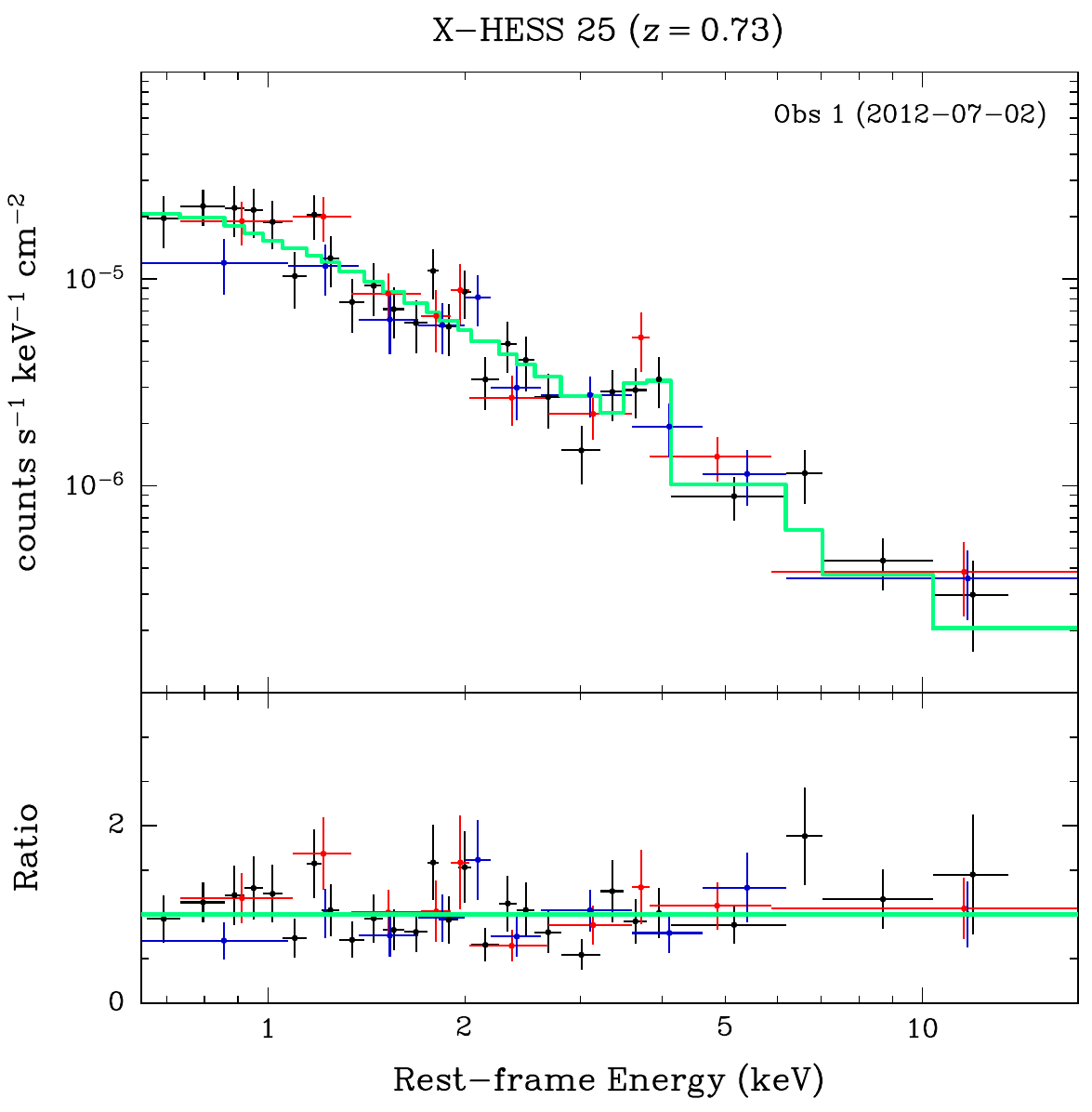}} &
     \subfloat{\includegraphics[width = 2.1in]{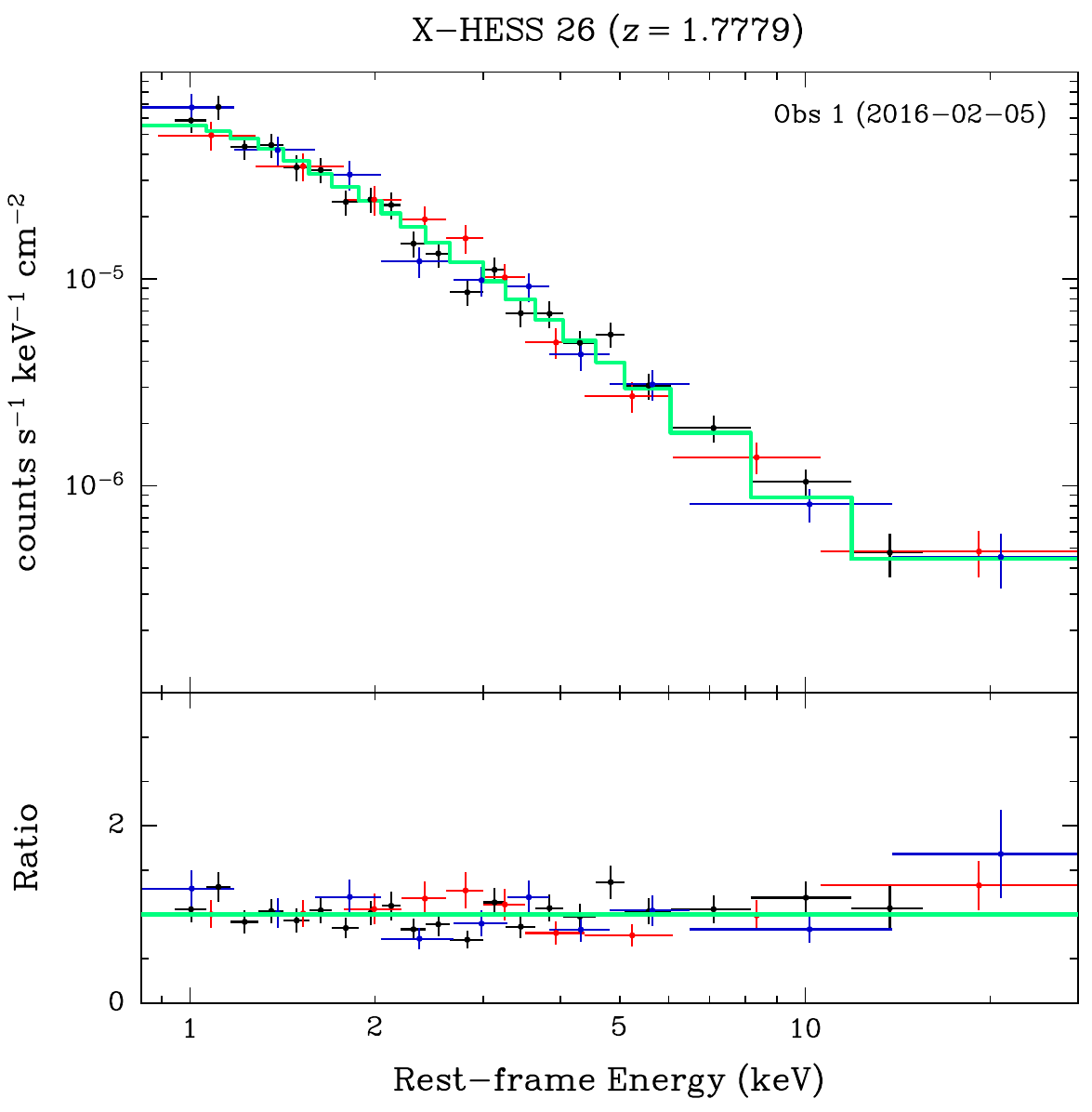}} \\
     \end{tabular}
\begin{minipage}{1.2\linewidth}
    \centering
    {Continuation of Fig. \ref{fig:xhess_spectra}.}
\end{minipage}
\end{figure}

\begin{figure}[h]
     \ContinuedFloat
     \centering
     \renewcommand{\arraystretch}{2}
     \begin{tabular}{ccc}
     \subfloat{\includegraphics[width = 2.1in]{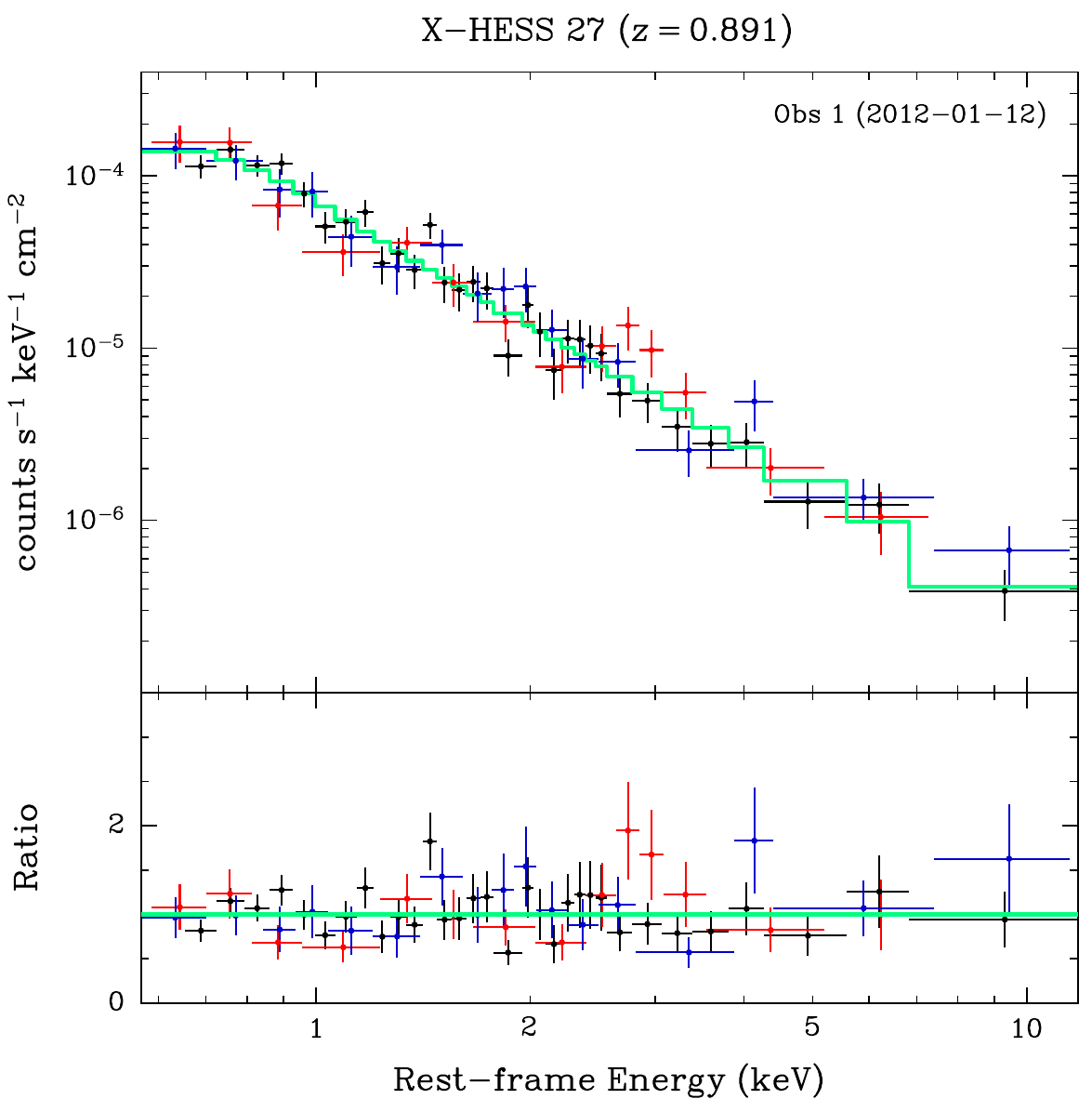}} &
     \subfloat{\includegraphics[width = 2.1in]{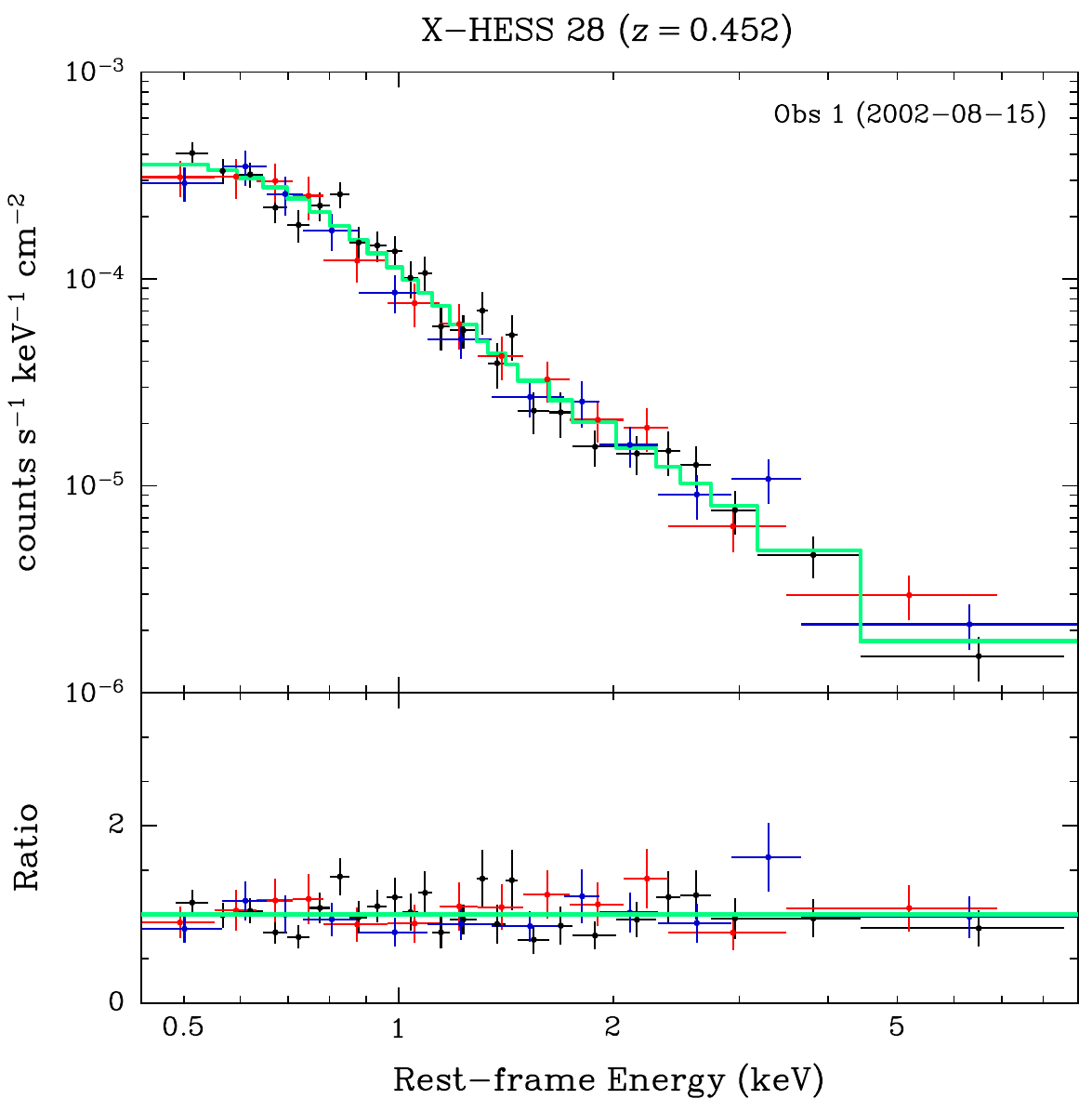}} &
     \subfloat{\includegraphics[width = 2.1in]{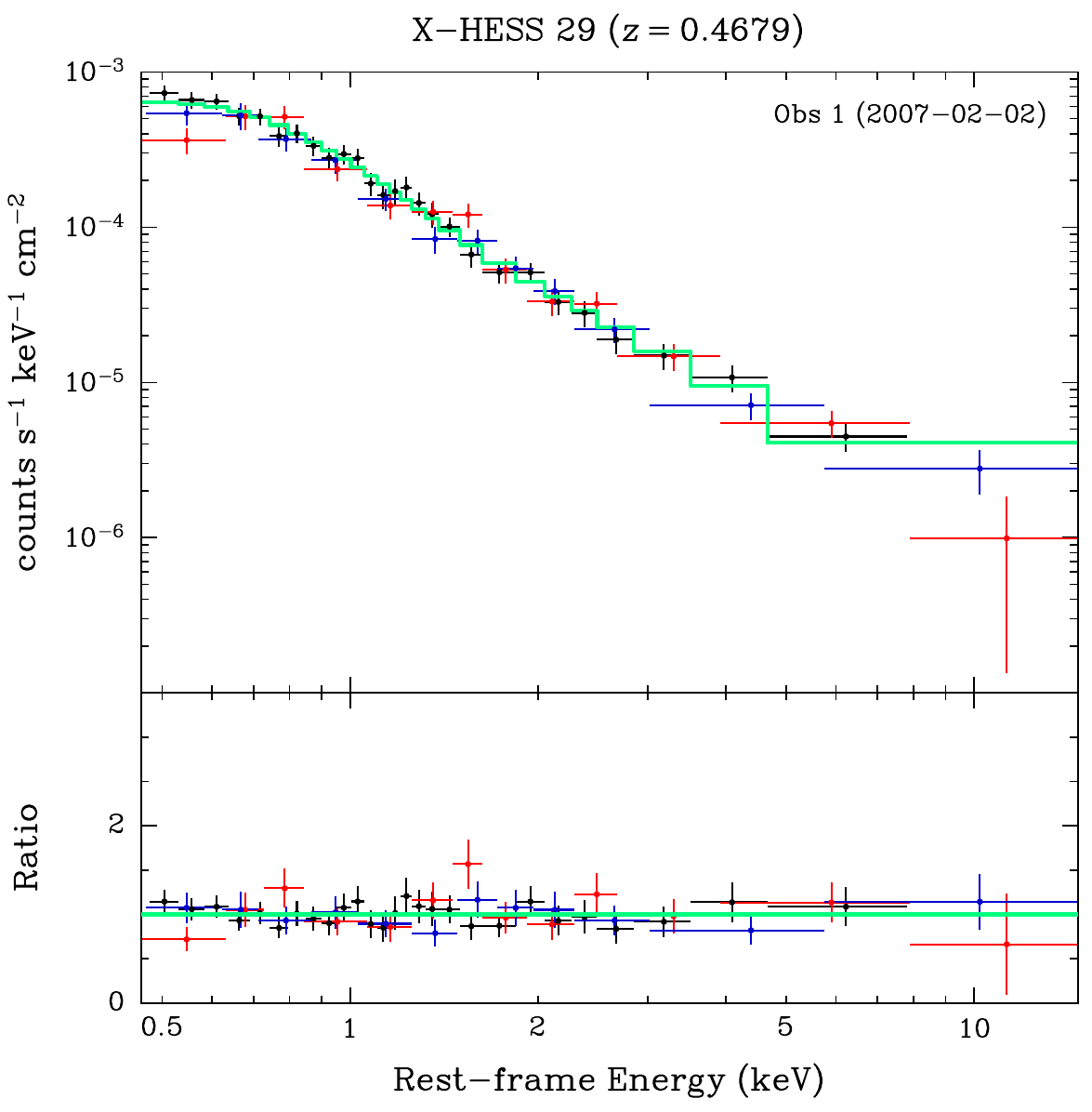}} \\
     \subfloat{\includegraphics[width = 2.1in]{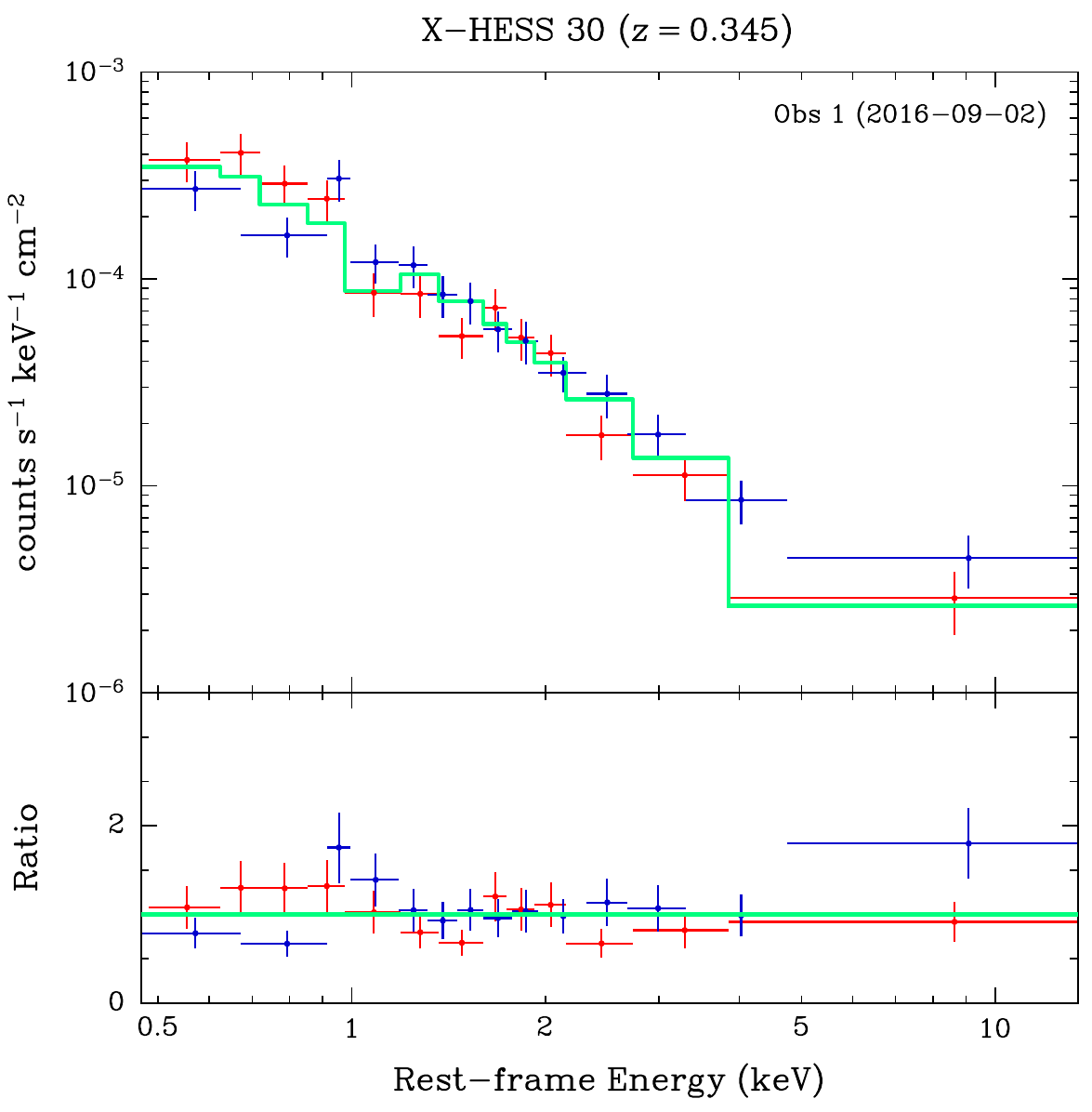}} &
     \subfloat{\includegraphics[width = 2.1in]{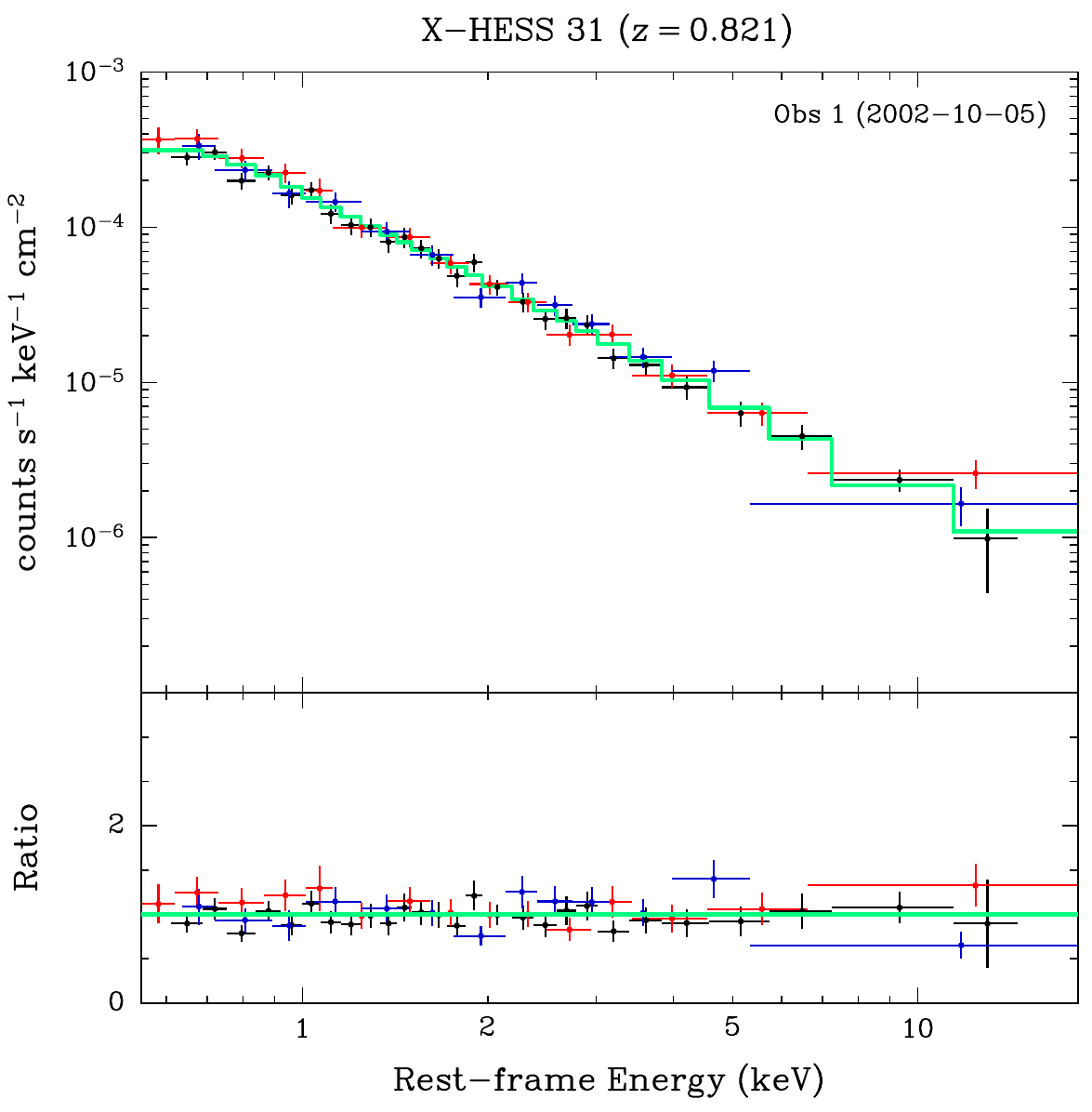}} &
     \subfloat{\includegraphics[width = 2.1in]{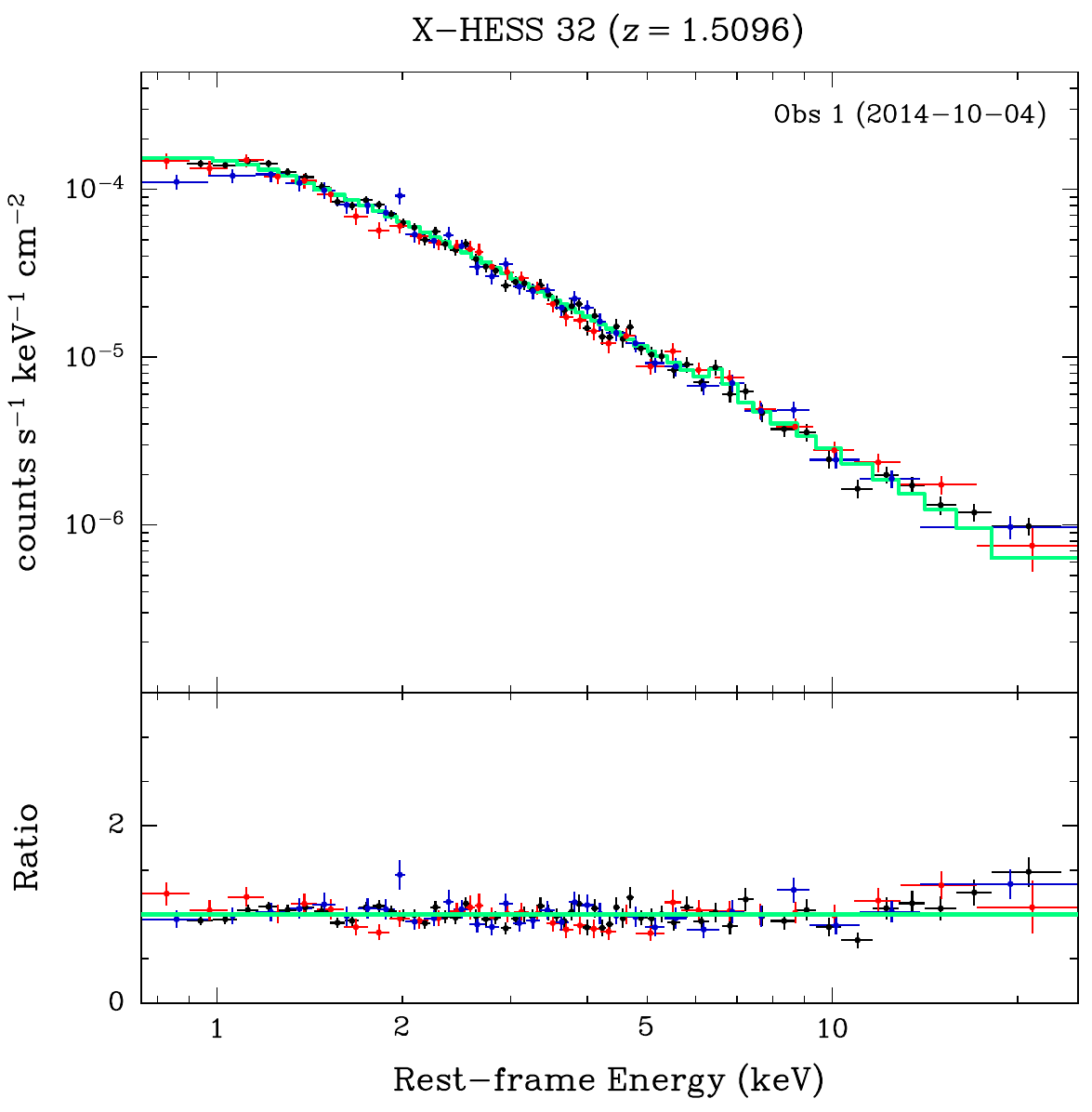}} \\
     \subfloat{\includegraphics[width = 2.1in]{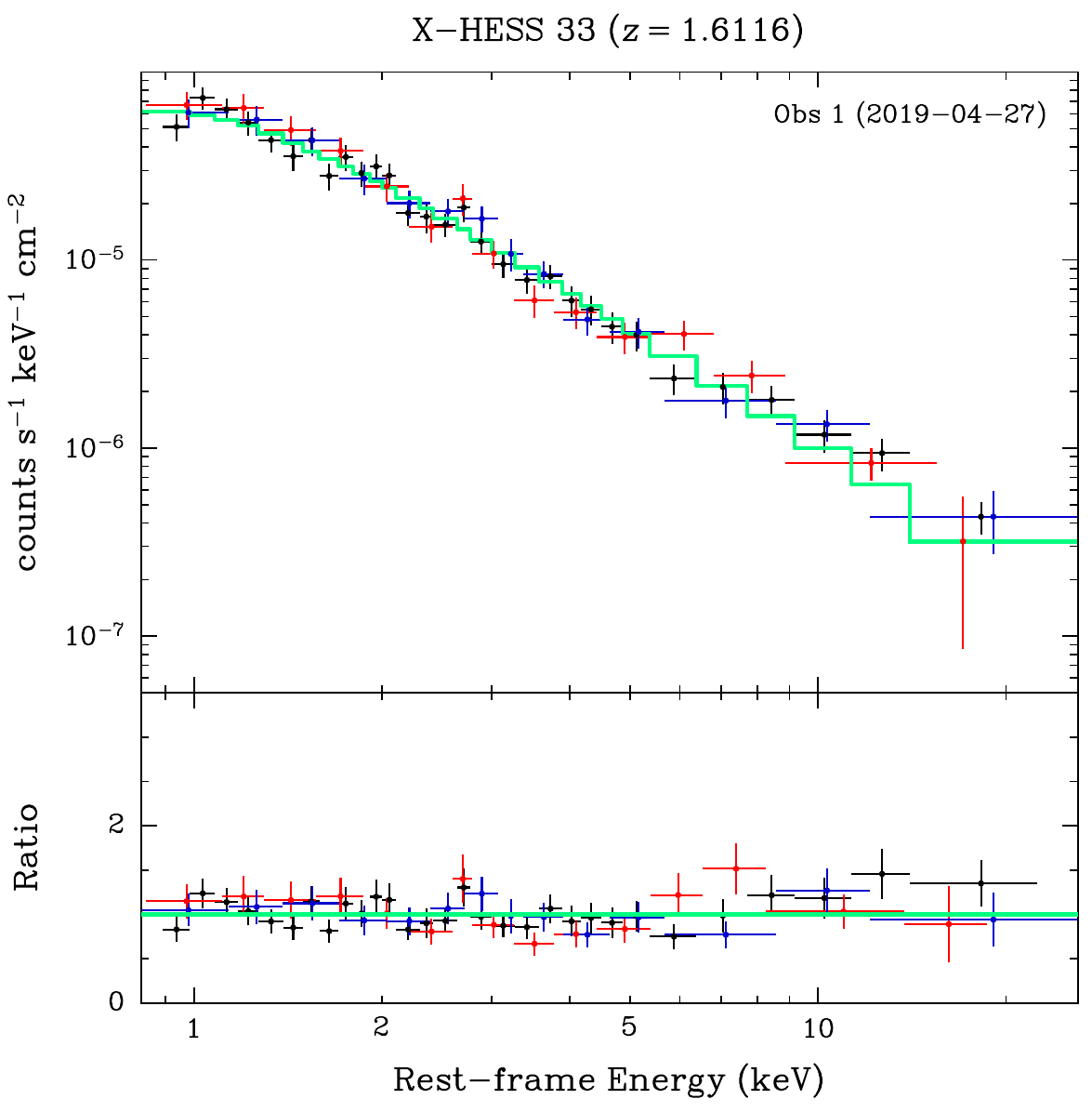}} &
     \subfloat{\includegraphics[width = 2.1in]{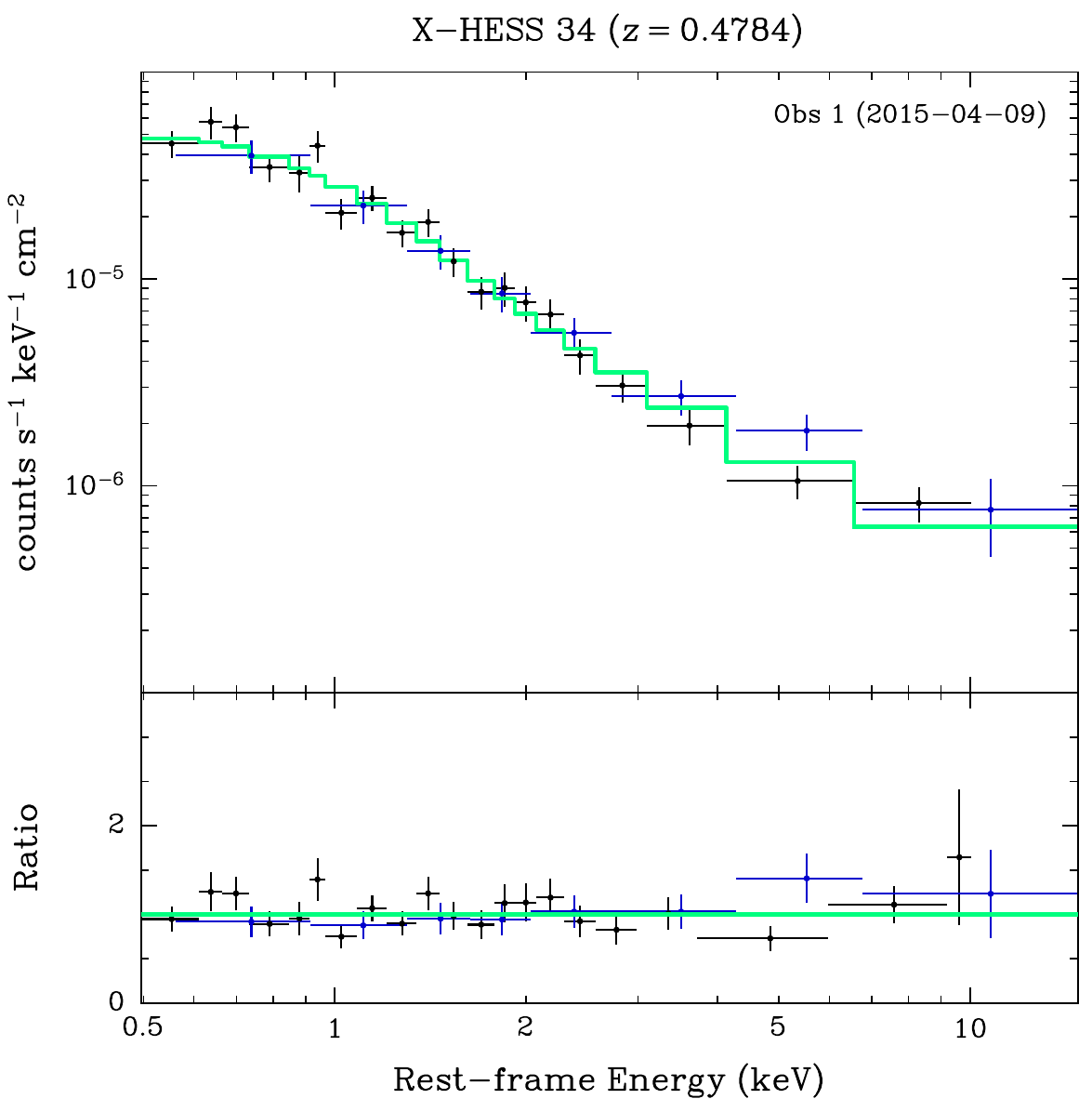}} &
     \subfloat{\includegraphics[width = 2.1in]{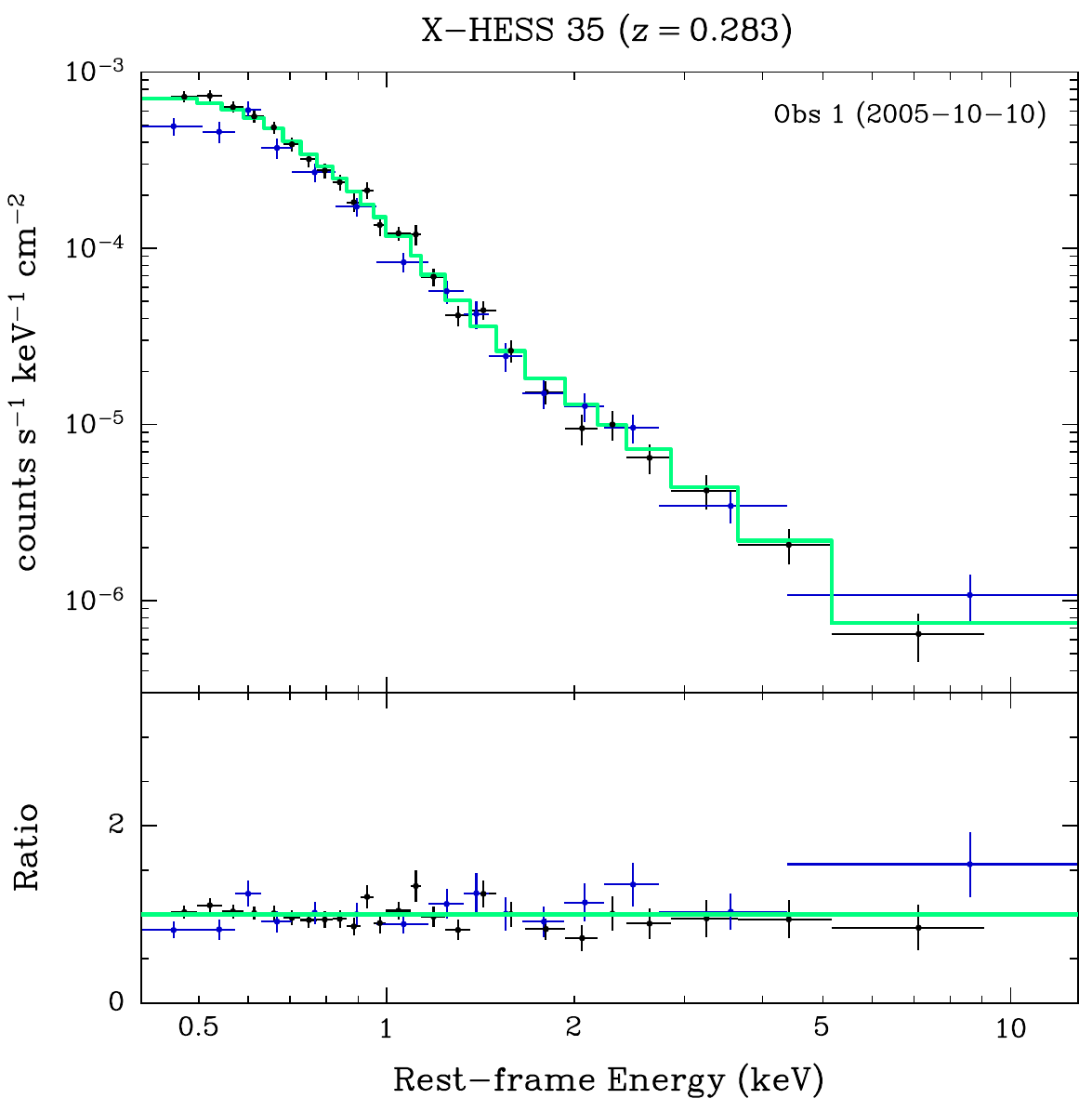}} \\
     \subfloat{\includegraphics[width = 2.1in]{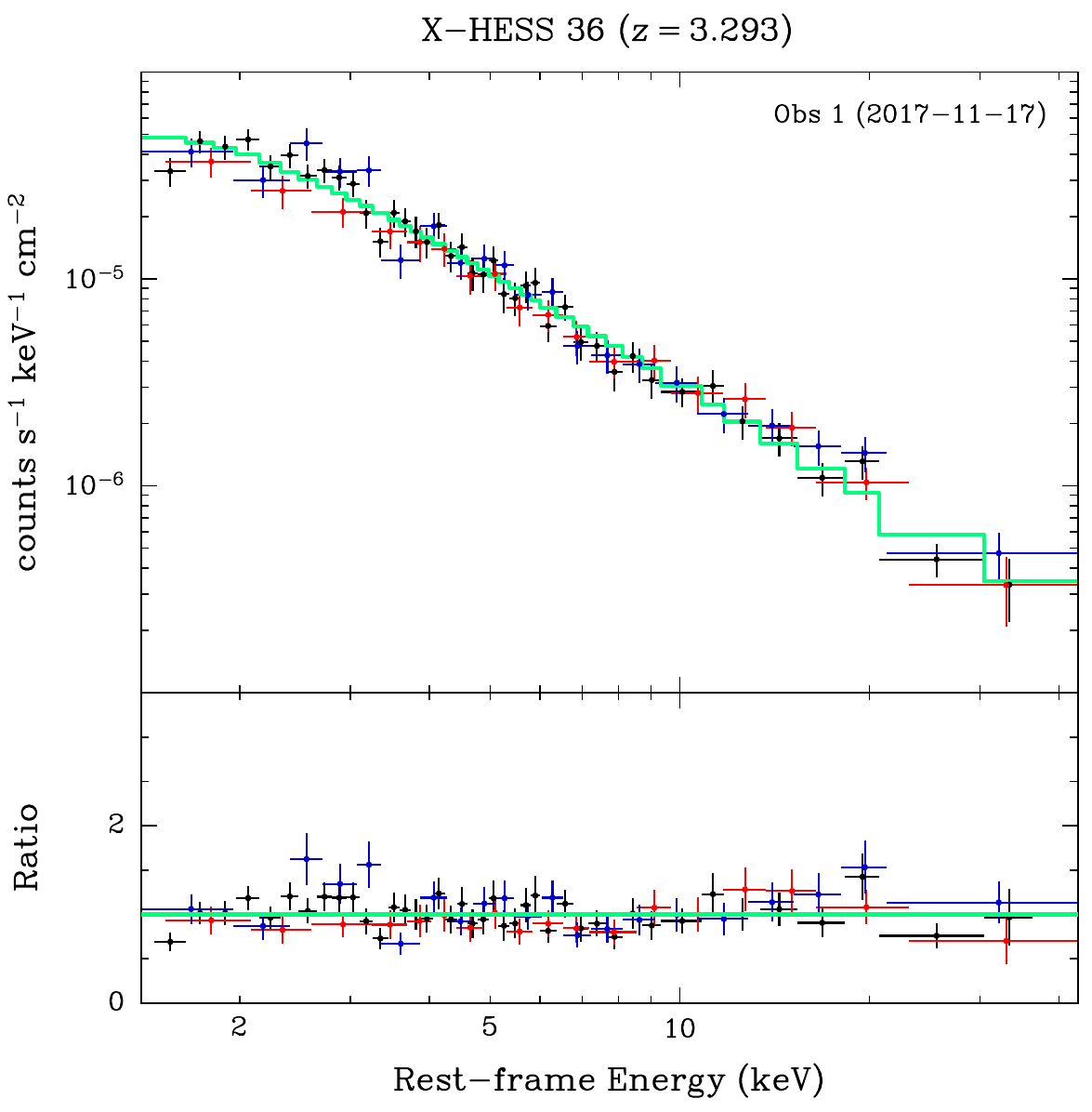}} &
     \subfloat{\includegraphics[width = 2.1in]{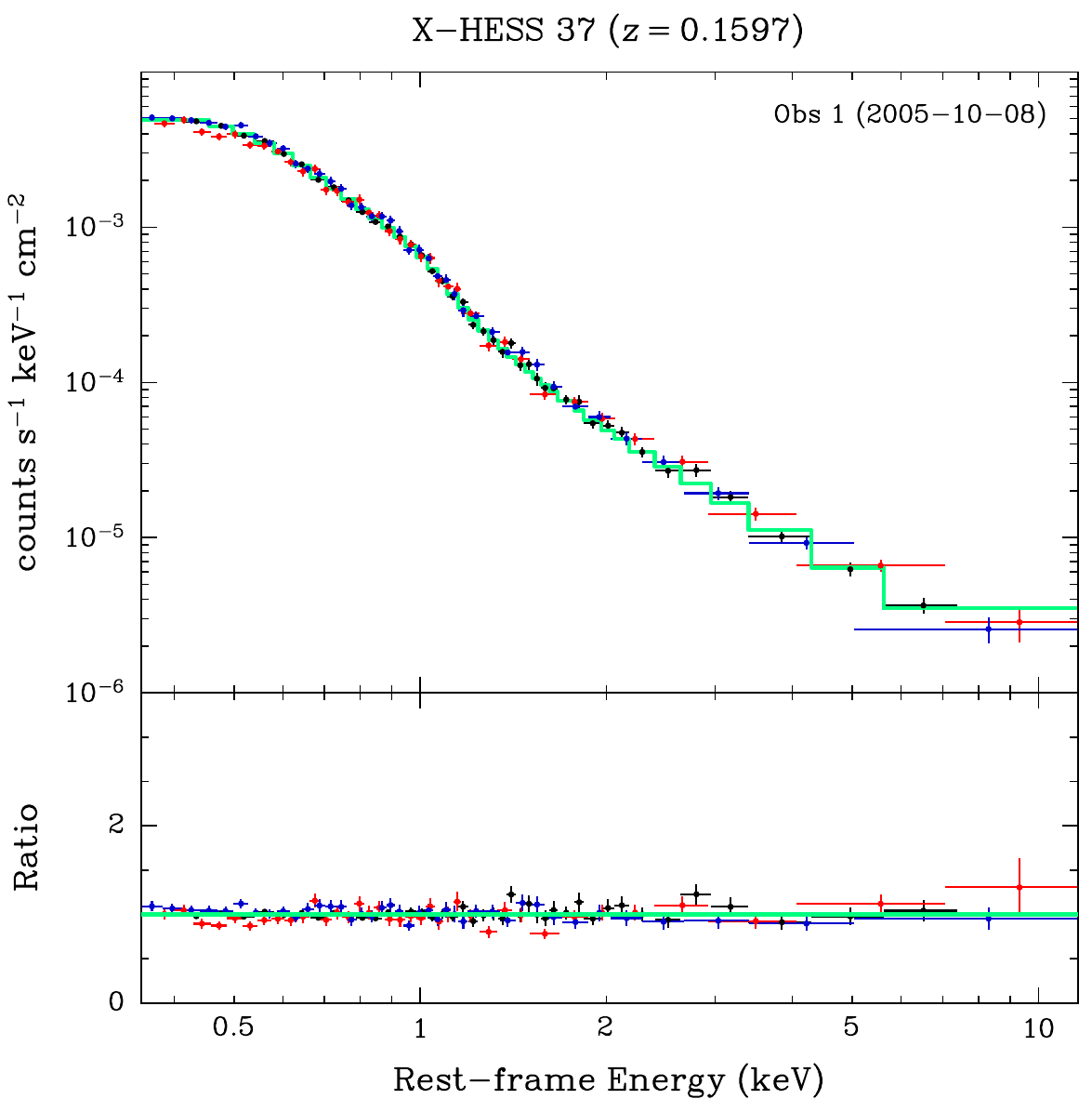}} &
     \subfloat{\includegraphics[width = 2.1in]{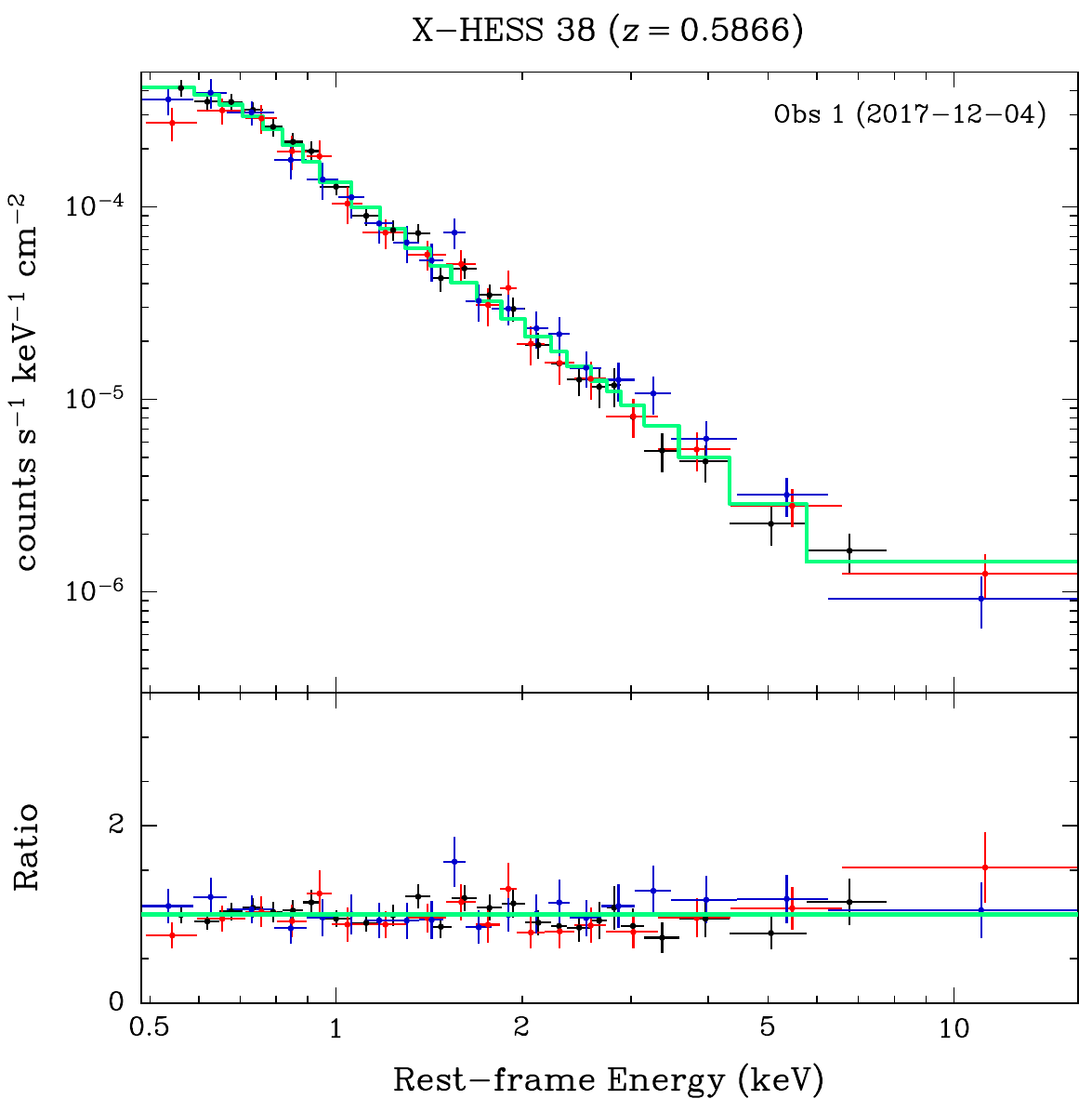}} \\
     \end{tabular}
\begin{minipage}{1.2\linewidth}
    \centering 
    {Continuation of Fig. \ref{fig:xhess_spectra}.}
\end{minipage}
\end{figure}

\begin{figure}[h]
     \ContinuedFloat
     \centering
     \renewcommand{\arraystretch}{2}
     \begin{tabular}{ccc}
     \subfloat{\includegraphics[width = 2.1in]{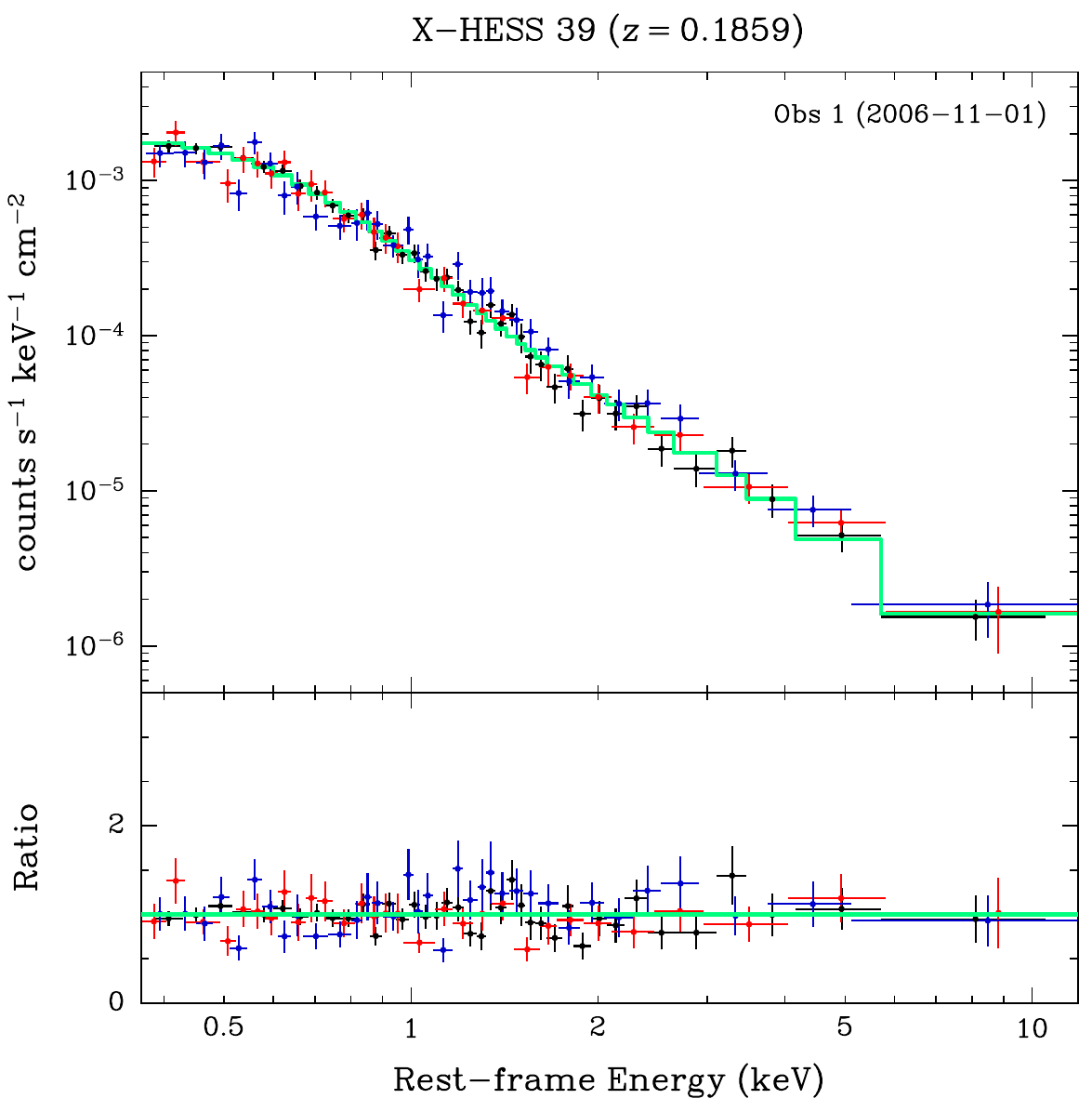}} &
     \subfloat{\includegraphics[width = 2.1in]{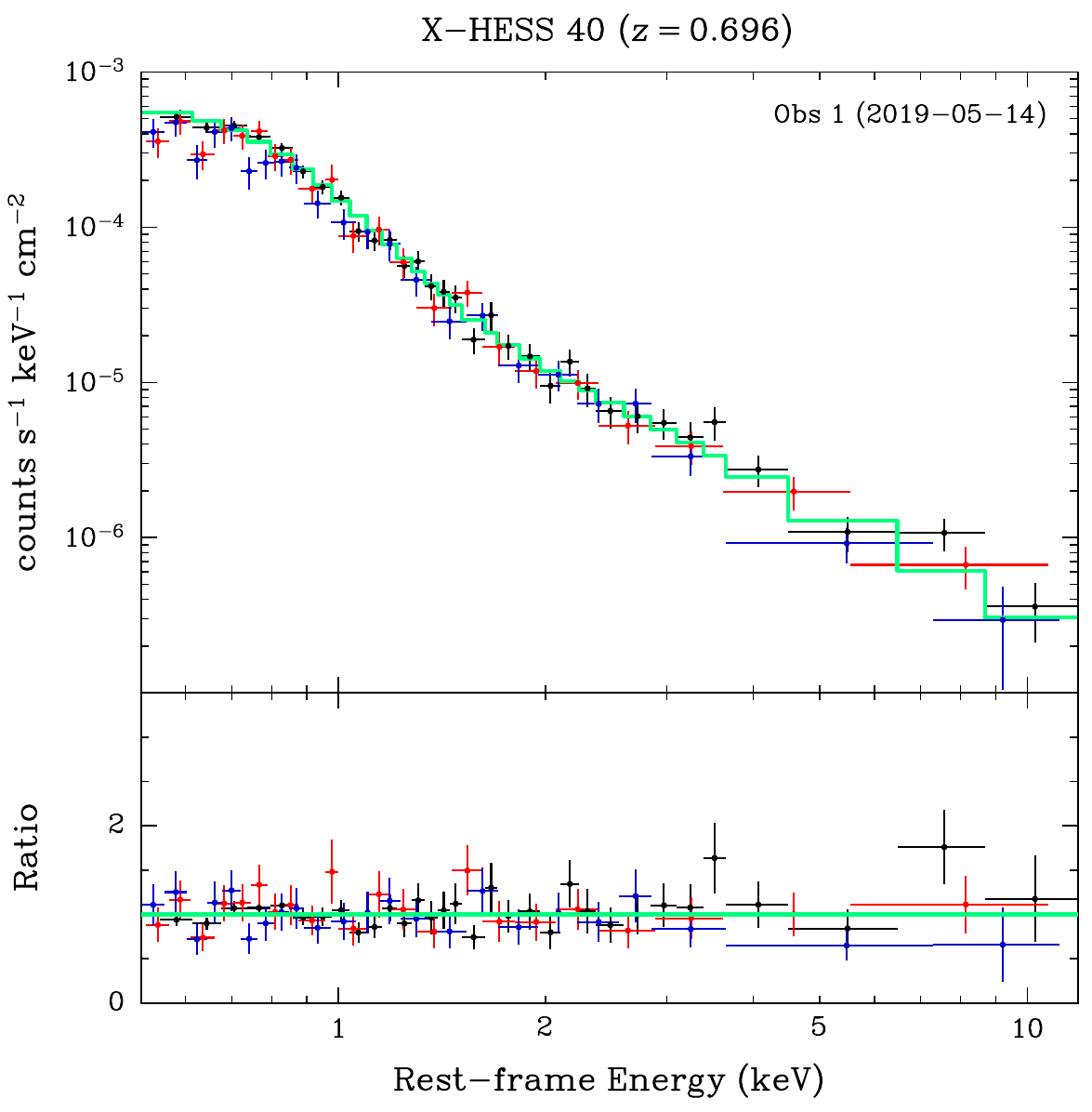}} &
     \subfloat{\includegraphics[width = 2.1in]{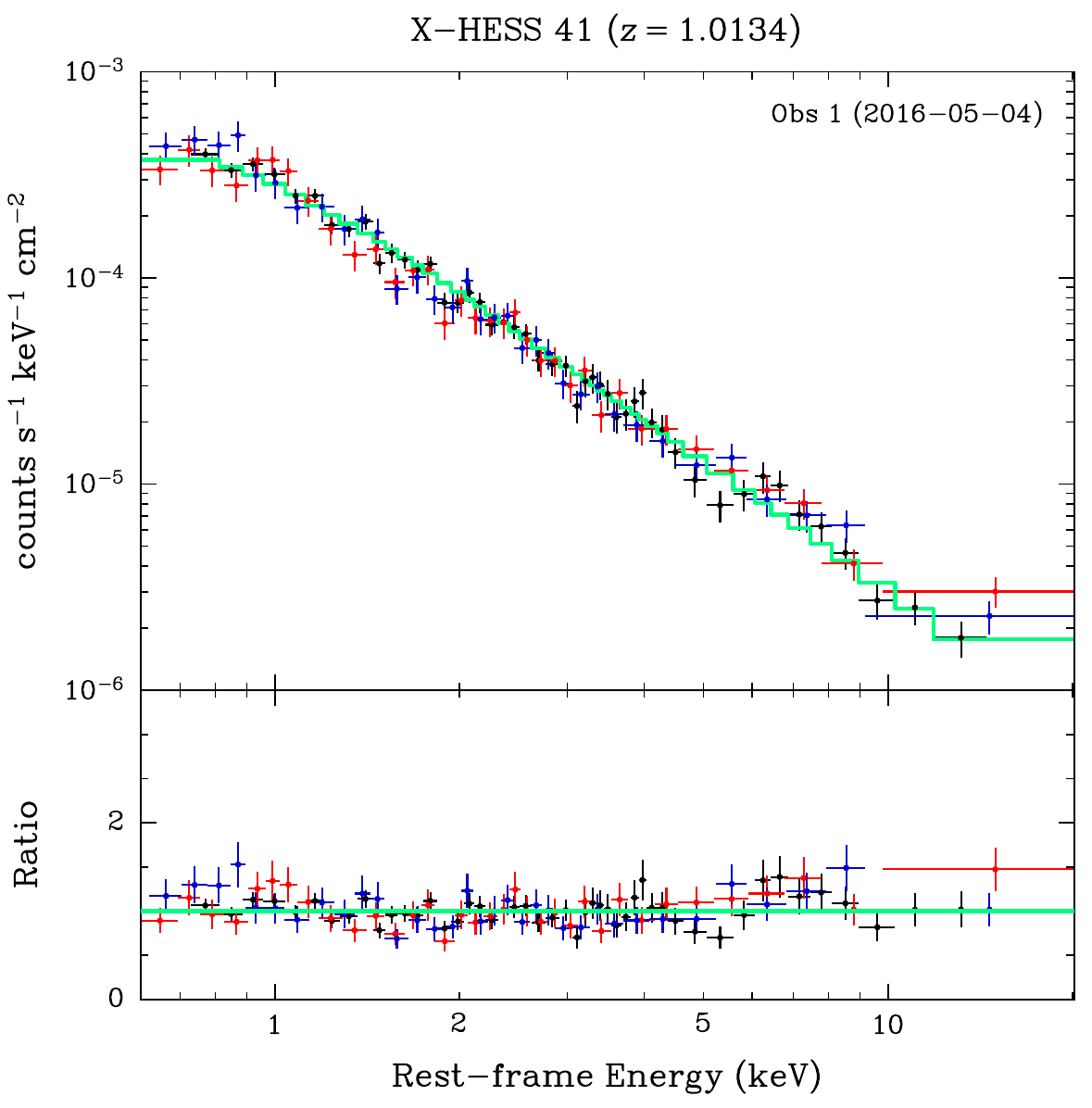}} \\
     \subfloat{\includegraphics[width = 2.1in]{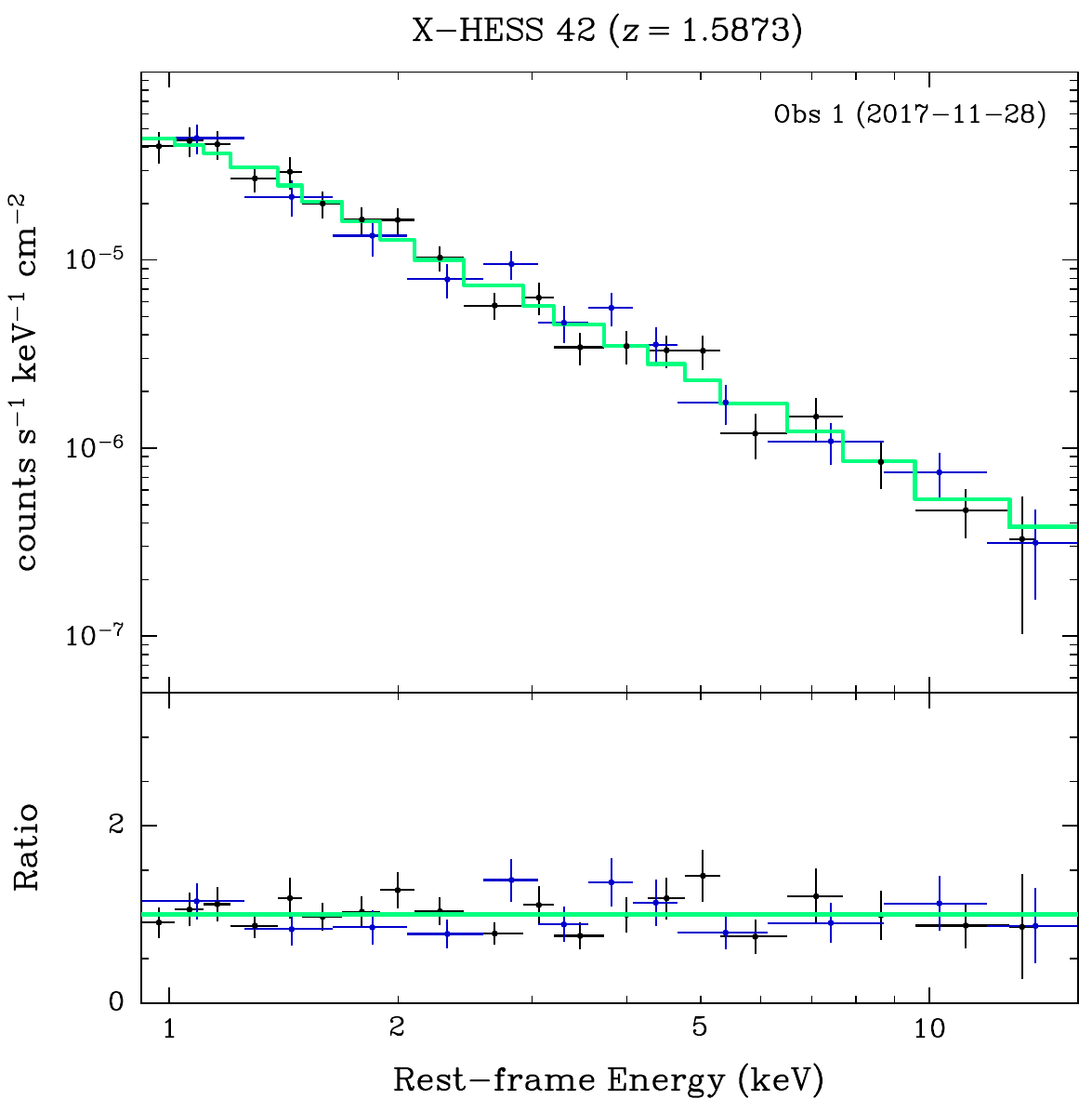}} &
     \subfloat{\includegraphics[width = 2.1in]{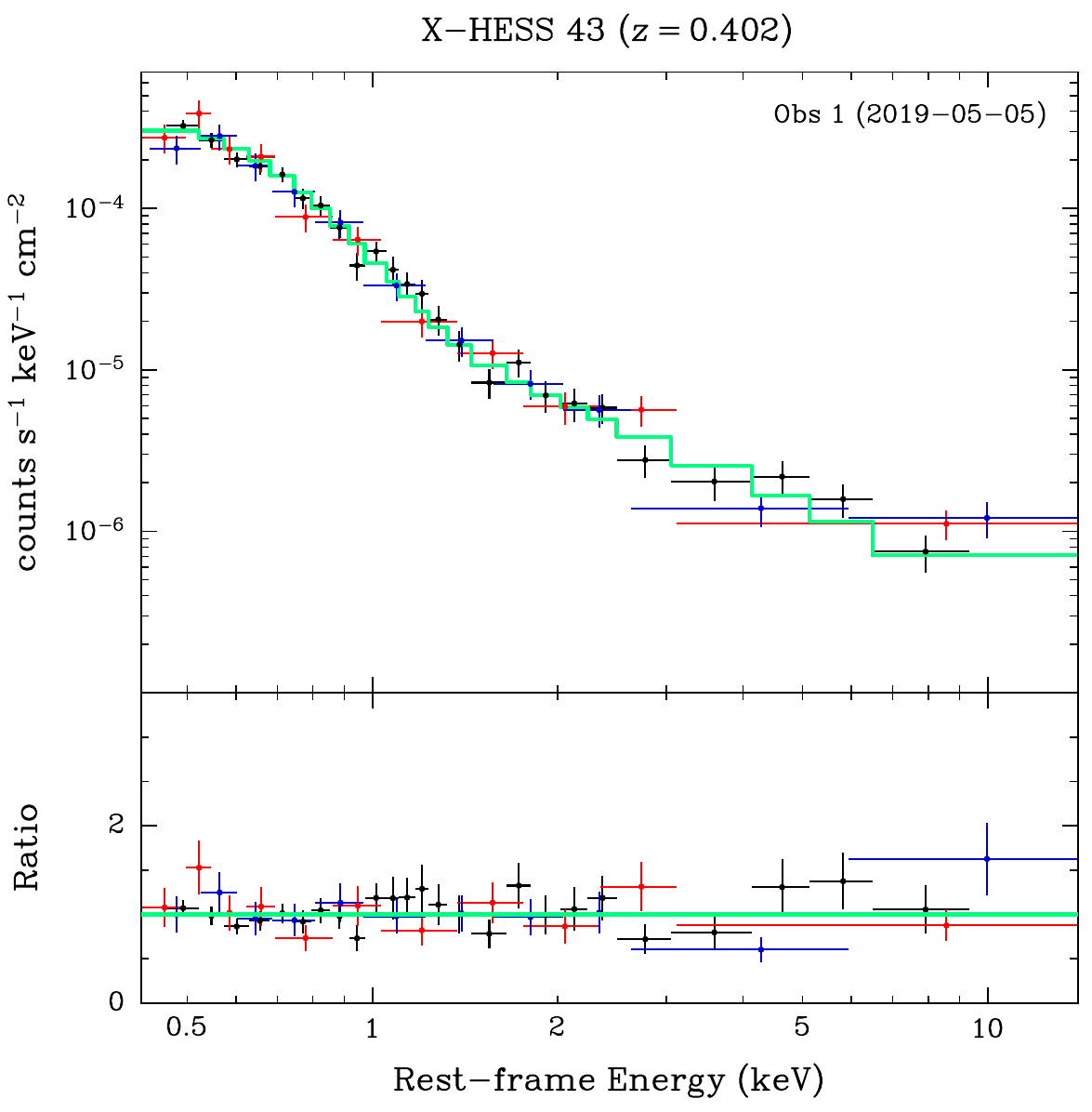}} &
     \subfloat{\includegraphics[width = 2.1in]{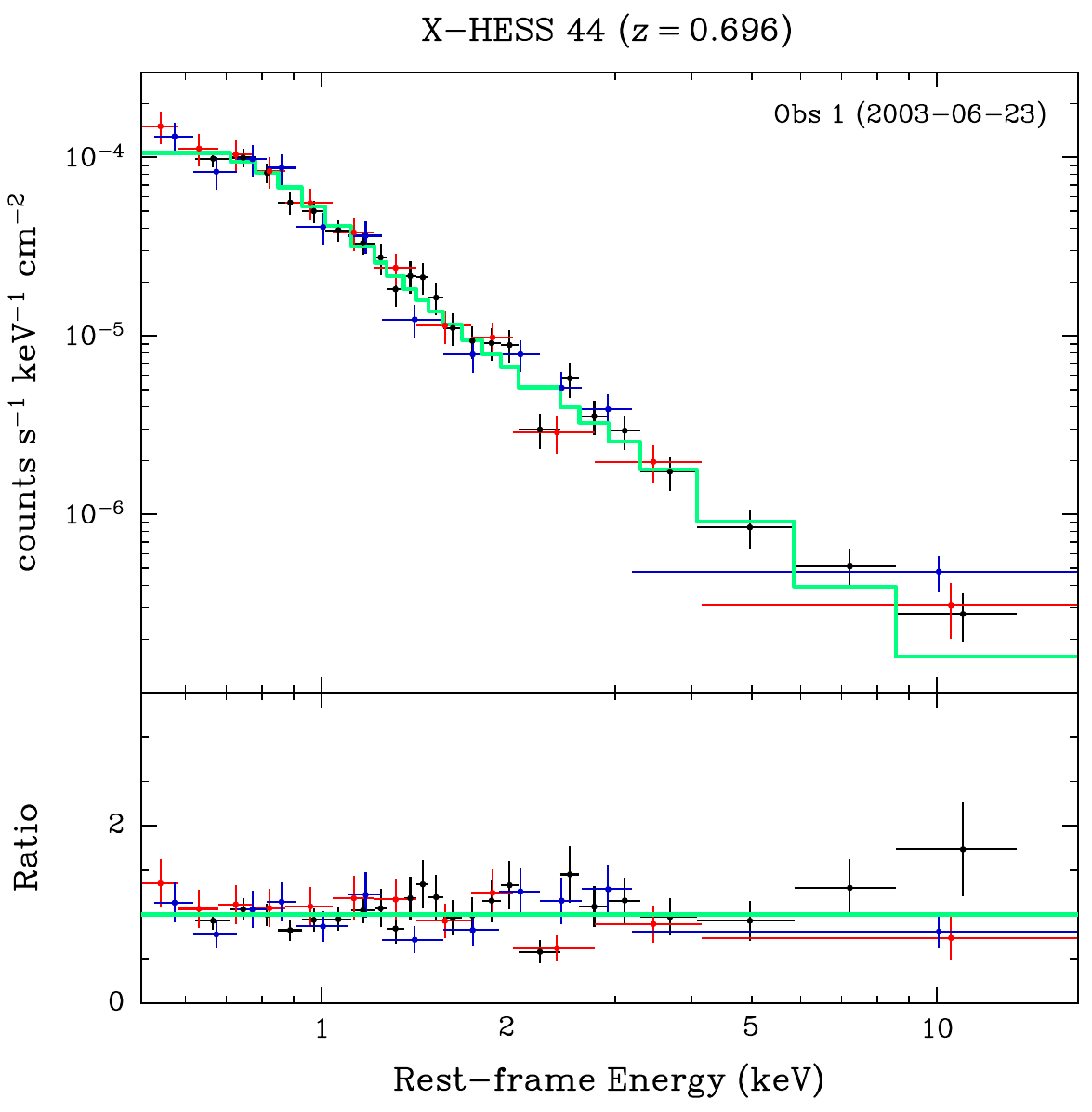}} \\
     \subfloat{\includegraphics[width = 2.1in]{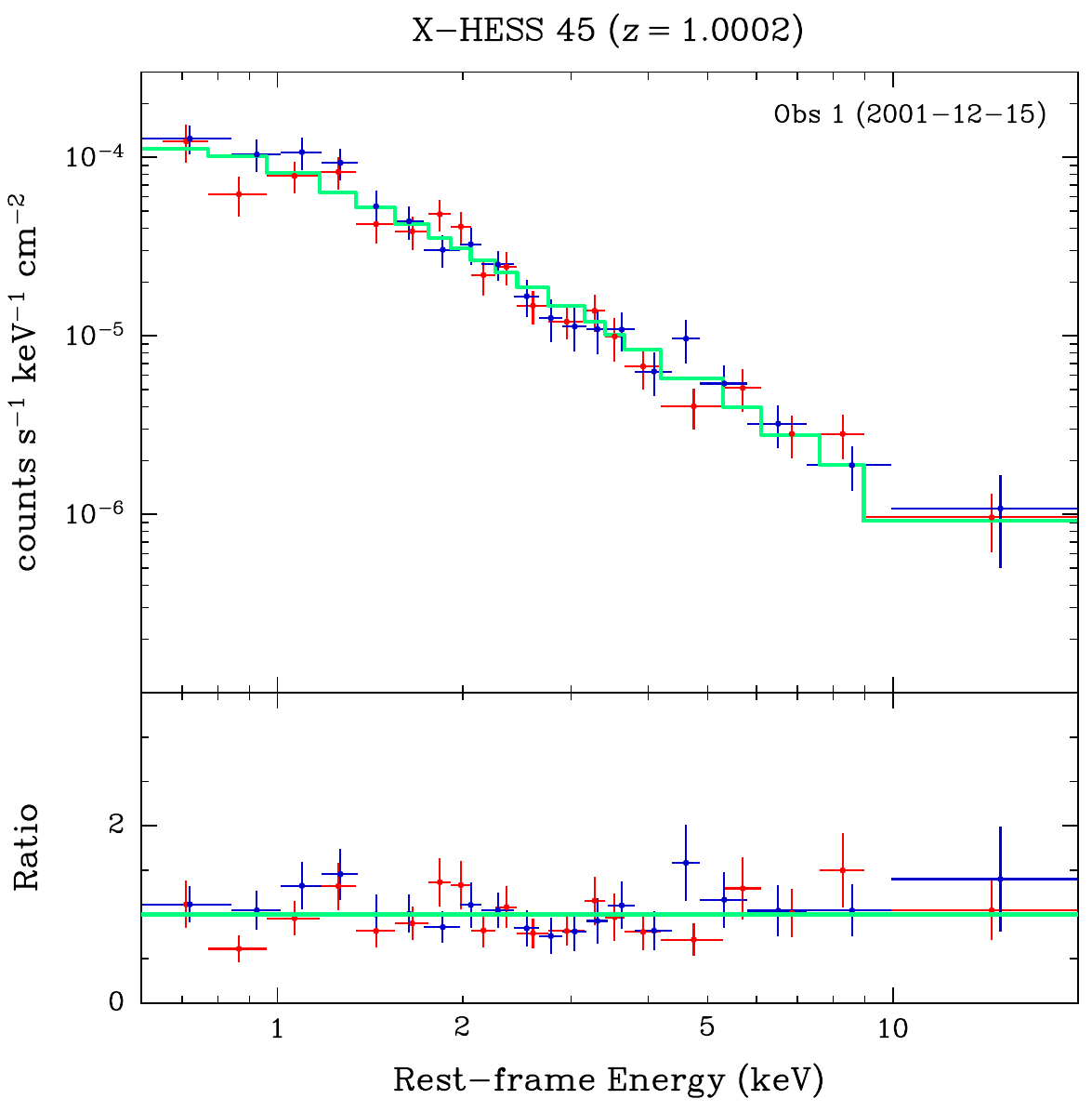}} &
     \subfloat{\includegraphics[width = 2.1in]{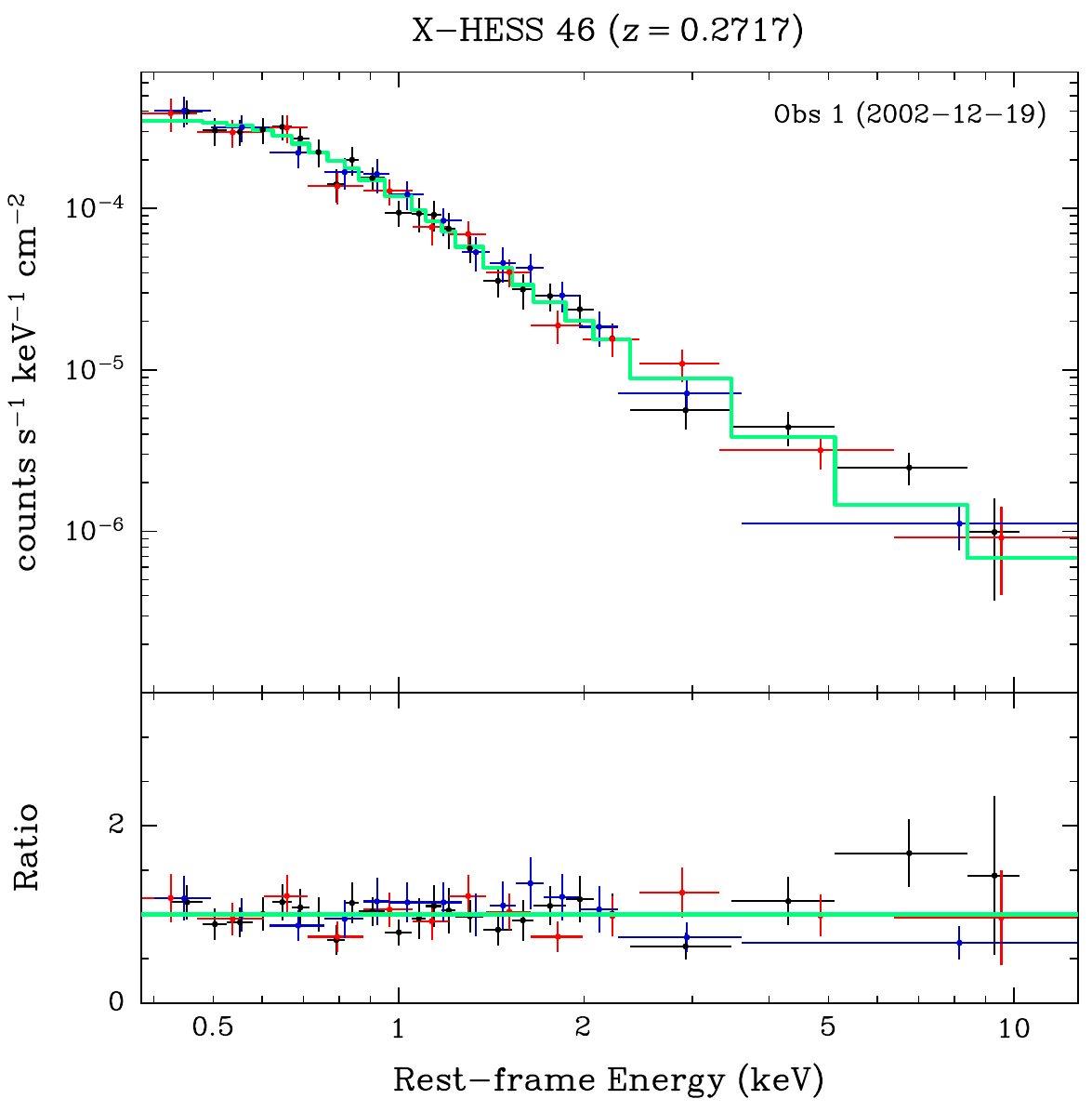}} &
     \subfloat{\includegraphics[width = 2.1in]{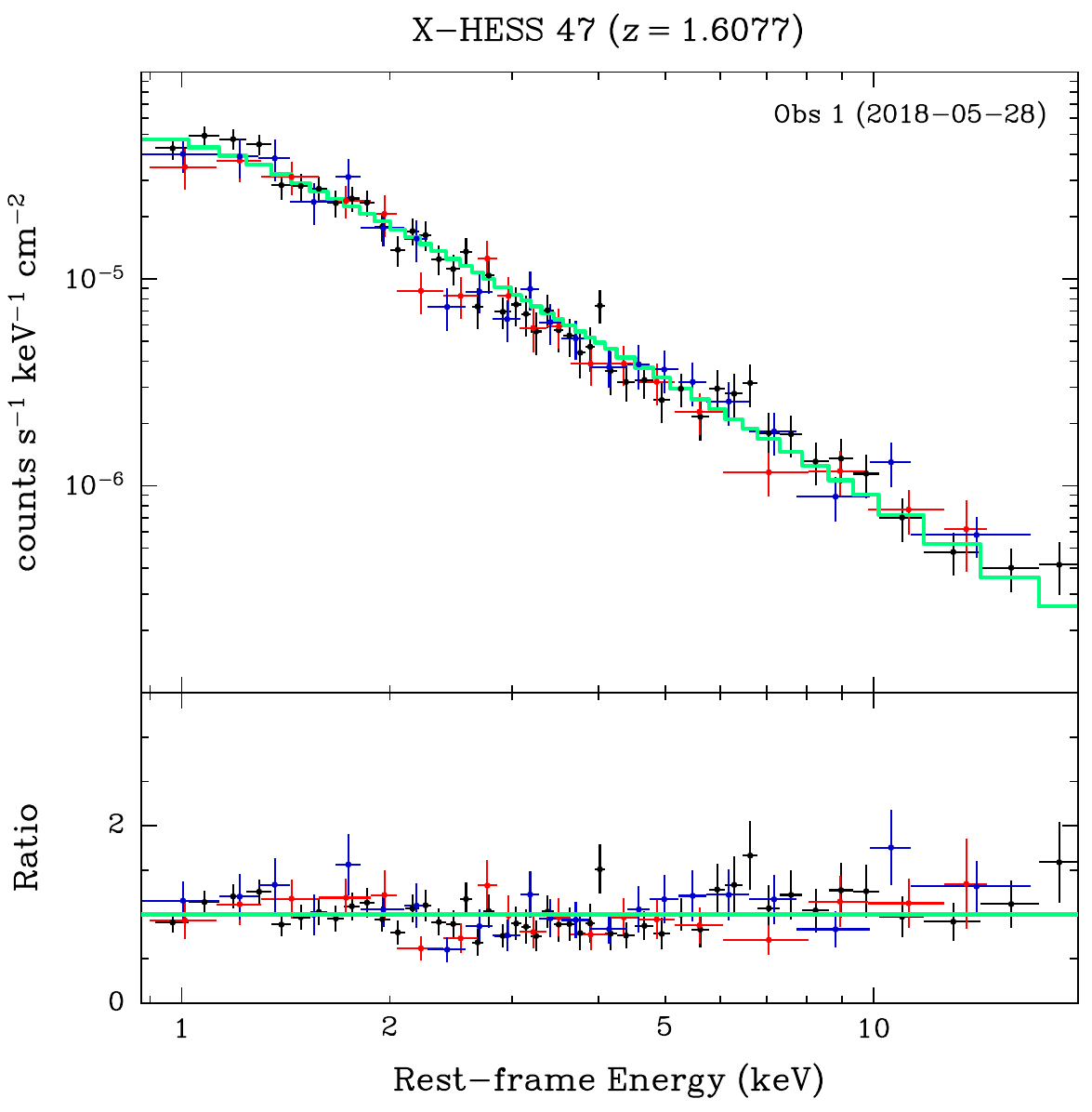}} \\
     \subfloat{\includegraphics[width = 2.1in]{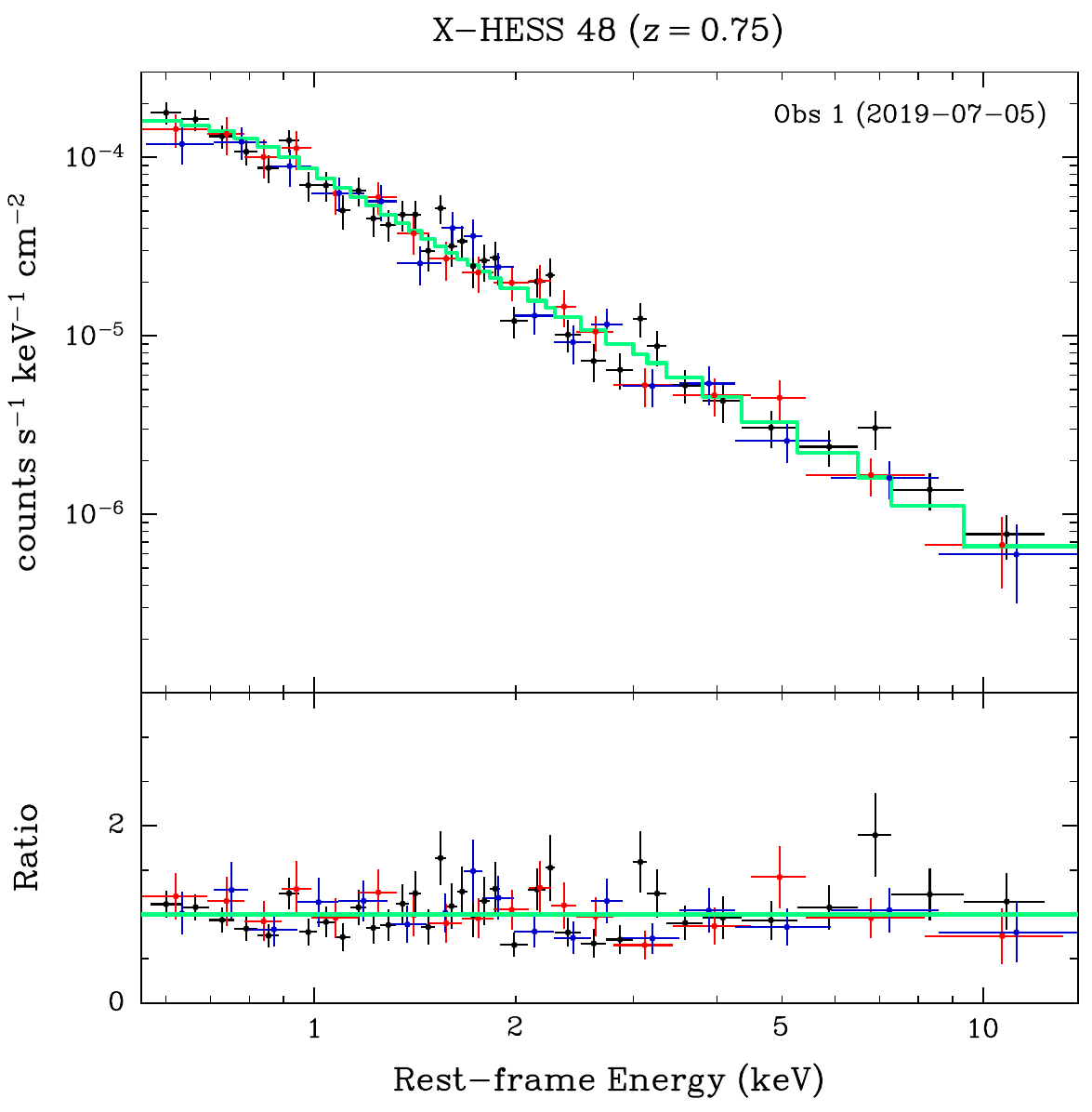}} &
     \subfloat{\includegraphics[width = 2.1in]{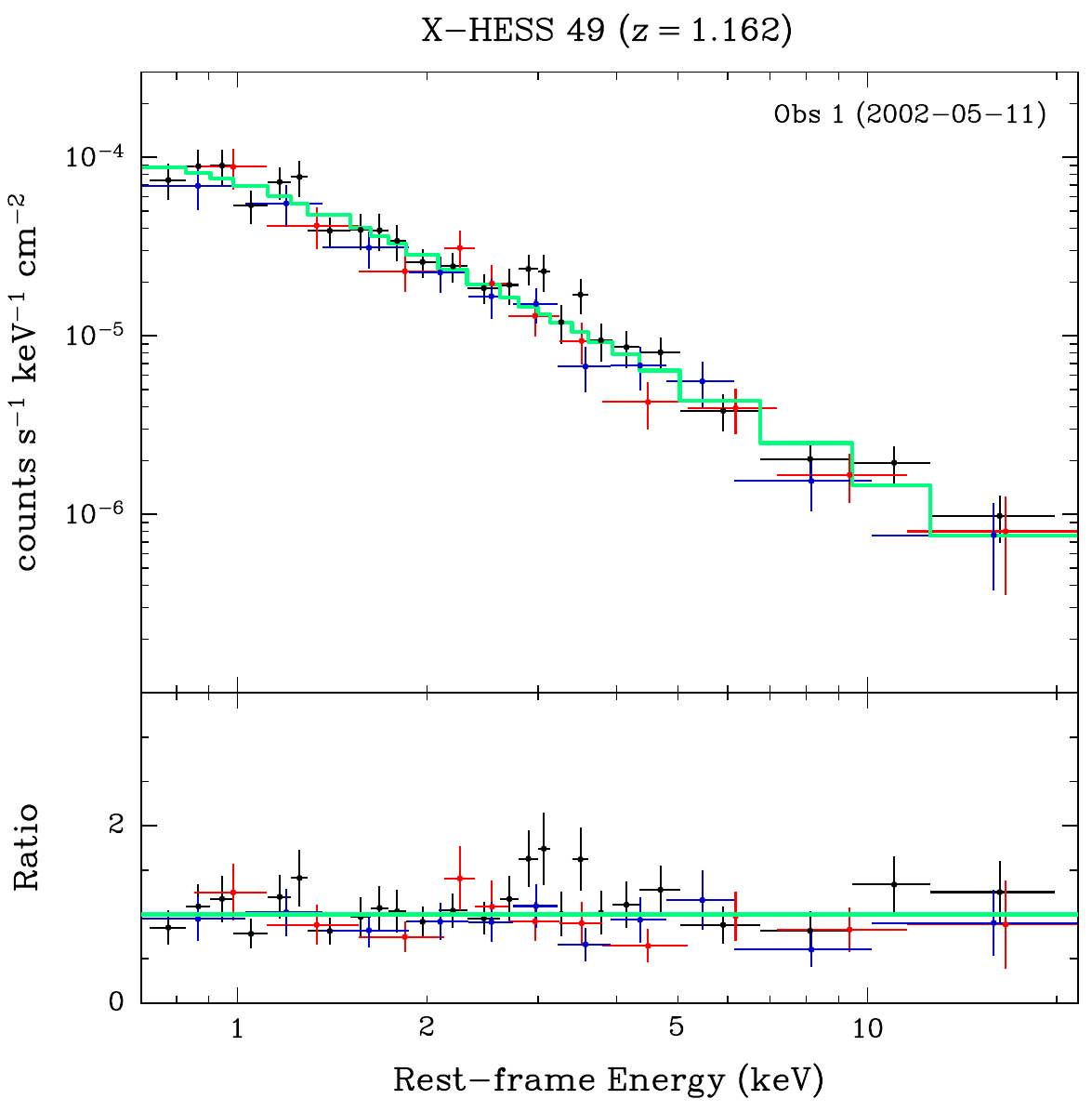}} &
     \subfloat{\includegraphics[width = 2.1in]{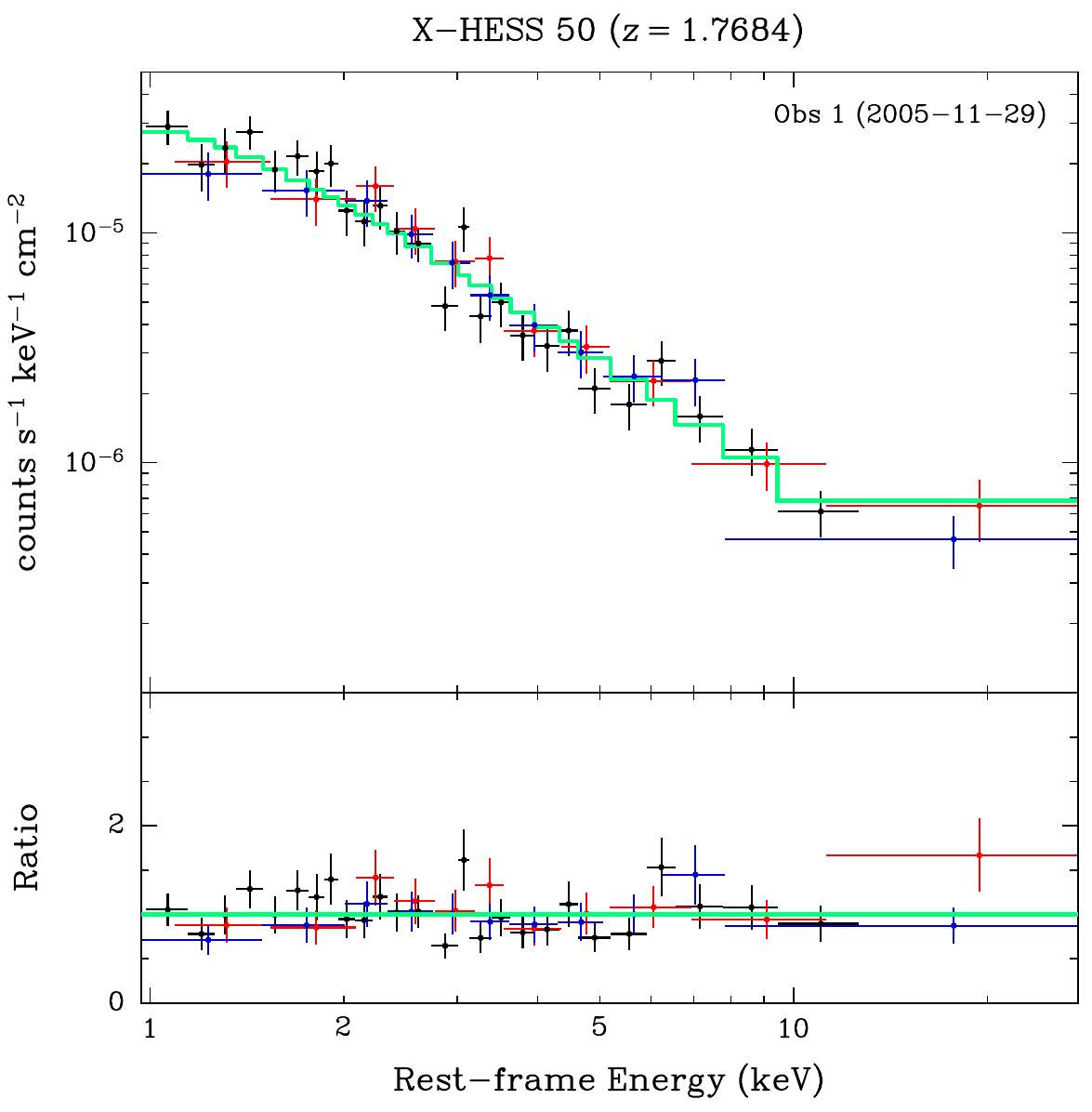}} \\
     \end{tabular}
\begin{minipage}{1.2\linewidth}
    \centering
    {Continuation of Fig. \ref{fig:xhess_spectra}.}
\end{minipage}
\end{figure}

\begin{figure}[h]
     \ContinuedFloat
     \centering
     \renewcommand{\arraystretch}{2}
     \begin{tabular}{ccc}
     \subfloat{\includegraphics[width = 2.1in]{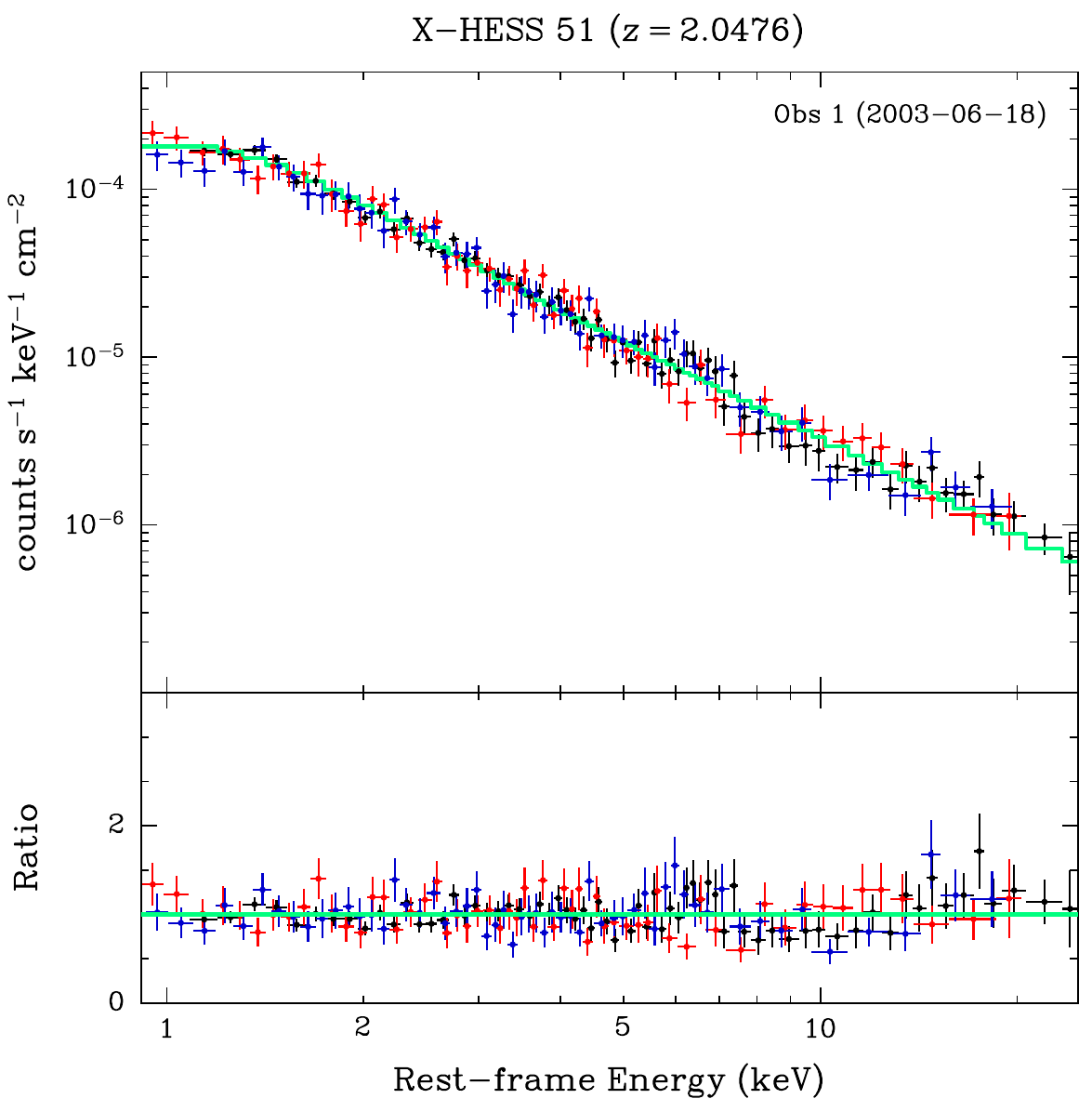}} &
     \subfloat{\includegraphics[width = 2.1in]{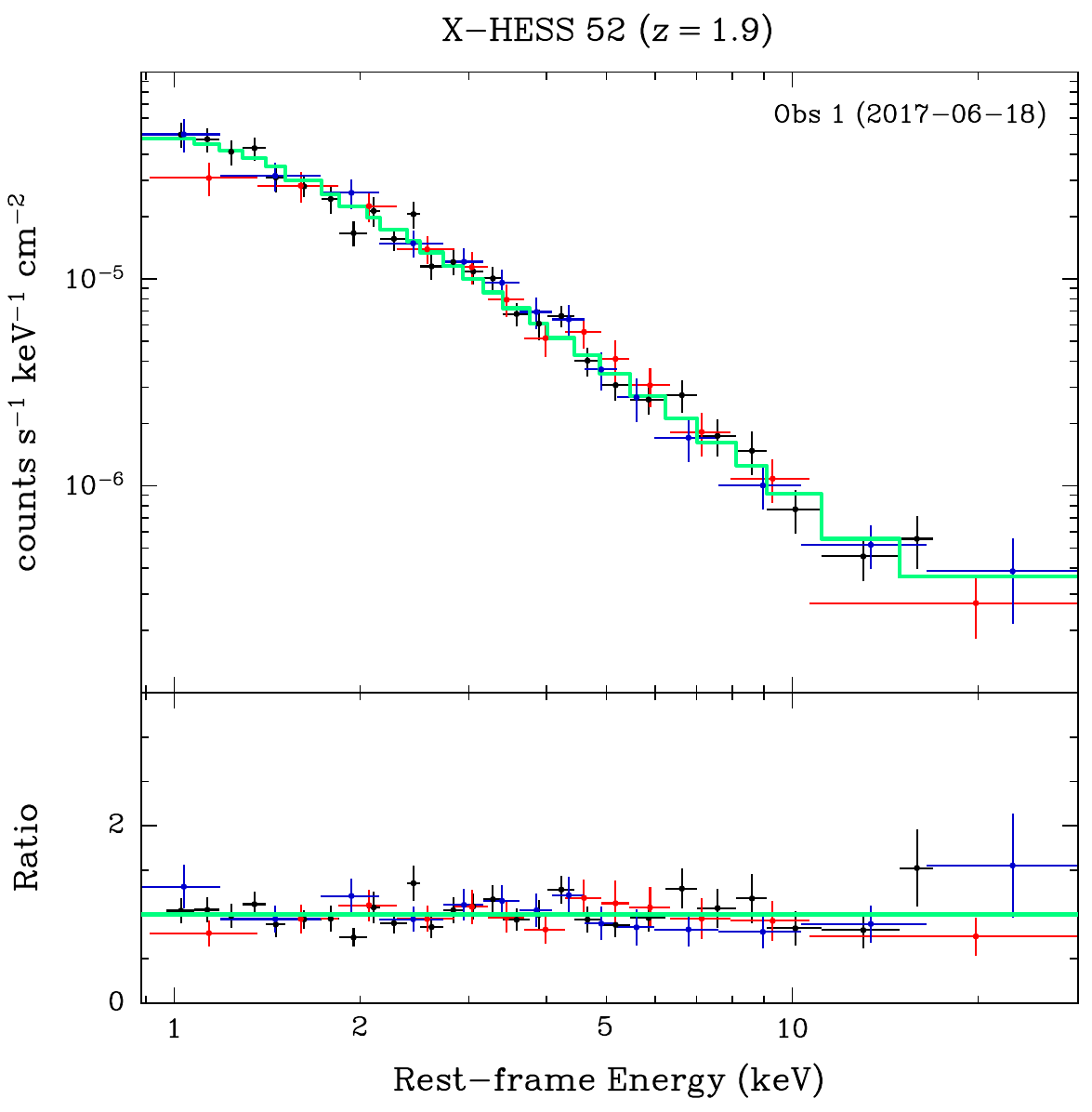}} &
     \subfloat{\includegraphics[width = 2.1in]{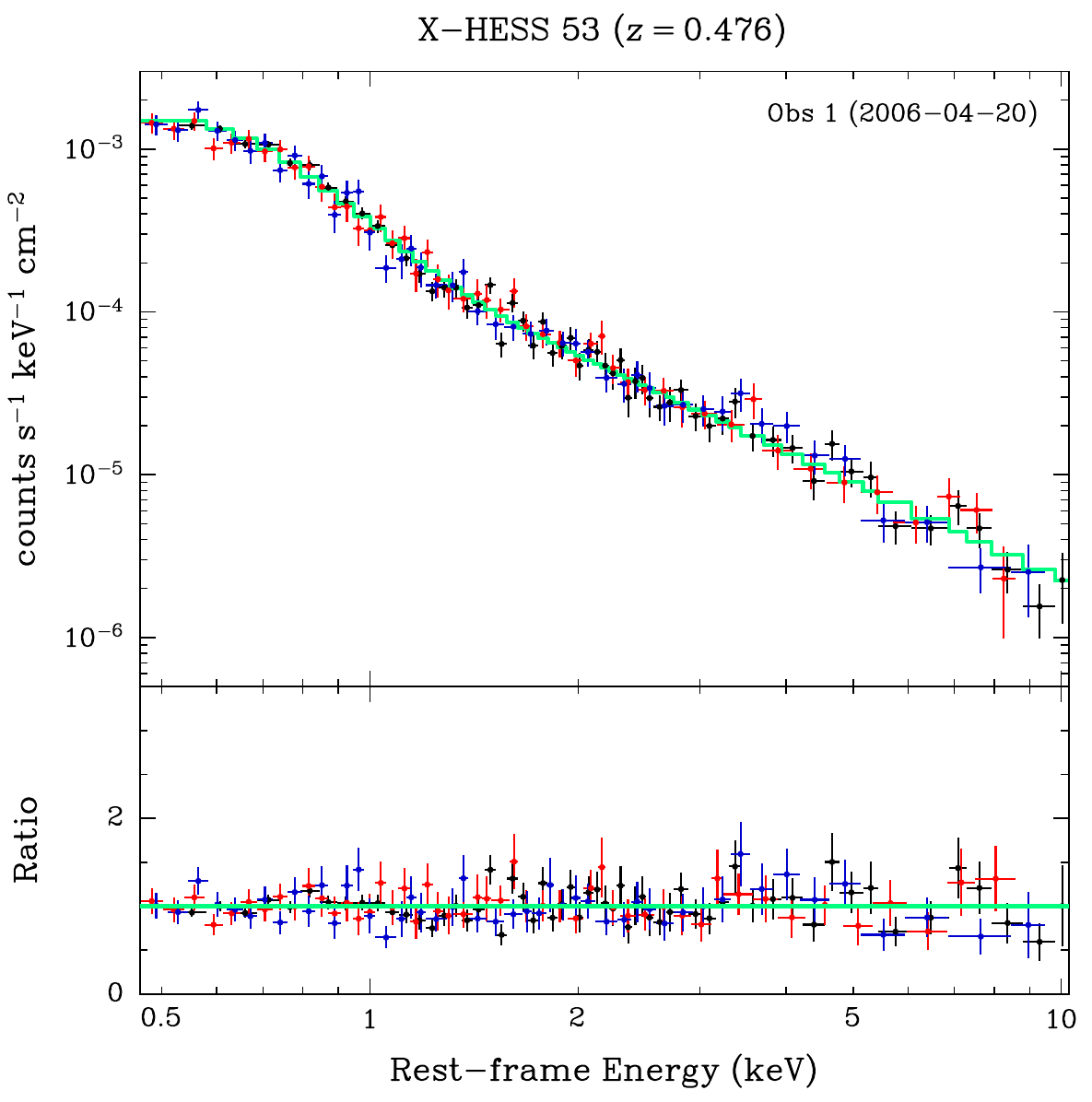}} \\
     \subfloat{\includegraphics[width = 2.1in]{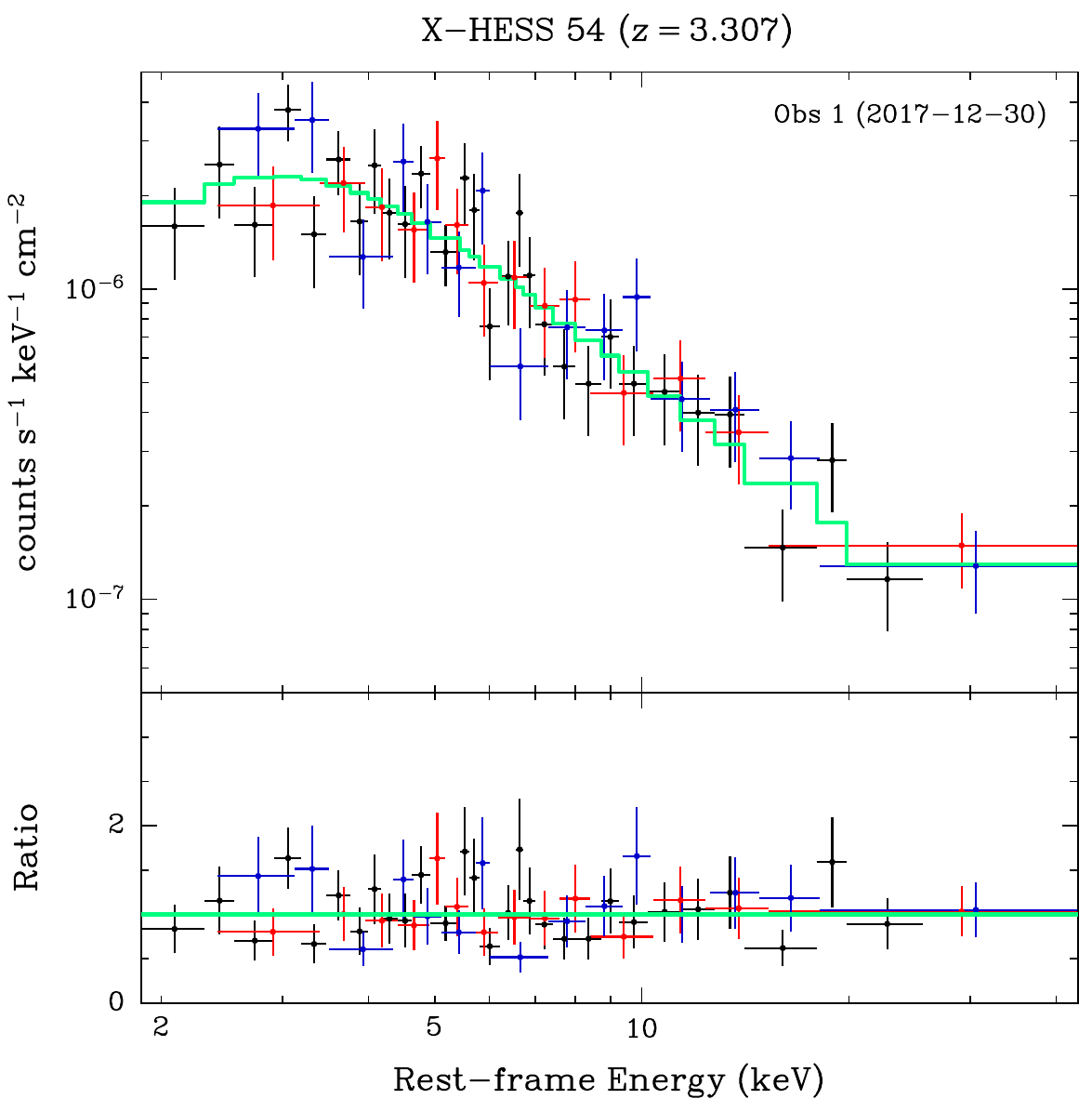}} &
     \subfloat{\includegraphics[width = 2.1in]{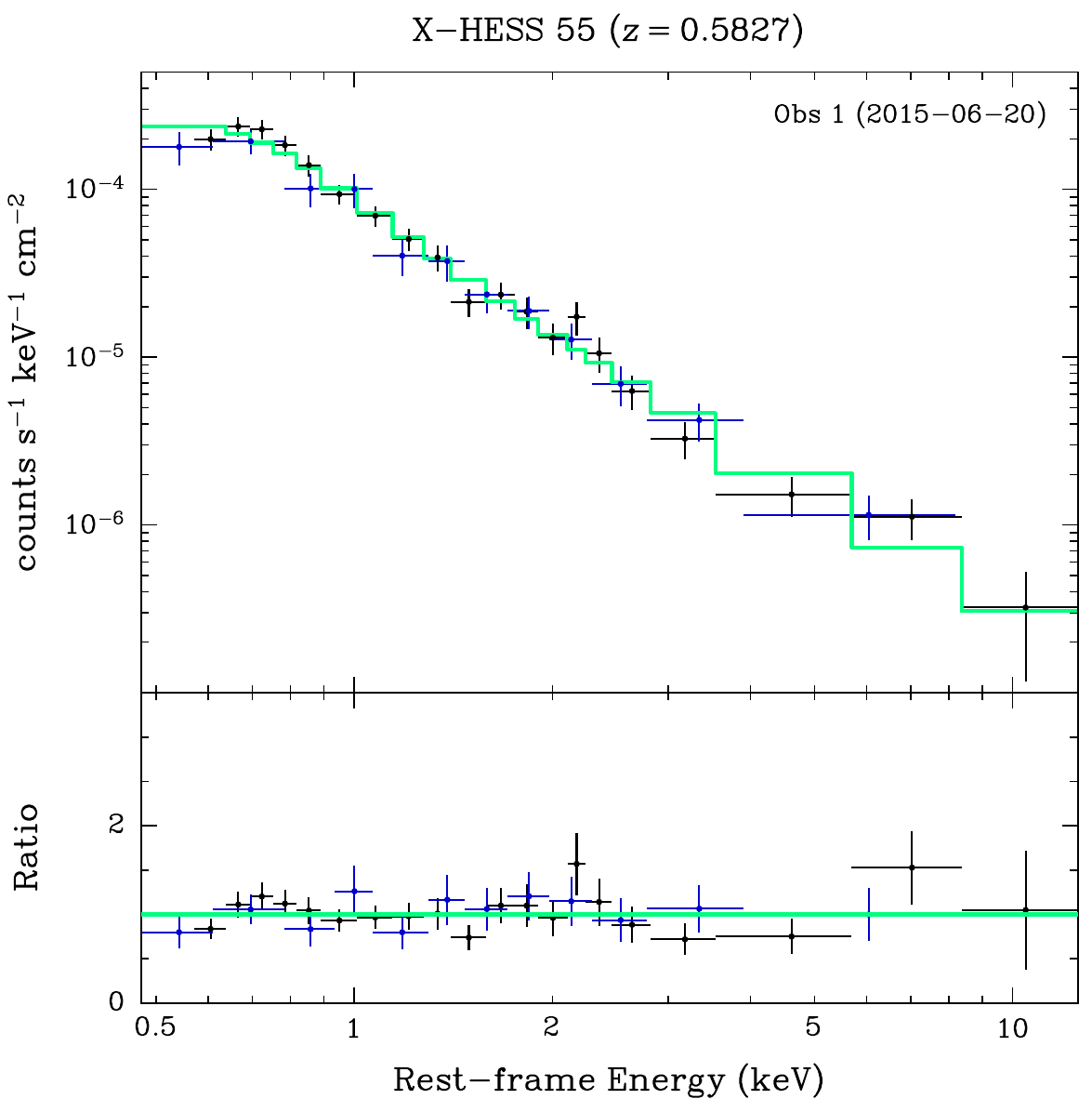}} &
     \subfloat{\includegraphics[width = 2.1in]{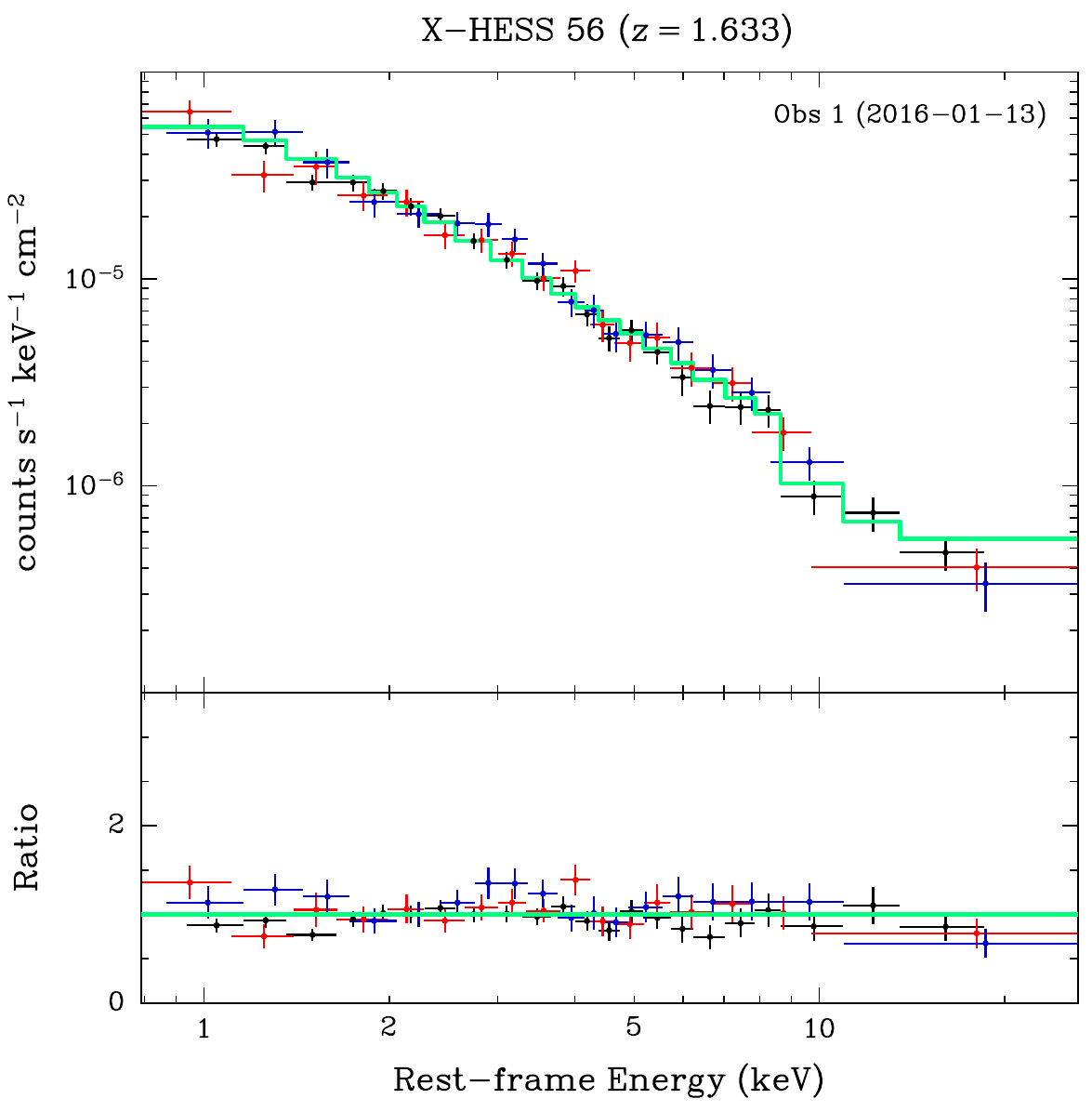}} \\
     \subfloat{\includegraphics[width = 2.1in]{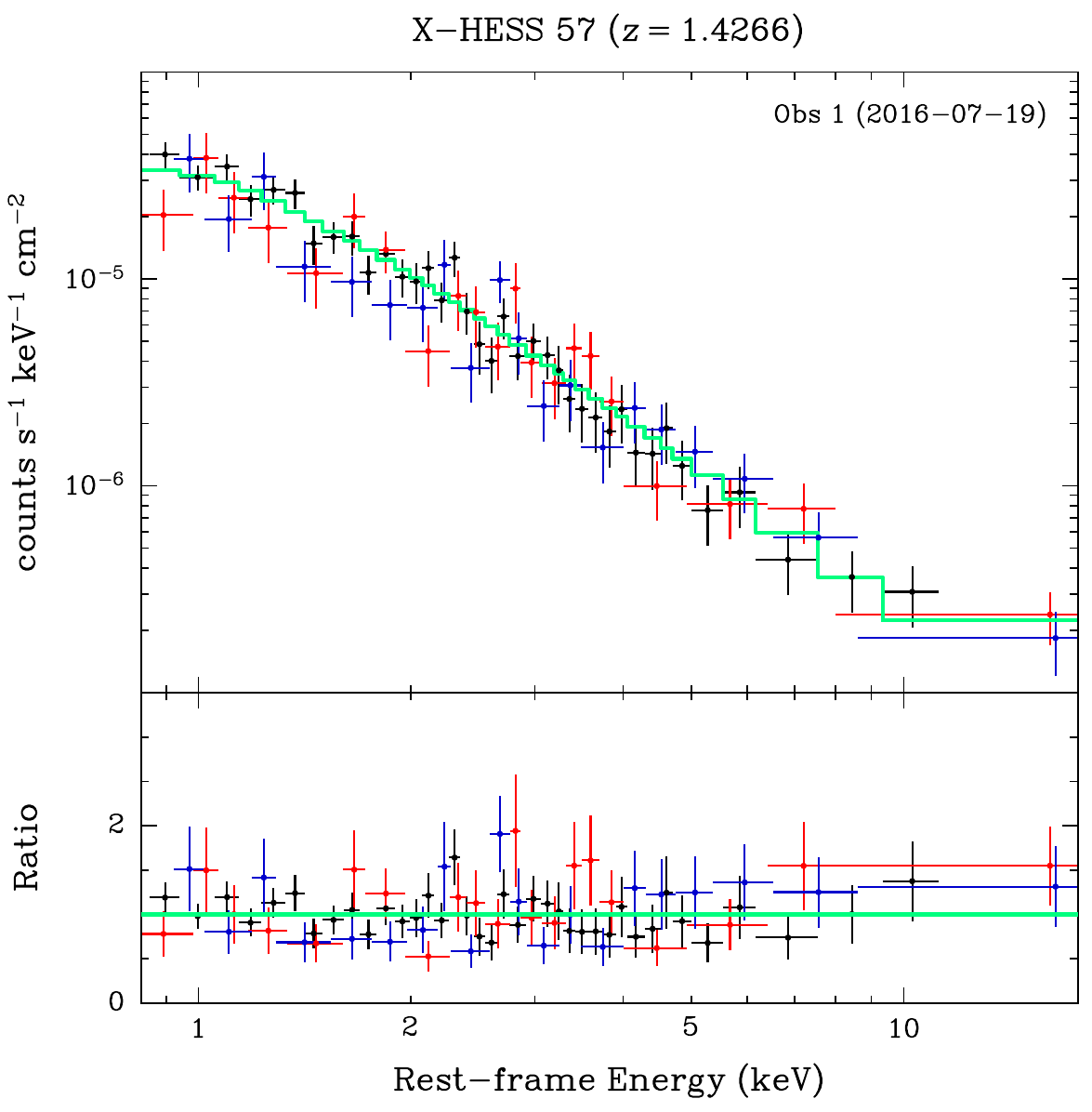}} &
     \subfloat{\includegraphics[width = 2.1in]{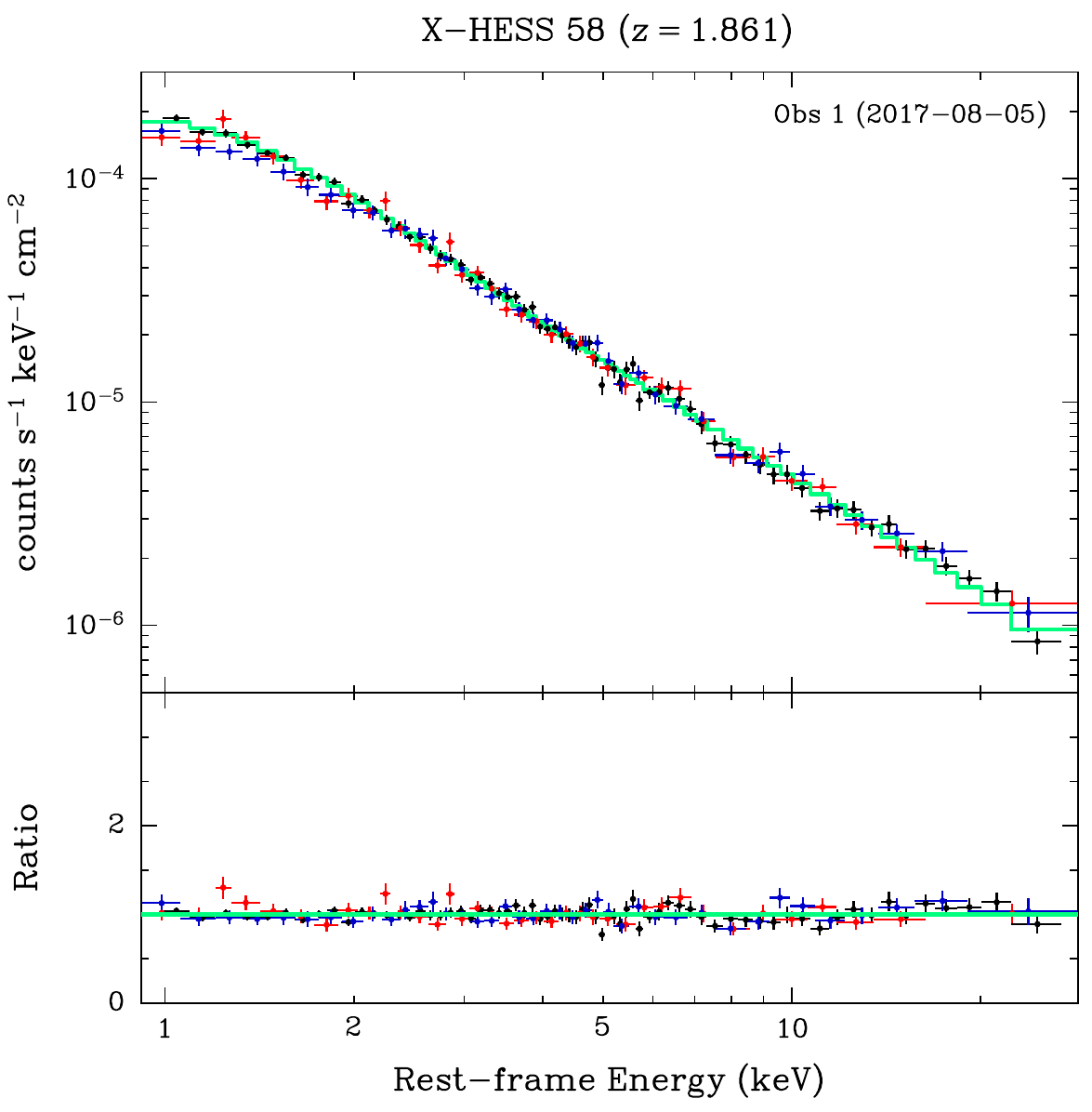}} &
     \subfloat{\includegraphics[width = 2.1in]{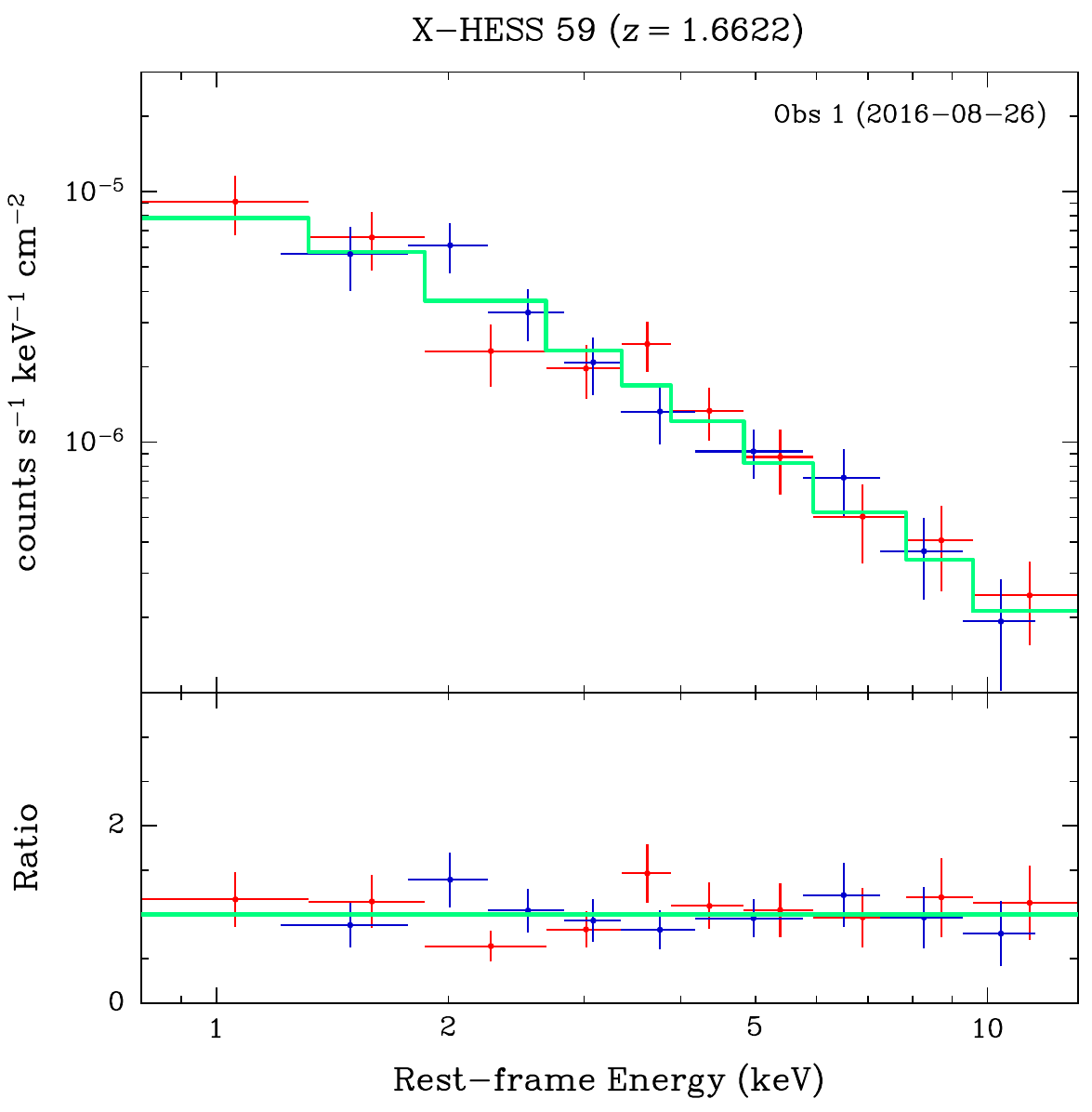}} \\
     \subfloat{\includegraphics[width = 2.1in]{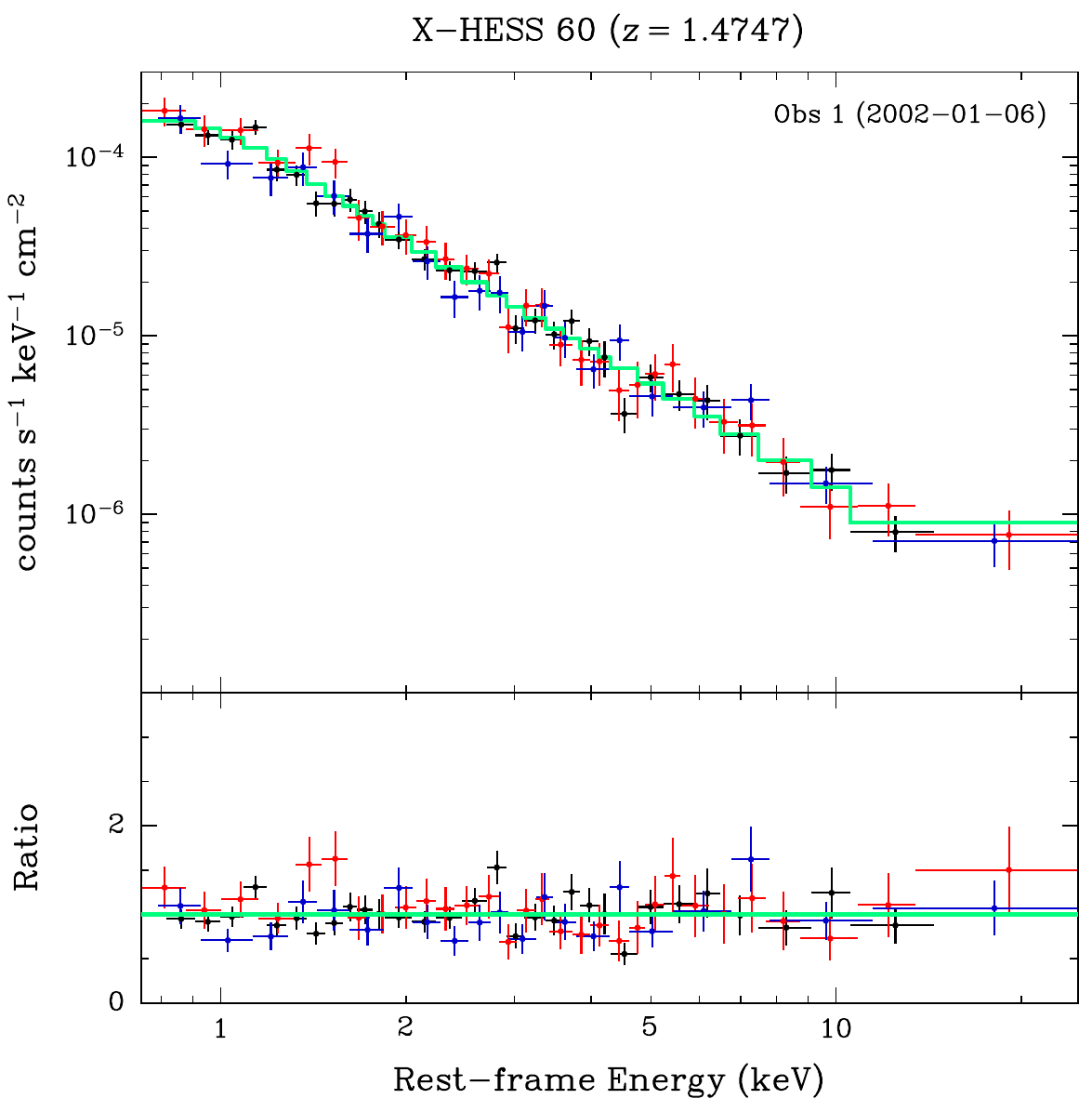}} &
     \subfloat{\includegraphics[width = 2.1in]{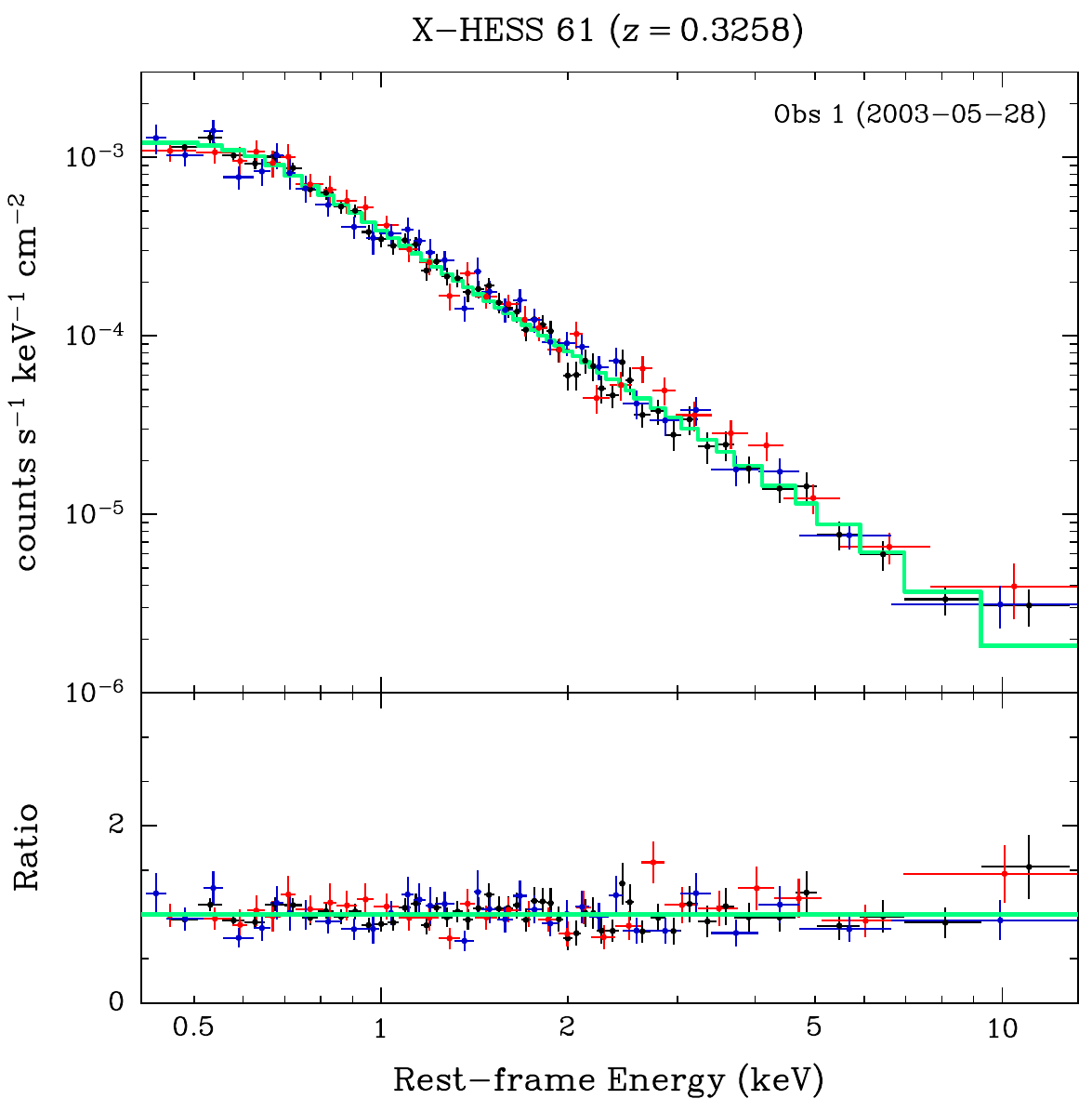}} &
      \\
     \end{tabular}
\begin{minipage}{1.2\linewidth}
    \centering 
    {Continuation of Fig. \ref{fig:xhess_spectra}.}
\end{minipage}
\end{figure}

\clearpage
\twocolumn
\normalsize
\restoregeometry
\renewcommand{\arraystretch}{1}

\end{document}